\documentclass[rmp,aps,twocolumn,nofootinbib]{revtex4}
\usepackage{graphicx}
\usepackage{graphics}
\usepackage{epsfig}
\begin{document}
\title{Failure Processes in Elastic Fiber Bundles}
\author{Srutarshi Pradhan}
\email{pradhan.srutarshi@ntnu.no}
\affiliation{Department of Physics, Norwegian University of Science and 
Technology, NO-7491 Trondheim, Norway}
\affiliation{SINTEF Petroleum Research, NO-7465 Trondheim, Norway}
\author{Alex Hansen}
\email{alex.hansen@ntnu.no}
\affiliation{Department of Physics, Norwegian University of Science and 
Technology, NO-7491 Trondheim, Norway}
\author{Bikas K. Chakrabarti}
\email{bikask.chakrabarti@saha.ac.in}
\affiliation{Theoretical Condensed Matter Physics Division and Centre for 
Applied Mathematics and Computational Science, Saha Institute of Nuclear 
Physics, 1/AF Bidhan Nagar, Kolkata 700064, India.}
\begin{abstract}
The fiber bundle model describes a
collection of elastic fibers under load.  The fibers
fail sucessively and for each failure, the load distribution among the
surviving fibers changes.  Even though very simple, this model captures
the essentials of failure processes in a large number of materials and
settings.  We present here a review of the fiber bundle model with different
load redistribution mechanisms from the point of view of statistics and
statistical physics rather than materials science, with a focus on
concepts such as criticality, universality and fluctuations.  We discuss the
fiber bundle model as a tool for understanding phenomena such as creep,
and fatigue, how it is used to describe the behavior of fiber reinforced
composites as well as modelling e.g.\ network failure, traffic jams and 
earthquake dynamics.     
\end{abstract}
\maketitle
\tableofcontents
\section{Introduction}

In materials science and engineering, a class of simple models, known as
fiber bundle models (FBM), has proven to be very effective in practical
applications such as fiber reinforced composites.  In this context,
such models have a history that goes back to the twenties \cite{p26}, and
they constitute today an elaborate toolbox for studying such materials,
rendering computer studies orders of magnitudes more efficient than
brute force methods.  Since the late eighties \cite{s89}, these models have
received increasing attention in the physics community due to their
deceivingly simple appearance coupled with an extraordinary richness of
behaviors.  As these models are just at the edge of what is possible
analytically and typically not being very challenging from a numerical
point of view so that extremely good statistics on large systems are
 available, they are perfect as model systems
for studying failure phenomena as a part of theoretical physics.

Fracture and material stability has for practical reasons interested humanity
ever since we started using tools:  our pottery should be able to withstand
handling, our huts should be able to withstand normal weather.  As science
took on the form we know today during the Renaissance, Leonardo da Vinci
studied five hundred years ago experimentally the strength of wires ---
fiber bundles --- as a function of their length \cite{lb01}.  Systematic
strength studies, but on beams, were also
pursued systematic by Galileo Galilei one hundred years later, as was done by
Edme Mariotte (of gas law fame) who pressurized vessels until they burst in
connection with the construction of a fountain at Versailles.  For some reason,
mainstream physics moved away from fracture and breakdown problems in the
nineteenth century, and it is only during the last twenty years that fracture
problems have been studied within physics proper.  The reason for this is
most probably the advent of the computer as a research tool, rendering problems
that were beyond the reach of systematic theoretical study now accessible.

If we were to single out the most important contribution from the physics
community with respect to fracture phenomena, it must be the focus on
{\it fluctuations\/} rather than averages.  What good is the knowledge
of the average behavior a system when faced with a single sample and this
being liable to breakdown given the right fluctuation?  This review, being
written by physicists, reflects this point of view, and hence, fluctuations
play an important role throughout it.

Even though we may trace the study of fiber bundles to Leonardo da Vinci,
their modern story starts with the already mentioned work by \textcite{p26}.
In 1945, Daniels published a seminal review cum research article on fiber
bundles which still today must be regarded as essential reading in the field
\cite{d45}.  In this paper, the fiber bundle model is treated as a problem
of statistics and the analysis is performed within this framework, rather
than treating it within materials science.  The fiber bundle is viewed
as a collection of elastic objects connected in parallel and clamped to a
medium that transmits forces between the fibers.  The elongation of a fiber is
linearly related to the force it carries up to a maximum value.  When this
value is reached, the fiber fails by no longer being able to carry any force.
The threshold value is assigned from some initially chosen probability 
distribution, and do not change thereafter.  When the fiber
fails, the force it carried is redistributed.  If the clamps deform under
loading, fibers closer to the just-failed fiber will absorb more of the
force compared to those further away.  If the clamps, on the other hand, are
rigid, the force is equally distributed to all the surviving fibers. Daniels
discussed this latter case.  A typical question posed and answered in this
paper would be the average strength of a bundle of $N$ fibers, but also
the variance of the average strength of the entire bundle.  The present
review takes the same point of view, discussing the fiber bundle model
as a {\it statistical model.\/}  Only in Section V we discuss the fiber
bundle model in the context of materials science with all the realism
of real materials considered.  However, we have not attempted to include
any discussions of the many experimental studies that have been performed
on systems where fiber bundles constitute the appropriate tool.  This is
beyond the scope of this statistical-physics based review.

After introducing (Section II) the fiber bundle model to our readers, in 
Section III, we present the {\it Equal Load Sharing Model,\/} which was
sketched just a few lines back.  This seemingly simple model is in fact
extremely rich.  For example, the load at which catastrophic failure occurs
is a {\it second order critical point\/} with essentially all the features
usually seen in systems displaying such behavior.  However in this case,
the system is analytically tractable.  In fact, we believe that the equal
load sharing fiber bundle may be an excellent system for teaching
second order phase transitions at the college level.  Under the heading
of Fluctuations, we discuss the burst distribution, i.e., the statistics
of simultaneously failing fibers during loading: When a fiber fails
and the force it was carrying is redistributed, one or more other fibers
may be driven above their failing thresholds.
In this equal load sharing model, the absolute rigidity of the bar
(transmitting forces among the fibers) suppresses the stress fluctuations
among  the fibers. As such, there is no apparent growth of the (fluctuation
correlation) length scale. Hence, although there are precise recursion
relations and their linearized solutions are available near the fixed point
(see Sec. III), no straight forward application  of the Renormalization
group techniques \cite{f74} has been made to extract the
exponents through length scaling.

In Section IV {\it Local Load Sharing\/} is discussed.  This bit of added
realism comes at the added cost that analytical treatment becomes much
more difficult.  There are, still, a number of analytical results in the
literature.  One may see intuitively how local load sharing complicates
the problem, since the relative positions of the fibers now become
important.  Under global load sharing, every surviving fibers gets
the same excess force and, hence, where they are do not matter.  There are
essentially three local load sharing models in the literature.  The first
one dictates that the {\it nearest surviving neighbors\/} of the failing
fiber absorb its load.  Then there are ``softer models" where the
redistribution follows a power law in the distance to the failing fiber.
Lastly, there is the model where the clamps holding the fibers are elastic
themselves, and this leads to non-equal redistribution of the forces.

Section V contains a review of the use of fiber bundle models in applications
such as materials science.  We discuss fatigue, thermal failure,
viscoelastic effects and precursors of global failure.  We then go on to 
review the large field of modeling
fiber reinforced composites.  Here fiber bundle models constitute the
starting point of the analysis, which, by its very nature, is rather complex
seen from the viewpoint of statistical physics.  Lastly, we review some
applications of fiber bundle models in connection with systems that initially
would seem quite far from the concept of a fiber bundle, such as traffic
jams.

We end this review by a summary with few concluding remarks in Section VI.

\section{Fiber bundle models}
Imagine a heavy load hanging from a rigid anchor point (say, at
the roof) by a rope or a bundle of fibers. If the load exceeds a threshold
value, the bundle fails. How does the failure proceeds in the bundle?
Unless all the fibers in the bundle have got identical breaking thresholds
(and that never happens in a real rope), the failure dynamics proceeds
in a typical collective load transfer way. One can assume that in this 
kind of situation the load is equally
shared  by all the intact fibers in the
bundle. However, the breaking threshold for each of the fibers being
different, some fibers fail before others and, consequently, the load
per surviving fiber increases as it gets redistributed and shared
equally by the rest. This increased load per fiber may induce further
breaking of some fibers and the avalanche continues, or it stops if
all the surviving fibers can withstand the redistributed load per
fiber. This collective or cooperative failure dynamics and the consequent 
avalanches
or bursts are typical for the failure in any many-body system. It
captures the essential features of failure during fracture propagation
(recorded by acoustic emissions), earthquake avalanches (main and aftershocks),
traffic jams (due to dynamic clustering), etc.

The model was first introduced in $1926$ by  \textcite{p26}
in the context of textile engineering. Since then it was modified
a little and investigated, mainly numerically, with various realistic
fiber threshold distributions by the engineering community \cite{d45,c57a,hp78,ps83}.
Starting from late eighties, physicists took interest in
the avalanche distribution in the model and in its dynamics \cite{s89,ng91,hh92,s92,gip93,khh97,zrsv97,asl97,kzh00,pbc02}.
A recursive dynamical equation was set up for the equal-load-sharing
version recently \cite{d99,pc01} and the dynamic critical behavior is now 
solved exactly \cite{pbc02,bpc03,ph07}.
In addition to the extensive numerical
results \cite{zd94,hh94} on the effect of short-range fluctuations 
(local load sharing
cases), some progress with analytical studies \cite{dl94,h85,hp91,gip93,khh97} 
have also been made.

There are a large number of experimental studies of various materials and
phenomena that have successfully been analyzed within the framework of
the fiber bundle model.  For example, \textcite{ls04} have used
the fiber bundle model to propose explanations for changes in
fibrious collagen and its relation to neuropathy in connection with diabetes.
\textcite{tl01} propose a method to monitor the structural integrity
of fiber-reinforced ceramic-matrix composites using electrical resistivity
measurements.  The basic idea here is that when the fibers in the composite
themselves fail rather than just the matrix in which they are embedded,
the structure is about to fail.  The individual fiber failures is recorded
through changes in the electrical conductivity of the material.  Acoustic
emission, the crackling sounds emitted by materials as they are loaded,
provide yet another example where fiber bundle models play and important
role, see e.g., \textcite{nhgs05}.

\section{Equal load sharing model}

\begin{figure}
\epsfysize=1.7in
\epsfbox{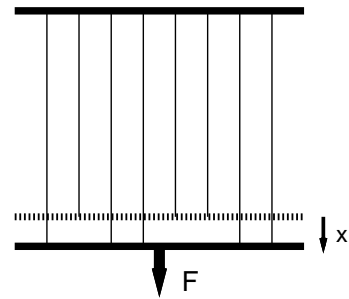}
\caption{A fiber bundle model having $N$ parallel fibers. The original position of the rigid platform on which force has been applied, is indicated. All the fibers are assumed to have the same elastic constant (normalized
to unity here) until breaking, while the breaking strength of these fibers
are
assumed to be randomly dispersed. As the bundle gets strained ($x > 0$),
some weaker fibers (having lower stress or strain capacity) fail, and the
number of intact fibers decrease from its starting value $N$.
Consequently, the stress on them (or the strain) increases due to the
redistribution or transfer of loads from the failing fibers. Some of these
fibers may fail further as  they can not support this extra redistributed
load. The process  stops if there is no further failure, and the
bundle will show nonlinear elastic response (although each fiber has linear
elastic behavior until breaking). Otherwise, the bundle fails when the
stress concentration on the  fibers (due to the dynamic stress
redistribution) becomes so high that none of the fibers can withstand that.
}
\label{fig:fbm-model}
\end{figure}
 The simplest  and the oldest version of the model
is the equal load sharing (ELS) model, in which the load 
previously carried by a failed fiber is shared equally by all the remaining 
intact fibers in the system. As the applied load is shared globally, 
this model is also known as global load sharing (GLS) model or democratic fiber
bundle model. 
Due to the consequent mean-field nature, some exact results 
could be extracted for this model and this  was demonstrated by 
\textcite{d45} in a classic work some sixty years ago. The typical
relaxation dynamics of this model has been solved recently which has clearly 
established a robust critical behavior \cite{pbc02,bpc03,ph07}.  
It may be mentioned at the outset that the ELS or GLS models do not allow
for spatial fluctuations (due to the absolute rigidity of the platform in
Fig. 1) and hence such models belong to the the mean field 
category of critical dynamics, see e.g., \textcite{s87}. Fluctuations in 
breaking
time or in avalanche  statistics (due to randomness in fiber strengths)
are of course possible in such models and are discussed in details in this
section.

A bundle can be loaded in two different ways: Strain controlled and 
force controlled. In the strain controlled method, at each step the whole 
bundle is stretched till the weakest fiber fails.  Clearly, when number of 
fibers $N$ is very large, strain is increased by infinitesimal amount at each
step until complete breakdown and therefore the process is  
considered as a \emph{quasi-static} way of loading. On the other hand, in 
the force controlled method, the external force (load) on the bundle is 
increased by same 
amount at each step until the breakdown. The basic difference between these 
two methods is that the  
first method ensures the failure of single fiber (weakest one among the 
intact fibers) at each loading step, while in the second method sometimes none 
of the fibers fail and sometimes more than one fail in one loading step.      

\begin{figure}
\epsfysize=2in
\epsfbox{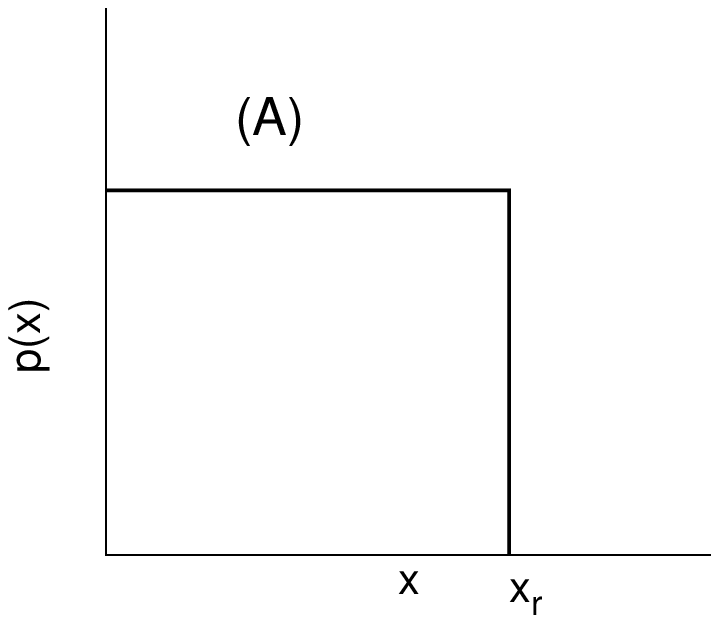}
\epsfysize=2in
\epsfbox{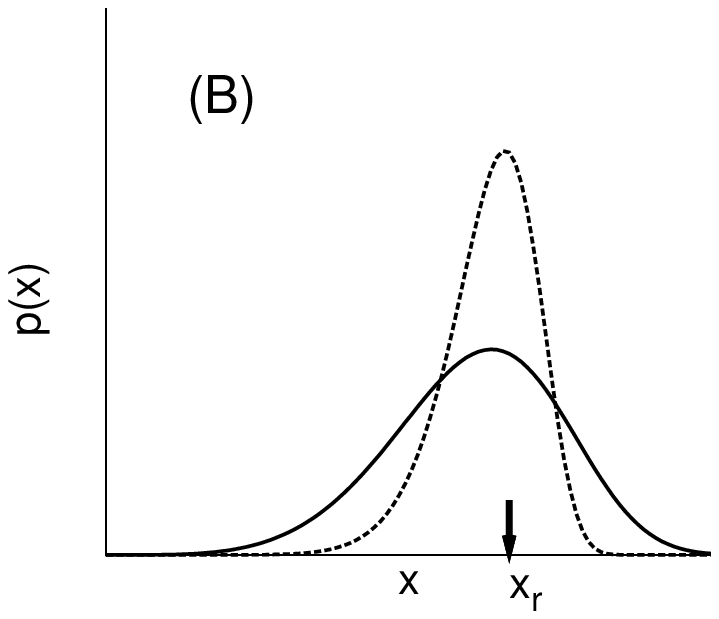}
\caption{The uniform distribution (A) and the Weibull distribution (B) with $\rho=5$ (solid line) and $\rho=10$ (dotted line).  }
\label{fig:distributions}
\end{figure}

Let $x$ denote the strain of the fibers in the bundle. Assuming
the fibers to be linearly elastic up to their respective failure point
(with unit
elastic constant), we can represent the stresses on each of the surviving
fibers by the same  quantity $x$. The strength (or threshold) of a fiber is 
usually determined by the stress value $x$ it can bear, and beyond 
which it fails.  We therefore denote the strength (threshold)
distribution of the fibers in the bundle by $p(x)$ and the corresponding 
cumulative distribution by $P(x)=\int_{0}^{x}p(y)dy$.
Two popular examples of threshold distributions
are the uniform distribution \begin{equation}
P(x)=\left\{ \begin{array}{cl}
x/x_{r} & \mbox{ for }0\leq x\leq x_{r}\\
1 & \mbox{ for }x>x_{r},\end{array}\right.\label{uniform}\end{equation}
 and the Weibull distribution \begin{equation}
P(x)=1-\exp(-(x/x_{r})^{\rho}).\label{Weibull}\end{equation}
 Here $x_{r}$ is a reference threshold, and the dimensionless number
$\rho$ is the Weibull index (Fig. \ref{fig:distributions}).

In the strain controlled loading, at a strain $x$, the total force on the
 bundle is $x$ times the number of intact fibers. The expected or average
force at this stage is therefore \cite{s89,hh92,s92}
 \begin{equation}
F(x)=N\, x\,(1-P(x)).\label{load}\end{equation}
The maximum $F_{c}$
of $F(x)$ corresponds to the value $x_{c}$ for which $dF/dx$ 
vanishes:
 \begin{equation}
1-P(x_{c})-x_{c}p(x_{c})=0.\label{critical-criteria}\end{equation}
Here the failure process is basically driven by \emph{fluctuations} and can be 
analyzed using extreme order statistics \cite{s89,hh92,s92,khh97}. 

In the force controlled method, if force $F$ is applied on a bundle having $N$ 
fibers,  when  the system reaches an equilibrium, the strain or effective 
stress $x$ is (see Fig. \ref{fig:fbm-model}) 
\begin{equation}
x(F) = \frac {F} {N\left[1- P (x)\right]}.
\label{strain} 
\end{equation}  
Therefore, at the equilibrium state, Eq. (\ref{load})  and Eq. (\ref{strain}) 
are identical. It is possible to construct recursive dynamics 
\cite{d99,pc01} of  the failure process  for a given load and the 
fixed-point solutions explore the \emph{average 
behavior} of the system  at the equilibrium state.

\subsection{Average behavior}

Fig. \ref{fig:fbm-model}  shows a static 
fiber bundle model in the ELS mode  where $N$ fibers are connected in parallel
 to each other 
(clamped at both ends) and a force is applied at one end.
At the first step
all fibers that cannot withstand the applied stress break. Then the
stress is redistributed on the surviving fibers, which compels further
fibers to break. This starts an iterative process that continues until
an equilibrium is reached, or all fibers fail. The average behavior is
 manifested when the initial load is  macroscopic (very large $N$). 

\subsubsection{Recursive breaking dynamics }

The breaking dynamics can be represented by recursion relations
 \cite{d99,pc01}  in
discrete steps. 
Let $N_t$ be the number of fibers  that survive after  step $t$, where $t$ 
indicates the number of stress redistribution steps.
Then one can write \cite{d99}
\begin{equation}
N_{t+1}= N\left[1-P\left(\frac{F}{N_t}\right)\right]. 
\end{equation}
 
Now  we introduce  $\sigma=F/N$, the applied stress and $U_{t}=N_t/N$, the 
 surviving fraction of total fibers. 
Then the effective stress  after $t$  step becomes 
$x_{t}=\frac{\sigma}{U_{t}}$ and after $t+1$  steps the 
surviving fraction of total fibers is $U_{t+1}=1-P(x_{t})$. Therefore we can 
construct the following  recursion relations \cite{pc01,pbc02}:
 \begin{equation}
x_{t+1}=\frac{\sigma}{1-P(x_{t})};x_{0}=\sigma\label{rec-x}
\end{equation}

\noindent and \begin{equation}
U_{t+1}=1-P(\sigma/U_{t});U_{0}=1.\label{rec-U}\end{equation}

At equilibrium  $U_{t+1}=U_{t}\equiv U^{*}$ and 
$x_{t+1}=x_{t}\equiv x^{*}$.
These equations (Eq. \ref{rec-x} and  Eq. \ref{rec-U}) can be solved at and 
around  the fixed points  for the
particular strength distribution $p(x)$. 

\subsubsection{Solution of the dynamics: Critical behavior}

Let us choose the uniform density of fiber strength distribution 
(Eq. \ref{uniform}) up to the cutoff $x_r=1$. 
Then the cumulative distribution becomes 
$P(\sigma/U_{t})=\sigma /U_{t}$.
Therefore  from Eq. (\ref{rec-x}) and Eq. (\ref{rec-U}) we can construct a pair 
of recursion relations 
 \begin{equation}
x_{t+1}=\frac{\sigma}{1-x_{t}}\label{rec-x-uniform}
\end{equation}
and 
\begin{equation}
U_{t+1}=1-\frac{\sigma}{U_{t}}.\label{rec-U-uniform}
\end{equation}
This nonlinear recursion equations are somewhat characteristic of the dynamics 
of fiber bundle models and such dynamics can be obtained in many different 
ways. For example, the failed fraction $1-U_{t+1}$ at step $t+1$ is given 
by the fraction $F/NU_{t} = \sigma/U_{t}$  of the load shared by the intact 
fibers at step $t$ and for the 
uniform distribution of thresholds (Fig. \ref{fig:distributions}A), one 
readily gets Eq. (\ref{rec-U-uniform}).

At the fixed point the above relations take the quadratic forms
\begin{equation}
x^{*^{2}}-x^{*}+\sigma=0
\label{fix-x-uniform}
\end{equation}
and
\begin{equation}
U^{*^{2}}-U^{*}+\sigma=0,
\label{fix-U-uniform}
\end{equation}
 with the solutions
\begin{equation}
x^{*}(\sigma)=\frac{1}{2}\pm(\sigma_{c}-\sigma)^{1/2}
\label{sol-x-uniform}
\end{equation}
and 
\begin{equation}
U^{*}(\sigma)=\frac{1}{2}\pm(\sigma_{c}-\sigma)^{1/2}.
\label{sol-U-uniform}
\end{equation}
Here $\sigma_{c}=\frac{1}{4} $ is the critical value of applied
stress beyond which the bundle fails completely. Clearly, for the effective 
stress (Eq. \ref{sol-x-uniform}) solution  with $(-)$ sign is the 
stable fixed point and  with $(+)$ sign is the unstable fixed point  whereas 
for fraction of unbroken fibers (Eq. \ref{sol-U-uniform}), it is just the 
opposite.       
Now the difference $U^{*}(\sigma)-U^{*}(\sigma_{c})$ behaves
like an order parameter signaling  partial failure of the bundle when it is 
non-zero (positive), although unlike conventional phase transitions it does 
not have a real-valued existence for $\sigma>\sigma_c$.   

\begin{equation}
O\equiv U^{*}(\sigma)-U^{*}(\sigma_{c})=(\sigma_{c}-\sigma)^{\alpha};\alpha=\frac{1}{2}.
\label{order-uniform}
\end{equation}

\begin{figure}
\epsfysize=2in
\epsfbox{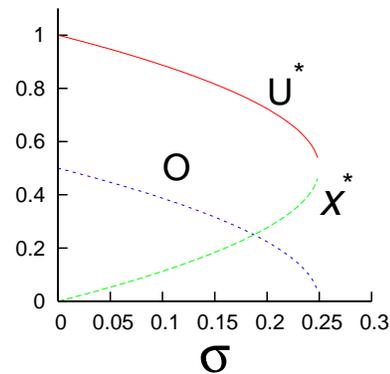}
\caption{Variation of effective stress ($x^*$), fraction of unbroken fibers ($U^*$) and the order parameter ($O$)  with the applied stress $\sigma$ for a bundle with uniform distribution (Eq. \ref{uniform}) of fiber strengths.  }
\label{fig:USO}
\end{figure}
Fig. \ref{fig:USO} shows the variation of $x^*$, $U^*$ and $O$ with the 
externally applied stress value $\sigma$. 
One can also obtain the breakdown susceptibility $\chi$,
defined as the change of $U^{*}(\sigma)$ due to an infinitesimal
increment of the applied stress $\sigma$: \begin{equation}
\chi=\left|\frac{dU^{*}(\sigma)}{d\sigma}\right|=\frac{1}{2}(\sigma_{c}-\sigma)^{-\beta};\beta=\frac{1}{2}.\label{sus-uniform}\end{equation}
Such a divergence in $\chi$
had already been reported in several studies \cite{zrsv97,zrsv99a,mgp00,d99}.

To study the dynamics away from criticality ($\sigma\rightarrow\sigma_{c}$
from below), the recursion relation (Eq. \ref{rec-U-uniform}) can be replaced
by a differential equation \begin{equation}
-\frac{dU}{dt}=\frac{U^{2}-U+\sigma}{U}.\label{diff-uniform}\end{equation}
Close to the fixed point, $U_{t}(\sigma)=U^{*}(\sigma)$
+ $\Delta{U}$, (where $\Delta{U}\rightarrow0$) and this gives \begin{equation}
\Delta{U}=U_{t}(\sigma)-U^{*}(\sigma)\approx\exp(-t/\tau),\end{equation}
 where $\tau=\frac{1}{2}\left[\frac{1}{2}(\sigma_{c}-\sigma)^{-1/2}+1\right]$.
Therefore, near the critical point: \begin{equation}
\tau\propto(\sigma_{c}-\sigma)^{-\theta};\theta=\frac{1}{2}.\label{tau-uniform}\end{equation}
At the critical point ($\sigma=\sigma_{c}$), a dynamic
critical behavior has been observed in the relaxation of the failure
process to the fixed point. From the recursion relation (Eq. \ref{rec-U-uniform})
it can be easily verified that the fraction $U_{t}(\sigma_{c})$
follows a simple power-law decay:

\begin{equation}
U_{t}=\frac{1}{2}(1+\frac{1}{t+1}),
\label{eq:fbm-omori}
\end{equation}
 starting from $U_{0}=1$. For large $t$ ($t\rightarrow\infty$),
this reduces to $U_{t}-1/2\propto t^{-\eta}$; $\eta=1$; indicating 
\emph{critical slowing down}  which is a robust characterization of the critical
state.

\subsubsection{Universality class of the model }

The critical properties, obtained above, are  for the 
uniform threshold distribution,
and the natural question is how general the results are. To check
the universality of the ELS model, two other types of fiber strength
distributions can be easily considered \cite{bpc03}: linearly increasing 
density distribution and linearly decreasing density distribution.

\begin{figure}[]
\resizebox{9cm}{!}{\rotatebox{0}{\includegraphics{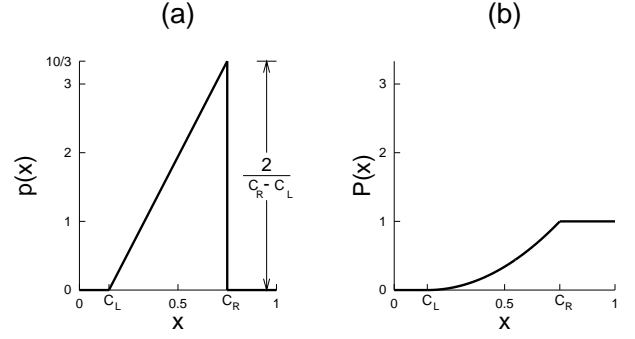}}}
\caption{(a) The density function $p(x)$ and (b) the
cumulative distribution $P(x)$ of random fiber strengths
$x$ distributed with linearly increasing density
in the interval $[C_L, C_R]$. In the particular instance shown
in the figure $C_L = 0.15$ and $C_R = 0.75$.}
\label{pratip_fig4}
\end{figure}

For linearly increasing density of fiber strengths in the 
interval $[C_L, C_R]$, the normalized density function and the
cumulative distribution are given by
(illustrated in Fig.~\ref{pratip_fig4}):

\begin{equation}
p(x) = \left \{ \begin{array}{l l}
 0, & ~0 \le x < C_L \\
 {{2(x - C_L)} \over {(C_R - C_L)^2}},
    & C_L \le x \le C_R \\
 0, & C_R < x
                \end{array}
       \right .
\label{eq:dens-lin}
\end{equation}

\noindent and
\begin{equation}
P(x) = \left \{ \begin{array}{l l}
 0, & ~0 \le x < C_L \\
 \left ({x - C_L} \over {C_R - C_L}
 \right )^2,
    & C_L \le x \le C_R \\
 1, & C_R < x .
                \end{array}
       \right .
\label{eq:prob-lin}
\end{equation}

\indent Now we introduce the transformed quantities:

\begin{equation}
\Gamma_0 = {\sigma \over {C_R - C_L}}, \hspace{0.5cm}
\Gamma_L = {C_L \over {C_R - C_L}}, \hspace{0.5cm}
\Gamma_t = {x_t \over {C_R - C_L}}.
\label{eq:trans-stress}
\end{equation}

\indent For an initial stress $C_L \leq \sigma \leq C_R$
(or, $\Gamma_L \leq \Gamma_0 \leq \Gamma_L+1$) along with the cumulative
distribution given by Eq.~(\ref{eq:prob-lin}), the recursion relations 
(Eq. \ref{rec-x} and Eq. \ref{rec-U}) appear as:
\begin{equation}
\Gamma_{t+1} = {\Gamma_0 \over {1 - (\Gamma_t - \Gamma_L)^2}}
\label{eq:stressrecur-lin}
\end{equation}

\noindent and
\begin{equation}
U_{t+1} = 1 - \left ({\Gamma_0 \over U_t} - \Gamma_L \right )^2 ,
 \hspace{1.0cm} U_0 = 1.
\label{eq:fracrecur-lin}
\end{equation}

\noindent The fixed point equations, (Eq. \ref{fix-x-uniform} and 
Eq. \ref{fix-U-uniform}), now assume cubic form:

\begin{equation}
\left (\Gamma^* \right )^3 - 2 \Gamma_L \left (\Gamma^* \right )^2
 + \left (\Gamma_L^2 - 1 \right ) \Gamma^* + \Gamma_0 = 0
\label{eq:stressfix-lin}
\end{equation}

\noindent where $\Gamma^* = x^* / (\sigma_R - \sigma_L)$, and

\begin{equation}
\left (U^* \right )^3 + \left ( \Gamma_L^2 - 1 \right ) \left (U^* \right )^2
 - \left (2 \Gamma_L \Gamma_0 \right ) U^* + \Gamma_0^2 = 0.
\label{eq:fracfix-lin}
\end{equation}
\noindent Consequently each of the recursions (Eq. \ref{eq:stressrecur-lin}
and Eq. \ref{eq:fracrecur-lin}) have three fixed points -- only one in
each case is found to be stable. For the redistributed stress the fixed
points are:

\begin{eqnarray}
\Gamma^*_1 & = & \frac{2}{3} \Gamma_L + 2 K_0 \cos\frac{\Phi}{3},
\label{eq:stressfix-lin1} \\
\Gamma^*_2 & = & \frac{2}{3} \Gamma_L - K_0 \cos\frac{\Phi}{3}
 + \sqrt{3} K_0 \sin\frac{\Phi}{3},
\label{eq:stressfix-lin2} \\
\Gamma^*_3 & = & \frac{2}{3} \Gamma_L - K_0 \cos\frac{\Phi}{3}
 - \sqrt{3} K_0 \sin\frac{\Phi}{3},
\label{eq:stressfix-lin3}
\end{eqnarray}

\noindent where
\begin{equation}
K_0 = \frac{1}{3} \sqrt{3 + \Gamma_L^2}
\label{eq:K-defn}
\end{equation}
\noindent and
\begin{equation}
\cos\Phi = {\Gamma_L \left ( 9 - \Gamma_L^2 \right ) - 27 \Gamma_0 / 2
            \over \left ( 3 + \Gamma_L^2 \right )^{3/2}}.
\label{eq:Phi-defn}
\end{equation}

\noindent Similarly, for the surviving fraction of fibers the fixed
points are:

\begin{eqnarray}
U^*_1 & = & {1 - \Gamma_L^2 \over 3} + 2 J_0 \cos\frac{\Theta}{3},
\label{eq:fracfix-lin1} \\
U^*_2 & = & {1 - \Gamma_L^2 \over 3} - J_0 \cos\frac{\Theta}{3}
 + \sqrt{3} J_0 \sin\frac{\Theta}{3},
\label{eq:fracfix-lin2} \\
U^*_3 & = & {1 - \Gamma_L^2 \over 3} - J_0 \cos\frac{\Theta}{3}
 - \sqrt{3} J_0 \sin\frac{\Theta}{3},
\label{eq:fracfix-lin3}
\end{eqnarray}
\noindent where
\begin{equation}
J_0 = \frac{1}{3} \sqrt{(\Gamma_L^2 - 1)^2 + 6 \Gamma_L \Gamma_0}
\label{eq:J-defn}
\end{equation}

\noindent and
\begin{equation}
\cos\Theta = {\left ( 1 - \Gamma_L^2 \right )
 \left [ \left ( \Gamma_L^2 - 1 \right )^2 + 9 \Gamma_L \Gamma_0 \right ]
 - 27 \Gamma_0^2 / 2
 \over \left [ \left ( \Gamma_L^2 - 1 \right )^2
               + 6 \Gamma_L \Gamma_0 \right ]^{3/2}}.
\label{eq:Theta-defn}
\end{equation}

\begin{figure}[]
\resizebox{9cm}{!}{\rotatebox{0}{\includegraphics{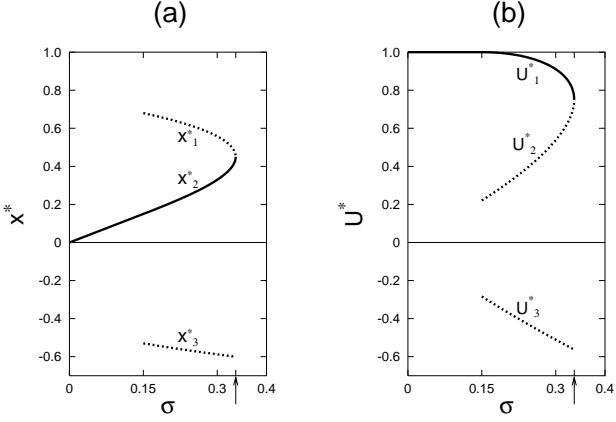}}}
\caption{The fixed points of (a) the redistributed stress,
and (b) the surviving fraction of fibers for the  distribution
of fiber strengths shown in Fig.~\ref{pratip_fig4}.
In each part of the figure the curve for the stable fixed points is shown
by a bold solid line and those for the unstable fixed points are shown
by bold broken lines. We have $C_L = 0.15$ and $C_R = 0.75$,
so that $\sigma_c = 0.3375$; the position of
the critical point is marked by an arrowhead.
For $\sigma \le C_L$ the fixed points are trivial: since there
are no broken fibers $x^* = \sigma$ and $U^* = U_0 = 1$.}
\label{pratip_fig5}
\end{figure}

\noindent Of these fixed points $\Gamma^*_2$ and $U^*_1$ are stable
whereas $\Gamma^*_1$, $\Gamma^*_3$ and $U^*_2$, $U^*_3$ are unstable
(Fig.~\ref{pratip_fig5}).

\indent The
discriminants of the cubic equations (Eq. \ref{eq:stressfix-lin} and
Eq. \ref{eq:fracfix-lin}) become zero at a critical value
$\sigma_c$ (or, $\Gamma_c$) of the initial
applied stress:

\begin{eqnarray}
\Gamma_c & = & {\sigma_c
                               \over {\sigma_R - \sigma_L}} \nonumber \\
~ & = & {2 \over 27} \left [ \Gamma_L \left ( 9 - \Gamma_L^2 \right )
                             + \left ( 3 + \Gamma_L^2 \right )^{3/2} \right ]
\label{eq:stresscrit-lin}
\end{eqnarray}

\noindent and then each of the quantities $\Gamma$ and $U$ have one
stable and one unstable fixed point. The critical point has
the trivial lower bound: $\sigma_c \geq C_L$.
\noindent The expression of $\Gamma_c$ in
Eq.~(\ref{eq:stresscrit-lin}) shows that it approaches the lower bound
as $\Gamma_L \to \infty$ which happens for finite values of $C_L$
and $C_R$ when $(C_R - C_L) \to 0$. It follows that
the upper bound for the critical point is also trivial: $\sigma_c \leq C_R$.
\noindent Also, at the critical point we get from Eq.~(\ref{eq:Phi-defn})
and Eq.~(\ref{eq:Theta-defn}):
\begin{equation}
\cos \Phi_{\rm crit} = \cos \Theta_{\rm crit} = -1
\label{eq:Phi-Theta-crit1}
\end{equation}
\noindent or,
\begin{equation}
\Phi_{\rm crit} = \Theta_{\rm crit} = \pi.
\label{eq:Phi-Theta-crit2}
\end{equation}

\indent The stable fixed points $\Gamma^*_2$ and $U_1^*$ are
positive real-valued
when $\Gamma_0 \leq \Gamma_c$; thus the fiber bundle
always reaches a state of mechanical equilibrium after partial failure
under an initial applied stress $\sigma_0 \leq \sigma_c$.
For $\sigma > \sigma_c$ (or, $\Gamma_0 >
\Gamma_c$), $\Gamma^*_2$ and $U_1^*$ are no longer
real-valued and the entire fiber bundle eventually breaks down.
The transition from the phase of partial failure to the phase of
total failure takes place when $\sigma$ just exceeds
$\sigma_c$ and the order parameter for this phase
transition is defined as in Eq.~(\ref{order-uniform}):
\begin{equation}
{O} \equiv U^*_1 - U^*_{\rm 1-crit}.
\label{eq:order-lin-defn}
\end{equation}

\noindent Close to the critical point but below it, we can write,
from Eq.~(\ref{eq:Theta-defn}) and  Eq.~(\ref{eq:Phi-Theta-crit2}), that:

\begin{eqnarray}
\pi - \Theta & \simeq & \sin \Theta \nonumber \\
~ & \simeq  & {3 \sqrt{3} \: \Gamma_c (3 + \Gamma_L^2)^{3/4}
 (\Gamma_c -\Gamma_0)^{1/2} \over
 [(\Gamma_L^2 - 1)^2 + 6 \Gamma_L \Gamma_c]^{3/2}}
\label{eq:Theta-nearcrit}
\end{eqnarray}

\noindent and the expressions for the fixed points in
Eq.~(\ref{eq:fracfix-lin1}) and Eq.~(\ref{eq:fracfix-lin2}) reduce
to the forms:

\begin{equation}
U^*_1 \simeq U^*_{\rm 1-crit} +
 {\Gamma_c (3 + \Gamma_L^2)^{3/4}
 \over (\Gamma_L^2 - 1)^2 + 6 \Gamma_L \Gamma_c}
 \left ( \Gamma_c - \Gamma_0 \right )^{1/2}
\label{eq:fracfix-lin1-nearcrit}
\end{equation}

\noindent and
\begin{equation}
U^*_2 \simeq U^*_{\rm 2-crit} -
 {\Gamma_c (3 + \Gamma_L^2)^{3/4}
 \over (\Gamma_L^2 - 1)^2 + 6 \Gamma_L \Gamma_c}
 \left ( \Gamma_c - \Gamma_0 \right )^{1/2},
\label{eq:fracfix-lin2-nearcrit}
\end{equation}

\noindent where
\begin{equation}
U^*_{1-{\rm crit}} = U^*_{2-{\rm crit}} = {1 - \Gamma_L^2 \over 3} +
 {1 \over 3} \sqrt{(\Gamma_L^2 - 1)^2 + 6 \Gamma_L \Gamma_c}
\label{eq:fracfix-lin12-crit}
\end{equation}

\noindent is the stable fixed point value of the surviving fraction
of fibers under the critical initial stress $\sigma_c$.
Therefore, following the definition of the order parameter in
Eq.~(\ref{eq:order-lin-defn}) we get from the above equation:

\begin{equation}
{O} = {\Gamma_c (3 + \Gamma_L^2)^{3/4}
 \over (\Gamma_L^2 - 1)^2 + 6 \Gamma_L \Gamma_c}
 \left ( \Gamma_c - \Gamma_0 \right )^{1/2},
 \hspace{1.0cm} \Gamma_0 \to \Gamma_c-.
\label{eq:order-lin-crit}
\end{equation}

\noindent On replacing the transformed variable $\Gamma_0$ by the original
$\sigma$, Eq.~(\ref{eq:order-lin-crit}) shows that the order parameter
goes to zero continuously following the same power-law as in
Eq.~(\ref{order-uniform})
for the previous case when $\sigma$ approaches its critical
value from below.

\indent Similarly the susceptibility diverges by the same power-law
as in Eq.~(\ref{sus-uniform}) on approaching the critical point
from below:

\begin{equation}
\chi = \left | {{\rm d} U^*_1 \over {\rm d}\sigma} \right |
     \propto \left (\Gamma_c - \Gamma_0 \right ) ^{-1/2},
 \hspace{1.0cm} \Gamma_0 \to \Gamma_c-.
\label{eq:suscept-lin-crit}
\end{equation}

\indent The critical dynamics of the fiber bundle is given by the
asymptotic closed form solution of the recursion
(Eq. \ref{eq:fracrecur-lin}) for $\Gamma_0 = \Gamma_c$:

\begin{equation}
U_t - U^*_{\rm 1-crit} \sim \left [
 {\left ( U^*_{\rm 1-crit} \right ) ^4
 \over 3 \left ( \Gamma_c \right ) ^2 - 2 \Gamma_L
 \Gamma_c U^*_{\rm 1-crit}} \right ] {1 \over t},
 \hspace{1.0cm} t \to \infty,
\label{eq:frac-lin-crit-dyn}
\end{equation}

\noindent where $\Gamma_c$ and $U^*_{\rm 1-crit}$
are given in Eq.~(\ref{eq:stresscrit-lin}) and
Eq.~(\ref{eq:fracfix-lin12-crit}) respectively.
This shows that the asymptotic relaxation of the surviving
fraction of fibers to its stable fixed point under the critical initial
stress has the same (inverse of step number) form as found in the case of 
uniform density of fiber strengths (Eq.~\ref{eq:fbm-omori}).

\begin{figure}[]
\resizebox{9cm}{!}{\rotatebox{0}{\includegraphics{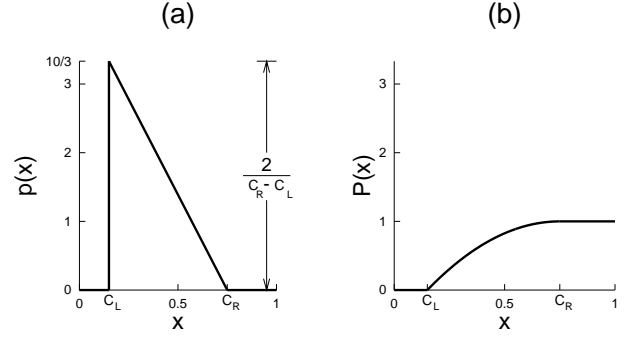}}}
\caption{(a) The density function $p(x)$ and (b) the
cumulative distribution $P(x)$ of fiber strengths
$x$ distributed with linearly decreasing density
in the interval $[C_L, C_R]$. Similar to the cases shown in
 Fig.~\ref{pratip_fig4} we have $C_L = 0.15$ and
$C_R = 0.75$ in this example also.}
\label{pratip_fig6}
\end{figure}

\indent We now consider
a fiber bundle
with a linearly decreasing density of fiber strengths in the interval
$[C_L, C_R]$.
The normalized density function and
cumulative distribution (illustrated in Fig.~\ref{pratip_fig6}) are:

\begin{equation}
p(x) = \left \{ \begin{array}{l c}
 0, & 0 \le x < C_L \\
 {{2(C_R - x)} \over {(C_R - C_L)^2}},
    & C_L \le x \le C_R \\
 0, & C_R < x
                \end{array}
       \right .
\label{eq:dens-lin'}
\end{equation}

\noindent and
\begin{equation}
P(x) = \left \{ \begin{array}{l c}
 0, & 0 \le x < C_L \\
 1 - \left ({{C_R - x} \over {C_R - C_L}}
 \right )^2,
    & C_L \le x \le C_R \\
 1, & C_R < x
                \end{array}
       \right .
\label{eq:prob-lin'}
\end{equation}

\indent With the transformed quantities defined in Eq.~(\ref{eq:trans-stress})
the recurrences (Eq. \ref{rec-x} and Eq. \ref{rec-U})
for $C_L \leq \sigma \leq C_R$ appear as:

\begin{equation}
\Gamma_{t+1} = {\Gamma_0 \over \left ( 1 + \Gamma_L - \Gamma_t \right )^2}
\label{eq:stressrecur-lin'}
\end{equation}

\noindent and
\begin{equation}
U_{t+1} = \left ( 1 + \Gamma_L - {\Gamma_0 \over U_t} \right )^2 ,
 \hspace{1.0cm} U_0 = 1.
\label{eq:fracrecur-lin'}
\end{equation}

\noindent The fixed point equations are again cubic:
\begin{equation}
\left (\Gamma^* \right )^3 - 2 \left ( 1 + \Gamma_L \right )
 \left (\Gamma^* \right )^2
 + \left ( 1 + \Gamma_L \right )^2 \Gamma^* - \Gamma_0 = 0,
\label{eq:stressfix-lin'}
\end{equation}

\begin{equation}
\left (U^* \right )^3 - \left ( 1 + \Gamma_L \right )^2 \left (U^* \right )^2
 + 2 \left ( 1 + \Gamma_L \right ) \Gamma_0 U^* - \Gamma_0^2 = 0
\label{eq:fracfix-lin'}
\end{equation}

\noindent and they have the following solutions:

\begin{eqnarray}
\Gamma^*_1 & = & {2 \over3} \left ( 1 + \Gamma_L \right )
 + 2 K_0' \cos\frac{\Phi'}{3},
\label{eq:stressfix-lin1'} \\
\Gamma^*_2 & = & {2 \over 3} \left ( 1 + \Gamma_L \right )
 - K_0' \cos\frac{\Phi'}{3} + \sqrt{3} K' \sin\frac{\Phi'}{3},
\label{eq:stressfix-lin2'} \\
\Gamma^*_3 & = & {2 \over 3} \left ( 1 + \Gamma_L \right )
 - K_0' \cos\frac{\Phi'}{3} - \sqrt{3} K' \sin\frac{\Phi'}{3},
\label{eq:stressfix-lin3'}
\end{eqnarray}

\noindent where
\begin{equation}
K_0' = {1 + \Gamma_L \over 3},
\label{eq:K'-defn}
\end{equation}

\begin{equation}
\cos \Phi' = {27 \Gamma_0 \over 2 \left ( 1 + \Gamma_L \right )^3} - 1
\label{eq:Phi'-defn}
\end{equation}

\noindent and

\begin{eqnarray}
U^*_1 & = & {(1 + \Gamma_L)^2 \over 3} + 2 J_0' \cos\frac{\Theta'}{3},
\label{eq:fracfix-lin1'} \\
U^*_2 & = & {(1 + \Gamma_L)^2 \over 3} - J_0' \cos\frac{\Theta'}{3}
 + \sqrt{3} J_0' \sin\frac{\Theta'}{3},
\label{eq:fracfix-lin2'} \\
U^*_3 & = & {(1 + \Gamma_L)^2 \over 3} - J_0' \cos\frac{\Theta'}{3}
 - \sqrt{3} J_0' \sin\frac{\Theta'}{3},
\label{eq:fracfix-lin3'}
\end{eqnarray}
\noindent where
\begin{equation}
J_0' = \frac{1}{3} \sqrt{(1 + \Gamma_L)^4 - 6 (1 + \Gamma_L) \Gamma_0},
\label{eq:J'-defn}
\end{equation}

\begin{equation}
\cos\Theta' = {\left ( 1 + \Gamma_L \right )^3 \left [
 \left ( 1 + \Gamma_L \right )^3 - 9 \Gamma_0 \right ] + 27 \Gamma^2 / 2
 \over \left [ \left ( 1 + \Gamma_L \right )^4
               - 6 \left ( 1 + \Gamma_L \right ) \Gamma_0 \right ]^{3/2}}.
\label{eq:Theta'-defn}
\end{equation}

\begin{figure}[]
\resizebox{9.5cm}{!}{\includegraphics{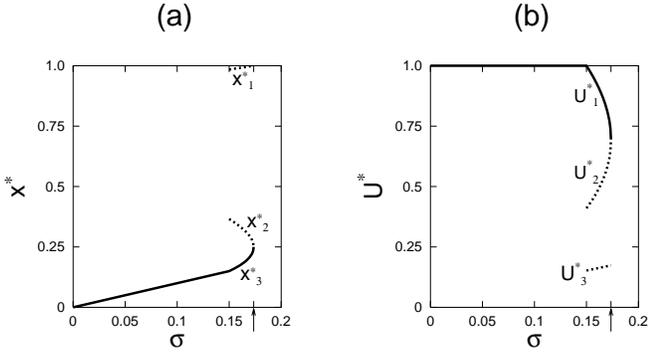}}
\caption{The fixed points of (a) the redistributed stress,
and (b) the surviving fraction of fibers for the  distribution
of fiber strengths shown in Fig.~\ref{pratip_fig6}. The curve for the stable
fixed points is shown by a bold solid line and those for the unstable
fixed points are shown by bold broken lines. In this example too
we have $C_L = 0.15$ and $C_R = 0.75$; here
$\sigma_c = 0.173611$, marked by an arrowhead. The
critical point is located lower than that in Fig.~\ref{pratip_fig5} due to
abundance of fibers of lower strengths compared to the previous case.}
\label{pratip_fig7}

\end{figure}

\noindent Here $\Gamma^*_3$ and $U^*_1$ are stable fixed points while
the rest are unstable (Fig.~\ref{pratip_fig7}).

\indent The discriminants of Eq.~(\ref{eq:stressfix-lin'}) and
Eq.~(\ref{eq:fracfix-lin'}) show that the critical applied stress in this
case, $\sigma_c'$ (or, $\Gamma_c'$),
is given by:
\begin{equation}
\Gamma_c'
 = {\sigma_c' \over C_R - C_L}
 = {4 \over 27} \left ( 1 + \Gamma_L \right ) ^3
\label{eq:stresscrit-lin'}
\end{equation}

\noindent or,
\begin{equation}
\sigma_c'
 = {4 C_R^3 \over 27 \left ( C_R - C_L \right )^2}.
\label{eq:stresscrit-lin'-alter}
\end{equation}

\noindent In order to satisfy the condition $\sigma_c'
\geq C_L$, it requires from Eq.~(\ref{eq:stresscrit-lin'-alter})
that $C_R \geq 3 C_L,$
\noindent which imposes an upper bound:
$\sigma_c' \le {C_R \over 3}$.

\indent Like before, for $\Gamma_0 \leq \Gamma_c'$
the stable fixed points are real-valued, which indicates that only partial
failure of the fiber bundle takes place before a state of mechanical
equilibrium is reached; for $\Gamma_0 > \Gamma_c'$
the fixed points are not real and a phase of total failure exists.
The order parameter $O$ of the transition is given by the
definition in Eq.~(\ref{eq:order-lin-defn}).

\indent For $\Gamma_0 = \Gamma_c'$ we get the
following properties from Eq.~(\ref{eq:Phi'-defn}),
Eq.~(\ref{eq:fracfix-lin1'}), Eq.~(\ref{eq:fracfix-lin2'}) and
Eq.~(\ref{eq:Theta'-defn}):

\begin{equation}
U^*_{1-{\rm crit}} = U^*_{2-{\rm crit}}
 = {4 \over 9} \left ( 1 + \Gamma_L \right )^2;
\label{eq:fracfix-lin12'-crit}
\end{equation}

\begin{equation}
\cos \Theta'_{\rm crit} = -1 \hspace{1.0cm}
 {\rm or}, \hspace{1.0cm} \Theta'_{\rm crit} = \pi
\label{eq:Theta'-crit}
\end{equation}

\noindent and
\begin{equation}
\cos \Phi'_{\rm crit} = 1 \hspace{1.0cm}
 {\rm or}, \hspace{1.0cm} \Phi'_{\rm crit} = 0.
\label{eq:Phi'-crit}
\end{equation}

\noindent Comparing Eq.~(\ref{eq:Phi'-crit}) and Eq.~(\ref{eq:Theta'-crit})
with Eq.~(\ref{eq:Phi-Theta-crit2}) we see that the critical values of
$\Theta$ and $\Theta'$ are the same whereas those of $\Phi$ and
$\Phi'$ differ by $\pi$ radians.

\indent Near the critical point, but below it, we get from
Eq.~(\ref{eq:fracfix-lin1'}) and Eq.~(\ref{eq:fracfix-lin2'}):

\begin{equation}
U^*_1 \simeq U^*_{\rm 1-crit}
 + {4 \over 3} \left (1 + \Gamma_L \right ) ^{1/2}
 \left ( \Gamma_c' - \Gamma_0 \right )^{1/2}
\label{eq:fracfix-lin1'-nearcrit}
\end{equation}

\noindent and
\begin{equation}
U^*_2 \simeq U^*_{\rm 2-crit}
 - {4 \over 3} \left (1 + \Gamma_L \right ) ^{1/2}
 \left ( \Gamma_c' - \Gamma_0 \right )^{1/2}.
\label{eq:fracfix-lin2'-nearcrit}
\end{equation}

\noindent Therefore, by the definition of the order parameter in
Eq.~(\ref{eq:order-lin-defn}) and that of the susceptibility in
(\ref{eq:suscept-lin-crit}), we get in this case ${O} \propto\left (
\Gamma_c' - \Gamma_0 \right )^{1/2}$ and
$\chi \propto \left ( \Gamma_c' - \Gamma_0 \right )^{-1/2}$,
$\Gamma_0 \to \Gamma_c' -$. These power laws have the
same exponents as the corresponding ones in the previous cases and
differ from those only in the critical point and the critical amplitude.

\indent At the critical point the asymptotic relaxation of the
surviving fraction of fibers to its stable fixed point [obtained as
an asymptotic solution to Eq. (\ref{eq:fracrecur-lin'})] is again found to be
a power law decay similar to Eq.~(\ref{eq:fbm-omori})
and Eq.~(\ref{eq:frac-lin-crit-dyn}):

\begin{eqnarray}
U_t - U^*_{\rm 1-crit} & \sim &
 {4 \over 3} {U^*_{\rm 1-crit} \over t}, \hspace{1.0cm} t \to \infty
 \nonumber \\
 & \sim & {16 \over 27} \left ( 1 + \Gamma_L \right )^2 {1 \over t}.
\label{eq:frac-lin-crit-dyn'}
\end{eqnarray}

\indent The two density functions,
Eq.~(\ref{eq:dens-lin}) and Eq.~(\ref{eq:dens-lin'}),
can be transformed from one to the other by a reflection on the
line $x = (C_L + C_R)/2$
(compare Fig.~\ref{pratip_fig4}(a) and Fig.~\ref{pratip_fig6}(a)).
But the fixed point equations and their
solutions do not have this symmetry. This is because
the density function $p(x)$ does not appear
directly in the recursion relations for the dynamics. It is the
cumulative distribution $P(x)$ which appears in
the recursion relations. Eq.~(\ref{eq:prob-lin}) and
Eq.~(\ref{eq:prob-lin'}) show that the cumulative distributions of
these two cases are not mutually symmetric about any value of
the threshold stress $x$
(compare Fig.~\ref{pratip_fig4}(b) and Fig.~\ref{pratip_fig6}(b)).
However a certain relation exists between the critical
values of the applied stress for a special case of these two models:
if $C_L = 0$, we get from Eq.~(\ref{eq:stresscrit-lin}) and
Eq.~(\ref{eq:stresscrit-lin'}) that $\sigma_c / C_R
 = \sqrt{4/27}$ and $\sigma_c' / C_R = 4/27$
respectively; therefore we have $\sigma_c' / C_R
 = \left ( \sigma_c / C_R \right )^2$.

\indent The critical behavior of the models discussed
in this section show that the power laws found here
are independent of the form of the cumulative distribution $P$.
The three threshold distributions studied here have a common
feature: the function $x^* \left [ 1 - P \left( x^* \right)
\right ]$ has a maximum which corresponds to the critical value of the initial
applied stress. All threshold distributions having this property
are therefore expected to lead to the same universality class as
the three studied here. If the threshold distribution does not
have this property we may not observe a phase transition at all.
For example, consider a fiber bundle model with
$P(x) = 1 - 1/x$,
$x \geq 1$.
Here $x^* \left [ 1 - P \left(x^* \right) \right ] = 1$
and the evolution of the fiber bundle is given by the recursion
relation $U_{t+1} = U_t/\sigma$ which implies that there is
no dynamics at all for $\sigma = 1$ and an exponential decay
to complete failure, $U_t = (\sigma)^{-t}$, for $\sigma > 1$.
There are no critical phenomena and therefore no phase transition.
However this general conclusion may not be true for finite-sized
bundles \cite{ms83}.

Thus the ELS fiber bundles (for different fiber threshold distributions)
show phase transition with a well defined order parameter which shows
similar power law variation on the way the critical point is approached.
For all the cases, discussed here, the susceptibility and relaxation time 
diverge following similar power laws and the failure processes show 
similar critical
slowing at the critical point. This suggests strongly that the critical
behavior is universal, which we now prove through general 
arguments \cite{hhp06}.

When an iteration is close to the fixed point, we have for the deviation
\begin{eqnarray}
\Delta U_{t+1} & =P\left(\frac{\sigma}{U^{*}}\right)-P\left(\frac{\sigma}{U^{*}+\Delta U_{t}}\right)=\Delta U_{t}\cdot\frac{\sigma}{U^{*2}}p(\sigma/U^{*}),\label{dev-1}\end{eqnarray}
to lowest order in $\Delta U_{t}$.
This guarantees an exponential relaxation to the fixed point, $\Delta U_{t}\propto e^{-t/\tau}$,
with parameter \begin{equation}
\tau=1\left/\ln\left(\frac{U^{*2}}{\sigma p(\sigma/U^{*})}\right)\right..\end{equation}
 Criticality is determined by the extremum condition (Eq. \ref{critical-criteria}),
which by the relation (Eq. \ref{rec-U}) takes the form \[
U_{c}^{2}=\sigma p(\sigma/U_{c}).\]
Thus $\tau=\infty$ at criticality. To study the relaxation at criticality
we must expand Eq. (\ref{dev-1}) to second order in $\Delta U_{t}$ since
to first order we simply get the useless equation $\Delta U_{t+1}=\Delta U_{t}$.
To second order we obtain 
\[
\Delta U_{t+1}=\Delta U_{t}-C \Delta U_{t}^{2},
\]
 with a positive constant $C$. This is satisfied by
 \[
\Delta U_{t}=\frac{1}{Ct}+\mathcal{O}(t^{-2}).
\]
 Hence in general the dominating critical behavior for the approach
to the fixed point is a power law with $\eta=1$. The values 
$\alpha=\beta=\theta=\frac{1}{2}$
can be shown to be consequences of the parabolic maximum of the load
curve at criticality. Thus all threshold distributions for which the
macroscopic strength function has a single parabolic maximum, is in
this universality class.

\begin{figure}
\epsfysize=2in
\epsfbox{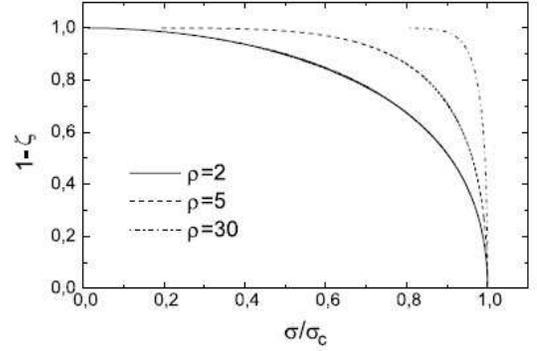}
\caption{Branching ratio as a function of applied stress for three different $\rho$ (Weibull index) [from \cite{mgp00}]. }
\label{fig:zeta}
\end{figure}
It is clear that at the critical stress value $\sigma_{c}$, ELS fiber bundles 
show phase transition from  partially broken state to completely broken state. 
What is the order of this phase transition?  \textcite{zrsv97,zrsv99a} 
considered the 
fraction of unbroken fibers as the order parameter and as it has a 
discontinuity at the critical stress value, they suggested, after a mean-field 
analysis, that it can be seen as a first-order phase 
transition similar to spinodal instability \cite{m94}. The additional reason 
for identifying the transition at $\sigma = \sigma_c$ as a first-order 
spinodal point had been \cite{kzh00} that in the presence of short-range 
interactions (as in LLS, see Section IV), the transition becomes discontinuous
 and first-order like. It is indeed hard to identify continuously changing 
order parameter there. We, however, believe the transition in ELS to be 
second-order.
Chronologically, a little later, a new parameter
was identified \cite{mgp00}: the branching ratio ($\zeta$), 
which  is defined as the probability of triggering further 
breaking given an individual failure. The branching ratio continuously 
approaches (Fig. \ref{fig:zeta}) the value $1$ at the critical stress ($\sigma_{c}$) starting from 
$0$ value (for very small $\sigma$). Also it shows a power law variation: 
 $1-\zeta \propto (\sigma_{c}-\sigma)^\beta$, with $\beta=1/2$. 
Therefore $1-\zeta$ acts as the order parameter showing a continuous 
transition at the critical point, signaling a second-order phase transition.
As mentioned earlier \textcite{pc01} and  \textcite{pbc02}   considered the 
difference between  the fraction of unbroken 
fibers at any $\sigma$ and at $\sigma_{c}$, as the order parameter ($O$): it
  shows a similar continuous variation with the applied stress: 
$O \propto (\sigma_{c}-\sigma)^\beta$, with $\beta=1/2$. Apart from this, 
the susceptibility and relaxation time diverge at the critical point following 
 power laws having universal exponent values \cite{pbc02,bpc03}. 
One may therefore conclude that at the critical point the ELS fiber bundles 
show a second-order phase transition with robust critical behavior as discussed
 here. 
             
\indent Finally we compare the ELS fiber bundle model studied
here with the mean-field Ising model. Though the order parameter
exponent (equal to $\frac{1}{2}$) of this model is identical to
that of the mean-field Ising model the two models are not in the
same universality class. The susceptibility in these models diverge
with critical exponents $\frac{1}{2}$ and $1$ respectively on
approaching the critical point. The dynamical critical exponents
are not the same either: in this fiber bundle model the surviving
fraction of fibers under the critical applied stress decays toward
its stable fixed point as $t^{-1}$, whereas the magnetization of the
mean-field Ising model at the critical temperature decays to zero
as $t^{-1/2}$.

\subsubsection{Relaxation behavior and critical amplitude ratio}

When an external load $F$ is applied to a fiber bundle, the iterative
failure process continues until all fibers fail, or an equilibrium
situation with a nonzero bundle strength is reached. Since the number
of fibers is finite, the number of steps, $t_{f}$, in this sequential
process is {\em finite}. Following \textcite{ph07}, we now determine 
how $t_{f}$ depends upon the applied stress $\sigma$.

\begin{figure}
\epsfysize=2in
\epsfbox{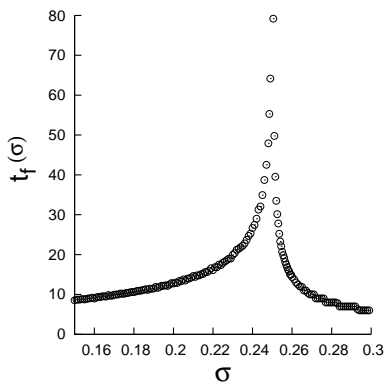}
\caption{Number of relaxation steps $t_{f}(\sigma)$
for a fiber bundle with a uniform threshold distribution (Eq. \ref{uniform}).
Here $\sigma_{c}=0.25$. The figure is based on 1000 samples, each
with $N=10^{6}$ fibers. }
\label{fig:relax-fig1_rmp}
\end{figure}

The state of the bundle can be characterized as \textit{pre-critical}
or \textit{post-critical} depending upon the stress value relative
to the critical stress $\sigma_{c}=F_{c}/N,$ above which the bundle
collapses completely. The function $t_{f}(\sigma)$ that we now focus
on, exhibits critical divergence when the critical point is approached
from \textit{either} side. As an example, we show in 
Fig. \ref{fig:relax-fig1_rmp} the $t_{f}(\sigma)$
obtained by simulation for a uniform threshold distribution.

We study the stepwise failure process in the bundle, when a fixed
external load $F=N\sigma$ is applied. Let $N_{t}$ be the number
of intact fibers at step no.\ $t$, with $N_{0}=N$. We want to determine
how $N_{t}$ decreases until the degradation process stops. 
When $N$ is a large number, we recall the basic recursion (Eq. \ref{rec-U}) to 
formulate the  breaking dynamics: 
\begin{equation}
U_{t+1}=1-P(\sigma/U_{t}),
\label{basic-rec}
\end{equation}
where $U_t = N_t /N $  is considered as a continuous variable.

\vspace{.2in}
\noindent \emph{(a) Post-critical relaxation}

\vspace{.2in}

We study first the post-critical situation, $\sigma>\sigma_{c}$,
with positive values of 
$\epsilon=\sigma-\sigma_{c}$,
 and start with the simplest one, uniform threshold distribution (Eq. \ref{uniform})  with  the critical point at $x_{c}=1/2$, $\sigma_{c}=1/4$. Then
the basic recursion relation (Eq. \ref{basic-rec}) takes the form 
\begin{equation}
U_{t+1}=1-\frac{\sigma}{U_{t}}=1-\frac{\frac{1}{4}+\epsilon}{U_{t}}.\label{Ut-uni}\end{equation}
 This nonlinear iteration can be transformed into a linear relation. 
We introduce first 
$U_{t}={\textstyle \frac{1}{2}}-y_{t}\sqrt{\epsilon},$
 into Eq. (\ref{Ut-uni}), with a result \begin{equation}
\frac{y_{t+1}-y_{t}}{1+y_{t}y_{t+1}}=2\sqrt{\epsilon}.\end{equation}
 Then we put 
$y_{t}=\tan v_{t}$, which gives \begin{equation}
2\sqrt{\epsilon}=\frac{\tan v_{t+1}-\tan v_{t}}{1+\tan v_{t+1}\;\tan v_{t}}=\tan(v_{t+1}-v_{t}).\end{equation}
 Hence we get $v_{t+1}-v_{t}=\tan^{-1}(2\sqrt{\epsilon})$, with solution
\begin{equation}
v_{t}=v_{0}+t\;\tan^{-1}(2\sqrt{\epsilon}).\end{equation}
 In the original variable the solution reads {\footnotesize \begin{eqnarray}
U_{t} & = & {\textstyle \frac{1}{2}}-\sqrt{\epsilon}\;\tan\left(\tan^{-1}(\frac{\frac{1}{2}-U_{0}}{\sqrt{\epsilon}})+\; t\;\tan^{-1}(2\sqrt{\epsilon})\right)\\
 & = & {\textstyle \frac{1}{2}}-\sqrt{\epsilon}\;\tan\left(-\tan^{-1}(1/2\sqrt{\epsilon})+\; t\;\tan^{-1}(2\sqrt{\epsilon})\right),\label{finaluni}\end{eqnarray}
 }where $U_{0}=1$ has been used.

\begin{figure}
\epsfbox{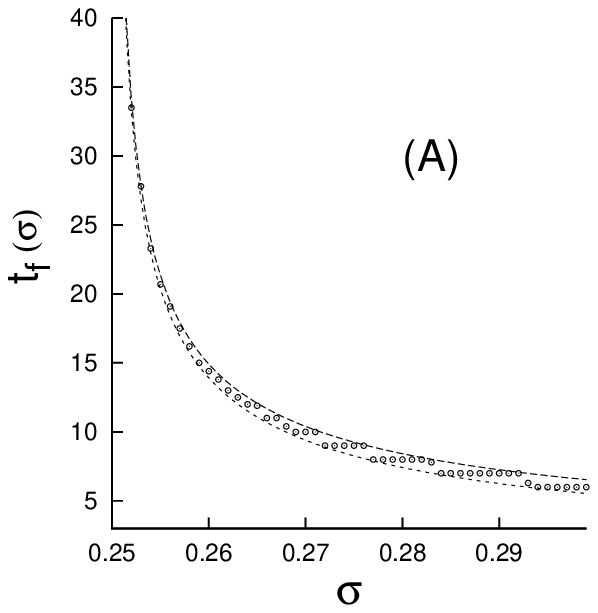}
\epsfbox{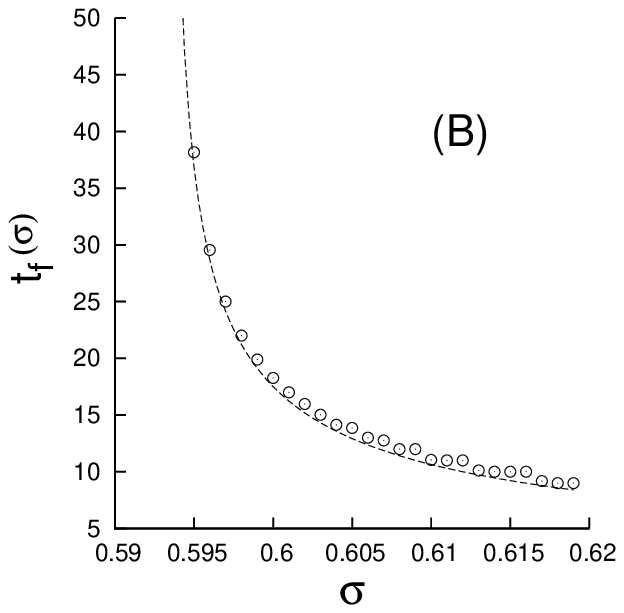}
\caption{Simulation results with post-critical stress
for (A) the uniform threshold distribution (Eq. \ref{uniform}), and (B)
the Weibull distribution (Eq. \ref{Weibull}) with index $5$.
 The graphs are based on
10000 samples with $N=10^{6}$ fibers in each bundle. Open circles
represent simulation data and dashed lines are the theoretical estimates
Eq. (\ref{tu}), Eq. (\ref{tl}) in (A) and Eq. (\ref{estimateW}) in (B). }
\label{fig:relax-fig2_rmp}
\end{figure}

Eq.\ (\ref{Ut-uni}) shows that when $U_{t}$ obtains a value in the
interval $(0,\sigma)$, the next iteration gives complete bundle failure.
Taking $U_{t}=\sigma$ as the penultimate value gives a lower bound,
$t_{f}^{l}$, for the number of iterations, while using $U_{t}=0$
in Eq. (\ref{finaluni}) gives an upper bound $t_{f}^{u}$. Adding unity
for the final iteration, Eq. (\ref{finaluni}) gives the bounds \begin{equation}
t_{f}^{u}(\sigma)=1+\frac{2\tan^{-1}(1/2\sqrt{\epsilon})}{\tan^{-1}(2\sqrt{\epsilon})},\label{tu}\end{equation}
 and \begin{equation}
t_{f}^{l}(\sigma)=1+\frac{\tan^{-1}((\frac{1}{4}-\epsilon)/\sqrt{\epsilon})+\tan^{-1}(1/2\sqrt{\epsilon})}{\tan^{-1}(2\sqrt{\epsilon})}.\label{tl}\end{equation}
Fig. \ref{fig:relax-fig2_rmp}A shows that these bounds nicely embrace 
the simulation results.

Note that both the upper and the lower bound behave as $\epsilon^{-\frac{1}{2}}$
for small $\epsilon$. A rough approximation near the critical point
is \begin{equation}
t_{f}(\sigma)\approx \kappa_{+}(\sigma-\sigma_{c})^{-\frac{1}{2}}.\end{equation}
 with $\kappa_{+}=\pi/2$.

Due to the inherent simplicity, uniform distribution is somewhat  easy to 
analyze. Therefore we now discuss how to
handle other distributions. Let us start with a Weibull
distribution (Eq. \ref{Weibull})  with index $5$.  The critical parameters for 
this case are $x_{c}=5^{-1/5}=0.72478$ and $\sigma_{c}=(5e)^{-1/5}=0.5933994$. 

\begin{figure}
\epsfysize=2.5in
\epsfbox{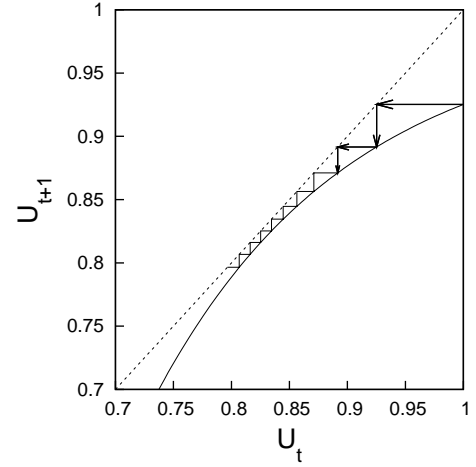}
\caption{  The iteration function $f(U)$ for the Weibull
distribution (Eq. \ref{Weibull}) with index $5$. Here $\sigma=0.6$, slightly 
greater than the critical value $\sigma_{c}=0.5933994.$} 
\label{fig:relax-fig3_rmp}
\end{figure}

The interesting values of the external stress are close to $\sigma_{c}$,
because for large super-critical stresses the bundle breaks down almost
immediately. For $\sigma$ slightly above $\sigma_{c}$ the iteration
function \begin{equation}
U_{t+1}=f(U_{t})=1-P(\sigma/U_{t})=e^{-(\sigma/U_{t})^{5}},\label{it-W}\end{equation}
 takes the form sketched in Fig. \ref{fig:relax-fig3_rmp}.

The iteration function is almost tangent to the reflection line $U_{t+1}=U_{t}$
and a long channel of width proportional to $\epsilon$ appears. The
dominating number of iterations occur within this channel (see Fig.  \ref{fig:relax-fig3_rmp}). The channel wall formed by the iteration function is almost parabolic
and is well approximated by a second-order expression \begin{equation}
U_{t+1}=U_{c}+(U_{t}-U_{c})+a(U_{t}-U_{c})^{2}+b(\sigma_{c}-\sigma).\label{quadr}\end{equation}
 Here $U_{c}=e^{-1/5}$ is the fixed point, $U_{t+1}=U_{t}$, of the
iteration at $\sigma=\sigma_{c}$. With $u=(U-U_{c})/b$ and $\epsilon=\sigma-\sigma_{c}$
 Eq. (\ref{quadr}) takes the form \begin{equation}
u_{t+1}-u_{t}=-Au_{t}^{2}-\epsilon,\end{equation}
 with $A=ab$. In the channel $u$ changes very slowly, so we may
treat the difference equation as a a differential equation: \begin{equation}
\frac{du}{dt}=-Au^{2}-\epsilon,\end{equation}
 with solution \begin{equation}
t\sqrt{A\epsilon}=-\tan^{-1}\left(u\sqrt{A/\epsilon}\right)+\mbox{ constant }.\end{equation}
 Thus \begin{equation}
t_{e}-t_{s}=(A\epsilon)^{-\frac{1}{2}}\left\{ \tan^{-1}(u_{s}\sqrt{A/\epsilon})-\tan^{-1}(u_{e}\sqrt{A/\epsilon})\right\} \label{tf-ti}\end{equation}
 is the number of iterations \textit{in the channel}, starting with
$u_{s}$, ending with $u_{e}$. This treatment is general and can
be applied to any threshold distribution near criticality. Although
the vast majority of the iterations occur in the channel, there are
a few iterations at the entrance and at the exit of the channel that
may require attention in special cases. The situation is similar to
type I intermittency in dynamical systems, but in our case the channel
is traversed merely once.

For the Weibull distribution the expansion (Eq. \ref{quadr}) has the
precise form\begin{eqnarray}
U_{t} & = & e^{-(\sigma/U)^{5}}\simeq e^{-1/5}+(U-U_{c})\nonumber \\
 &  & -{\textstyle \frac{5}{2}}e^{1/5}(U-U_{c})^{2}-5^{1/5}(\sigma-\sigma_{c}),\label{b1}\end{eqnarray}
 where $U_{c}=e^{-1/5}$, $a={\textstyle \frac{5}{2}}e^{1/5}$, $b=5^{1/5}$
and $A={\textstyle \frac{5}{2}}(5e)^{1/5}$. For completeness we must
also consider the number of iteration to reach the entrance to the
channel. It is not meaningful to use the quadratic approximation (Eq. \ref{b1})
where it is not monotonously increasing, i.e. for $U>U_{m}=U_{c}+1/(2a)=\frac{6}{5}e^{-1/5}\simeq0.98$.
Thus we take $U_{s}=U_{m}$ as the entrance to the channel, and add
one extra iteration to arrive from $U_{0}=1$ to the channel entrance.
(Numerical evidence for this extra step: For $\sigma=\sigma_{c}$
the iteration (Eq. \ref{it-W}) starts as follows: $U_{0}=1.00$, $U_{1}=0.93$,
$U_{2}=0.90$, while using the quadratic function with $U_{0}=U_{m}=0.98$
as the initial value, we get after one step $U_{1}=0.90$, approximately
the same value that the exact iteration reaches after two steps.)
With $U_{e}=0$ we obtain from Eq. (\ref{tf-ti}), in the Weibull case,
the estimate\begin{eqnarray}
t_{f} & = & 1+(A\epsilon)^{-1/2}\left\{ \tan^{-1}(e^{-1/5}\sqrt{A/\epsilon}\,/5b)\right.\nonumber \\
 &  & \left.+\tan^{-1}(e^{-1/5}\sqrt{A/\epsilon}\,/b)\right\} ,\label{estimateW}\end{eqnarray}
 with $A=\frac{5}{2}(5e)^{1/5}$ and $b=5^{1/5}$. 

Near the critical point Eq. (\ref{estimateW}) has the asymptotic form
\begin{equation}
t_{f}\approx\pi(A\epsilon)^{-1/2}=\kappa_{+}(\sigma-\sigma_{c})^{-1/2},\label{div}\end{equation}
 with $\kappa_{+}=\pi(2/5)^{1/2}(5e)^{-1/10}$. 
The critical
index is the same as for the uniform threshold distribution. The theoretical
estimates give an excellent representation of the simulation data
(see Fig. \ref{fig:relax-fig2_rmp}B).

\vskip.2in
\noindent \emph{(b) Pre-critical relaxation}
\vskip.2in

We now assume the external stress to be pre-critical, $\sigma<\sigma_{c}$,
and introduce the positive parameter 
$\varepsilon=\sigma_{c}-\sigma$
 to characterize the deviation from the critical point. Starting with
uniform threshold distribution and introducing 
$U_{t}={\textstyle \frac{1}{2}}+\sqrt{\varepsilon}/z_{t}$
 and $\sigma=\frac{1}{4}-\varepsilon$ into Eq. (\ref{Ut-uni}),
 one gets \begin{equation}
2\sqrt{\varepsilon}=\frac{z_{t+1}-z_{t}}{1-z_{t+1}\; z_{t}}.\end{equation}
 In this case we put
$z_{t}=\tanh w_{t}$,  
which gives \begin{equation}
2\sqrt{\varepsilon}=\frac{\tanh w_{t+1}-\tanh w_{t}}{1-\tanh w_{t+1}\;\tanh w_{t}}=\tanh(w_{t+1}-w_{t}).\end{equation}
 Thus $w_{t+1}-w_{t}=\tanh^{-1}(2\sqrt{\varepsilon})$, i.e. \begin{equation}
w_{t}=w_{0}+t\;\tanh^{-1}(2\sqrt{\varepsilon}).\end{equation}
 Starting with $U_{0}=1$, we obtain $z_{0}=2\sqrt{\varepsilon}$
and hence \begin{equation}
w_{t}=(1+t)\;\tanh^{-1}(2\sqrt{\varepsilon}).\end{equation}
 This corresponds to \begin{equation}
U_{t}={\textstyle \frac{1}{2}}+\frac{\sqrt{\varepsilon}}{\tanh\left\{ (1+t)\tanh^{-1}(2\sqrt{\varepsilon})\right\} }\label{ntor}\end{equation}
 in the original variable.

\begin{figure}
\epsfysize=2.5in
\epsfbox{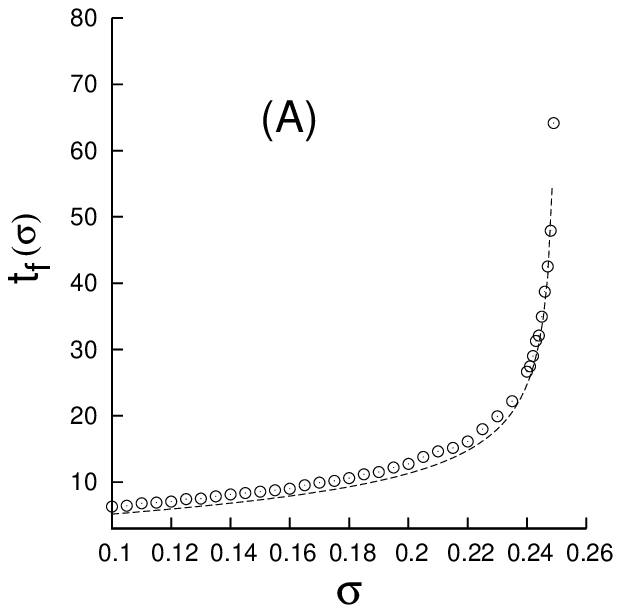}
\epsfysize=2.5in
\epsfbox{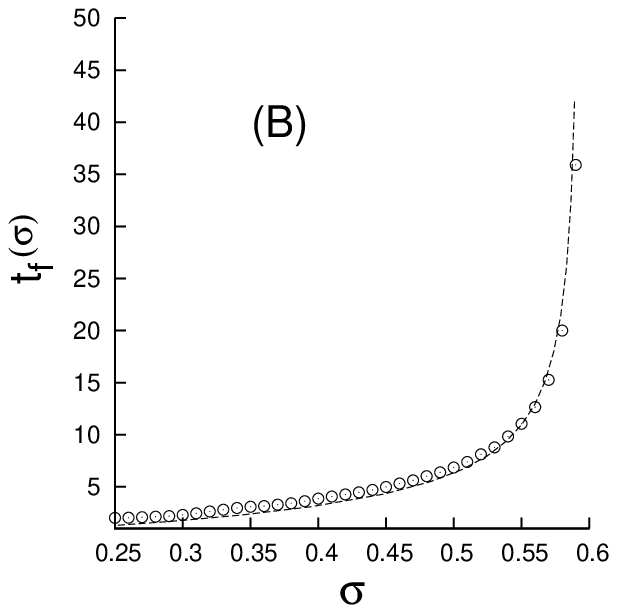}
\caption{ Simulation results with pre-critical stress
for (A) the uniform threshold distribution (Eq. \ref{uniform}), and (B) the 
Weibull distribution (\ref{Weibull}) with index $5$. The graphs are based on 
$10000$ samples with $N=10^{6}$
fibers in each bundle. Open circles represent simulation data and
the dotted lines are the theoretical estimates, Eq.\ (\ref{subnorm})
in (A) and Eq.\ (\ref{par1}-\ref{par2}) in (B).}
\label{fig:relax-fig4_rmp}
\end{figure}

Apparently $U_{t}$ reaches a fixed point $U^{*}=\frac{1}{2}+\sqrt{\varepsilon}$
after an infinite number of iterations. However, for a  bundle with
 \textit{finite} number of fibers,  only a finite number
of steps is needed for the iteration to arrive at a fixed point $N^{*}$ which 
is approximately \cite{pbc02,ph07}  
 \begin{equation}
N^{*}=\frac{N}{2}\left(1-\sqrt{1-4\sigma}\right)+\frac{1}{2}\left(1+(1-4\sigma)^{-1/2}\right).\label{app}\end{equation}
As a consequence,
we can use \begin{equation}
U_{t}=\frac{N^{*}}{N}=\frac{1}{2}+\sqrt{\varepsilon}+\frac{1}{4N}\left(2+\varepsilon^{-1/2}\right)\end{equation}
 as the final value in Eq. (\ref{ntor}). Consequently we obtain the following
estimate for the number of iterations to reach this value: \begin{equation}
t_{f}(\sigma)=-1+\frac{\coth^{-1}\left\{ 1+(1+2\sqrt{\varepsilon})/4N\varepsilon\right\} }{\tanh^{-1}(2\sqrt{\varepsilon})}.\label{subnorm}\end{equation}
 Fig. \ref{fig:relax-fig4_rmp}A shows that the simulation data are well 
approximated by the analytic formula (Eq. \ref{subnorm}).

For very large $N$ Eq. (\ref{subnorm}) is approximated by \begin{equation}
t_{f}=\frac{\ln(N)}{4}\;\varepsilon^{-1/2}=\kappa_{-}(\sigma_{c}-\sigma)^{-1/2}.\end{equation}
 with $\kappa_{-}=\ln(N)/4$. The critical behavior is again characterized
by a square root divergence.

Again we use the Weibull distribution (Eq. \ref{Weibull})  as an example threshold distribution.
In principle, the
iteration, \begin{equation}
U_{t+1}=1-P(\sigma/U_{t}),\end{equation}
 will reach a fixed point $U^{*}$ after infinite many steps. The
deviation from the fixed point, $U_{t}-U^{*}$, will decrease exponentially
near the fixed point: \begin{equation}
U_{t}-U^{*}\propto e^{-t/\tau},\label{exp}\end{equation}
 with \begin{equation}
\tau=1/\ln\left\{ U^{*2}\sigma^{-1}/p(\sigma/U^{*})\right\} .\end{equation}

For the Weibull threshold distribution with index $=5$  \begin{equation}
p(\sigma/U^{*})=5(\sigma/U^{*})^{4}\;\exp\left(-(\sigma/U^{*})^{5}\right)=5\sigma^{4}/U^{*3},\end{equation}
 and thus \begin{equation}
\tau=1/\ln(U^{*5}/5\sigma^{5}).\end{equation}
If we allow ourselves to use the exponential
formula (Eq. \ref{exp}) all the way from $U_{0}=1$, we obtain \begin{equation}
U_{t}-U^{*}=(1-U^{*})e^{-t/\tau}.\label{fixW}\end{equation}

For a finite number $N$ of fibers the iteration will stop after a
finite number of steps. It is a reasonable supposition to assume that
the iteration stops when $N_{t}-N^{*}$ is of the order $1$. This
corresponds to take the left-hand side of Eq. (\ref{fixW}) equal to $1/N$.
The corresponding number of iterations is then given by \begin{equation}
t_{f}=\tau\;\ln\left(N(1-U^{*})\right)\label{tGensub}\end{equation}
 in general, and \begin{equation}
t_{f}=\frac{\ln\left(N(1-U^{*})\right)}{\ln(U^{*5}/5\sigma^{5})}\label{tWsub}\end{equation}
 in the Weibull case. Solving the Weibull iteration $U^{*}=\exp(-(\sigma/U^{*})^{5})$
with respect to $\sigma$ and inserting into Eq. (\ref{tWsub}), we obtain
\begin{eqnarray}
t_{f} & = & -\frac{\ln\left\{ N(1-U^{*})\right\} }{\ln\left\{ 5(-\ln U^{*})\right\} }\label{par1}\\
\sigma & = & U^{*}(-\ln U^{*})^{1/5}.\label{par2}\end{eqnarray}
 These two equations represent the function $t(\sigma)$ on parameter
form, with $U^{*}$ running from $U_{c}=e^{-1/5}$ to $U^{*}=1$.

For $U^{*}=U_{c}=e^{-1/5}$ Eq. (\ref{par1}) shows that $t_{f}$ is infinite,
as it should be. To investigate the critical neighborhood we put 
$U^{*}=U_{c}(1+\Delta U)$
with $\Delta U$ small, to obtain to lowest order \begin{eqnarray}
t_{f} & = & \frac{\ln(N)}{5\Delta U}\label{57}\\
\sigma_{c}-\sigma & = & \frac{5}{2}\sigma(\Delta U)^{2}\label{58}\end{eqnarray}
 The combination of Eq. (\ref{57}) and Eq. (\ref{58}) gives, once more,
the square root divergence \begin{equation}
t_{f}(\sigma)\simeq \kappa_{-}(\sigma_{c}-\sigma)^{-1/2},\label{a}\end{equation}
 now with the magnitude \begin{equation}
\kappa_{-}=10^{-1/2}(5e)^{-1/10}\ln(N).\label{b}\end{equation}
Simulation results for the pre-critical Weibull distribution are shown
in Fig. \ref{fig:relax-fig4_rmp}B. which shows good agreement with the 
analytic solution (Eqs. \ref{par1}- \ref{par2}). 

For a general threshold distribution the divergence and its amplitude
are most easily deduced by expanding both the load curve $\sigma=x[1-P(x)]$
and the characteristic time $\tau$ around the critical threshold
$x_{c}$. To lowest contributing order in $x_{c}-x$ we find \begin{equation}
\sigma=\sigma_{c}-\frac{1}{2}[2p(x_{c})+x_{c}p^{\prime}(x_{c})]+(x_{c}-x)^{2}\label{y}\end{equation}
 and \begin{equation}
\tau=\frac{x_{c}p(x_{c})}{2p(x_{c})+x_{c}^{2}p^{\prime}(x_{c})}(x_{c}-x).\label{x}\end{equation}
 Inserting for $(x_{c}-x)$ from the equation above, and using (\ref{tWsub}),
we find \begin{equation}
t_{f}=\kappa_{-}(\sigma_c -\sigma) ^{-1/2}\label{xx}\end{equation}
 with

\begin{equation}
\kappa_{-}=x_{c}p(x_{c})[4p(x_{c})+2x_{c}p^{\prime}(x_{c})]^{-1/2}\ln(N).\label{generalk}\end{equation}

To show how the magnitude of the amplitude $\kappa_{-}$
depends on the form of the threshold distribution, we consider a Weibull
distribution \begin{equation}
P(x)=1-e^{(x/a)^{\rho}}\label{Weibullk}\end{equation}
 with varying coefficient $\rho$, and constant average strength. With
$a=\Gamma(1+1/\rho)$ the average strength $\langle x\rangle$ equals
unity, and the width takes the value \begin{equation}
w=\left(\langle x^{2}\rangle-\langle x\rangle^{2}\right)^{\frac{1}{2}}=\left(\Gamma(1+2/\rho)/\Gamma^{2}(1+1/\rho)-1\right)^{\frac{1}{2}}.\end{equation}
 Here $\Gamma$ is the Gamma function. Using the power series expansion
$\Gamma(1+z)=1-0.577z+0.989z^{2}+\ldots$ we see how the width decreases
with increasing $\rho$: \begin{equation}
w\simeq\frac{1.52}{\rho}.\label{width}\end{equation}

For the Weibull distribution (Eq. \ref{Weibullk}) we use Eq. (\ref{generalk})
to calculate the amplitude $\kappa_{-}$, with the result \begin{equation}
\kappa_{-}=(\Gamma(1+1/\rho)/2\rho)^{\frac{1}{2}}(\rho e)^{-1/2\rho}\ln(N)\simeq(2\rho)^{-\frac{1}{2}}\;\ln(N),\label{KWeibullk}\end{equation}
 the last expression for large $\rho$. Comparison between Eq. (\ref{width})
and Eq. (\ref{KWeibullk}) shows that for narrow distributions

\begin{equation}
\kappa_{-}\propto\sqrt{w}.\label{Kwidth}\end{equation}

That narrow distributions give small amplitudes could be expected:
Many fibers with strengths of almost the same magnitude will tend
to break simultaneously, hence the relaxation process goes quicker.
\vskip.2in
\noindent \emph{(c) Universality of critical amplitude ratio}
\vskip.2in
As function of the initial stress $\sigma$ the number of relaxation 
steps, $t_{f}(\sigma)$,
shows a divergence $\left|\sigma-\sigma_{c}\right|^{-1/2}$ at the
critical point, both on the pre-critical and post-critical side. This
is a generic result, valid for a general probability distribution
of the individual fiber strength thresholds. On the post-critical
side $t_{f}(\sigma)$ is independent of the system size $N$ for large
$N$. On the pre-critical side there is, however, a weak (logarithmic)
$N$-dependence, as witnessed by Eqs. ($46$), ($47$) and ($55$).
Note that the critical amplitude ratio takes the same value $\kappa_{-}/\kappa_{+}=\ln(N)/2\pi$
for the uniform and the Weibull distributions. This shows
the universal nature of the  critical amplitude ratio, independent of 
the threshold distribution. Note the difference with normal
critical phenomena \cite{a76}  due to the
appearance
of the $\ln(N)$ in this amplitude ratio here.

\subsubsection{Non-linear stress-strain behavior}

Fiber bundle model captures correctly the  non-linear elastic behavior  
in ELS mode \cite{s89,pbc02}. In case of 
strain controlled loading, using the theory of extreme order statistics, 
it has been shown \cite{s89} that ELS bundles shows non-linear 
stress-strain behavior 
after an initial linear part up to which no fiber fails. Similar non-linear 
behavior is seen in the force controlled loading case as well. Moreover, from 
the recursive failure dynamics, the amount of stress drop at the breaking point
can be calculated exactly \cite{pbc02}. 
To demonstrate the scenario we consider an ELS bundle with  uniform fiber
strength distribution, having a low cutoff $C_L$, such that for stresses 
below the low cutoff, none of the fibers fail. Hence, until failure of
any of the fibers, the bundle shows linear
elastic behavior. As soon as the fibers start to fail, the stress-strain
relationship becomes non-linear. This non-linearity can be easily calculated 
in the ELS model, using Eq. (\ref{rec-U}) for the failure dynamics of the model.

Fibers  are here assumed to be  elastic, each having unit force constant,
with their breaking strengths (thresholds) distributed uniformly
within the interval $[C_{L},1]$ : \begin{equation}
p(x)=\left\{ \begin{array}{cc}
0, & 0\leq x\leq C_{L}\\
\frac{1}{1-C_{L}}, & C_{L}<x\leq1\end{array}\right..\label{nonlinear-dist}\end{equation}
For an applied stress $\sigma\leq C_{L}$ none of the fibers break,
though they are elongated by an amount $\varepsilon = x =\sigma$.
The dynamics of breaking starts when applied stress $\sigma$ becomes
greater than $C_{L}$. For $\sigma>C_{L}$, the basic recursion relation
 (Eq. \ref{rec-U}) takes the form:
 \begin{equation}
U_{t+1}=\frac{1}{1-C_{L}}\left[1-\frac{\sigma}{U_{t}}\right],\label{nonlinear-1}\end{equation}
which has stable fixed points: \begin{equation}
U^{*}(\sigma)=\frac{1}{2(1-C_{L})}\left[1+\left(1-\frac{\sigma}{\sigma_{c}}\right)^{1/2}\right].\label{nonlinear-2}\end{equation}
The model now has a critical point $\sigma_{c}=1/[4(1-C_{L})]$
beyond which the bundle fails completely.
At each fixed point, there
will be an equilibrium elongation $\varepsilon(\sigma)$ and a corresponding
stress $S =U^{*}\varepsilon(\sigma)$ develops in the system (bundle).
From Eq. (\ref{nonlinear-1}), one gets (for $\sigma>C_{L}$)
\begin{equation}
U^{*}(\sigma)=\frac{1-x^{*}}{1-C_{L}};   x^{*} = \frac{\sigma} {U^{*}}.
\label{nonlinear-3}
\end{equation}
Also, from the force balance condition, at each fixed-point 
$\varepsilon(\sigma)=x^{*}$. 
Therefore, the  stress-strain relation for the ELS model finally becomes: \begin{equation}
S=\left\{ \begin{array}{cc}
\varepsilon, & 0\leq\sigma\leq C_{L}\\
\varepsilon(1-\varepsilon)/(1-C_{L}), & C_{L}\leq\sigma\leq\sigma_{c}\\
0, & \sigma>\sigma_{c}\end{array}\right..\label{stress-strain}\end{equation}

\begin{figure}
\includegraphics[width=7cm,height=5cm]{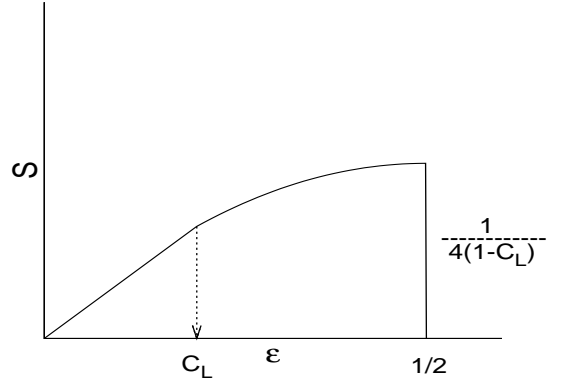}
\caption{ The stress-strain
curve for an ELS bundle having uniform fiber strength distribution with a low 
cutoff $C_L = 0.2$.}
\label{fig:stress-strain}
\end{figure}
The stress-strain relation in a ELS bundle  is shown in 
Fig. \ref{fig:stress-strain}, where
the initial linear region has unit slope (the force constant
of each fiber). This Hooke's region for the  stress $S$ continues up to
the strain value $\varepsilon=C_{L}$, until which no one of the
fibers breaks. After this, nonlinearity appears due to the failure
of a few of the fibers and the consequent decrease of $U^{*}(\sigma)$.
It finally drops to zero discontinuously by an amount $x_{c}^{*}U^{*}(\sigma_{c})=1/[4(1-C_{L})]$
at the breaking point $\sigma=\sigma_{c}$ or $\varepsilon=x_{c}^{*}=1/2$ for the bundle. It may be noted that in this model the internal stress $x_{c}^{*}$ is 
universally equal to $1/2$, independent of $C_{L}$ at the failure point 
$\sigma=\sigma_{c}$.

\subsubsection{Effect of a low cutoff: Instant failure situation}

A low cutoff in the fiber threshold distribution excludes the presence
of very weak fibers in a bundle. The weaker fibers mainly reduces
the strength of a bundle. But in practice we always try to
build stronger and stronger materials (ropes, cables etc.) from the
fibrous elements. Therefore this situation (exclusion of weaker fibers)
is very realistic. In this section we discuss the effect \cite{ph05} of a 
low cutoff on the failure properties of ELS  bundles.  

We follow the weakest fiber breaking approach \cite{d45,hh92}:
The applied load is tuned in such a way that only the weakest fiber
(among the intact fibers) will fail after each step of loading. We
first find the extreme condition when the whole bundle fails instantly
after the first fiber ruptures. As the strength thresholds of $N$
fibers are uniformly distributed between $C_{L}$ and $1$ 
(Eq. \ref{nonlinear-dist}), the
weakest fiber fails at a stress $C_{L}$ (for large $N$). After
this single fiber failure, the load will be redistributed within intact
fibers resulting a global stress $x_{f}=NC_{L}/(N-1)$.
Now, the number of intact fibers having strength threshold below $x_{f}$
is

\begin{equation}
NP(x_{f})=N\int_{C_{L}}^{x_{f}}p(y)dy =\frac{N(x_{f}-C_{L})}{(1-C_{L})}.
\label{Ad-1}
\end{equation}

Stress redistribution can break at least another fiber if $NP(x_{f})\geq1$
and this `second' failure will trigger another failure and so on.
Thus the successive breaking of fibers cannot be stopped till the
complete collapse of the bundle. Clearly, there cannot be any fixed
point (critical point) for such `instant failure' situation. Putting
the value of $x_{f}$ we get

\begin{equation}
\frac{N(\frac{NC_{L}}{N-1}-C_{L})}{(1-C_{L})}\geq1;\label{Ad-2}\end{equation}
which gives \begin{equation}
C_{L}\geq\frac{(N-1)}{(2N-1)}.\label{Ad-3}\end{equation}
In the large $N$ limit the above condition can be written as $C_{L}\geq1/2$.
Therefore, the condition to get a fixed point in the failure process
is $C_{L}<1/2$.

\begin{figure}
\includegraphics[width=6cm,height=5cm]{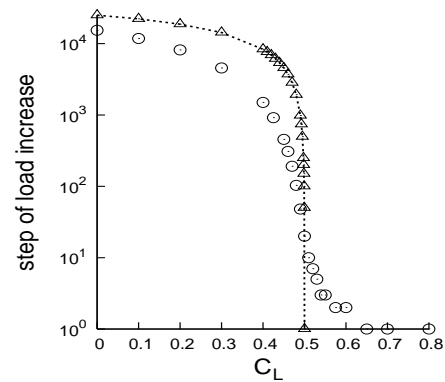}
\caption{ 
The number of steps of load
increase (till final failure) is plotted against $C_{L}$ for
a ELS model having $50000$ fibers. The dotted line represents the
analytic form (eqn. \ref{step no}), triangles are the simulated data
for a strictly uniform strength distribution, and the circles represent
the data (averages are taken for $5000$ samples) for a uniform on
average distribution. 
}
\label{fig:els-step}
\end{figure}

We can also calculate how many steps are required to attain the final
catastrophic failure for $C_{L}<1/2$. Let us assume that we
have to increase the external load $n$ times before the final failure.
At each step of such load increment only one fiber fails. Then after
$n$ step the following condition should be fulfilled to have a catastrophic
failure:

\begin{equation}
N\int_{x_{i}}^{x_{i}[1+1/(N-n)]}p(y)dy\geq1.\label{uni-step-1}\end{equation}
where \begin{equation}
x_{i}=C_{L}+\frac{n(1-C_{L})}{N}\label{uni-step-2}\end{equation}
The solution gives \begin{equation}
n=\frac{N}{2}\left(1-\frac{C_{L}}{1-C_{L}}\right).\label{step no}\end{equation}
The above equation suggests that at $C_{L}=1/2$, $n=0$.
But in reality we have to put the external load once to break the
weakest fiber of the bundle. Therefore, $n=1$ for $C_{L}\geq1/2$
(Fig. \ref{fig:els-step}). To check the validity of the above calculation we take `strictly
uniform' and uniform on average distributions of fiber strength. In
our `strictly uniform' distribution the strength of the $k$-th fiber
(among $N$ fibers) is $C_{L}+(1-C_{L})k/N$. We can see
in Fig. \ref{fig:els-step}  that the `strictly uniform distribution' exactly 
obeys the
analytic formula (Eq. \ref{step no}) but the uniform on average distribution
shows slight disagreement which comes from the fluctuation in the
distribution function for a finite system size. This fluctuation will
disappear in the limit $N\rightarrow\infty$ where we expect perfect
agreement.

\subsection{Fluctuations}

If the contribution to breakdown phenomena in materials science by
statistical physics were to be expressed in one word, that word would
have to be ``fluctuations".  In the context of fiber bundles, this concept
refers to the effects of the fibers each having properties that are
statistically distributed around some mean, which cannot be reproduced
by substituting the fiber bundle by an equivalent one where each fiber
is identical to all the others.

Intuitively, it is not difficult to accept that fluctuations must play
an important role in the breakdown properties of fiber bundles --- or in
fracture and breakdown phenomena in general. A plane ride in turbulent
weather compared to one in smooth weather is a reminder of this.

Closely connected to the question of fluctuations is that of phase
transitions and criticality \cite{s87}.  Leaving the fiber bundles for a
moment, consider a fluid whose temperature is slowly raised.  At a
well-defined temperature determined by the surrounding pressure, the
fluid starts to boil.  Each gas bubble that rise to the surface is due
a fluctuation being larger than a well-defined size for which the bubble grows
rather than shrinks away.  At a particular pressure the character of the
boiling changes character. There is no longer any size that determined whether
a nascent bubble grows or shrinks.  There are bubbles of all sizes.  At
this particular point, the system is critical and undergoes a second
order phase transitions. The boiling process at other pressures signals a
first order transition.

A brittle material under stress develops microcracks.  These appear where
the material is weak or where the local stress field is high.  As the stress
increases, more and more microcracks accumulate until either one or a
few microcracks go unstable and grow to macroscopic dimensions causing
failure.  The spatial fluctuations of the local material properties
cause the appearance of microcracks.  Their subsequent growth accentuate
these initial fluctuations, but in a highly complex manner due to interactions
between the growing cracks.  There are similarities between this scenario
and a first order transition \cite{zrsv97}.  On the other hand, stable
mode I crack growth as studied experimentally by M{\aa}l{\o}y and
Schmittbuhl \cite{ms97,ms01} shows all the signs of the advancing crack
front showing a dynamics compatible with being at a critical point.

We now turn to the global load sharing fiber bundle model in light of the
preceeding remarks.

\subsubsection{Burst distribution for continuous load increase}

When a fiber ruptures somewhere, the stress on the intact 
fibers increases.
This may in turn trigger further fiber failures, which can produce
 bursts (avalanches) that either lead to a stable situation or to breakdown
of the whole bundle. A burst is usually defined as the amount 
or number ($\Delta$) of simultaneous fiber failure during loading.  
One may study 
the  distribution $D(\Delta)$ of the bursts appear during the entire failure 
process until the complete breakdown of the bundle. 

The property of the fiber bundle model of interest in the present context,
is the fluctuation driven {\it burst distribution.\/}  In order to define this 
property, we 
again consider a finite bundle containing $N$ elastic fibers whose strength 
thresholds are picked randomly from a probability density $p(x)$. Let $x_k$ be 
the ordered sequence of failure thresholds: $x_1 \leq x_2 \leq ...\leq x_N$. 
Then the the external load or force $F$ on the bundle  (Eq. \ref{load})  at 
the point where $k$th fiber is about to fail can be written as:
\begin{equation}
F_k = (N+1-k) x_k,
\label{load-HH}   
\end{equation}   
where elastic constant of the fibers is set equal to unity as before.
Note that the sequence of external loads $F_k$ is not monotonously increasing.
This may be readily seen from Eq.\ (\ref{load-HH});
the total load is the product
of a monotonously increasing {\it fluctuating\/} quantity $x_k$ and a
monotonously {\it decreasing\/} quantity $(N+1-k)$.  Suppose now that our
control parameter is the total load $F$,  and that $k-1$ fibers have broken.
In order to be in this situation, $F>F_k > F_j$ for all $j < k$.  The latter
inequality ensures that the situation we are studying is not unstable.
We increase $F$ until it reaches $F_k$, at which fiber $k$ breaks.  If now
$F_{k+1} \le F_k$, then fiber $k+1$ will also break without the external
load $F$ being further increased.  The same may be true for $F_{k+2}$ and
so on until the $(k+\Delta-1)$th bond breaks. Thus, $F_{k+j} \le F_k$ for
$j < \Delta$.  If now $F_{k+\Delta}>F_k$, the burst of breaking bonds
then stops at this point, and we have experienced a burst event of size
$\Delta$.

\begin{figure}
\centering
\includegraphics[width=7cm]{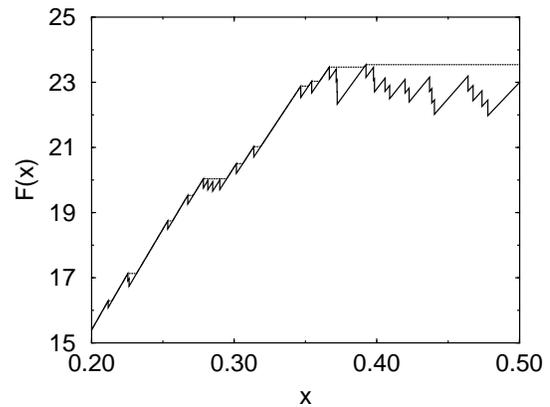}
\caption{The solid curve indicates the total force $F(x)$ as a
function of $x$.  However, when
our control parameter is $F$ rather than $x$, the system will follow the
dotted line, Eq.~(\ref{lfm}).  The bursts
are the horizontal parts of $F_{ph}(x)$. Here $N=100$.}
\label{khh_rmp}
\end{figure}

The total force $F$ expressed as a function of elongation $x$, is shown
in Fig.~\ref{khh_rmp}.  When the control parameter is elongation $x$,
the solid curve is followed.  However, when the force $F$ is the control
parameter, the broken line, given by
\begin{equation}
\label{lfm}
F_{ph}={\rm LMF}\ F(x)\;,
\end{equation}
where LFM designates the {\it least monotonic function.\/}

\vspace{.1in}
\noindent\emph{(a) Generic case}
\vspace{.1in}

\begin{figure}
\centering
\includegraphics[width=6cm]{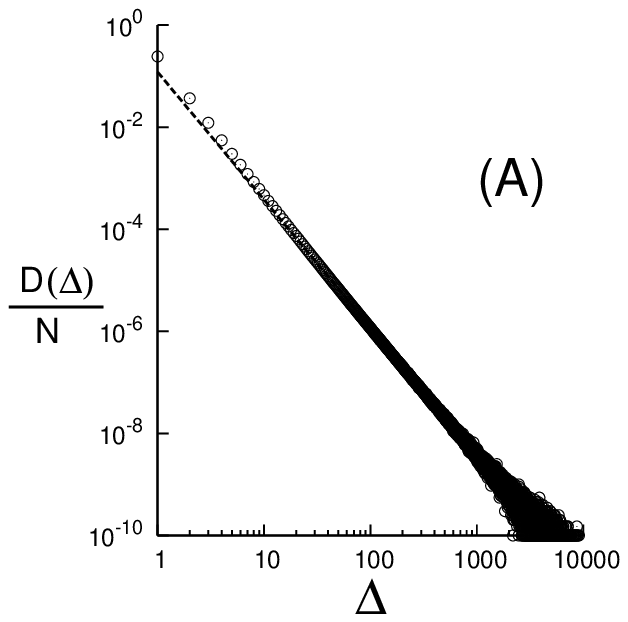}

\includegraphics[width=6cm]{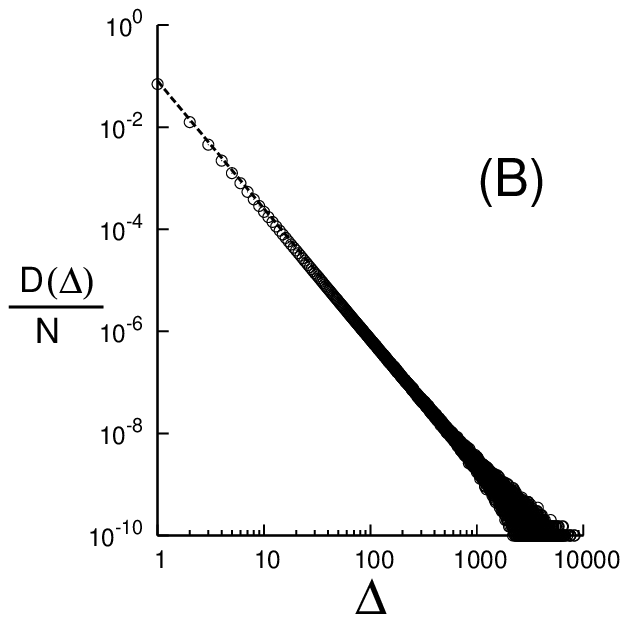}
\caption{The burst distribution $D(\Delta)/N$
for the uniform distribution (A) and the Weibull distribution with
index 5 (B). The dotted lines represent the power law with exponent
$\xi=5/2$. Both figures are based on $20000$ samples of bundles each
with $N=10^{6}$ fibers.}
\label{ah-fig1}
\end{figure}

It was shown in  \textcite{hh92}
that the average number of burst events
of size $\Delta$ per fiber, $D(\Delta)/N$, follows a power law of the form
\begin{equation}
\label{eq9}
D(\Delta)/N = C \Delta^{-\xi}
\end{equation}
in the limit $N \to \infty$.  Here,
\begin{equation}
\label{eq10}
\textstyle \xi={5\over 2}
\end{equation}
is the {\it universal\/} burst exponent.  The value (Eq. \ref{eq10})
is, under very mild assumptions, independent of the threshold distribution
$P(x)$: the probability density needs to have a quadratic maximum somewhere in
the interval $x_{\min} < x < x_{\max}$.
We demonstrate this in Fig.~\ref{ah-fig1}.
The prefactor $C$ in Eq. (\ref{eq9}) is given by
\begin{equation}
\label{eq11}
C={{x_c p(x_c)^2}\over {\sqrt{2 \pi}[x_c p'(x_c)+2p(x_c)]}}\;,
\end{equation}
where $x_c$ is the solution of the equation
\begin{equation}
\label{eq12}
x_c p(x_c)=1-P(x_c)\;,
\end{equation}
and is the value of $x$ for which the characteristics has a maximum.
Eqs.\ (\ref{eq9}) to (\ref{eq12}) were derived in 
\textcite{hh92} using combinatorial arguments.  We will, however, in the
following, take an alternative route based on a mapping between
the global load sharing model and a Brownian process
 \cite{s92,hh94}. Before we explain this mapping we quote, for later 
comparison, the pertinent results of the Hemmer-Hansen analysis \cite{hh92}:
The probability  $\Phi(\Delta,x)$ that a burst
event at elongation $x$ will have the size $\Delta$ is
\begin{equation}
\label{eq13}
\Phi(\Delta,x) = {{\Delta^{\Delta-1}}\over{\Delta !}}\
{{m(x)}\over{1-m(x)}} \left[[1-m(x)] e^{m(x)-1}\right]^\Delta \;,
\end{equation}
where
\begin{equation}
m(x)=1-{{xp(x)}\over{1-P(x)}}\;.
\label{eq14}
\end{equation}
Note in particular that by Eq. (\ref{eq12}) $m(x_c)=0$.
Let us now assume that we do not load the fiber bundle until complete
collapse, i.e., until  $x = x_c$, but stop at a value
$x_s < x_c$.  We may then ask for
$D(\Delta,x_s)/N$, the expected number of burst events of size $\Delta$
during the breakdown process that occurs between $x=0$ and $x=x_s$.
This is given by the integral
\begin{eqnarray}
\label{eq15}
&{{D(\Delta,x_s)}\over N} = \int_0^{x_s} p(x) dx\ \Phi(\Delta,x)=\nonumber\\
&{{\Delta^{-3/2}}\over
{\sqrt{2\pi}}}\ \int_0^{x_s} dx\ p(x) {{m(x)}\over{1-m(x)}}
\left[ [1-m(x)]e^{m(x)}\right]^\Delta\;,\\
\nonumber
\end{eqnarray}
where  on the right-hand side the Stirling approximation $\Delta !
\approx \sqrt{2\pi} \Delta^{\Delta+1/2} e^{-\Delta}$
for large $\Delta$ has been used.
The integrand in Eq.(\ref{eq15}) is strongly peaked near $x=x_c$.  We therefore
expand it to second  order in $y=x_c-x$ to find
\begin{eqnarray}
                                            & {{D(\Delta,x_s)}\over N} =
{{\Delta^{-3/2}}\over {\sqrt{2\pi}}}\ p(x_c)m'(x_c)\nonumber\\
& \int_{x_c-x_s}^{\infty}
dy\ y e^{-m'(x_c)^2 y^2\Delta /2}\;,\nonumber\\
\label{eq16}
\end{eqnarray}
where we have extended the upper integration limit
to $\infty$.  We may do this integral to get
\begin{equation}
\label{eq17}
{{D(\Delta,x_s)}\over N}=C\Delta^{-5/2}\ e^{-m'(x_c)^2 \Delta
(x_c-x_s)^2/2}\;,
\end{equation}
where $C$ is defined by Eq.~(\ref{eq11}).

We may write (\ref{eq17}) in scaling form,
\begin{equation}
\label{eq18}
{{D(\Delta,x_s)}\over N}=\Delta^{-\xi} G(\Delta, x_s) = \Delta^{-\xi}
G\left(\Delta^\mu (x_c-x_s)\right)\;,
\end{equation}
where
\begin{equation}
\label{eq19}
G(y)=C\ e^{-m'(x_c)^2y^2/2}\;.
\end{equation}
In particular $G(y)$ tends to the constant $C$ for $y \to 0$.
Two universal critical exponents appear, $\xi=5/2$,
Eq.~(\ref{eq10}), and
\begin{equation}
\label{eq20}
\mu=\textstyle {1\over 2}\;.
\end{equation}

It is, thus, in the above sense, that the fracture process of the fiber
bundle approaches a critical point at total breakdown:  The
distribution of burst events follows a power law with an upper cutoff that
diverges as the bundle approaches total failure.

In \textcite{s92} and later on in full detail in \textcite{hh94}, the burst
distribution (Eq. \ref{eq9}) was
derived from the assumption that $F_k$ may be directly interpreted as a biased
random walk.  The precise nature of this random walk is elucidated below.
It is a peculiar asymmetric walk with variable step length. In the limit
$N\to\infty$ and continuous time variable $k/N\to t$ and
$\Delta k/N\to \delta t >0$, this random walk may be mapped onto a continuous
Brownian process. Such Brownian processes have been studied by \textcite{pt73},
 \textcite{ds85} and  \textcite{d89}
in connection with the distribution of the strength $S$ of
fiber bundles. We will in the following derive Eq.~(\ref{eq15})
by means of a biased random-walk model with variable step length.
We find this an interesting example of
universality in statistical physics: The asymptotic behavior of one
model is found by using a different model with the same asymptotic
behavior as the first one, but which is simpler to solve before the
continuum limit is taken.

Under increasing load the variation of the force per fiber, $f=F/N$,
will consist of a systematic
nonfluctuating part, given by the average load-elongation
characteristics, with a small fluctuation of
order $1/\sqrt{N}$ superimposed.

The precise value of the force fluctuation depends upon whether one studies
the force $f(x)$ at given elongation, or the force $f_k$
at which fiber number $k$ breaks.  Let us for both quantities calculate the
variance of $f$, $\sigma^2_f$, starting with the constant-$k$ ensemble.

The force per fiber when the $k$th fiber is
about to break is, by $F_k=(N+1-k) x_k$,
\begin{equation}
\label{eq21}
f_k = [1 - P(\overline{x}_k)] x_k \; ,
\end{equation}
where $x_k$ is the elongation when the $k$th fiber breaks, and we have
defined $\overline{x}_k$ by
\begin{equation}
\label{eq22}
P(\overline{x}_k) = {k \over {N+1}}.
\end{equation}
For large $N$ $\overline{x}_k$ is essentially the average value of $x_k$.
For a fixed $k$ the variance of $f_k$ is by Eq. (\ref{eq21}) given by the
variance of $x_k$:
\begin{equation}
\label{23}
\sigma_f^2(k) = [1-P(x_k)]^2 \sigma_{x_k}^2 \; ,
\end{equation}
and we seek therefore the probability $\varphi(x) dx$ that the $k$th
threshold in the ordered threshold
sequence lies in the interval $(x,x+dx)$. This probability equals
\begin{eqnarray}
&\varphi(x)\ dx=\nonumber\\
& {{N!}\over{(k-1)!(N-k)!}}P(x)^{k-1} [1-P(x)]^{N-k} p(x) dx\;.\nonumber\\
\label{eq24}
\end{eqnarray}
For large $k$ and $N$, and using Eq.~(\ref{eq22}), this is close to the
Gaussian distribution
\begin{eqnarray}
\label{eq25}
&\varphi(x)\ dx=
\left({{Np( \overline{x}_k )^2}
\over{2\pi P(\overline{x}_k )[1-P(\overline{x}_k )]}}\right)^{1/2}\nonumber\\
&e^{-Np(\overline{x}_k)^2(x-\overline{x}_k)^2
/2P(\overline{x}_k )[1-P(\overline{x}_k )]}\ dx\;.\\
\nonumber
\end{eqnarray}
This gives the variance of $x_k$ and thus of $f_k$,
\begin{equation}
\label{eq26}
\sigma_f^2(k) = {{P(\overline{x}_k)
[1-P(\overline{x}_k)]^3}\over {Np(\overline{x}_k)^2}} .
\end{equation}

Let us now compare with the force fluctuation at constant elongation.
The force per fiber is the following function of elongation $x$,
\begin{equation}
\label{eq27}
f(x)=N^{-1}\sum_{i=1}^Nx\Theta(t_i-x)\;,
\end{equation}
where $t_i$ is the breakdown threshold for
the $i$th fiber, and $\Theta(t)$ is
the Heavyside function.  This gives
immediately the average force
\begin{equation}
\label{eq28}
\langle f\rangle_x = x[1-P(x)]\;,
\end{equation}
i.e., the characteristics,  as well as the variance
\begin{equation}
\label{eq29}
\sigma^2_f(x)={{x^2P(x)[1-P(x)]}\over N}\;.
\end{equation}

Although the two types of force fluctuations have different variances,
in both cases $\sigma \propto 1/\sqrt{N}$.

The nonmonoticities of the force $f$ within the fluctuation zone produce
bursts. Since the fluctuations are so small for large $N$, one can treat
the burst events {\it locally.\/}

We now consider the force sequence $F_k$ as a stochastic process.
Since we seek the {\it asymptotic\/} burst distribution,
we are interested in the behavior after many steps of the process.
It is convenient, however, to start with the one-step process.

Let us determine the probability distribution of the force increase
$\Delta F = F_{k+1}-F_k$ between two consecutive bursts, the first
one taking place at elongation $x_k$ with $F_k=(N-k+1)x_k$.

Since $\Delta F = (N-k)(x_{k+1}-x_k)-x_k$, we have
\begin{equation}
\label{eq30}
\Delta F \ge - x_k\;.
\end{equation}
The probability to find the $k+1$'th threshold in $(x_{k+1},x_{k+1}+dx_{k+1})$
for given $x_k$,
\begin{equation}
(N-k-1) {{[1-P(x_{k+1})]^{N-k-2}}\over{[1-P(x_k)]}} p(x_{k+1}) dx_{k+1}\;,
\label{eq31}
\end{equation}
gives directly by using the connection
$x_{k+1} = x_k +(\Delta F(x_k)+x_k)/(N-k)$,
the probability density $\rho (\Delta F;x_k)$ of $\Delta F$
\begin{eqnarray}
\label{eq32}
&\rho (\Delta F;x_k) =\nonumber\\
&{{N-k-1}\over{N-k}}
{{[1-P(x_k+{{\Delta F+x_k}\over{N-k}})]^{N-k-2}}\over{[1-P(x_k)]^{N-k-1}}}\cr
& p(x_k+{{\Delta F+x_k}\over{N-k}})\;.\\
\nonumber
\end{eqnarray}
For large $N-k$ this simplifies to
\begin{eqnarray}
\label{eq33}
&\rho(\Delta F;x_k)\nonumber\\
&=\cases{{p(x_k)}\over{1-P(x_k)} \exp\left[-{{(\Delta F+x_k)p(x_k[)}\over
{1-P(x_k)]}}\right]
            & for $\Delta F\ge -x_k$,\cr
0           & for $\Delta F<   -x_k$.\cr}\nonumber\\
\end{eqnarray}
This one-dimensional random walk
is asymmetric in more than one way. First of all, it has nonzero bias
\begin{equation}
\langle \Delta F \rangle (x_k) = {{1-P(x_k)-x_kp(x_k)}\over{p(x_k)}}\;.
\label{eq34}
\end{equation}
In addition the probability distribution around this average is very
asymmetric.

The variance is easily determined,
\begin{equation}
\label{eq35}
\sigma_{\Delta F}^2 (x_k) = \left[{{1- P(x_k)}\over{p(x_k)}}\right]^2 \;.
\end{equation}

The Brownian-motion limit of a one-dimensional random walk is
completely determined by the first and second moments
of the single-step probability distribution. The results just obtained
enables us therefore to select an  ``ordinary''
biased random walk with constant step length $a$
which has the same Brownian motion limit as the burst process.

We imagine having a one-dimensional
random walk along the $z$-axis with a constant bias.
Each step is of length $a$. Let the probability to take a step in the
negative $z$ direction be $q$, and let $p$ be the probability to take a
step in the positive $z$ direction.
The walk is biased when $p$ is different from
$q$. The probability distribution of the position $z_1$ after one step
has the average
\begin{equation}
\label{eq36}
\langle z_1 \rangle = a(p-q)\;,
\end{equation}
and variance
\begin{equation}
\label{eq37}
\sigma_1^2= 4pq a^2 \; .
\end{equation}
Elimination of $a$ yields
\begin{equation}
\label{eq38}
{{p-q}\over{2\sqrt{pq}}} = {{\langle z_1 \rangle}\over{\sigma_1}}\;.
\end{equation}
Since $p+q=1$, the bias parameters are determined.

After $k$ steps a Gaussian distribution,
\begin{equation}
\label{eq39}
{e^{-[z_k-z_0-ka(p-q)]^2/8pqa^2 k}\over{\sqrt{8\pi pqa^2k}}}
\end{equation}
is approached when $k$ increases.

The two processes will have the same asymptotic behavior when we make the
identification
\begin{equation}
\label{eq40}
{{p-q}\over{2\sqrt{pq}}} = {{\langle \Delta F \rangle}\over{\sigma_F}} =
1- {{x_k p(x_k)}\over{1-P(x_k)}} = m(x_k)\;,
\end{equation}
where $m(x)$ is defined by Eq.~(\ref{eq14}). When the bias is small,
both $p$ and $q$ are close to 1/2, and we have to lowest order
\begin{eqnarray}
\label{eq41}
p=\textstyle {1\over 2}[1+m( x )]\;,\nonumber\\
q=\textstyle {1\over 2}[1-m( x )]\;.\\
\nonumber
\end{eqnarray}
We have now made the promised mapping between
the fiber bundle problem and a random walk with a constant bias.
A constant bias may be used since bursts can be treated locally.

The next step is to calculate the burst distribution for such
a biased random walk.
Since this biased random walk by construction has the same asymptotic
behavior in the limit $N\to \infty$ as the original fiber bundle problem,
the two burst distributions will asymptotically be the same.

In terms of the biased random walk, a burst event of size $\Delta$ at
``time" $k$ may be defined as follows:  (i) $z_{k+i} < z_k$ for $0<i<\Delta$
and $z_{k+\Delta} \ge z_k$.  (ii) Furthermore, to ensure that we are not
counting burst event {\it inside\/} other burst events, the condition
$z_k>z_j$ for $k>j$ is necessary.

The first condition is in fact a special case of the ``Gambler's ruin"
problem \cite{f66}.  A gambler plays a series of independent games against
a bank with infinite resources.  In each game, the gambler either looses or
wins one Euro, and the probability that the bank wins is $p=(1+B)/2$, while
the probability that the gambler wins is $q=(1-B)/2$.  If the gambler
starts out with a capital of $z$ Rs, the probability that she is ruined
after precisely $\Delta$ games is
\begin{equation}
\label{eq42}
\pi(z,\Delta)=
{z\over\Delta} {\Delta \choose {{\Delta\over 2}-{z\over 2}}} p^{(\Delta-z)/2}
q^{(\Delta+z)/2}\;.
\end{equation}
The probability that condition i is fulfilled for a
biased random walk burst of size $\Delta$ is then
\begin{equation}
\label{eq43}
{\textstyle {1\over 2}}\pi(z=1,\Delta)=
{{\Delta^{-3/2}}\over{\sqrt{2\pi}}}\
\sqrt{{{1-B}\over{1+B}}}\ (1-B^2)^{\Delta/2}\;,
\end{equation}
where we have assumed that $\Delta >> 1$.

The probability that a biased random walker returns at least once to the
origin is \cite{f66} $1-|p-q|=1-B$.  The probability that condition (ii),
namely that $z_j>z_k$ for all $j<k$ is fulfilled, is then simply $1-(1-B)
=B$, and we have that the probability for having a burst of size $\Delta$
happening at ``time" $k$ is
\begin{equation}
\label{eq44}
\Phi_{RW}(\Delta,B)=
{\textstyle {1\over 2}} B \pi(z=1,\Delta)
={{\Delta^{-3/2}}\over{\sqrt{2\pi}}}\ B\ e^{-B^2\Delta/2}\;,
\end{equation}
where we in addition have assumed that $B<<1$.

Returning to the fiber bundle model, the bias $B=m(x )$.
When $x$ is close to $x_c$, we have $B=m'(x_c)y$
where $y=(x_c- x )$.  Thus, the probability to have a burst of
size $\Delta$ between $y$ and $y+dy$ is
\begin{eqnarray}
\label{eq45}
&\Phi_{RW}(\Delta,m'(x_c)y)p(x_c) dy\nonumber\\ & =
{{\Delta^{-3/2}}\over{\sqrt{2\pi}}}\ p(x_c)\ m'(x_c)\
e^{-m'(x_c)y^2\Delta/2}\;.\nonumber\\
\end{eqnarray}
Thus, the cumulative burst distribution up the elongation $x_s$ is
\begin{equation}
\label{eq46}
\int_{x_c- x_s}^\infty \Phi_{RW}(\Delta,m'(x_c)y)p(x_c) dy\;.
\end{equation}

Comparing this expression to Eq.~(\ref{eq16}), we see that they are
identical. This completes the derivation of
the asymptotic burst distribution via the mapping
between the random-walk problem and the burst process.

\vspace{.1in}
\noindent\emph{(b) Special cases}
\vspace{.1in}

The burst distribution given in Eq.~(\ref{eq15}) is valid when the
threshold distribution has a parabolic maximum inside the interval
of the thresholds.  We now consider threshold distributions that do
{\it not\/} reach their maximum at the boundaries of the interval
\cite{khh97}.
Model examples of such threshold distributions are
\begin{equation}
P(x)=\left\{ \begin{array}{ll}
0 & \mbox{for }\; x \leq x_0 \\
1-[1+(x-x_0)/x_r]^{-\alpha_0} & \mbox{for }\; x > x_0
\end{array} \right.
\label{brede}
\end{equation}
Here $\alpha_0$ and $x_0$ are positive parameters, and $x_r$ is a reference
quantity which we for simplicity put equal to unity in the following.
These distributions are all characterized by diverging moments.  When
$\alpha_0 \le 1$, even the first moment --- the mean ---
as well as all other moments diverge.  This class of threshold distributions
are rich enough to exhibit several qualitatively different burst
distributions.

\begin{figure}
\centering
\includegraphics[width=6cm]{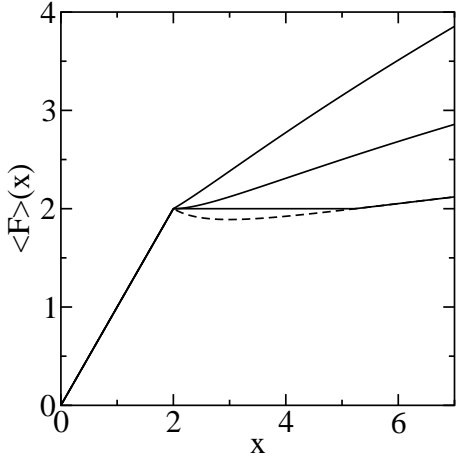}
\caption{The macroscopic bundle strength $\langle F\rangle(x)$ for the 
distribution (Eq. \ref{brede}), with $x_0=2x_r$, and for
$\alpha_0 = \frac{1}{3}$ (upper curve), $\frac{1}{2}$ (middle curve), and
$\frac{2}{3}$ (lower curve).  The broken part of the $\alpha_0=2/3$-curve
is unstable and the macroscopic bundle strength will follow the solid
line.}
\label{fig-3b-6}
\end{figure}

The corresponding macroscopic bundle strength per fiber is
\begin{equation}
\frac{\langle F\rangle(x)}{N} = \left\{\begin{array}{ll}
x & \mbox{for}\;\;\;x\leq x_0  \\
\frac{x}{(1+x-x_0)^{\alpha_0}} &\mbox{for} \;\;\; x>x_0
\end{array} \right.
\label{macro}
\end{equation}
In Fig.~\ref{fig-3b-6} the corresponding macroscopic force curves
$\langle F\rangle (x)$ are sketched.  We note that when $\alpha_0\to 1$,
the plateau in Eq.\ (\ref{macro}) becomes infinitely wide.

The distribution of burst sizes is given by Eq.~(\ref{eq15}).
In the present case the function $m(x)$ takes the form
\begin{equation}
m(x) = \frac{xp(x)}{1-P(x)} = \frac{\alpha_0 x}{1+x-x_0}.
\label{af}
\end{equation}
A simple special case is $x_0=1$, corresponding to
\[ p(x) =
\alpha_0 \; x^{-\alpha_0 -1} \hspace{1cm} \mbox{for} \;\; x \geq 1,\]
 since then the function (\ref{af}) is independent of $f$:
\[ m(x) = \alpha_0.\]
This gives at once
\begin{eqnarray}
\frac{D(\Delta)}{N} & = &  \frac{1-\alpha_0}{\alpha_0} \;
\frac{\Delta^{\Delta - 1}}{\Delta !} \left[\alpha_0 e^{-\alpha_0}\right]^{\Delta}\nonumber\\
 & \simeq & \frac{1-\alpha_0}{\alpha_0\sqrt{2\pi}}\;\Delta^{-\frac{3}{2}}
\left[\alpha_0 e^{1-\alpha_0}\right]^{\Delta}.
\label{one}
\end{eqnarray}

In other cases it is advantageous to change
integration variable in Eq.~(\ref{eq15})
from $x$ to $m$:
\begin{eqnarray}
&\frac{D(\Delta)}{N}
 = \frac{\Delta^{\Delta -1}}{e^{\Delta}\Delta !}\;
\frac{1}{\alpha_0^{\alpha_0 -1}(1-x_0)^{\alpha_0}}\nonumber\\
& \int\limits_{\alpha_0 x_0}^{\alpha_0}
(\alpha_0 - m)^{\alpha_0 -1}(1-m)m^{-1}\left(me^{1-m}\right)^{\Delta}\;dm.
\nonumber\\
\label{integral}
\end{eqnarray}
The asymptotics for large $\Delta$,
beyond the  $\Delta^{-\frac{3}{2}}$
dependence of the prefactor, is determined by the $\Delta$-dependent
factor in the integrand.  The maximum of $me^{1-m}$ is unity, obtained for
$m = 1$, and the
asymptotics depends crucially on whether $m=1$ falls outside the range
of integration, or inside
(including the border). If the maximum falls
inside the range of integration the
$D(\Delta) \propto \Delta^{-\frac{5}{2}}$ dependence
remains. A special case of this is $\alpha_0 = 1$, for which  the
maximum of the integrand is located at the integration limit and the
macroscopic force has a ``quadratic'' maximum {\it at infinity.\/}
Another special case is  $\alpha_0 x_0=1$ (and $\alpha_0 <1$), for which
again the standard asymptotics $\Delta^{-\frac{5}{2}}$ is valid. In this
instance the macroscopic force has a quadratic {\it minimum\/} at
$x=x_0$ (see Fig.~\ref{fig-3b-6}
for $\alpha_0=1/2$), and  critical behavior arises
just as well from a minimum as from a maximum.

In the remaining cases, in which $m=1$ is not within the range of
integration in Eq.~(\ref{integral}), the burst distribution is always a
power law with an exponential cut-off,
\begin{equation}
\label{eq-3b-100}
\frac{D(\Delta)}{N} \simeq \Delta^{-\xi} \; A^{\Delta}\;.
\end{equation}
Here $\xi$ and $A$ depend
on the parameter values $x_0$ and $\alpha_0$, however.
This is easy to understand. Since
\begin{equation}
\frac{dm(x)}{dx} = \frac{\alpha_0 (1-x_0)}{(1+x-x_0)^2},
\end{equation}
we see that $m(x)$ is a monotonically
decreasing function for $x_0>1$, so that
the maximum of $me^{1-m}$ is obtained at the lower limit $x=x_0$, where
$m=\alpha_0 x_0$. The asymptotics
\begin{equation}
D(\Delta) \propto \Delta^{-\frac{5}{2}} \left(\alpha_0 x_0e^{1-\alpha_0 x_0}
\right)^{\Delta}
\label{two}
\end{equation}
follows.

This is true merely for $\alpha_0 x_0 < 1$, however. For $\alpha_0 x_0 > 1$
the macroscopic force $\langle F\rangle(x)$ {\it decreases\/}
near $x=x_0$ so that a macroscopic burst takes place at a force
$x_0$ per fiber, and stabilization
is obtained
at a larger elongation $x_1$ (Fig.~\ref{fig-3b-6}).
The subsequent bursts have an asymptotics
\begin{equation}
D(\Delta) \propto \Delta^{-\frac{5}{2}} \left(a(f_1) e^{1-m(x_1)}
\right)^{\Delta},
\end{equation}
determined by the neighborhood of $x=x_1$.
For $t_0<1$, the maximum of $me^{1-m}$ is obtained at $x=\infty$, leading
to the asymptotics
\begin{equation}
D(\Delta) \propto \Delta^{-\frac{3}{2}-\alpha_0} \left(\alpha_0 e^{1-\alpha_0}
\right)^{\Delta},
\end{equation}
reflecting the power-law behavior of the integrand at infinity.

\begin{table}
\label{ah-tab1}
\begin{center}
\begin{tabular}{cc}
\hline
Parameters & Asymptotics \\
\hline
$0 \leq x_0<1, \alpha_0 < 1$ &
$\Delta^{-\frac{3}{2}-\alpha_0} (\alpha e^{1-\alpha_0})^{\Delta}$ \\
$0\leq x_0 < 1, \alpha_0 =1$ & $\Delta^{-\frac{5}{2}}$ \\
$x_0 = 1, \alpha_0<1$ & $\Delta^{-\frac{3}{2}}(\alpha_0 e^{1-\alpha_0})^{\Delta}$\\
$1<x_0<\alpha_0 ^{-1}$
& $\Delta^{-\frac{5}{2}}(\alpha_0 x_0 e^{1-\alpha_0 x_0})^{\Delta}$ \\
$1 < x_0 =\alpha_0 ^{-1}$ & $\Delta^{-\frac{5}{2}}$ \\
$1<\alpha_0 ^{-1}<x_0$ & $\Delta^{-\frac{5}{2}} e^{-\Delta /\Delta_0}$ \\
\hline
\end{tabular}
\end{center}
\caption{\label{tab1} Asymptotic behavior of the burst distribution for
strong threshold distributions in the ELS model.}
\end{table}

The results are summarized in Table \ref{tab1}.
Note that the $x_0=1$ result (\ref{one})  cannot be obtained by
putting $x_0=1$ in Eq.~(\ref{two})
since in (\ref{integral}) the order of the limits
$\Delta \rightarrow \infty$ and $x_0 \rightarrow 1$ is crucial.

\vspace{.1in}
\noindent\emph{(c) Crossover behavior}
\vspace{.1in}

\begin{figure}
\centering
\includegraphics[width=6cm]{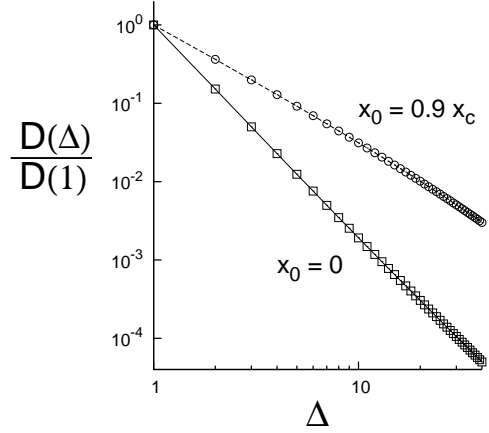}
\caption{The distribution of bursts for thresholds uniformly distributed
in an interval $(x_{0},x_{c})$, with $x_{0}=0$ and with 
$x_{0}=0.9x_{c}$.
The figure is based on 50~000 samples, each containing $N=10^{6}$ fibers. 
The exponents of the distributions are $\xi=5/2$ and $\xi=3/2$ respectively.}
\label{fig-3b-2}
\end{figure}

When all the bursts are recorded for the entire failure process, we
have seen that the burst distribution $D(\Delta)$ follows the asymptotic
power law $D\propto\Delta^{-5/2}$. If we just sample bursts that
occur near the breakdown point, a different behavior is seen.
As an illustration we consider the uniform threshold distribution, and
compare the complete burst distribution with what one gets when one
samples merely burst from breaking fibers in the threshold interval
$(0.9x_{c},x_{c})$. Fig.~\ref{fig-3b-2} shows clearly that in the latter case
a different power law is seen.

This observation may be of practical importance, as it gives a
criterion for the imminence of catastrophic failure \cite{phh05}. 
This proposal has so far not been tested experimentally.  However,  it 
is enticing to note the recent observation by Kawamura of a crossover
behavior in the magnitude distribution of earthquakes before large
earthquakes appear \cite{k06}. We return to this result in
Section V.C and Fig.~ \ref{fig:kawamura}.


We introduce the following notation in Eq.~(\ref{eq15}),
\begin{eqnarray}
\label{eq-3b-1}
&\frac{D(\Delta)}{N}=
\frac{\Delta^{\Delta-1}e^{-\Delta}}{\Delta!}\nonumber\\
&\int_{0}^{x_{c}}p(x)r(x)[1-r(x)]^{\Delta-1}
\exp\left[\Delta\, r(x)\right]dx\;,\nonumber\\
\end{eqnarray}
where
\begin{equation}
r(x)=1-\frac{x\, p(x)}{Q(x)}=
\frac{1}{Q(x)}\;\frac{d}{dx}\left[x\, Q(x)\right]\;,
\label{eq-3b-8}
\end{equation}
and $Q(x)=\int_x^\infty p(x)\ dx$.
We note that $r(x)$ vanishes at the point $x_{c}$.
If we have a situation in which the weakest fiber has its threshold
$x_{0}$ just a little below the critical value $x_{c}$, the contribution
to the integral in the expression (Eq. \ref{eq-3b-1}) for the burst distribution
will come from a small neighborhood of $x_{c}$. Since $r(x)$ vanishes
at $x_{c}${\small ,} it is small here, and we may in this narrow
interval approximate the $\Delta$-dependent factors in Eq. (\ref{eq-3b-1})
as follows
\begin{eqnarray}
&(1-r)^{\Delta}\, e^{\Delta\, r}  
= \exp\left[\Delta(\ln(1-r)+r)\right]\nonumber \\
 & =  \exp[-\Delta(r^{2}/2+{\mathcal{O}}(r^{3}))]
\approx\exp\left[-\Delta r(x)^{2}/2\right]\nonumber\\
\label{eq-3b-6}
\end{eqnarray}
We also have
\begin{equation}
\label{eq-3b-9}
r(x)\approx r'(x_{c})(x-x_{c})\;.
\end{equation}

Inserting everything into Eq.~(\ref{eq-3b-1}), we obtain to
dominating order
\begin{eqnarray}
\frac{D(\Delta)}{N} & 
= & \frac{\Delta^{\Delta-1}\, 
e^{-\Delta}}{\Delta!}\int_{x_{0}}^{x_{c}}p(x_{c})\; r'(x_{c})(x-x_{c})\nonumber\\
& \times & e^{-\Delta\, r'(x_{c})^{2}(x-x_{c})^{2}/2}\; dx\nonumber \\
 & = & \frac{\Delta^{\Delta-2}\, 
e^{-\Delta}p(x_{c})}{\left|r'(x_{c})\right|\Delta!}
\left[e^{-\Delta\, 
r'(x_{c})^{2}(x-x_{c})^{2}/2}\right]_{x_{0}}^{x_{c}}\nonumber \\
& = & \frac{\Delta^{\Delta-2}\, e^{-\Delta}}{\Delta!}
\frac{p(x_{c})}{\left|r'(x_{c})\right|}
\left[1-e^{-\Delta/\Delta_{c}}\right]\;,
\label{eq-3b-15}
\end{eqnarray}
with
\begin{equation}
\Delta_{c}=\frac{2}{r'(x_{c})^{2}(x_{c}-x_{0})^{2}}\;.
\label{eq-3b-16}
\end{equation}

By use of the Stirling approximation
$\Delta!\simeq\Delta^{\Delta}e^{-\Delta}\sqrt{2\pi\Delta}$,
the burst distribution (Eq. \ref{eq-3b-15}) may be written as
\begin{equation}
\frac{D(\Delta)}{N}=C\Delta^{-5/2}\left(1-e^{-\Delta/\Delta_{c}}\right)\;,
\label{eq-3b-12}
\end{equation}
with a nonzero constant
\begin{equation}
\label{eq-3b-13}
C=(2\pi)^{-1/2}p(x_{c})/\left|r'(x_{c})\right|\;.
\end{equation}
We can see from Eq. (\ref{eq-3b-12}) that there is a crossover at a burst
length around $\Delta_{c}$,
\begin{equation}
\label{eq-3b-14}
\frac{D(\Delta)}{N}\propto\left\{ \begin{array}{cl}
\Delta^{-3/2} & \mbox{ for }\Delta\ll\Delta_{c}\\
\Delta^{-5/2} & \mbox{ for }\Delta\gg\Delta_{c}\end{array}\right.\;.
\end{equation}

\begin{figure}
\centering
\includegraphics[width=7cm]{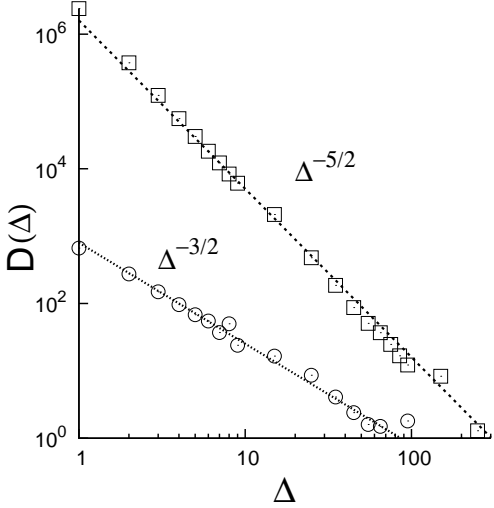}
\caption{The distribution of bursts
for the uniform threshold distribution for a single fiber bundle with
$N=10^{7}$ fibers. Results with $x_{0}=0$, i.e., when all bursts
are recorded, are shown as squares and data for bursts near the
critical point ($x_{0}=0.9x_{c}$) are shown by circles.}
\label{fig3b-3}
\end{figure}

We have thus shown the existence of a crossover from the generic asymptotic
behavior $D\propto\Delta^{-5/2}$ to the power law $D\propto\Delta^{-3/2}$
near criticality, i.e., near global breakdown. The crossover is a
universal phenomenon, independent of the threshold distribution $p(x)$.


The simulation results we have shown so far are based on {\it averaging\/}
over a large number of samples. For applications it is important that
crossover signal can be seen also in a single sample. We show in 
Fig. \ref{fig3b-3}
that equally clear crossover behavior is seen in a {\it single\/}
fiber bundle when $N$ is large enough. Also, as a practical tool
one must sample finite intervals ($x_{i}$, $x_{f}$) during the fracture
process. The crossover will be observed when the interval is close
to the failure point \cite{phh05,phh06}.

\begin{figure}
\includegraphics[width=2.5in,height=2.4in]{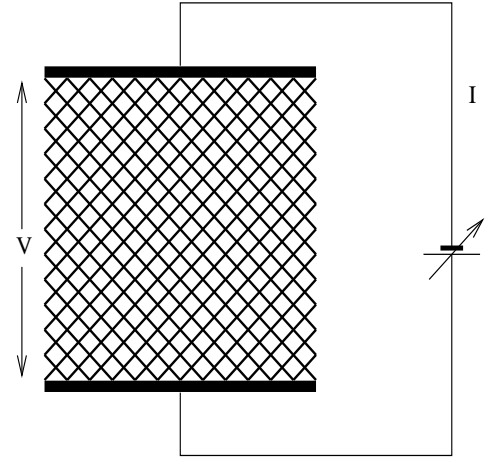} 

\caption{ A fuse model of size $100\times 100$.  Each bond is a fuse with a 
burn-out threshold $t$ drawn from a probability distribution $p(t)$.
}
\label{fig:fusenet}
\end{figure}

The ELS fiber bundle model is a simple model in that it is analytically
tractable.  A step up in complexity from the ELS fiber bundle model,
is the {\it random fuse model\/} \cite{hr90}.  While
resisting most analytical treatments, this model retains
{\it computational\/} tractability.  The fuse model consists of a
lattice in which each bond is a fuse, i.e., an ohmic resistor as long as
the electric current it carries is below a threshold value. If the
threshold is exceeded, the fuse burns out irreversibly. The threshold
$t$ of each bond is drawn from an uncorrelated distribution $p(t)$.
The lattice is placed between electrical bus bars and an increasing current
is passed through it. The lattice is a two-dimensional square one placed
at $45{}^{\circ}$ with regards to the bus bars, and the Kirchhoff equations
are solved numerically at each node assuming that all fuses have the same
resistance.  We show the model in Fig. \ref{fig:fusenet}.  The ELS fiber bundle
model may be interpreted as a mean-field version of the random fuse model
\cite{zrsv97}.  Hence, the random fuse model may be used as a testing
ground for results (see Table \ref{els-tab}) found with the ELS fiber bundle 
model to explore their robustness when other effect not present in the fiber 
bundle model enter.

\begin{figure}
\centering
\includegraphics[width=7cm]{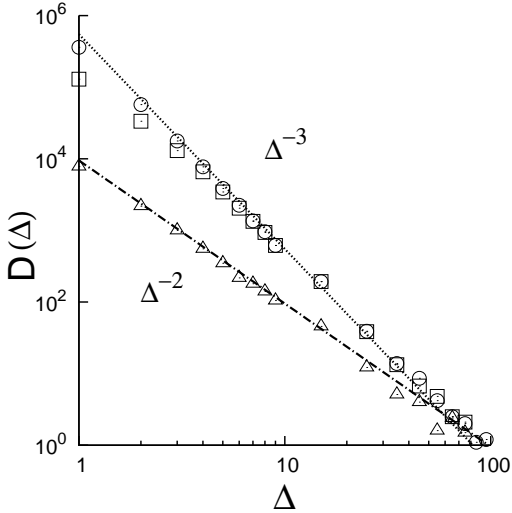}
\caption{ The burst distribution based on 300 samples random fuse lattices
of size $100\times100$. The threshold $t$ are uniformly distributed on the
unit interval. On the average, catastrophic
failure sets in after $2097$ fuses have blown. The circles denote
the burst distribution measured throughout the entire breakdown process.
The squares denote the burst distribution based on bursts appearing
after the first $1000$ fuses have blown. The triangles denote the
burst distribution after $2090$ fuses have blown. The two straight
lines indicate power laws with exponents $\xi=3$ and $\xi=2$, respectively.}
\label{fig:fuse-cross}
\end{figure}

To test the crossover phenomenon in a more complex situation than
for ELS fiber bundle model, we consider the random fuse model \cite{phh06}.
When one records all the bursts in the random fuse model, the distribution
follows a power law $D(\Delta)\propto\Delta^{-\xi}$ with $\xi\approx 3$,
which is consistent with the value reported in recent studies.
We show the histogram in Fig. \ref{fig:fuse-cross}.
With a system size of $100\times100$,
$2097$ fuses blow on the average before catastrophic failure sets
in. When measuring the burst distribution only after the first $2090$
fuses have blown, a different power law is found, this time with $\xi=2$.
After $1000$ blown fuses, on the other hand, $\xi$ remains the same as for the 
histogram recording the entire failure process (Fig. \ref{fig:fuse-cross}).

\begin{figure}
\centering
\includegraphics[width=6cm]{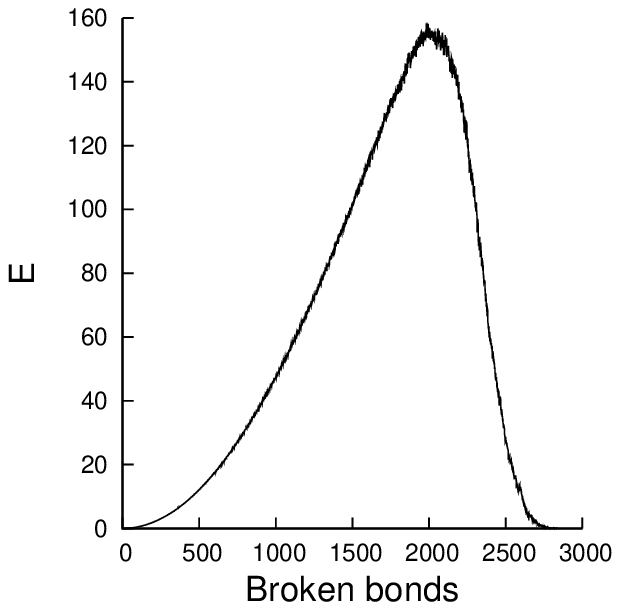}
\vskip.2in
\includegraphics[width=6cm]{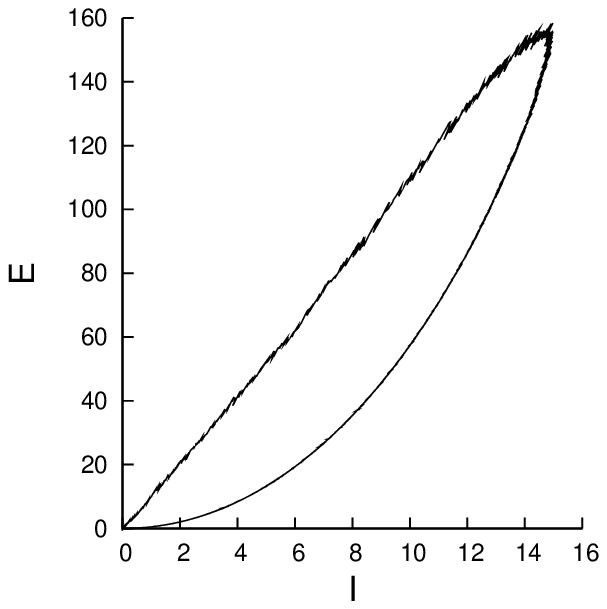}
\caption{ Power dissipation
$E$ as a function of the number of broken bonds (upper) and as a function
of the total current $I$ flowing in the fuse model (lower).
}
\label{fig:fig10-11}
\end{figure}

\begin{figure}
\centering
\includegraphics[width=7cm]{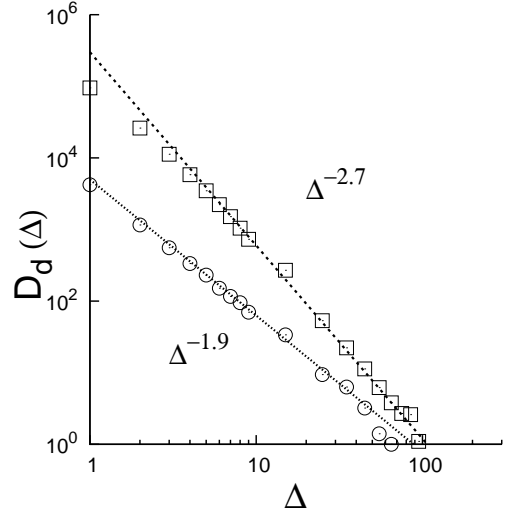}
\caption{The power dissipation avalanche histogram  $D_{d}(\Delta)$  for
the fuse model. The slopes of the two straight lines are $-2.7$ and $-1.9$, 
respectively. The circles show the histogram of avalanches recorded after 1000 
fuses has blown, whereas the squares show the histogram recorded after $2090$ 
fuses have blown.  This is close to catastrophic failure.}
\label{fig:fig12}
\end{figure}

In Fig. \ref{fig:fig10-11},
we show the power dissipation $E$ in the network as
a function of the number of blown fuses and as a function of the total
current. The dissipation is given as the product of the voltage drop
across the network $V$ times the total current that flows through
it. The breakdown process starts by following the lower curve, and
follows the upper curve returning to the origin. It is interesting
to note the linearity of the unstable branch of this curve. In Fig \ref{fig:fig12}, we record the avalanche distribution for power dissipation,
$D_{d}(\Delta)$.

\vskip.1in

Recording, as before, the avalanche distribution throughout the entire
process as well as recording only close to the point at which the
system catastrophically fails, result in two power laws, with exponents
{\small $\xi=2.7$} and {\small $\xi=1.9$,} respectively. It is interesting
to note that in this case there is not a difference of unity between
the two exponents. The power dissipation in the fuse model corresponds
to the stored elastic energy in a network of elastic elements. Hence,
the power dissipation avalanche histogram would in the mechanical
system correspond to the released energy. Such a mechanical system
could serve as a simple model for earthquakes.

\begin{figure}
\centering
\includegraphics[width=5cm]{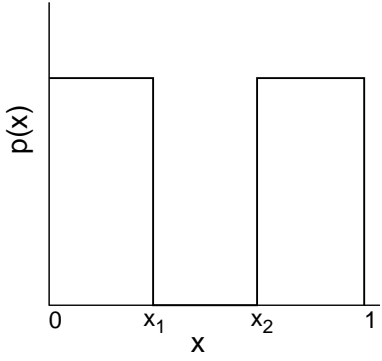}
\caption{Mixed uniform distribution.}
\label{ah-dutta1}
\end{figure}

\textcite{dd07a} studied the critical behavior of a 
bundle of fibers under global load sharing scheme with threshold
strength chosen randomly from a distribution which is uniform, but
discontinuous.  The form of the distribution is
\begin{eqnarray}
p(x) &=& \frac {1}{1-(x_{2}-x_{1})}~~~~~~
0<x\leq x_{1}\nonumber\nonumber\\
&=&0~~~~~~~~~~~~~~~~~~~~~~x_{1}<x<x_{2}\nonumber\\
~~~~~~~~~~&=&\frac {1}{1-(x_{2}-x_{1})}~~~~~x_{2}\leq x \leq 1
\label{dd1}
\end{eqnarray}
where $x_2-x_1$ is the gap in the threshold distribution as
shown in Fig.\ \ref{ah-dutta1}.

Here, a fraction $f$ of the fibers belong to the weaker section
$(0<x\leq x_1)$ and the remaining to the stronger section
$(x_2\leq x\leq 1)$. The condition of uniformity
of the distribution demands
\begin{equation}
x_{1} = \frac{f}{1-f}(1-x_{2})\;,
\label{dd2}
\end{equation}
so that fixing $x_1$ and $f$ immediately settles the value of $x_2$.

\begin{figure}
\centering
\includegraphics[width=7cm, angle=-90]{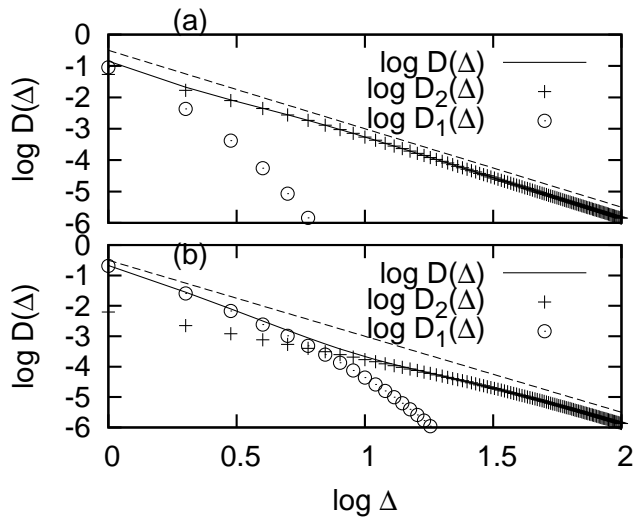}
\caption{Total avalanche size distribution $D(\Delta)$, $D_1(\Delta)$ and
$D_2(\Delta)$
obtained by numerical integration of Eqs.\ (\ref{dd5}) and (\ref{dd6}).
Figure (a) corresponds to
$x_1=0.08,x_2=0.28$ and $f=0.1$ and Figure (b) to
$x_1=0.25$, $x_2=0.42$ and $f=0.3$. As $x_2\to 0.5$,
the nonuniversal region increases in the small $\Delta$ region, whereas
$\Delta \propto \Delta^{-5/2}$ for large $\Delta$.
The dotted line has a slope of -5/2. From \textcite{dd07a}.}
\label{ah-dutta2}
\end{figure}

To study the dynamics of this model, once again the recursive equation approach
was used. Importantly, the redistributed stress must cross $x_2$
for the complete failure of the bundle to take place.
When the external load is such that the redistributed stress at an instant
$t$, i.e., $x(t)$ is greater than $x_2$,
the fixed point solution has the form
\begin{equation}
U^* = \frac{1}{2(1-(x_{2}-x_{1}))} 
 \left[1 + \sqrt{1-\frac{\sigma}{\sigma_{c}}}\right]\;,
\label{dd3}
\end{equation}
so that the critical stress $\sigma_c$ is
\begin{equation}
\sigma_{c} = \frac {1} {4[1-(x_{2}-x_{1})]}
\label{dd4}
\end{equation}
and the redistributed stress at the critical point is found to be $1/2$ as in 
the uniform distribution \cite{phh05}.
This immediately restricts the value of $x_2$ to be less than $1/2$ and
therefore, $x_2=0.5$ is defined as the critical distribution in this model.
On the other hand, uniformity condition, Equation (\ref{dd2}),
puts another restriction,namely $x_1<f$.
Interestingly, the critical stress is a function of the gap
$x_2-x_1$ and reduces to one fourth when the gap goes to zero.
The exponents related to the order parameter and susceptibility stick to their
mean field values.
However, the existence of a forbidden region shows a prominent signature in
the avalanche size distribution of the mixed model.
The expression for the total avalanche size distribution $D(\Delta)$
in this model includes two terms, one due to the contribution from thresholds
between $0$ to $x_1$ and the other from the stronger section of
fibers.
Hence, the total avalanche size $D(\Delta)$ is
$D(\Delta)/N=D_1(\Delta)+D_2(\Delta)$ where
\begin{eqnarray}
&{D_1(\Delta)}=\frac{\Delta^{\Delta-1}}{\Delta!}\frac{1}
{1-x_2+x_1}\int_0^{x_1}dx
(\frac{1-x_2+x_1-2x}{x})\nonumber\\
&\left[\frac{x}{1-x_2+x_1-x}\times
{\rm exp}(-\frac{x}{1-x_2+x_1-x})\right]^{\Delta}\nonumber\\
\label{dd5}
\end{eqnarray}
 and
\begin{eqnarray}
D_2(\Delta)=\frac{\Delta^{\Delta-1}}{\Delta!}
\frac{1}{1-x_2+x_1}\int_{x_2}^
{0.5}dx(\frac{1-2x}{x})\nonumber\\
\times\left[\frac{x}{1-x}
{\rm exp}(-\frac{x}{1-x})\right]^{\Delta}\;.
\label{dd6}
\end{eqnarray}
The leading behavior of $D_1(\Delta)$ is given by \cite{dd07a}
\begin{equation}
D_1(\Delta)=\Delta^{-5/2}e^{(1-x_m)\Delta} x_m^{\Delta}
\label{dd6-1}
\end{equation}
where
\begin{equation}
x_m=\frac{x_1}{1-x_2}
\label{dd8}
\end{equation}
which clearly indicates a rapid fall of the contribution of weaker fibers.
On the other hand, $D_2(\Delta)$ resembles the imminent failure behavior
studied by \textcite{phh05}, where the avalanche size exponent
shows a crossover
from $5/2$ to $3/2$ as $x_2\to0.5$. For the mixed model, the total
avalanche size distribution $D(\Delta)$ shows a nonuniversal behavior
for small $\Delta$ values, though eventually there is a crossover to the
universal mean field value. The most fascinating observation is the following:
though the gap in the distribution is always present, nonuniversality is only
prominent in the limit $x_2\to 0.5$.
Divakaran and Dutta showed that this nonuniversal
behavior stems from the avalanche of fibers in the weak section and only
in the vicinity of the critical distribution,
the contribution of $D_1(\Delta)$ overcomes $D_2(\Delta)$.
Otherwise, the faster fall of $D_1(\Delta)$ and large value of $D_2(\Delta)$
together force the avalanche size exponent to be $5/2$. It is to
be noted that the nonuniversal behavior is most prominent at a critical
distribution where the avalanche size exponent crosses over to $3/2$ in the
asymptotic limit.
The typical behavior of $D(\Delta)$ is shown in Fig.\ \ref{ah-dutta2}
for two different
distributions highlighting the increase in nonuniversal region
as $x_2$ approaches 0.5. However, for many discontinuities in the threshold 
distributions, avalanche size distribution shows a nonuniversal, non power-law 
behavior \cite{dd08} for small-size avalanches, although the large avalanches 
still exhibit similar crossover behavior as we discuss here. 

Divakaran and Dutta also
looked at the model where fibers from two different Weibull distributions are
mixed \cite{dd07b}.
Though an interesting variation of the critical stress with the mixing parameter
was obtained using a  probabilistic method introduced by
\textcite{mgp00},
there is no  deviation in the avalanche size exponent.
In a recent paper, \textcite{hkpk08} studied the infinite gap
limit of the discontinuity model. Here, they
considered a fraction $\alpha_{inf}$ of the fibers having infinite 
threshold strength
mixed with fibers having threshold chosen from a distribution $p(x)$.
They observed a critical fraction $\alpha_c$ such that
for $\alpha_{inf}>\alpha_c$, the avalanche size exponent
switches from the well known mean field exponent $\xi =5/2$ to a lower value 
$\xi =9/4$.
It was also showed
that such a behavior is observed for those distributions where the
macroscopic constitutive behavior has a maxima and a point of inflexion.
It is also claimed that below a critical gap, the Hidalgo et al.\ model reduces
to the discontinuity model of Divakaran and Dutta.
\textcite{kn08} studied the global load sharing fiber bundle model
in a wedge-shaped geometry.  That is, the fibers are connected to two rigid
blocks placed at an angle with respect to each other.  The fibers are loaded
by rotating the blocks with respect to each other, resulting in a linear
loading gradient on them.  In the limit of a threshold distribution tending
towards zero width, the fibers break in an orderly fashion according to the 
load,
and hence, position in the wedge.  As the width is increased, a process zone
--- i.e., a zone where some fibers fail whereas others stay intact ---
develops.  When the width of the threshold distribution is wide enough, the
process zone spans the entire bundle.  In this limit a burst size
exponent $\xi=5/2$ is recovered.  However, with a narrower distribution so
that a well-defined process zone smaller than the size of the bundle develops,
the burst exponent $\xi=2.0$ is found.

\subsubsection{Burst distribution for discrete load increase}

When the bundle
is stretched continuously from zero, fluctuation plays crucial role and 
the generic result is a power
law \cite{hh92} $ D(\Delta)\propto \Delta^{-\xi}$, for large $\Delta$, with $\xi=5/2$.
However, experiments may be performed in a different manner, where
the load is increased in finite steps of size $\delta$ . The value of the
exponent increases  \cite{pbc02,hp07} then to $3$: 
$D(\Delta)\propto \Delta^{-3}$.
The basic reason for the difference in the power laws is that increasing
the external load in steps reduces the fluctuations in the force.
The derivation (see Section III.B.1) of the asymptotic size 
distribution $D(\Delta)\propto \Delta^{-5/2}$
of avalanches, corresponding to stretching by infinitesimal steps,
shows the importance of force fluctuations \cite{hh92}. An effective
reduction of the fluctuations requires that the size $\delta$ of
the load increase is large enough so that a considerable number of
fibers break in each step.

Here is an analytic derivation, following \textcite{hp07},  how to calculate the burst distribution
in such situation. For the uniform distribution of thresholds 
(Eq. \ref{uniform}), the load curve is parabolic, 
\begin{equation}
\langle F\rangle=N\; x\;(1-x),\label{loaduniform}
\end{equation}
so that the expected critical load equals $F_{c}=N/4$. With a sufficiently 
large $\delta$ we may use the macroscopic load
equation (Eq. \ref{loaduniform}) to determine the number of fibers broken
in each step. The load values are $m\delta$, with $m$ taking the
values $m=0,1,2,\ldots,N/4\delta$ for the uniform threshold distribution.
By Eq. (\ref{loaduniform}) the threshold value corresponding to the load
$m\delta$ is \begin{equation}
x_{m}={\textstyle \frac{1}{2}}\left(1-\sqrt{1-4m\delta/N}\right).\end{equation}
 The expected number of fibers broken when the load is increased from
$m\delta$ to $(m+1)\delta$ is close to \begin{equation}
\Delta=Ndx_{m}/dm=\delta/\sqrt{1-4m\delta/N}.\end{equation}
 Here the minimum number of $\Delta$ is $\delta$, obtained in the first
load increase. The integral over all $m$ from $0$ to $N/4\delta$
yields a total number $N/2$ of broken fibers, as expected, since
the remaining one-half of the fibers burst in one final avalanche.

The number of avalanches of size between $\Delta$ and $\Delta+d\Delta$, $D(\Delta)\; d\Delta$,
is given by the corresponding interval of the counting variable $m$:
$D(\Delta)\; d\Delta=dm$.
Since \begin{equation}
\frac{d\Delta}{dm}=\frac{2\delta^{2}}{N}(1-4m\delta/N)^{-3/2}=\frac{2}{N\delta}\; \Delta^{3},\end{equation}
 we obtain the following distribution \cite{pbc02,hp07} of avalanche sizes: 
\begin{equation}
D(\Delta)=\frac{dm}{d\Delta}={\textstyle \frac{1}{2}}N\delta\; \Delta^{-3},\hspace{1cm}(\Delta \geq\delta).\label{Guniform}\end{equation}
 For consistency, one may estimate the total number of bursts by integrating
$D(\Delta)$ from $\Delta=\delta$ to $\infty$, with the result $N/4\delta$,
as expected.

\begin{figure}
\epsfysize=2in
\epsfbox{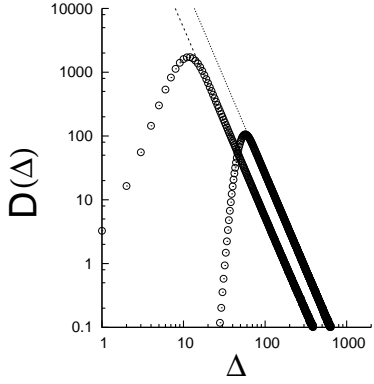}
\caption{ Avalanche size distribution for the uniform threshold
distribution (Eq. \ref{uniform}) when the load is increased in steps
of $\delta=10$ and $\delta=50$ (upper curve). The dotted lines show
the theoretical asymptotics (Eq. \ref{Guniform}) for $\delta=10$ and
$\delta=50$. The figure is based on $10000$ samples with $N=10^{6}$
fibers in the bundle. }
\label{fig:els-per-RMP1}
\end{figure}

Fig. \ref{fig:els-per-RMP1}  shows that the theoretical power 
law (Eq. \ref{Guniform}) fits
the simulation results perfectly for sufficiently large $\Delta$. The
simulation records also a few bursts of magnitude less than $\delta$
because there is a nonzero probability to have bundles with considerably
fewer fibers than the average in a threshold interval. However, these
events will be of no importance for the asymptotic power law in the
size distribution. 

\begin{figure}
\epsfysize=2in
\epsfbox{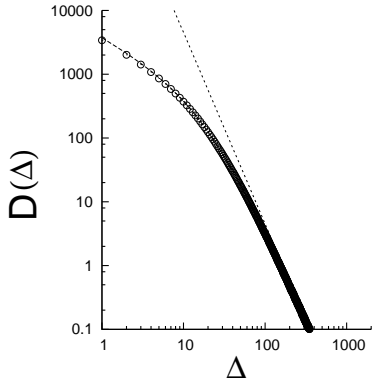}
\caption{Avalanche size distribution for the Weibull distribution
(\ref{Weibull}) with index $5$. Open circles represent simulation data, 
dashed lines
are analytic expressions (Eq. \ref{D_Weibull}) and
dotted line is the asymptotic power law with exponent $-3$. The
load is increased in steps of $\delta=20$. The figure is based on
$10000$ samples of bundles with $N=10^{6}$ fibers. }
\label{fig:els-per-RMP2}
\end{figure}

In order to see whether the asymptotic exponent value  $\xi=3$  is general,
simulations for another threshold distribution has been performed,
the Weibull distribution (Eq. \ref{Weibull}) with index $5$, which confirms 
 similar asymptotic behavior (Fig. \ref{fig:els-per-RMP2}).   

For a general threshold distribution $P(x)$ a load interval $\delta$
and a threshold interval are connected via the load equation $\langle F\rangle=N\; x\;(1-P(x))$.
Since $d\langle F\rangle/dx=N[1-P(x)-xp(x)]$, an increase $\delta$
in the load corresponds to an interval \begin{equation}
dx=\frac{\delta}{N[1-P(x)-xp(x)]}\label{dx}\end{equation}
 of fiber thresholds. The expected number of fibers broken by this
load increase is therefore \begin{equation}
\Delta=N\; p(x)\;dx=\frac{p(x)}{1-P(x)-xp(x)}\;\delta.\label{n}\end{equation}
 Note that this number diverges at the critical point, i.e.\ at the
maximum of the load curve, as expected.

Following the similar method, as in case of uniform distribution, we can 
 determine \cite{hp07} the asymptotic distribution for large
$\Delta$: \begin{equation}
D(\Delta)\simeq C\; \Delta^{-3},\label{Dasymp}\end{equation}
 with a nonzero constant \begin{equation}
C=N\delta\;\frac{p(x_{c})^{2}}{2p(x_{c})+x_{c}p'(x_{c})},\end{equation}
where we  have used that at criticality $1-P(x_{c})=x_{c}p(x_{c})$. Thus
the asymptotic  exponent value  $\xi=3$  is universal.

For the Weibull distribution considered in Fig. \ref{fig:els-per-RMP2} we obtain \begin{equation}
D(\Delta)=N\delta\; \Delta^{-3}\;\frac{25x^{9}\; e^{-x^{5}}}{4+5x^{5}},\hspace{.2cm}\mbox{ and }\hspace{.2cm}\Delta=\frac{5\delta\; x^{4}}{1-5x^{5}}.
\label{D_Weibull}
\end{equation}
 This burst distribution must be given on parameter form, the elimination
of $x$ cannot be done explicitly. The critical point is at $x=5^{-1/5}$
and the asymptotics is given by Eq. \ref{Dasymp}, with $C=N\delta(625e)^{-1/5}$.

If we let the load increase $\delta$ shrink to zero, we must recover
the asymptotic $D(\Delta)\propto \Delta^{-5/2}$ power law valid for continuous
load increase. Thus, as function of $\delta$, there must be a crossover
from one behavior to the other. It is to be expected that for $\delta\ll1$
the $D(\Delta)\propto \Delta^{-5/2}$ asymptotics is seen, and when $\delta\gg1$
the $D(\Delta)\propto \Delta^{-3}$ asymptotics is seen.

\subsubsection{Energy bursts in fiber bundle model }

So far we have discussed in detail the statistical distribution of the \textit{size} of avalanches in fiber bundles \cite{hh92,phh05,hhp06,rkh06}.  Sometimes the
 avalanches cause a sudden internal stress redistribution in the material, 
and are accompanied by a rapid release of mechanical energy. A useful
experimental technique to monitor the energy release is to measure
the acoustic emissions (AE), the elastically radiated waves produced in
the bursts \cite{ppvac94,ggbc97,s91a,f91,dmp91}. Experimental 
observations suggest that AE signals follow  power law distributions. What is 
the origin of such power laws? Can we explain it through a general scheme of 
fluctuation guided breaking dynamics that has been demonstrated well in 
 ELS fiber bundle model?
  
We now determine the statistics of the energies released \cite{ph08} in 
fiber bundle avalanches. As the fibers obey Hooke's law, the energy stored
in a single fiber at elongation $x$ equals $\frac{1}{2}x^{2}$, where
we for simplicity have set the elasticity constant equal to unity.
The individual thresholds $x_{i}$ are assumed to be independent
random variables with the same cumulative distribution function $P(x)$
and a corresponding density function $p(x)$.

\vspace{.1in}
\noindent \emph{(a) Energy statistics}
\vspace{.1in}

Let us characterize a burst by the number $\Delta$ of fibers that 
fail, and by the lowest threshold value $x$ among the $\Delta$ failed fibers.
The threshold value $x_{{\rm max}}$ of the strongest fiber in the
burst can be estimated to be \begin{equation}
x_{{\rm max}}\simeq x+\frac{\Delta}{Np(x)},\label{xmax}\end{equation}
 since the expected number of fibers with thresholds in an interval
$\delta x$ is given by the threshold distribution function as $N\; p(x)\;\delta x$.
The last term in (Eq. \ref{xmax}) is of the order $1/N$, so for a very
large bundle the differences in threshold values among the failed
fibers in one burst are negligible. Hence the energy released in a
burst of size $\Delta$ that starts with a fiber with threshold $x$ is
given as \begin{equation}
E={\textstyle \frac{1}{2}}\; \Delta\; x^{2}.\label{En}\end{equation}

Following  \textcite{hh92}
the expected number of bursts of size $\Delta$, starting at
a fiber with a threshold value in the interval $(x,x+dx)$, is 
\begin{eqnarray}
f(\Delta,x)\; dx & = & N\frac{\Delta^{\Delta-1}}{n!}\;\frac{1-P(x)-xp(x)}{x}\nonumber\\
& \times & X(x)^{\Delta}\; e^{-\Delta X(x)}\; dx,
\end{eqnarray}
where  \begin{equation}
X(x)=\frac{x\; p(x)}{1-P(x)}.\end{equation}
The expected number of bursts with energies less than $E$ is therefore
\begin{equation}
G(E)=\sum_{\Delta}\int\limits _{0}^{\sqrt{2E/\Delta}}f(\Delta,x)\; dx,\end{equation}
 with a corresponding energy density \begin{equation}
g(E)=\frac{dG}{dE}=\sum_{\Delta}(2E\Delta)^{-1/2}\; f(\Delta,\sqrt{2E/\Delta}).\end{equation}
 Explicitly, \begin{equation}
g(E)=N\sum_{\Delta}g_{\Delta}(E),\label{g}\end{equation}
 with 
\begin{eqnarray}
g_{\Delta}(E)& = & \frac{\Delta^{\Delta-1}}{2E\; \Delta!}(1-P(s)-sp(s))\nonumber\\
& \times & \left[\frac{sp(s)}{1-P(s)}\exp\left(-\frac{sp(s)}{1-P(s)}\right)\right]^{\Delta}.\label{gn}\end{eqnarray}
Here $s\equiv\sqrt{2E/\Delta}.\label{s}$
With a critical threshold value $x_{c}$, it follows from (Eq. \ref{En})
that a burst energy $E$ can only be obtained if $\Delta$ is sufficiently
large, $\Delta\geq2E/x_{c}^{2}.$ Thus the sum over $n$ starts with 
$\Delta=1+[2E/x_{c}^{2}]$, where $[a]$ denotes the integer part of $a$.

\vspace{.1in}
\noindent \emph{(b) High energy asymptotics}
\vspace{.1in}

Bursts with high energies correspond to bursts in which many fibers
rupture. In this range we use Stirling's approximation for the
factorial $\Delta!$, replace $1+[2E/x_{c}^{2}]$ by $2E/x_{c}^{2}$, and
replace the summation over $\Delta$ by an integration. Thus

\begin{eqnarray}
g(E) & \simeq & \frac{N}{2E^{3/2}\pi^{1/2}}\int\limits _{2E/x_{c}^{2}}^{\infty}\frac{e^{\Delta}}{\Delta^{3/2}}\;(1-P(s)-sp(s))\nonumber \\
 & \times & \left[\frac{sp(s)}{1-P(s)}\exp\left(-\frac{sp(s)}{1-P(s)}\right)\right]^{\Delta}\; d\Delta.\label{integral1}\end{eqnarray}
By changing integration
variable from $\Delta$ to $s$ we obtain {\begin{eqnarray}
g(E) & \simeq & \frac{N}{2E^{3/2}\pi^{1/2}}\int\limits _{0}^{x_{c}}(1-P(s)-sp(s))\nonumber\\ & \times & \left[\frac{sp(s)}{1-P(s)}\exp\left(1-\frac{sp(s)}{1-P(s)}\right)\right]^{\Delta}\; ds\nonumber \\
 & = & \frac{N}{2E^{3/2}\pi^{1/2}}\int\limits _{0}^{x_{c}}(1-P(s)-sp(s))e^{-Eh(s)}\; ds,\nonumber\\
\label{integral2}\end{eqnarray}
}  with \begin{equation}
h(s)\equiv\left[-\frac{1-P(s)-sp(s)}{1-P(s)}+\ln\frac{1-P(s)}{sp(s)}\right]\frac{2}{s^{2}}.\label{h}\end{equation}

For large $E$ the integral (Eq. \ref{integral2}) is dominated by the
integration range near the minimum of $h(s)$. At the upper limit
$s=x_{c}$ we have $h(x_{c})=0$, since $1-P(x_{c})=x_{c}p(x_{c})$.
 This is also a minimum of $h(s)$, having quadratic form,
\begin{equation}
h(s)\simeq\left(\frac{2p(x_{c})+x_{c}p'(x_{c})}{x_{c}^{2}p(x_{c}}\right)^{2}\;(x_{c}-s)^{2}.\end{equation}
 Inserting these expressions into (Eq. \ref{integral2}) and integrating,
we obtain the following asymptotic expression, \begin{equation}
g(E)\simeq N\;\frac{C}{E^{5/2}} \sim E^{-\xi_e}, \label{as}\end{equation}
 where \begin{equation}
C=\frac{x_{c}^{4}p(x_{c})^{2}}{4\pi^{1/2}\,[2p(x_{c})+x_{c}p'(x_{c})]}.\label{C}\end{equation}

\begin{figure}
\includegraphics[width=2.5in,height=2.4in]{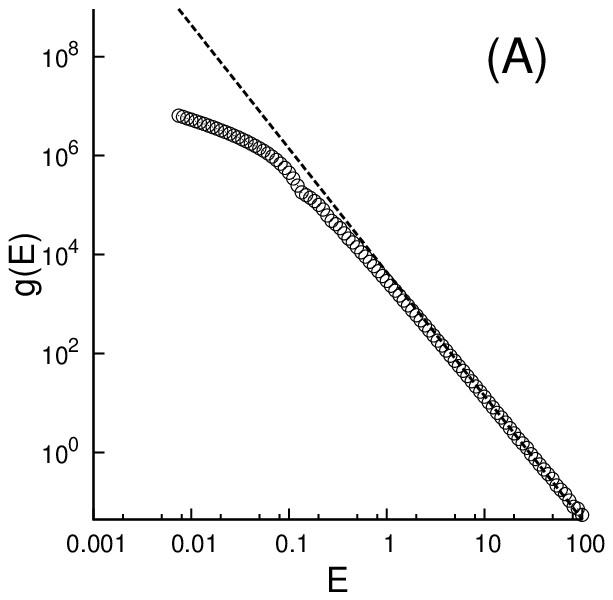} 

\includegraphics[width=2.5in,height=2.4in]{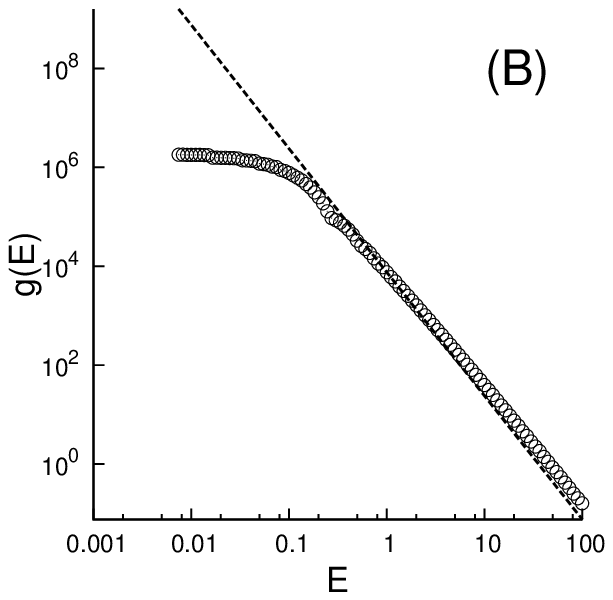} 

\caption{Simulation results for $g(E)$ characterizing
energy bursts in fiber bundles with (A) the uniform threshold distribution
(Eq. \ref{uniform}) and (B) the Weibull distribution (Eq. \ref{Weibull})
of index $2$. The graphs are based on 1000 samples with $N=10^{6}$
fibers in each bundle. Open circles represent simulation data, and
dashed lines are the theoretical results (Eqs. \ref{as} - \ref{C}) for
the asymptotics. }
\label{fig:energy_bursts_high}
\end{figure}

In Fig. \ref{fig:energy_bursts_high} we compare the theoretical formula with simulations for
the uniform distribution (Eq. \ref{uniform}), 
which corresponds to $x_{c}=\frac{1}{2}$, and $C=2^{-7}\pi^{-1/2}$,
and for the Weibull distribution (Eq. \ref{Weibull})  with index $\rho=2$,
 which corresponds to $x_{c}=2^{-1/2}$ and $C=2^{-5}(2\pi e)^{-1/2}$.

The corresponding asymptotics (Eq. \ref{as}) are also exhibited in Fig.  \ref{fig:energy_bursts_high}.
For both threshold distributions the agreement between the theoretical
asymptotics and the simulation results is very satisfactory. The exponent
$-5/2$ in the energy burst distribution is clearly universal. Note
that the asymptotic distribution of the burst magnitudes $\Delta$ is governed
by the same exponent \cite{hh92}.

\vspace{.1in}
\noindent \emph{(c) Low-energy behavior}
\vspace{.1in}

\begin{figure}
\includegraphics[width=2.5in,height=2.4in]{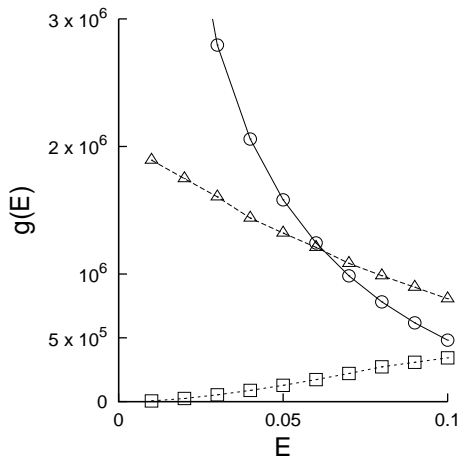} 

\caption{ Simulation results for the burst distribution
$g(E)$, in the low-energy regime, for the uniform threshold distribution
(circles), the Weibull distribution with $\rho=2$ (triangles) and Weibull
distribution with $\rho=5$ (squares). The graphs are based on 1000 samples
with $N=10^{6}$ fibers in each bundle. }
\label{fig:energy_bursts_low}
\end{figure}

The low-energy behavior of the burst distribution is by no means 
universal: $g(E)$ may diverge, vanish or stay constant as $E\rightarrow0$,
depending on the nature of the threshold distribution. In Fig.  \ref{fig:energy_bursts_low}
we exhibit simulation results for the low-energy part of $g(E)$ for
the uniform distribution and the Weibull distributions of index $2$
and index $5$.

We see that $g(E)$ approaches a finite limit in the Weibull $\rho=2$
case, approaches zero for Weibull $\rho=5$ and apparently diverges in
the uniform case. All this is easily understood, since bursts with
low energy predominantly correspond to single fiber bursts ($\Delta=1$,
\textit{i.e.} $E=x^{2}/2$) and to fibers with low threshold values.
The number of bursts with energy less than $E$ therefore corresponds
to the number of bursts with $x<\sqrt{2E}$, which is close to $N\; P(\sqrt{2E})$.
This gives \begin{equation}
g(E)\simeq N\;\frac{p(\sqrt{2E})}{\sqrt{2E}}\hspace{10mm}\mbox{when }E\rightarrow0.\label{Enull}\end{equation}
 For the uniform distribution $g(E)$ should therefore diverge as
$(2E)^{-1/2}$ for $E\rightarrow0$. The simulation results in Fig.  \ref{fig:energy_bursts_low}
are consistent with this divergence. For the Weibull Distribution  of index 2, 
on the other hand, (Eq. \ref{Enull}) gives $g(E)\rightarrow2N$ when 
$E\rightarrow0$,
a value in agreement with simulation results in the figure. Note that
for a Weibull distribution of index $\rho$, the low-energy behavior
is $g(E)\propto E^{(\rho-2)/2}$. Thus the Weibull with $\rho=2$ is a borderline
case between divergence and vanishing of the low-energy density.
The same lowest-order results can be obtained from the general expression
(Eq. \ref{g}), which also can provide more detailed low-energy expansions.

For high energies the energy density obeys a power law
with exponent $-5/2$. This asymptotic behavior is universal, independent
of the threshold distribution. A similar  power law dependence is found in some 
experimental observations on acoustic emission
studies \cite{ppvac94,ggbc97} of loaded composite materials.
In contrast, the low-energy behavior of $g(E)$ depends crucially on
the distribution of the breakdown thresholds in the bundle. $g(E)$
may diverge, vanish or stay constant for $E\rightarrow0$.

\begingroup
\squeezetable
\begin{table}
\label{els-tab}
\begin{center}
\begin{tabular}{ccc}
\hline
Exponent for & Value & Comment \\
\hline
Order parameter $(\alpha)$ &
$1/2$ &
$-$ \\
\\

Breakdown susceptibility $(\beta)$ &
$1/2$ &
$-$ \\
\\

Relaxation time $(\theta)$ &
$1/2$ &
amplitude ratio $= \ln N/2\pi$  \\
\\

Avalanche size distribution $(\xi)$ &
$3$ &
discrete load increase \\
 &
$5/2$ &
continuous load increase \\
\\

Energy burst distribution $(\xi _e)$ &
$5/2$ &
in the asymptotic limit.\\ 
&
&
For the low energy limit,\\
&
&
 distribution is non universal. \\
\hline
\end{tabular}
\end{center}
\caption{\label{els-tab} Exponents for order parameter ($O$), breakdown 
susceptibility ($\chi$), relaxation time ($\tau$), avalanche size distribution 
$D(\Delta)$ and energy burst distribution $g(E)$ in the ELS model.}
\end{table}
\endgroup

\section{Local load sharing model}

So far we have studied fiber bundles where the force once carried by
a failing fiber is spread equally among all the surviving fibers.  This
may often be a very good approximation.  However, intuitively it is
natural that fibers being closer to a failing fiber feel more of an effect
than fibers further away --- an effect reminiscent of stress enhancement
around cracks.  We will in this section discuss three classes of models
where there are local effects in how the forces carried by failed fibers
are distributed.  We start with the most extreme where the forces are
totally absorbed by the nearest surviving fibers.  We then move on to
models where the stress is distributed according to a power law in the
distance from the failing fiber, and lastly a model where we assume
the {\it clamps\/} the fibers are attached to are soft and therefore
deform due to the loading of the fibers --- as can be seen in pulling on the
hairs on one's arm.

\subsection{Stress alleviation by nearest neighbors}

\label{Sec:4A}

\begin{figure}
\centering
\includegraphics[width=4cm]{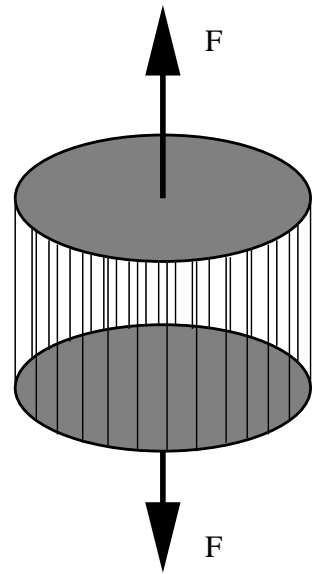}
\caption{A fiber bundle with periodic boundary conditions.
The externally applied force $F$ is the control parameter.}
\label{ah-fig4a-0}
\end{figure}

The extreme form for local load redistribution is that all extra stresses
caused by a fiber failure are taken up by the  {\it nearest-neighbor\/}
surviving fibers \cite{hp81,ps83,h85,kp87,hp91,dl94}.
The simplest geometry is one-dimensional, so that the $N$
fibers are ordered linearly, with or without periodic boundary
conditions. In this case precisely {\it two\/} fibers, one on each side, take
up, and divide equally, the extra stress, see Fig.~\ref{ah-fig4a-0}.
When the strength thresholds take only two values, the bundle strength
distribution has been found analytically \cite{h85,hp91,dl94}.

At a total force $F$ on the bundle the force on a fiber
surrounded by $n_l$ previously failed fibers on the left-hand side,
and $n_r$ on the right-hand side, is then
\begin{equation}
\frac{F}{N}
\left(1+{\textstyle \frac{1}{2}}(n_l+n_r)\right) = f(2+n_l+n_r)\;.
\label{HP}
\end{equation}
Here
\begin{equation}
f=\frac{F}{2N}
\label{xa}
\end{equation}
is one-half the force-per-fiber, is a convenient variable to use as the
driving {\it force parameter.\/}

\begin{figure}
\centering
\includegraphics[width=6cm]{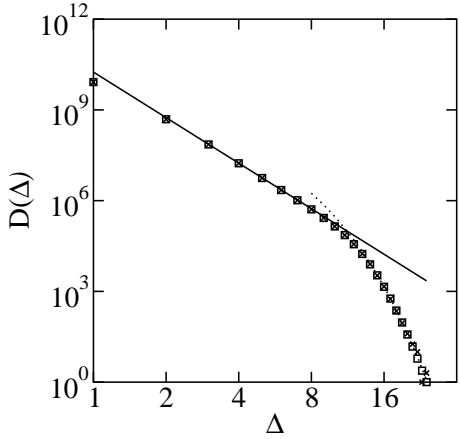}
\caption{Burst distribution in local model as found numerically
for 4 000 000 samples with $N=20 000$ fibers (crosses), and calculated from
Eq. (\ref{850}) (boxes).  The straight line shows the power law
$\Delta^{-5}$ and the broken curve the function $\exp(-\Delta/\Delta_0)$
with $\Delta_0 =1.1$. Note the small value of $\Delta_0$.}
\label{ah-fig4a-1}
\end{figure}

\textcite{zd94} and \textcite{hh94b}
studied numerically
the burst distribution in the local load-sharing model.
In Fig.~\ref{ah-fig4a-1} we exhibit simulation
results similar to those first appearing in  \textcite{hh94} using threshold
strengths randomly distributed on the unit interval.
Again a power-law distribution
seems to appear. However, the burst exponent $\xi$ seems much larger
than in the global lad-sharing model, Eq.~\ref{eq10},
\begin{equation}
\xi \simeq 5\;.
\label{eq:4a-6}
\end{equation}
Thus the relative frequency of long (non-fatal) bursts is
considerably reduced.

It was concluded from the  numerics that systems with local load-sharing
are {\it not\/} in the universality class of fiber bundles with global load
redistribution.

\textcite{khh97} set out to analytically calculate the burst
distribution in the local load-sharing model, finding the surprising
result that there is no power law distribution of bursts at all; it is
exponential.

\textcite{khh97}
did not find the burst distribution for general threshold
distribution. Rather, they limited their study to the uniform threshold
distribution in the force parameter $f$, given by
\begin{equation}
\label{eq-4a-1}
P(f) = \left \{ \begin{array}{ll}
f& \hspace{3mm} \mbox{for } 0 \leq f < 1 \\
1 & \hspace{3mm} \mbox{for } f \geq 1.
\end{array} \right.
\end{equation}

Bursts in the local and the global models have  different characters.
In the local model a burst develops  with one failure
acting as the seed. If many neighboring fibers have failed, the load on the
fibers on each side is high, and if they burst the load on the new neighbors
will be even higher, etc. In this way a weak region in the bundle may be
responsible for the failure of the whole bundle. For a large number $N$ of
fibers the probability of a weak region somewhere is higher. This
hints in a qualitative way that
the maximum load the bundle are able to carry
does not increase proportional to $N$, but slower than linear.

The result of a calculation based on combinatorics \cite{khh97}
was the burst distribution
\begin{eqnarray}
&D(\Delta) = \int\limits_0^{1/(\Delta +2)} \sum_{n=1}^N \sum_{L_1=\Delta}^{M(f)}\sum_{L_2=0}^{M(f)-L_1} \frac{P_f(n,L_1;f)}{S(L_1;f)}p(L_1,\Delta;f)\nonumber\\
&P_f(N-n-1,L_2;f)[1-(L_1+L_2+2)f]\;df.\nonumber\\\label{850}
\end{eqnarray}
where $S(l;f)$ is the probability that a selected region of $l$ consecutive
fibers have all failed whereas the two fibers at both boundaries are still
intact at force parameter $y$, $p(l,a;f)df $ is the probability
that a force increase from $f$ to $f+df$ leads to a burst of length $l$ and
magnitude $a$. $P_f(n,L;f)$ is the probability at force parameter $f$
that among the $n$ first fibers there is no fatal burst, and that the last $L$
fibers of these have all failed.

The load distribution rule (Eq. \ref{HP}) implies that burst of size
$\Delta$  does necessarily
lead to a complete breakdown of the whole bundle if the external force
is too high, i.e., if $x$ exceeds a critical value $ x_{\max}$.
Since here a fiber can at most take a load
of unity, we have
\begin{equation} 
f_{\max} = \frac{1}{\Delta + 2}\;.
\label{max}
\end{equation}

Let us  now attempt to find an simple
estimate for the maximal force per fiber
that the fiber bundle can tolerate. In order to do that we assume that the
fatal burst occurs in a region where no fibers have previously failed
so that the burst has
the same magnitude and length. We know that a single burst of length
$\Delta = f^{-1}-2$ is fatal, Eq.~(\ref{max}), so our criterion is
simply
\begin{equation}
D(f^{-1}-2) = 1\;.
\label{fatal}
\end{equation}

If we take into account that the two fibers adjacent to the burst should
hold, and ignore the rest of the bundle, the gap distribution would be
\begin{eqnarray}
&N^{-1} D(\Delta) \approx
\int\limits_0^{1/(\Delta+2)}[1-(2+\Delta)f]^2p(\Delta,
\Delta;f)\;df\nonumber\\
&=\frac{2p(\Delta,\Delta)}{\Delta(\Delta+1)(\Delta+2)^{\Delta+1}}.\nonumber\\
\end{eqnarray}
With the abbreviation
\[   R_{\Delta} = \frac{p(\Delta,\Delta)}{(\Delta -1) !},\]
we have
\begin{eqnarray}
&D(\Delta)/N \approx \frac{2(\Delta+2)!}{\Delta^2(\Delta+1)^2
(\Delta+2)^{\Delta+2}}
R_{\Delta}\nonumber\\
&\simeq \frac{\sqrt{8\pi(\Delta +2)}\;}{\Delta^2(\Delta+1)^2}e^{-\Delta-2}
\;R_{\Delta},\nonumber\\
\end{eqnarray}
using Stirling's formula.

Taking logarithms we have
\begin{eqnarray}
&\ln D(\Delta)-\ln N = -
(\Delta + 2)\left[1+ \frac{\ln R_{\Delta}}{\Delta+2} + {\cal O}
\left(\frac{\ln \Delta}{\Delta}\right)
\right]\nonumber\\
& \simeq - (\Delta +2)\;,\nonumber\\
\label{950}
\end{eqnarray}
using that
\begin{equation}
\lim_{n\rightarrow \infty} R_n^{1/n} = 1
\label{conv}
\end{equation}
for $R_{\Delta}$ when $\Delta$ is large.

The failure criterion (Eq. \ref{fatal}) then takes the form
\begin{equation}
\label{912}
\ln N \simeq \frac{1}{f}\;.
\end{equation}
Since $f=F/2N$ we have the following estimate for the maximum force $F$
that the fiber bundle
can tolerate before complete failure:
\begin{equation}
F \simeq \frac{2N}{\ln  N}.
\label{Ffatal}
\end{equation}
Due to the assumption that the fatal burst occurs in a region with no
previously failed fibers, the numerical prefactor is an overestimate.
The size dependence
\begin{equation}
F \propto \frac{N}{\ln  N}.
\label{lFfatal}
\end{equation}
shows that the maximum load the
fiber bundle can carry does not increase proportionally
to the number of fibers, but slower.
This is to be expected since the probability of
finding somewhere a stretch of
weak fibers that start a fatal burst increases
when the number of fibers increases.

The $N/\ln N$ dependence agrees with a previous estimate by 
\textcite{zd95} for a uniform threshold distribution.  The bimodal distribution
used in \cite{hp91,dl94} also shows this behavior.

The burst distribution Eq.~(\ref{950}) is exponential.  The probability
of a single burst zipping through the fiber bundle grows with the
system size $N$.  This contrasts strongly with the global load sharing
model whose strength (maximum force it can sustain) grows linearly with $N$,
and whose burst distribution follows a universal power law.  If the latter
behavior is reminiscent of a second order transition with a critical point,
the local load sharing model behaves more as if moving towards a first order
phase transition.

\begin{figure}
\includegraphics[width=6cm,height=5cm]{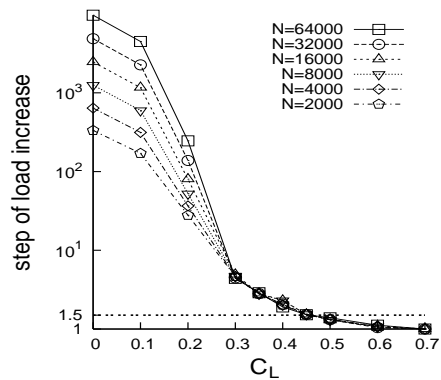}
\caption{ 
Numerical estimate of
the upper bound of $C_{L}$ in LLS model: For $C_{L}\geq0.5$
the average step values go below $1.5$, i.e., the bundle fails at
one step in most of the realizations. 
}
\label{fig:lls-step}
\end{figure}

Let us discuss briefly the effect of a low cutoff (in fiber strength 
distribution) on the failure properties of  LLS model.  As in case of ELS 
model (see Section III.A), we consider  uniform fiber 
threshold distribution
having a low cutoff $C_{L}$ (Eq. \ref{nonlinear-dist}). We  present a 
probabilistic argument to determine the upper limit of $C_{L}$, beyond which
the whole bundle fails at once. Following the weakest fiber breaking
approach the first fiber fails at an applied stress $C_{L}$
(for large $N$). As we are using periodic boundary conditions, the
$n_c$ nearest neighbors ($n_c$ is the coordination number) bear the
terminal stress of the failing fiber and their stress value rises
to $x_{f}=C_{L}(1+1/n_c)$. Now, the number of nearest neighbors
(intact) having strength threshold below $x_{f}$ is $(nn)_{fail}=n_c P(x_{f})$
(see Eq. \ref{Ad-1}). Putting the value of $P(x_{f})$ and
$x_{f}$ we finally get \begin{equation}
(nn)_{fail}=\frac{(C_{L})}{(1-C_{L})}.\label{eq:lls-prob}\end{equation}
If $(nn)_{fail}\geq1$, then at least another fiber fails and this
is likely to trigger a cascade of failure events resulting complete collapse
of the bundle. Therefore, to avoid the `instant failure' situation we
must have $(nn)_{fail}<1$, from which we get the upper bound of $C_{L}$:
$C_{L}<\frac{1}{2}.$
As the above condition does not depend on the coordination
number $n$, at any dimension the whole bundle is likely to collapse
at once for $C_{L}\geq1/2$. It should be mentioned that LLS
model should behave almost like ELS model at the limit of infinite
dimensions and therefore the identical bound (of $C_{L}$) in
both the cases is not surprising. A numerical study \cite{ph05} confirms
(Fig. \ref{fig:lls-step}) the above analytic argument in one
dimension. When average step value goes below $1.5$, one step failure
is the dominating mode then. One can find the extreme limit of
$C_{L}$ when all the
nearest neighbors fail after the weakest fiber breaks. Then the LLS
bundle collapses instantly for sure. Setting $(nn)_{fail}=n_c$ one gets
the condition  $C_{L}\geq n_c /(1+n_c)$, where stress level of all the
nearest neighbors crosses the upper cutoff 1 of the strength distribution.
Clearly such failure is very rapid (like a chain reaction) and does not
depend on the shape of the strength distributions, except for the upper
cutoff. Also as $n_c$ increases (ELS limit), $C_{L}$ for instant failure
assumes the trivial value 1. Similar sudden failure in FBM
has been discussed by  \textcite{mgp01} in the context of 
a `one sided load transfer' model.

\begin{figure}
\includegraphics[width=6cm,height=5cm]{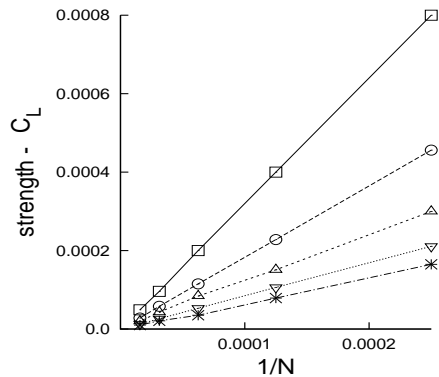}
\caption{ 
The strength -- $C_{L}$
is plotted against $1/N$ for different $C_{L}$ values: $0.3$
(square), $0.35$ (circle), $0.4$ (up triangle), $0.45$ (down triangle),
$0.5$ (star). All the straight lines approach $0$ value as $N\rightarrow\infty$.
}
\label{fig:lls-strength}
\end{figure}

The local load sharing (LLS) scheme introduces stress enhancement
around the failed fiber, which accelerate damage evolution. Therefore,
a few isolated cracks can drive the system toward complete failure
through growth and coalescence. The LLS model shows zero strength
(for fiber threshold distributions starting from zero value) at the
limit $N\rightarrow\infty$, following a logarithmic dependence on
the system size ($N$) \cite{s80,gip93,pc03a}. 
Now for threshold distributions having a low cutoff ($C_{L}$), the ultimate
strength of the bundle cannot be less than $C_{L}$. For such
a uniform distribution (Eq. \ref{nonlinear-dist}),  numerical simulations
show (Fig. \ref{fig:lls-strength}) that as $C_{L}$ increases the quantity 
(strength-$C_{L}$)
approaches zero following straight lines with $1/N$, but the slope
gradually decreases --which suggests that the system size dependence
of the strength gradually becomes weaker.

\begin{figure}
\centering
\includegraphics[width=9cm]{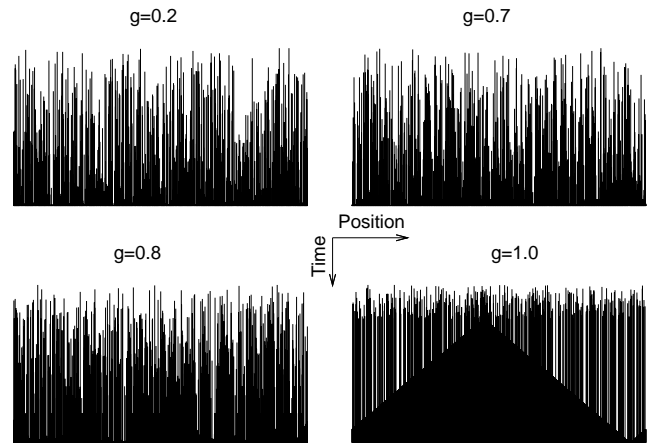}
\caption{Space time diagram of breaking sequences of the interpolating
fiber bundle model of \textcite{pch05}. Black lines represent broken fibers 
and white regions are unbroken parts of the bundle. As the interpolation
parameter $g$ is increased, there is increasing localization of the failing
fibers.}
\label{ah-fig24}
\end{figure}

In \textcite{hh94} a model interpolating between the global
load sharing fiber bundle and a variant of a local load sharing model was
introduced and studied. \textcite{k04} and \textcite{pch05}
followed up this work.  In the model studied by  \textcite{pch05}
a fraction $g$ of the load a failing fiber carries would be distributed among
its surviving neighbors and a fraction $1-g$ among all surviving fibers.
Hence, for $g=1$ the model would be purely local load sharing, whereas for
$g=0$ it would be purely global load sharing.  We show in Fig.\ \ref{ah-fig24}
space time diagrams of the one-dimensional version of the model for different
values of $g$.  For increasing values of $g$, there is increasing localization.

Both \textcite{k04} and \textcite{pch05} found a phase
transition when interpolating between the global
load sharing model and the local load sharing model discussed earlier in this
section.  In the one-dimensional model shown in Fig.\ \ref{ah-fig24}, the
critical value of $g$, is $g_c=0.79\pm0.01$ for a flat threshold distribution
\cite{pch05}.

\subsection{Intermediate load-sharing models}
\label{Sec:4B}

A crucial mechanism in brittle fracture is the stress enhancement that occurs
at crack tips.  The stress field has a $1/\sqrt{r}$ singularity, where $r$
is the distance to the crack tip, in this region.  It is the interplay between
fracture growth due to this singularity and due to weak spots in the material
that drive the development of the fracture process \cite{hr90}.  Clearly, there
is a cutoff in the stress field as $r\to 0$.  This may be caused by
non-linearities in the material constitutive relations or by microstructure
in the material such as the presence of crystallites.

\textcite{hmkh02} has introduced a fiber bundle that contains
a power-law dependence on the distance from a failing fiber on the force
redistribution in order to model the stress singularity seen around crack
tips.  The fiber bundle is implemented as a regular two-dimensional grid of
parallel fibers clamped between two stiff blocks.  Assuming that fiber $j$
has just failed, a force transfer function
\begin{equation}
\label{forcetransferhidalgo}
F(r_{i,j},\gamma) = \frac{Z}{r_{ij}^\gamma}\;,
\end{equation}
where
\begin{equation}
\label{normalizationhidalgo}
\frac{1}{Z} = \sum_{i\in I} \frac{1}{r_{ij}^{\gamma}}\;,
\end{equation}
and $I$ is the set of intact fibers, redistributes the forces. $\gamma$ is
treated as a parameter on the unit interval.  There are two limiting cases,
$\gamma\to 0$ which recovers the global load-sharing fiber bundle and
$\gamma\to\infty$ which recovers the local load-sharing model with
nearest-neighbor stress alleviation (Section \ref{Sec:4A}). The load
increase on fiber $i$ is hence given by
\begin{equation}
\label{stresshidalgo}
f_i\to f_i+\sum_{j\in B} f_j\ F(r_{ij},\gamma)\;,
\end{equation}
where $B$ is the set of failed fibers up to that point.

\begin{figure}
\resizebox{7cm}{!}{\rotatebox{-90}{\includegraphics{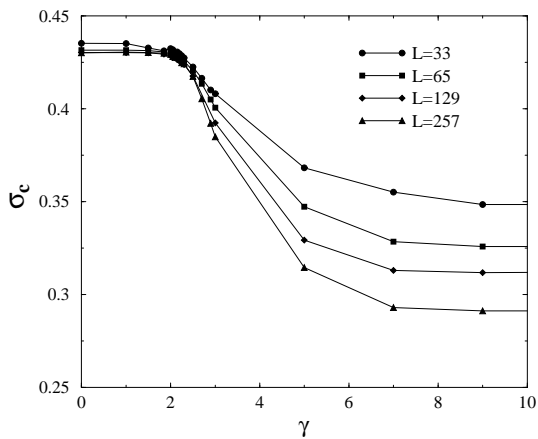}}}
\caption{Strength at failure, $\sigma_c$ as a function of $\gamma$
in the variable range fiber bundle of \textcite{hmkh02}.}
\label{ah-fig18}
\end{figure}

\begin{figure}
\resizebox{7cm}{!}{\rotatebox{-90}{\includegraphics{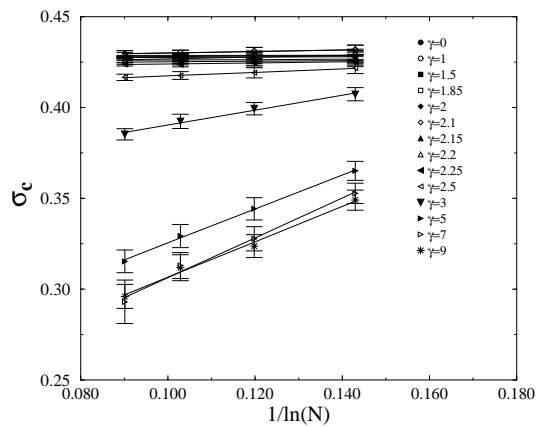}}}
\caption{Strength at failure, $\sigma_c$ as a function of the number
of fibers $N$ for different range exponents $\gamma$
in the variable range fiber bundle of \textcite{hmkh02}.}
\label{ah-fig19}
\end{figure}

This model is too complex for analytical treatment and numerical simulations
must be invoked.  Around $\gamma=2.0$ there is a transition in behavior between
essentially global load sharing and local load sharing as described in
Section \ref{Sec:4A}: For $\gamma < 2.0$, the maximum sustainable force
scales the number of fibers in the bundle at the outset,
$N$, whereas for $\gamma > 2.0$, a $N/\ln N$-behavior is observed as in
Eq.~(\ref{lFfatal}).  This is seen in Figs.\ \ref{ah-fig18} and
\ref{ah-fig19}.

\begin{figure}
\includegraphics[width=8cm]{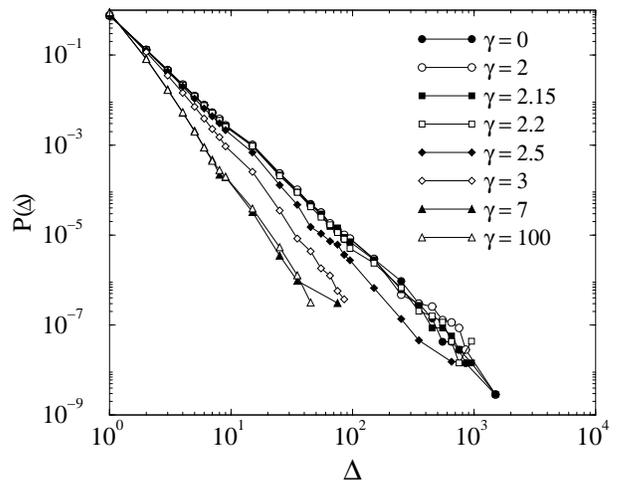}
\caption{Burst size distribution for different range exponents $\gamma$
in the variable range fiber bundle of \textcite{hmkh02}. The
data points where $\gamma \le 2.0$ can be fitted to a power law with
exponent $\xi=5/2$.  For larger $\gamma$, the slope becomes steeper
and resembling the behavior seen in the local load-sharing model, see
Fig.\ \ref{ah-fig4a-1}.}
\label{ah-fig20}
\end{figure}

The burst distribution shows a power-law distribution with exponent $\xi=5/2$
again signaling global load sharing behavior for smaller values of $\gamma$.
As $\gamma$ is increased, deviations from this behavior is seen.  This must
be interpreted as a crossover towards local load sharing behavior as
described in the previous section, see Fig.\ \ref{ah-fig20}.


\begin{figure}
\centering
\includegraphics[width=7cm]{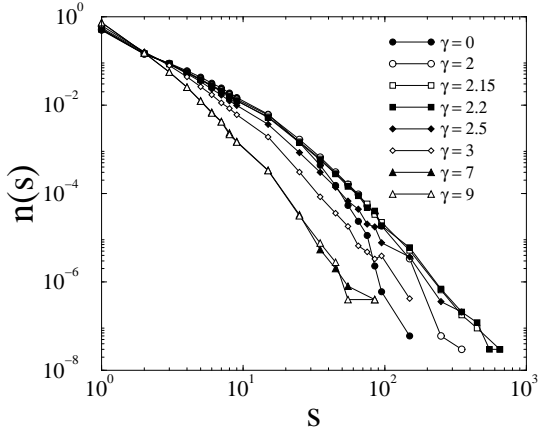}
\caption{Size distribution of clusters of broken bonds at collapse
in the variable range fiber bundle of \textcite{hmkh02}. There
is no evident power law behavior.}
\label{ah-fig21}
\end{figure}

Lastly, the structure of the clusters of failed fibers at breakdown is studied.
In the global load sharing model implemented in two dimensions does not yield
anything particular. There is a percolation transition in the cluster size
distribution when the relative density of failed fibers reaches the
percolation threshold, but this has no particular significance in the
evolution of the model. \textcite{hmkh02} find the cluster distribution
having two distinct behaviors, depending on whether $\gamma$ is smaller
than or larger than 2.  There is no clear power law behavior, see Fig.\
\ref{ah-fig21}.  

\textcite{hzh08} have studied an anisotropic version of this model.  The
force transfer function (\ref{forcetransferhidalgo}) is in this work
generalized to
\begin{equation}
F(r_{i,j},\gamma)=\frac{Z}
{(\alpha\Delta x_{i,j}^2+(1-\alpha)\Delta y_{i,j}^2))^{\gamma/2}}\;,
\label{forecetransferhidalgoani}
\end{equation}
where $\alpha$ is an anisotropy parameter.  The behavior of this model 
turns out to be quite similar to that found in the isotropic model. 

\begin{figure}
\centering
\includegraphics[width=8cm]{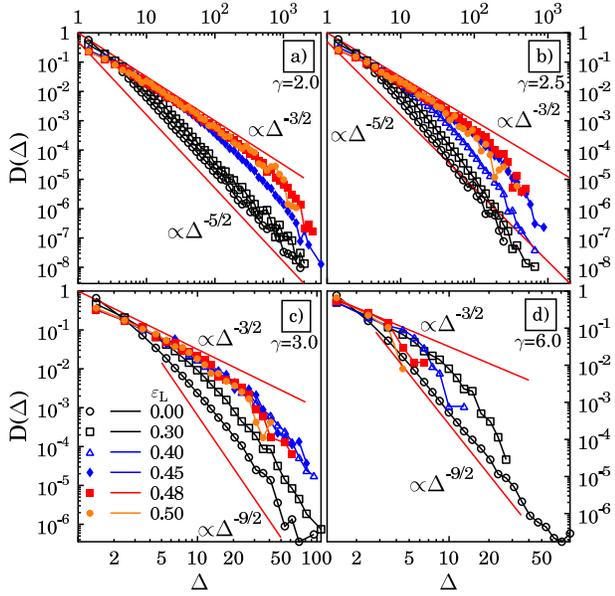}
\caption{Burst size distributions in the variable range fiber bundle
model for different $\gamma$ and $\epsilon_L$ values: a) $\gamma=2.0$,
b) $\gamma=2.5$, c) $\gamma=3.0$ and d) $\gamma=6.0$.
From \textcite{rkh06}.}
\label{ah-fig22}
\end{figure}

\textcite{rkh06} introduced a low cutoff in the
threshold distribution \cite{pbc02} in the variable range fiber
bundle model of \textcite{hmkh02}. They studied the
burst distribution as a function of $\gamma$ the
burst distribution as a function of $\gamma$ and cutoff in the
failure thresholds in terms of deformation, $\epsilon_L$.  Fig.\ \ref{ah-fig22}
summarizes their findings:  For the explored values of $\gamma$ in the range
$2.0 \le \gamma \le 6.0$, a crossover from burst exponent $\xi=5/2$ to  
$\xi=3/2$ is seen
for small $\epsilon_L$, whereas for larger $\gamma\approx 6.0$, the burst
distribution may be fitted to a value $\xi=9/2$ for small $\epsilon_L$.

\textcite{ng91} and \textcite{ngdpt94} introduced a fiber bundle model
where the fibers are hierarchically organized.  The fibers are paired
two by two.  Each of these pairs is seen as an ``order one" fiber.  This
pairing is repeated for the order one fibers, creating order two fibers
and so on.  Within each sub-bundle, the fibers are subject to equal load
sharing.  If the cumulative threshold distribution for the fibers at order
zero is $P_0(x)=P(x)$, then the threshold distribution at level one is
\begin{equation}
\label{levelonegabrielov}
P_1(x)=P_0(x)[2P_0(2x)-P_0(x)]\;,
\end{equation}
a result which is readily generalized to any level.  Even though the
starting point here is global load sharing, the approximation introduced
by treating the fibers at each level as fibers with given thresholds,
leads to the introduction of spatial load dependence in the model.  To
our knowledge, bursts have not been studied within this framework.

\subsection{Elastic medium anchoring}
\label{Sec:4C}

In this Section we generalize the fiber bundle problem to include more
realistically  the elastic response of the surfaces to which the fibers are
attached.  So far, these have been assumed to be infinitely stiff for
the equal load sharing model, or their response has been modeled  as very
soft, but in a fairly unrealistic way in the local load sharing models, see
Section \ref{Sec:4A}.  We will end up with a description that is somewhat
related to the models of the previous Section, in particular the model of
\textcite{hmkh02}.   In \textcite{bhs02}, a realistic model for the
elastic response of the clamps was studied.  The model was presented in
the context of  the failure of weldings. In this language, the two clamps
were seen as elastic media glued together at a common interface.

Without loss of generality, one of the media may be assumed to be infinitely
stiff whereas the other is soft.  When a force is applied to a given fiber, the
soft clamp responds by a deformation falling off inversely to the distance from
the loaded fiber.  Hence, the problem becomes one of solving the response
of the surface with respect to a given loading of the fibers.  Fibers exceeding
their maximally sustainable load fail, and the forces and deformations must be
recalculated.
The two clamps can be pulled apart by controlling (fixing) either the
applied force or the {\it displacement.\/} The displacement is defined
as the change in the distance between two points, one in each clamp
positioned far from the interface. The line connecting
these points is perpendicular to the average position of the
interface. In our case, the pulling is accomplished by controlling the
displacement. As the displacement is increased slowly, fibers
will fail, eventually ripping the two surfaces apart.

We now concretize these ideas in a model.  It  consists of a
two-dimensional square $L\times L$ lattices with periodic boundary
conditions. The lower one represents the hard, stiff surface and the
upper one the elastic surface. The nodes of the two lattices are
matched, (i.e., there is no relative lateral displacement). The
fibers are modelled as in the previous sections: elastic up to a
threshold value which has been individually chosen for each fiber
from some threshold distribution.  The spacing between the fibers
is $a$ in both the $x$ and $y$ directions.  The force that each fiber
is carrying is transferred over an area of size $a^2$ to the soft clamp:
As the two clamps are
separated by controlling the displacement of the hard clamp relative to
the zero level, $D$,
the forces carried by the fibers increases from zero.
When the force carried by a fiber reaches its breaking threshold, it breaks
irreversibly and the forces redistribute themselves through the
deformation of the soft clamp.  Hence, the fibers are broken
one by one until the two clamps are no longer in mechanical contact.
The force, $f_i$, carried by the $i$th fiber is given by
\begin{equation}
\label{M1}
f_i = -k(u_i-D)\;,
\end{equation}
where $k$ is the spring constant
and $u_i$ is the deformation of the elastic clamp at site
$i$. All unbroken fibers have $k=1$ while a broken fiber has
$k=0$. The quantity $(u_i-D)$ is, therefore, the length, and since $k=1$,
also the force carried by fiber $i$.
The deformation of the soft clamp is described by the coupled system of
equations,
\begin{equation}
\label{M2}
u_i=\sum_{j} G_{i,j} f_j\;,
\end{equation}
where the elastic Green function, $G_{i,j}$ is given by \cite{ll58,j85}
\begin{equation}
\label{M3}
G_{i,j} = \frac{1-s^2}{\pi e a^2}\ \int_{-a/2}^{+a/2}\int_{-a/2}^{+a/2}\
\frac{dx'\ dy'}{|(x-x',y-y')|}\;.
\end{equation}
In this equation, $s$ is the Poisson ratio, $e$ the elastic constant,
and $|{\vec i}-{\vec j}|$ the distance between sites $i$ and $j$. The
indices $i$ and $j$ run over all $L^2$ sites.  The integration over the
area $a^2$ is done to average the force from the fibers over this area.
As remarked by \textcite{bhs02}, the
Green function  (Eq. \ref{M3}) applies for a medium occupying the infinite half
space. However, with a judicious choice of elastic constants, it may
be used for a finite medium if its range is small compared to $L$, the
size of the system.

By combining Eq. (\ref{M1}) and Eq. (\ref{M2}), one obtains
\begin{equation}
\label{M4}
({\bf I} +{\bf K G}) {\vec f} = {\bf K} {\vec D}\;,
\end{equation}
where matrix-vector notation is used. ${\bf I}$ is the $L^2\times
L^2$ identity matrix, and ${\bf G}$ is the Green function represented
as an $L^2\times L^2$ dense matrix. The constant vector ${\vec D}$ is
$L^2$ dimensional. The {\it diagonal\/} matrix ${\bf K}$ is also
$L^2\times L^2$. Its matrix elements are either 1, for unbroken
fibers, or 0 for broken ones.

Once Eq.~(\ref{M4}) is solved for the force, ${\vec f}$,
Eq.~(\ref{M2}) yields the deformations of the elastic clamp.

Eq.~(\ref{M4}) is of the familiar form ${\bf A}{\vec x}={\vec
b}$.  Since the Green function connects all nodes to all other nodes,
the $L^2\times L^2$ matrix ${\bf A}$ is dense which puts severe limits
on the size of the system that may be studied.

The simulation proceeds as follows: One starts with all springs present,
each with its stochastic breakdown threshold. The two media are
then pulled apart, the forces calculated using the Conjugate
Gradient algorithm (CG) \cite{bh88,ptvf92}, and the fiber
which is the nearest to its threshold is broken, i.e., the
matrix element corresponding to it in the matrix ${\bf K}$ is
zeroed. Then the new forces are calculated, a new fiber broken and so
on until all fibers have failed.

However, there are two problems that render the simulation of large
systems extremely difficult from a numerical point of view.
The first is that since ${\bf G}$ is
$L^2\times L^2$ {\it dense\/} matrix, the number of operations per CG
iteration scales like $L^4$. Even more serious is the fact that as the
system evolves and springs are broken, the matrix $({\bf I}+k{\bf G})$
becomes ill-conditioned.
To overcome the problematic $L^4$ scaling of the algorithm, the
matrix-vector multiplications are done in Fourier space since the
Green function is diagonal in this space.  Symbolically, these multiplications
may be written as follows,
\begin{equation}
\label{M5}
({\bf I} + {\bf K} {\bf F^{-1}F}{\bf G}){\bf F^{-1}F} {\vec f} = {\bf
K} {\vec D}\;,
\end{equation}
where ${\bf F}$ is the FFT (fast Fourier transform) operator and
${\bf F^{-1}}$ its inverse
(${\bf F^{-1}F}=1$). Since ${\bf I}$ and ${\bf K}$ are diagonal,
operations involving them are performed in real space. With this
formulation, the number of operations/iteration in the CG algorithm
now scales like $L^2\ln(L)$ rather than $L^4$.

To overcome the ill-conditioning of the matrix $({\bf I}+k{\bf G})$
we need
to precondition the matrix \cite{bh88,bhn86}. This means that instead of
solving Eq.~(\ref{M5}), one solves the equivalent problem
\begin{equation}
\label{M6}
{\bf Q}({\bf I} + {\bf K} {\bf F^{-1}F}{\bf G}){\bf F^{-1}F} {\vec f}
= {\bf Q} {\bf K} {\vec D}\;,
\end{equation}
where we simply have multiplied both sides by the arbitrary, positive
definite preconditioning matrix ${\bf Q}$. Clearly, the ideal choice
is ${\bf Q_{0}}=({\bf I} + {\bf K}{\bf G})^{-1}$ which would always
solve the problem in one iteration. Since this is not possible in
general, we look for a form for ${\bf Q}$ which satisfies the
following two conditions: (1) As close as possible to ${\bf Q_{0}}$,
and (2) fast to calculate. The choice of a good ${\bf Q}$ is further
complicated by the fact that as the system evolves and fibers are
broken, corresponding matrix elements of ${\bf K}$ are set to
zero. So, the matrix $({\bf I} + {\bf K}{\bf G})$ evolves from the
initial form $({\bf I} + {\bf G})$ to the final one ${\bf I}$.
\textcite{bhs02} did not find a fixed ${\bf Q}$
that worked throughout the entire breakdown process.
They therefore chose the form
\begin{equation}
\label{M7}
{\bf Q}={\bf I}-({\bf K}{\bf G})+({\bf K}{\bf G})({\bf K}{\bf G})
-({\bf K}{\bf G})({\bf K}{\bf G})({\bf K}{\bf G})+ ...
\end{equation}
which is the Taylor series expansion of ${\bf Q_{0}}=({\bf
I} + {\bf K}{\bf G})^{-1}$. For best performance, the number of terms
kept in the expansion is left as a parameter since it depends on the
physical parameters of the system. It is important to emphasize the
following points. (a) As fibers are broken, the preconditioning
matrix evolves with the ill-conditioned matrix and, therefore, remains
a good approximation of its inverse throughout the breaking
process. (b) All matrix multiplications involving ${\bf G}$ are done
using FFTs. (c) The calculation of ${\bf Q}$ can be easily organized
so that it scales like $n L^2 \ln(L)$ where $n$ is the number of terms
kept in the Taylor expansion, equation (\ref{M7}).
The result is a stable accelerated algorithm which scales
essentially as the volume of the system.

\begin{figure}
\centering
\includegraphics[width=6cm,angle=0]{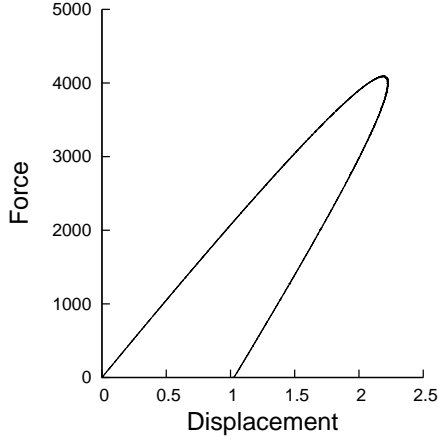}
\caption{Force-displacement curve, $128\times 128$ systems with $e=10$.}
\label{ah-fig5}
\end{figure}

\begin{figure}
\centering
\includegraphics[width=6cm,angle=0]{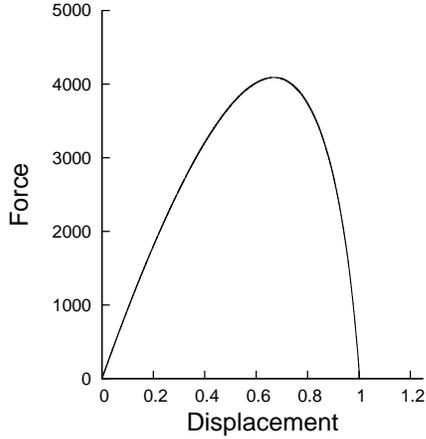}
\caption{Force-displacement curve, $128\times 128$ systems with $e=100$.}
\label{ah-fig6}
\end{figure}

Fig.~\ref{ah-fig5} shows the force-displacement curve for a system of
size $128\times 128$ and
elastic constant $e=10$.  Whether we control the applied force, $F$,
or the displacement, $D$, the system will eventually suffer
catastrophic collapse.  However, this is not so when $e=100$ as shown
in Fig.~\ref{ah-fig6}.  In this case, only controlling the force will lead
to catastrophic failure.  In the limit when $e\to\infty$, the model
becomes the equal load-sharing fiber bundle model,
where $F=(1-D)D$.  In this limit there are no spatial correlations and
the force instability is due to the decreasing total elastic
constant of the system making the force on each surviving bond
increase faster than the typical spread of threshold values. No such
effect exists when controlling displacement $D$.  However, when the
elastic constant, $e$, is small, spatial correlations in the form of
localization, where fibers that are close in space have a tendency to
fail consecutively, do develop, and these are responsible for the displacement
instability which is seen in Fig.~\ref{ah-fig5}.

\begin{figure}
\centering
\includegraphics[width=6cm,angle=0]{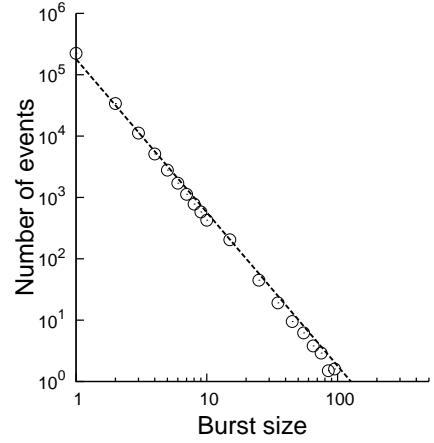}
\caption{Burst distribution for $128\times 128$, $e=10$.  The slope of
the straight line is $-2.5$.}
\label{ah-fig7}
\end{figure}

\begin{figure}
\centering
\includegraphics[width=6cm,angle=0]{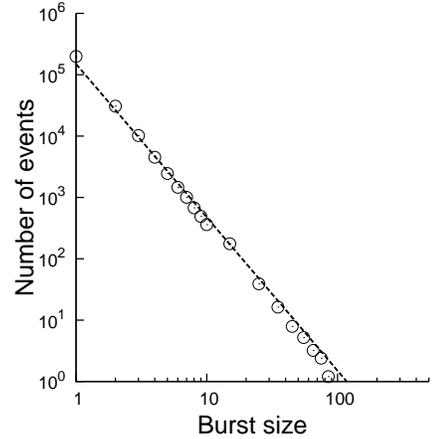}
\caption{Burst distribution for $128\times 128$, $e=100$.  The slope of
the straight line is $-2.5$.}
\label{ah-fig8}
\end{figure}

We now turn to the study of the burst distribution.
Figs.~\ref{ah-fig7} and \ref{ah-fig8} show the burst distribution for
$e=10$ and 100. In both cases we find that the burst distribution
follows a power law with an exponent $\xi=2.6\pm0.1$.
It was argued in Ref.\ \cite{bhs02} that the value of $\xi$ in this case
is indeed $5/2$ as in the global load sharing model.  These two figures should
be compared with Fig.\ \ref{ah-fig20} showing the burst distribution
in the variable range fiber bundle model of Hidalgo et al., \cite{hmkh02},
where $\xi=5/2$ is recovered as long as the range exponent $\gamma$ is small,
rendering the forces long range among the fibers.

\begin{figure}
\centering
\includegraphics[width=6cm,angle=0]{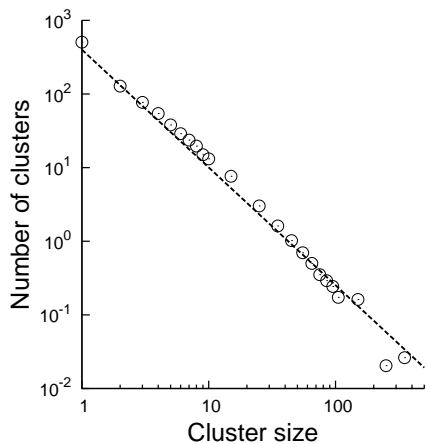}
\caption{Area distribution of zones where glue has failed for systems of
size $128\times 128$ and elastic constant $e=10$.  The straight line
is a least square fit and indicates a power law with exponent $-1.6$.}
\label{ah-fig10}
\end{figure}

As the failure process proceeds, there is an increasing competition
between local failure due to stress enhancement and local failure due
to local weakness of material.  When the displacement, $D$, is the
control parameter and $e$ is sufficiently small (for example $e=10$),
catastrophic failure eventually occurs due to localization. The onset
of this localization, i.e., the catastrophic regime, occurs when
the two mechanisms are equally important. This may be due to self-organized
criticality \cite{btw87} occurring at this
point.  In order to test whether this is the case,
\textcite{bhs02} measured
the size distribution of broken bond clusters at the point when $D$
reaches its maximum point on the $F-D$ characteristics, {\it i.e.} the
onset of localization and catastrophic failure.  The analysis was
performed using a Hoshen-Kopelman algorithm \cite{sa94}.  The result is
shown the
in Fig.~\ref{ah-fig10}, for 56 disorder realizations, $L=128$ and
$e=10$. The result is consistent with a power law distribution with
exponent $-1.6$, and consequently with self organization.  If this process
were in the universality class of percolation, the exponent would have been
2.05.  Hence, we are dealing with a new universality class in this system.
This behavior should be contrasted to the one seen in the variable range
fiber bundle
model studied by \textcite{hmkh02}
where no power law distribution was found  (see Fig.\ \ref{ah-fig21} in Section \ref{Sec:4B}).

\section{Fiber bundles in material science and other applications}
\label{Sec:5}

The aim of this Section is to demonstrate how the fiber bundle model
may be used as a tool for studying both important phenomena occuring in
materials such as fatigue and for studying important classes of materials,
notably fiber reinforced composites.  A general review has recently been
written by \textcite{mb09} on this subject.  We also
discuss applications of fiber bundle models in other contexts, ranging from
traffic modeling to earthquake dynamics.

\subsection{Time dependent failure: Fatigue or creep phenomena}

Materials may undergo time dependent deformation under steady load. 
Sometimes when a load is applied,
though surviving at first stage, the system fails after long time. This 
type of failure is referred as the fatigue-failure or creep rupture 
\cite{l93,c94}. The
fatigue-failure is basically a thermally activated process \cite{pt83}
and originates at the atomic level of the fibers where the molecules
are in random thermal vibrations. Eventually a molecule acquires sufficient
thermal energy to overcome the local energy barrier and slips relative
to other molecules. The frequency of such events is greatly enhanced
by increases in temperature, stress and impurity level. After a molecular
slip or rupture, neighboring molecules become overloaded and the failure
rate increases. These molecular failures accumulate locally and produce
microcracks within the material. Also micro-cracks can grow at the
crack-tips with time due to chemical diffusion in the atmosphere \cite{l93}
which helps the growth of fractures. These failures nucleate
around the defects in the solid and the failure behavior and its statistics 
therefore crucially depends on the disorder or impurity distribution
within the sample. The system then fails under a stress less than
its normal strength ($\sigma_{c}$) and the failure time ($\tau$)
depends on both, the applied load and the impurity level. 

Fatigue-failure in fiber bundle model was first studied by 
  \textcite{c56,c57a} considering different classes of fibers
and several breakdown rules. The probabilistic analysis gives the
lifetime distribution under various loading conditions: Constant load,
loads proportional to time and periodic loads. The `time dependent
fatigue' and `cycle dependent fatigue' both are addressed in this work
introducing the concept of `memory' effect, i.e., the load history
can affect the failure of fibers. Such time dependent failure
in fiber bundles have been considerably extended and generalized by
Phoenix et al. \cite{p78,p79,pt83,np01}
for equal load sharing (ELS) and local load sharing (LLS) bundles. 
The approximate fatigue life time distributions
have been achieved through probabilistic analysis introducing power
law breakdown rule and exponential breakdown rule at the molecular
level \cite{pt83}.

When dry fibers are replced by viscoelastic elements having time dependent 
deformation properties, fiber bundle model exhibits creep behavior 
\cite{hkh02,khhp03} in terms of 
the macroscopic response under constant external load. There exists a critical 
load (or stress) below which the deformation attains a constant value 
(infinite lifetime) and above the critical load, deformation increases 
monotonically resulting global failure (finite lifetime). Another extension of
classical fiber bundle model, the continuous damage model \cite{hkh01,khhp03}, 
captures similar
 creep behavior assuming that a fiber can fail more than once and at each 
failure its stiffness gets reduced by a constant factor.
In both of these models the lifetime of the bundle diverges at the critical 
load following robust power law variation with the applied load.   

Also, few experiments \cite{pm93,bc01,kcfaszh07} have been performed
to observe the failure time of materials and its statistics.
The effect of thermal activation and disordered
noise on the failure have recently been measured for material breakdown
and approximate fatigue behavior has been obtained 
\cite{gsc99,scg01,gcgzs02,r00,pc03b,kcfaszh07} using fiber bundle models.

In this section, we discuss several approaches of achieving fatigue-failure
behavior in equal load sharing fiber bundle models. The approaches
differ basically in the way how time-dependence has been incorporated in the 
failure process.

\subsubsection{Thermally induced failure in fiber bundles}

The influence of  noise on macroscopic failure in fiber bundle model,
  has been studied numerically \cite{gsc99,cgs01,scg01,gcgzs02},
 using both disorder noise and thermal noise. 
The strength of each fiber is characterized by a {\it critical
stress} $x_i^{(c)}$ which is  a random variable that
follows a normal distribution of mean $x^{(c)}$ and variance
$KT_d$:
\begin{equation}
x_i^{(c)}= x^{(c)}+ N_d(KT_d),
\label{thermal_eq1}
\end{equation}
where $K$ is the Boltzmann constant and  $N_d$ is  the disorder noise. 
Again each fiber is subjected to an additive time dependent random 
stress $\Delta
x_i(t)$ which follows a zero  mean normal distribution of variance
$KT$:
\begin{equation}
\Delta x_i(t)= N_T(t,KT).
\label{thermal_eq2}
\end{equation}
Here $N_T$ is the thermal noise.

\begin{figure}
\includegraphics[width=6cm,height=6cm]{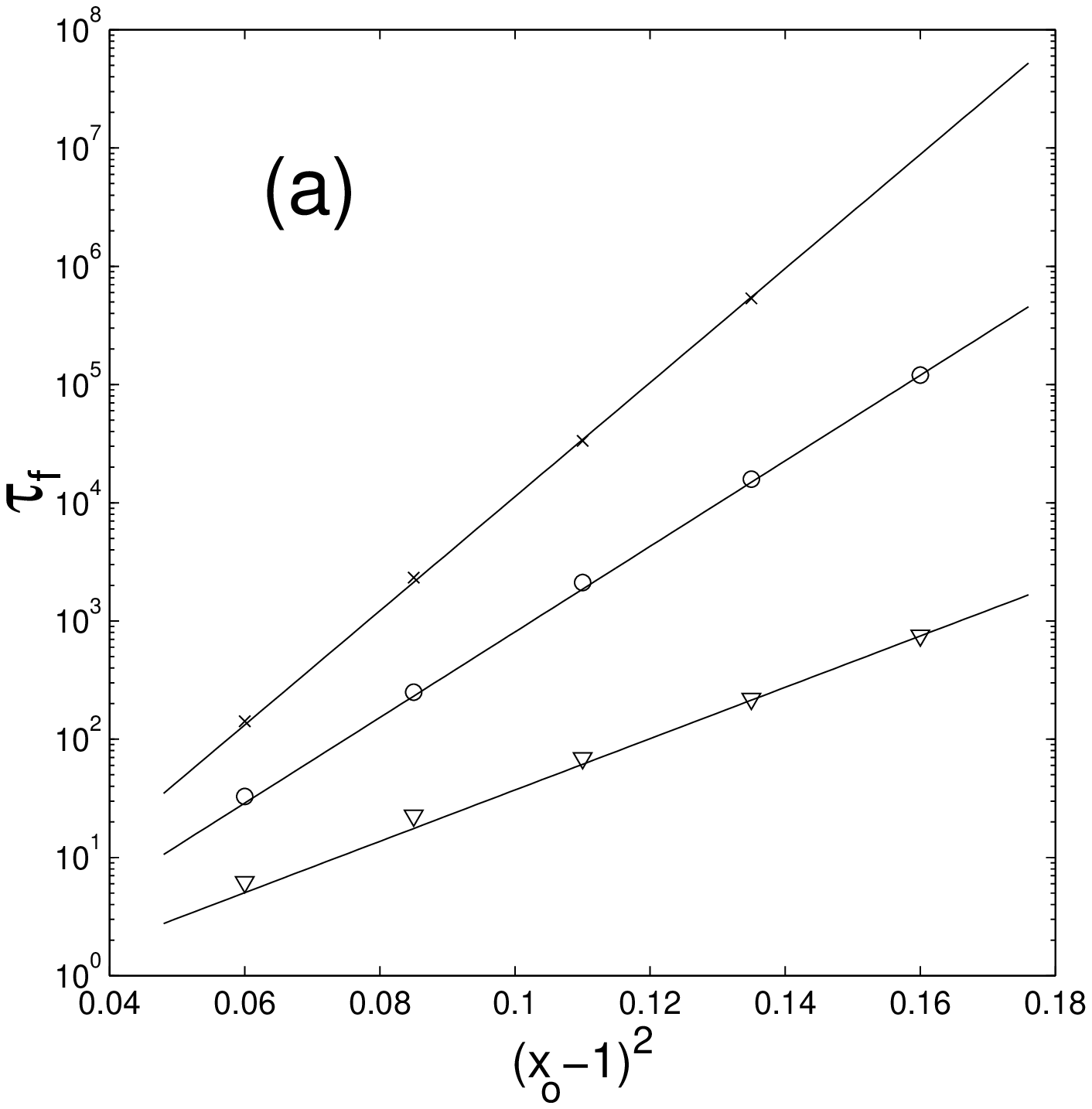}
\includegraphics[width=6cm,height=6cm]{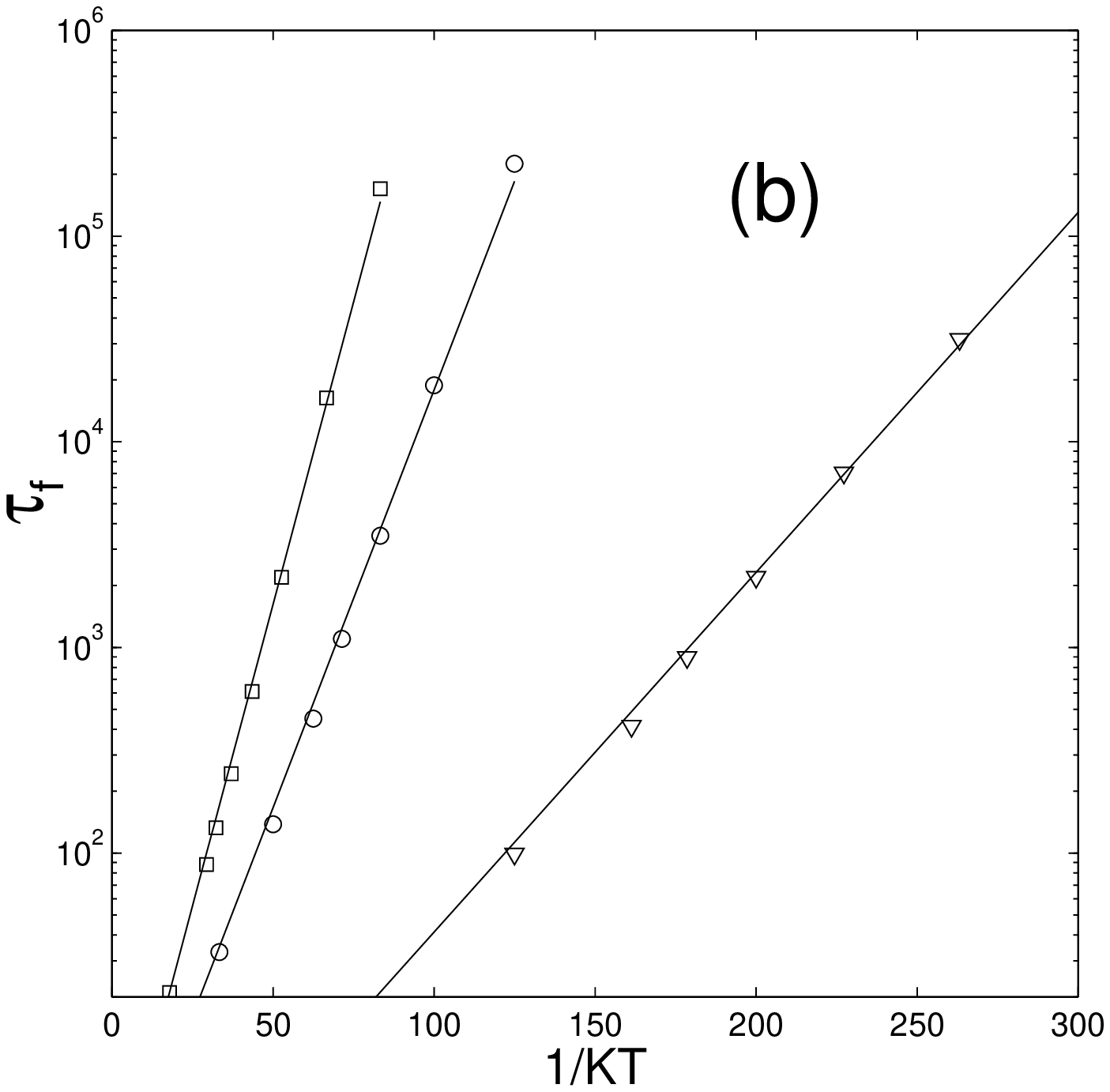}
\caption{
 Failure time $\tau _f$ of an homogeneous bundle ($KT_d=0$) in a
creep test.
 (a) $\tau _f$ as a function of  the normalized force
$(1-x_0)^2$ for several values of thermal noise variance $KT$:
 (cross) $KT=0.0045$; (circle) $KT=0.006$; (triangle) $KT=0.01$.
(b) $\tau _f$ as a function of $1/KT$ for several values of
 $x_0$: (box) $x_0=0.45$; (circle) $x_0=0.54$; (triangle) $x_0=0.7$. 
 Continuous lines in (a) and (b) are  the
fits with Eq. (\ref{thermal_eq4}) and Eq. (\ref{thermal_eq5}), respectively.
 Adapted from  \textcite{cgs01}. }
\label{thermal1}
\end{figure}

 Due to the equal load sharing scheme, if a number $n(t)$ of
fibers are broken at time $t$ after force $F$ is applied on the bundle, 
the local force on each of the remaining fibers will be:
\begin{equation}
x_i(t)=\frac {x_0N}{N-n(t)}+\Delta x_i(t),
\label{thermal_eq3}
\end{equation}
where $N$ is the total number of fibers in the intact bundle and $x_0=F/N$ is 
the initial force per fiber.
If $KT=0$, the model reduces to the static one.
In that case the applied force is  increased linearly 
from zero to the critical value $F_c$ above which the whole
bundle breaks. Therefore, at  a constant force $F$, the
bundle  breaks in a single avalanche only if $F>F_c$, otherwise it will never 
break. 
If $KT \ne 0$ then the system can break at an applied force $F<F_c$ due to the
thermal effect. Such thermal failure of the model has been studied 
numerically as a function of $x_0$, $KT$ and $KT_d$ \cite{gsc99,cgs01}.

\begin{figure}
\includegraphics[width=6cm,height=6cm]{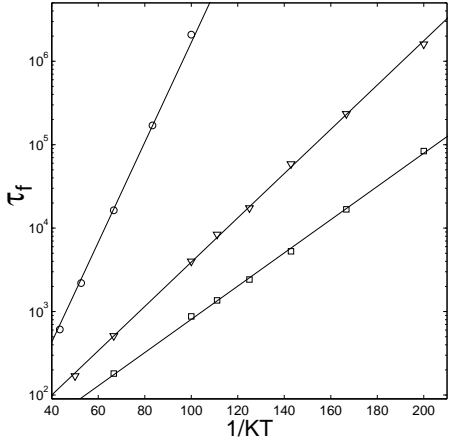}
\caption{Failure time $\tau _f$ of an heterogeneous bundle ($KT_d\neq 0$)
in creep test. The life time $\tau _f$ is plotted as a function 
of $1/KT$ at  $x_0=0.45$. Different symbols correspond to  different values of 
 $ KT_d$: (Circle) $KT_d=0$; (triangle) $KT_d=0.02$; (box) $KT_d=0.04$. 
Adapted from \textcite{cgs01}.}
\label{thermal2}
\end{figure}

 When   $KT\ne 0$, the failure time $\tau _f $ as a function of 
$(1-x_0)^2$ follows an exponential law  for any fixed value of $KT_d$:
\begin{equation}
\tau _f \sim \exp \left[\alpha (1-x_0) ^2\right],
\label{thermal_eq4}
\end{equation}
where $\alpha$ is a fitting parameter [Fig. \ref{thermal1}(a)], which is a 
function of $KT$.  At constant stress, the failure time  depends on thermal 
noise $KT$ as:
\begin{equation}
\tau _f \sim \tau _0 \exp \left( \frac A{KT}\right),
\label{thermal_eq5}
\end{equation}
where A is a function of $x$ [Fig. \ref{thermal1}(b)]. Similar result has also 
been observed in case of a heterogeneous fiber bundle [Fig. \ref{thermal2}]. 

 One can compare these results with
 Pomeau's theory \cite{p92} for the failure time of solids: 
\begin{equation}
\tau _f =\tau _o \exp{\left( \alpha_g \frac{\Gamma_s
^dY^{(d-1)}}{KT_{eff} \ P_s^{(2d-2)}}\right) }, 
\label{thermal_eq6}
\end{equation}
where $P_s$ is the imposed stress, $\tau _o$ is a constant, $\Gamma_s $ the 
surface energy, $Y$
the Young modulus,  $\alpha_g$ is a
constant which depends on the geometry, $T_{eff}$ is an effective
temperature and $d$ the dimensionality of the system.  This theory is 
based on the physical argument that thermal activation of
 micro cracks \cite{p92,gf91} is responsible for the macroscopic
failure of the material. It can be noted that the functional dependence of 
$\tau _f$ on stress is different for fiber bundle model and for solids. 
The main reason of which is the different geometry of the stress distribution 
in the fiber bundle and the solids.

These numerical studies suggest that disorder noise amplifies the effect of 
thermal noise and  reduces the dependence of $\tau _f$ on the
temperature and this can  explain the recent
experimental observations  on
microcrystals \cite{pm93}, gels \cite{bkpdm98} and macroscopic
composite materials \cite{ggc99}. 

The numerical observations described above have been confirmed later through an 
analytic investigation by \textcite{r00}. In a homogeneous (no disorder in 
fiber strengths) fiber bundle model, force or stress on each fiber is 
\begin {equation}
x = x_0 + \eta, 
\end{equation}  
where $x_0 = F/N$ and $\eta$ is the random noise with a Gaussian distribution

\begin{equation}
p(\eta) = \frac 1{\sqrt{2 \pi KT}} \exp{\left(-\frac{ \eta^2} {2KT}\right)}, 
\end{equation}
with zero mean and variance $KT$. Now the probability that one fiber survives after $t$ time step is 

\begin {equation}
p_1 (t) = [1-P(1-x_0)]^t , 
\end {equation}
where $P$ is the cumulative probability. Then the probability that the entire 
bundle can survive after $t$ time step is 

\begin {equation}
p_N (t) = [1-P(1-x_0)]^{Nt} . 
\end {equation}
Therefore the average failure time is 
\begin {equation}
 \langle\tau_1\rangle = \frac {-1}{N \ln [1-P(1-x_0)]} . 
\end {equation}
After the first fiber breaks, the situation remains the same with a smaller 
bundle and larger stress. Thus the average failure time after $i-1$ broken 
fiber is 
\begin {equation}
 \langle\tau_i\rangle = \frac {-1}{(N-i)\ln{\{1-P[1-Nx_0 /(N-i)]\}}} . 
\end {equation}

Now  the total failure time can be obtained by taking the sum over all $i$ as
\begin {equation}
 \langle\tau_f\rangle =\sum_{i=1}^N   \frac {-1}{(N-i)\ln{\{1-P[1-Nx_0 /(N-i)]\}}} . 
\end {equation}
 
When $N$ is large, one can replace the sum by a continuous integral:
\begin {eqnarray*}
\langle\tau_f\rangle &  = &  N^{-1}\int_0^N   \frac {-N}{(N-y)\ln{\{1-P[1-Nx_0 /(N-y)]\}}}dy\\
         &  = &  \int_{x_0}^\infty \frac {-1}{\ln{[1-P(1-z)]}} \frac {dz}{z}.
\end {eqnarray*}

To achieve a closed form equation it has been considered that the above sum 
is dominated by the time required for breaking the first fiber when $x_0$ is 
much smaller than the maximum load $x_c$ and when $KT<<1$. Then $P$ can be 
considered to be much smaller than $1$. Now the derivative of the average time 
with respect to $x_0$ gives    
 
\begin {equation}
\frac {\partial \langle\tau_f\rangle}{\partial x_0} =  \frac {1}{x_0 \ln{[1-P(1-x_0 )]}} \approx \frac {1}{x_0 P(1-x_0)} . 
\end {equation}
When $(1-x_0)^2 \gg KT$ the error function can be expanded as 

\begin {equation}
P(1-x_0) =  \frac {\sqrt{KT} \exp{\left( -\frac{(1-x_0)^2}{2KT}\right)  } }{\sqrt{2\pi} (1-x_0)} [1+ \mathcal{O}(KT)] . 
\end {equation}
Finally taking into account the dominating terms, one gets 
\begin {equation}
\langle\tau _f\rangle =  \frac {\sqrt{2 \pi KT}}{x_0} \exp{\left( -\frac{(1-x_0)^2}{2KT}\right) } . 
\end {equation}
These analytic expressions are identical to what have been observed earlier
 \cite{gsc99,scg01,gcgzs02,cgs01}  in numerical simulations. A similar 
analysis \cite{r00}  shows that when fiber 
strengths are distributed (heterogeneous case), the average first failure time 
can be expressed as 
   
\begin {equation}
\langle\tau _1\rangle =  \frac {\sqrt{2 \pi}}{N} \frac {(1-x_0)}{\sqrt{K(T+\Theta)}} \exp{\left( -\frac{(1-x_0)^2}{2K(T+\Theta)}\right) } ,
\label{roux_disorder} 
\end {equation}
where $\Theta$ is an effective temperature that is added to the temperature $T$
due to the disorder. Hence, the disorder leads to an effective temperature
$T_{eff}=T+\Theta$, and this is what 
\textcite{scg01} proposed: time independent heterogeneities 
of the system modify the effective temperature. 

\textcite{pcs02} estimate the total time to failure for the thermally 
activated fiber bundle model and find the same behavior as in 
Eq. (\ref{roux_disorder}): The disorder adds a constant
to the temperature of the fiber bundle.  In \textcite{gvsc06}, these results 
are generalized to the two-dimensional fuse model.

\subsubsection{Noise induced failure in fiber bundles}

Not only the temperature, but several other factors  
can result in fatigue-failure in materials: Weather effects, 
chemical effects etc. \cite{l93,c94}. Recently, 
there has been an attempt \cite{pc03b} to incorporate all the noise effects 
through a single 
parameter $KT$ ($K$ is Boltzmann constant) which can directly 
influence the failure probability of the individual elements.
Such a failure probability $p(\sigma,KT$) at any applied stress $\sigma$, 
induced
by a non-zero noise $KT$, has been 
formulated as:
 \begin{equation}
p(\sigma,KT)=\left\{ \begin{array}{cc}
\frac{\sigma}{x}\exp\left[-\frac{1}{KT}\left(\frac{x}{\sigma}-1\right)\right], & 0\leq\sigma\leq x\\ 
1, & \sigma>x\end{array}\right\} ,\label{fat-prob}\end{equation}
where $x$ is the failure strength of an element.
Clearly, the failure probability increases as $\sigma$ and $KT$ increases. 
Without any noise ($KT=0$) the model is
trivial: The bundle remains intact for stress $\sigma<\sigma_{c}$
and it fails completely for $\sigma\geq\sigma_{c}$ where $\sigma_c$ is the 
critical stress value. 

\begin{figure}
\includegraphics[width=6cm,height=5cm]{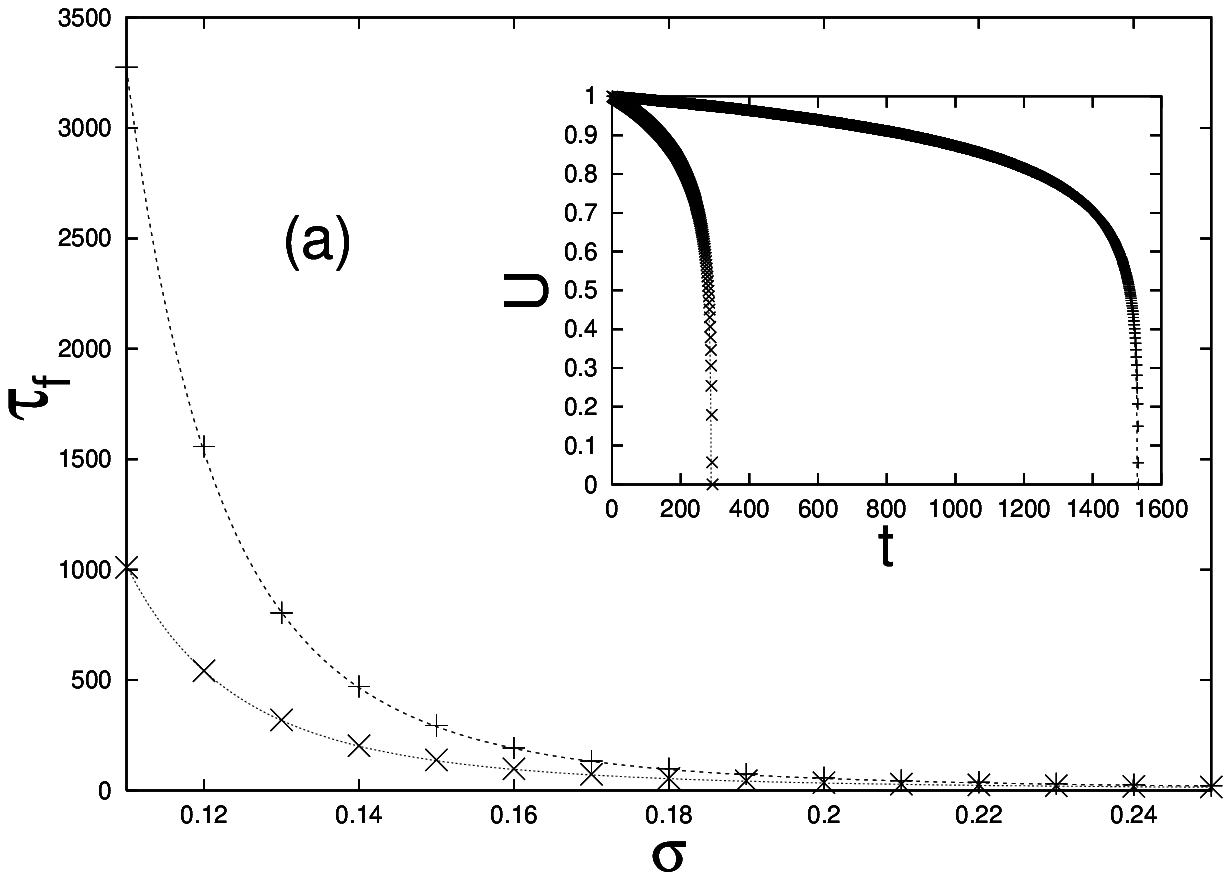}

\includegraphics[width=6cm,height=5cm]{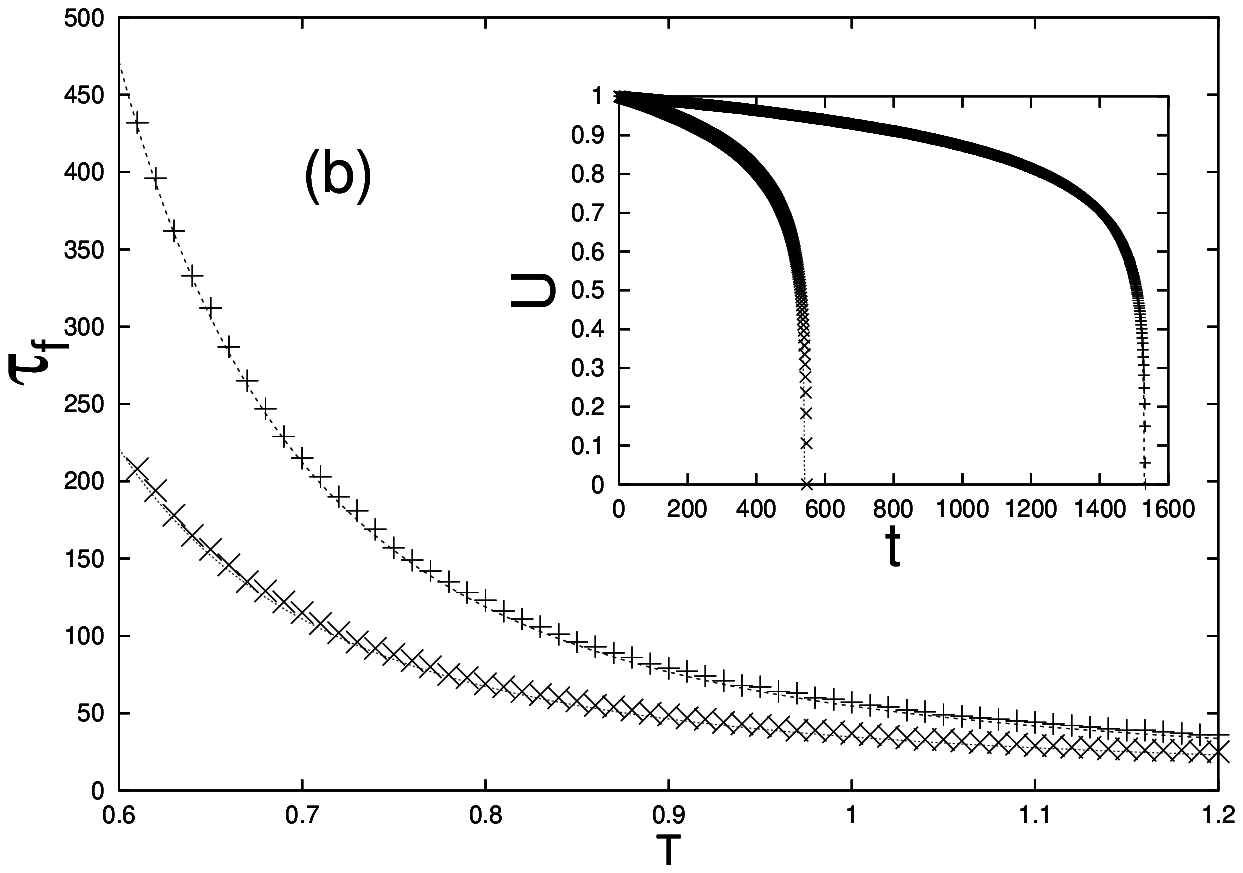}
\caption{The simulation
results showing variation of average failure time $\tau_f$ against
(a) external stress $\sigma$ and (b) against noise $T$ ($K=1$ here), for a 
homogeneous bundle containing $N=10^{5}$ fibers. The insets show 
the variation of the fraction $U$ of unbroken fibers with time $t$ for
different $T$ values [1.2 (cross) and 1.0 (plus)] in (a) and different $\sigma$
values [0.15 (cross) and 0.12 (plus)] in (b). The dotted and dashed
lines represent the theoretical result Eq. (\ref{fat-time2}). }
\label{fig:fat_homo}
\end{figure}
At $KT\neq0$ and under any stress $\sigma$ ($<\sigma_{c}$),
some fibers (weaker fibers) fail and the total load has to be supported
by the surviving fibers, which in turn enhances their stress value,
inducing further failure. The bundle therefore fails at $\sigma<\sigma_{c}$ 
after a finite time $\tau_f$.

In case of homogeneous bundle (all fibers have the same strength $x$), the critical stress value for the bundle is $\sigma_c=x$. Then the  
time dynamics at an applied stress $\sigma$ can be written \cite{pc03b}  as
 \begin{equation}
U_{t+1}=U_{t}\left[1-p\left(\frac{\sigma}{U_{t}},KT\right)\right],\label{fat-recur}
\end{equation}
where $U_{t}$ is the fraction of total fibers remains intact after time 
step $t$.
In the continuum limit, we can write the above recursion relation in a differential form \begin{equation}
-\frac{dU}{dt}=\frac{\sigma}{\sigma_{c}}\exp\left[-\frac{1}{KT}\left(\frac{\sigma_{c}}{\sigma}U-1\right)\right].\label{fat-diff}\end{equation}
The solution gives the failure time 
\begin{equation}
 \tau_f=\int_{0}^{\tau_f}dt=\frac{\sigma_{c}}{\sigma}\exp\left(-\frac{1}{KT}\right)\int_{0}^{1}\exp\left[\frac{1}{KT}\left(\frac{\sigma_{c}}{\sigma}\right)U\right]dU.\label{fat-time1}\end{equation}
Hence, for $\sigma<\sigma_{c}$ \begin{equation}
\tau_f =KT\exp\left(-\frac{1}{KT}\right)\left[\exp\left(\frac{\sigma_{c}}{\sigma KT}\right)-1\right].\label{fat-time2}\end{equation}
Again, for $\sigma\geq\sigma_{c}$, one gets $U_{t+1}=0$ from
Eq. (\ref{fat-recur}), giving $\tau_f=0$. For small $KT$ and as $\sigma\rightarrow\sigma_{c}$,
$\tau_f\simeq KT\exp\left[\left(\sigma_{c}/\sigma-1\right)/KT\right]$.
These results agree qualitatively
with the recent experimental observations \cite{bc01,gcgzs02}.

\begin{figure}
\includegraphics[width=5cm,height=6cm, angle=-90]{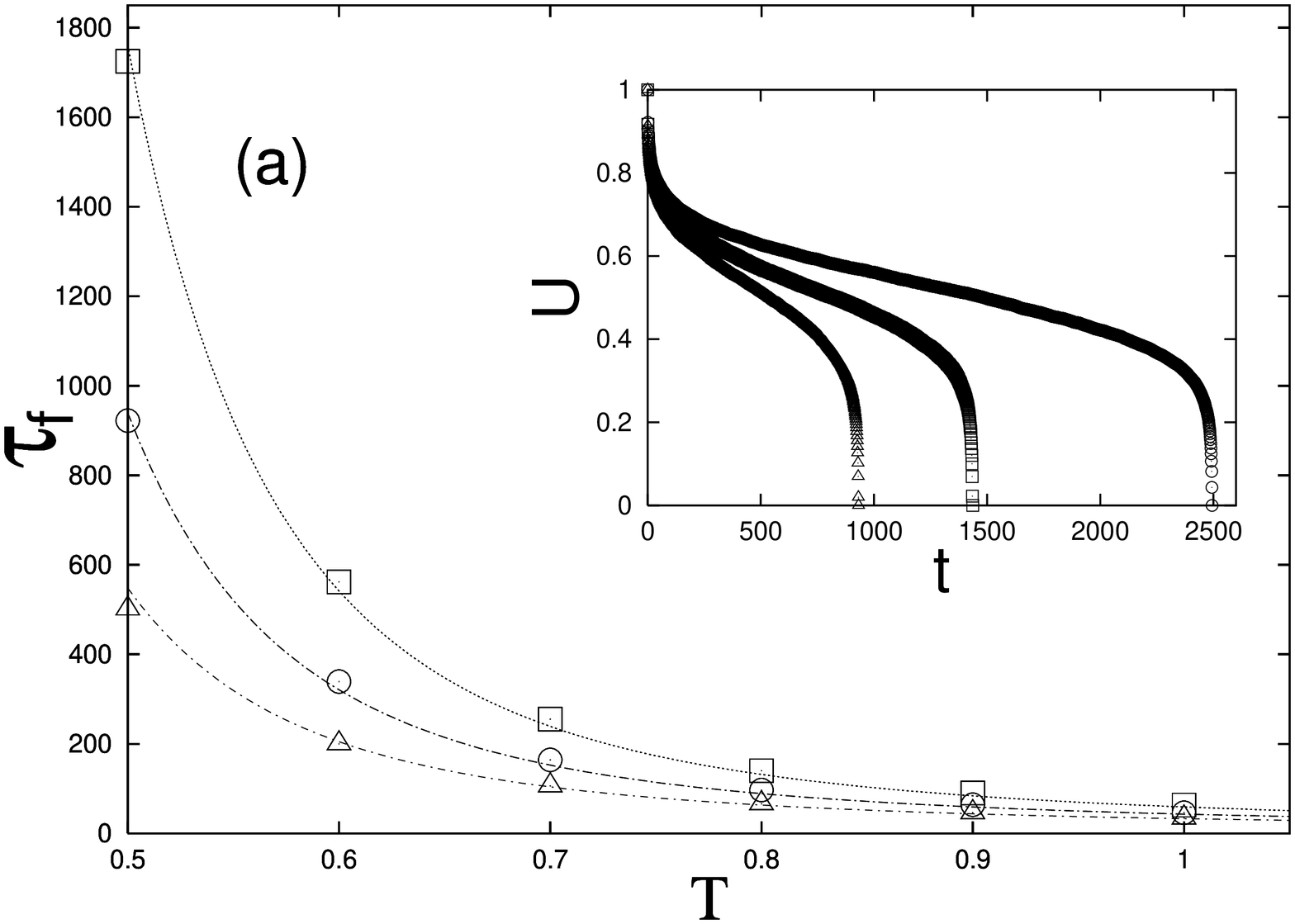}
\vskip.3in
\includegraphics[width=5cm,height=6cm,angle=-90]{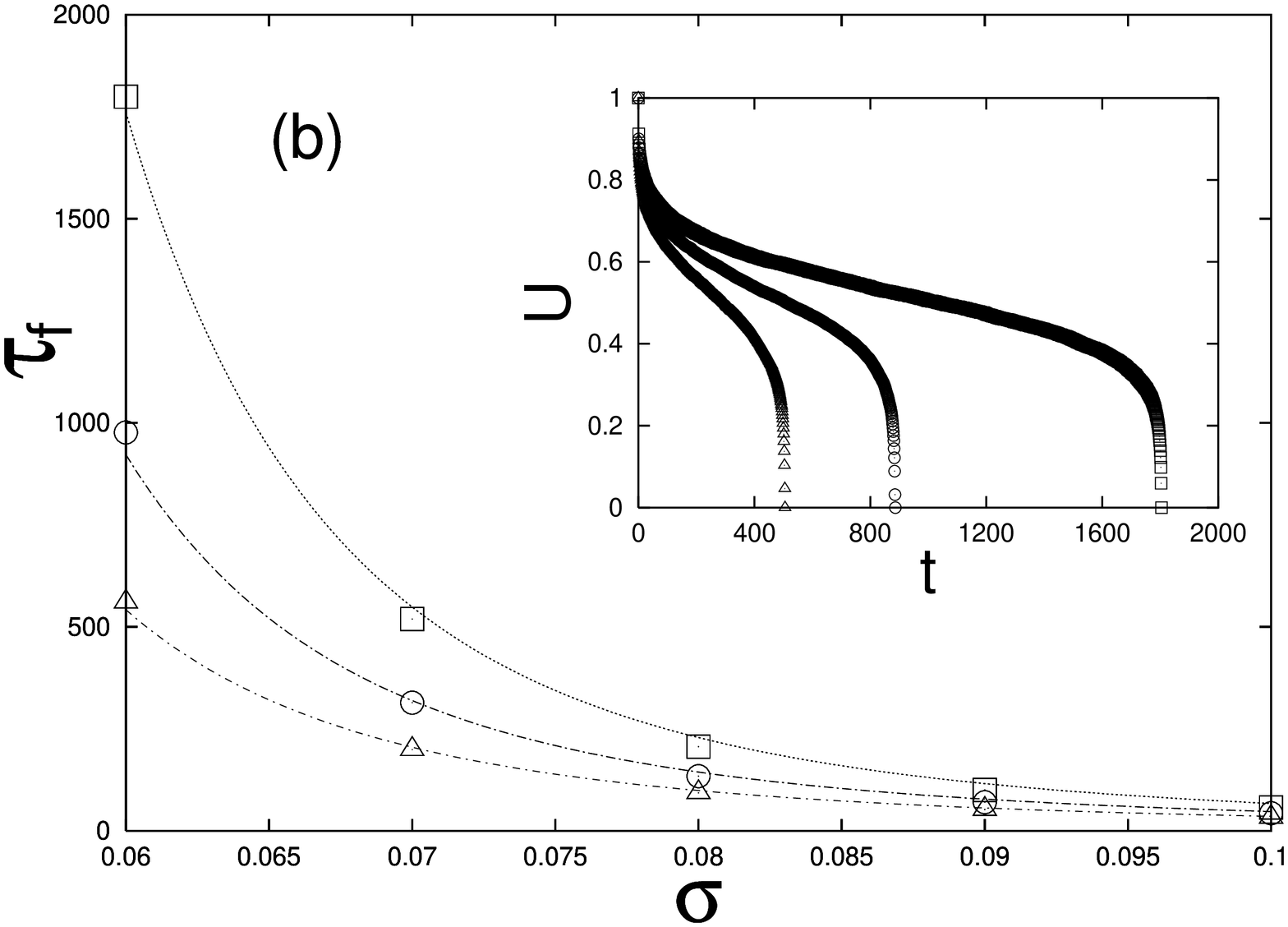}
\vspace{.3in}
\caption{  
Simulation results of fatigue behavior: (a) average failure time $\tau_f$ vs.
noise $T$ ($K=1$ here) for three different stress values and 
(b) $\tau_f$ vs. $\sigma$ for three different noise values.
The bundle contains $N=10^{5}$ fibers with uniformly distributed strength 
thresholds. The time variations of fraction of surviving fibers
are shown in the insets. The dotted
lines in (a) and (b) correspond to the fit with expression
 (Eq. \ref{eq:fati-nonhomo}) where
$\sigma_{c}\simeq0.245$  (exact value=$1/4$ \cite{pbc02,bpc03}).}
\label{fig:fat_hetero}
\end{figure}

In order to investigate the fatigue behavior in heterogeneous
fiber bundles, uniform distribution of fiber strengths has been 
considered \cite{pc03b}.
The noise-induced failure probability has the similar form 
: $p(\sigma,KT)=\exp\left[-\frac{1}{KT}\left(\frac{x}{\sigma}-1\right)\right]$
for $0<\sigma\leq x$ and $p(\sigma,KT)=1$ for $\sigma> x$,
where $x$ denotes the strength of the individual fibers
in the bundle. Now it is difficult to tackle the problem analytically. 
However,  Monte Carlo simulations \cite{pc03b} 
show (Fig. \ref{fig:fat_hetero}) the 
variations of average failure time ($\tau_f$) with noise ($KT$) and stress 
level ($\sigma$):
\begin{equation}
\tau_f =KT\exp\left(-\frac{1}{KT}\right)\left[\exp\left(\frac{\sigma_{c}}{\sigma KT}+\frac{1}{KT}\right)-1\right],\label{fat-het-time}
\label{eq:fati-nonhomo}
\end{equation}
 where, $\sigma_c$ is the critical stress. This phenomenological 
form Eq. (\ref{fat-het-time}) is indeed
very close to the analytic result Eq. (\ref{fat-time1}) for the homogeneous
fiber bundle.

\textcite{np01}  considered the breaking dynamics in a fiber bundle
where each fiber has a failure probability $p$ determined by the loading
time $t$ and the load $\sigma$ on it as $p(t,\sigma)= 1-exp[-t\sigma^\rho]$,
 and analyzed the life time $(t_f)$ distribution. For $1/2 \le \rho \le 1$,
 ELS and LLS models have identical Gaussian distribution for $t_f$. For 
$\rho>1$,  LLS shows extreme  statistics, while ELS gives Gaussian behavior, 
see e.g.,  \textcite{cs97}. \textcite{yki08} also studied similar 
$t_f$ distribution and their scaling property with load and temperature
variations.

\subsubsection{Creep rupture in viscoelastic fiber bundles}

\begin{figure}
\includegraphics[width=5cm,height=4cm]{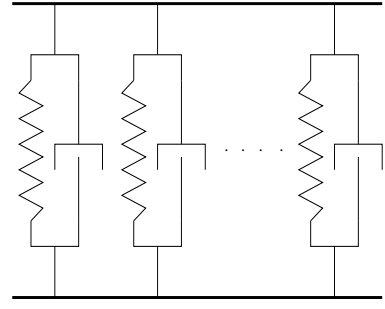}
\caption{The viscoelastic fiber bundle: intact
fibers are modeled by Kelvin-Voigt elements which consists of a spring and
a dashpot in parallel. From \textcite{khhp03}. }
\label{fig:kelvin} 
\end{figure}
Creep behavior has been achieved \cite{hkh02,khhp03} in a bundle of 
viscoelastic fibers, 
where a fiber is modeled by a Kelvin-Voigt element  
(see Fig.\ \ref{fig:kelvin}) and results in
the constitutive stress-strain relation
 \begin{eqnarray}
\sigma_{{\rm o}}=\beta_0\dot{\varepsilon}+Y\varepsilon.\label{eq:visco}
\end{eqnarray}
 Here $\sigma_0$ is the applied stress, $\varepsilon$ is the corresponding
 strain, $\beta_0$ denotes the damping coefficient, and $Y$ is the Young
modulus of the fibers. 

In the equal load sharing mode, the time evolution of the system
under a steady external stress $\sigma_{o}$ can be  described by
the equation
 \begin{eqnarray}
\frac{\sigma_{{\rm o}}}{1-P(\varepsilon)}=\beta_0\dot{\varepsilon}+Y\varepsilon.\label{eq:eom}
\end{eqnarray}

As one can expect intuitively, there is a critical load $\sigma_c$  for the 
system and  Eq.\ (\ref{eq:eom}) suggests two distinct
regimes  depending on the value of the external
load $\sigma_{{\rm o}}$: When $\sigma_{{\rm o}}$ is below the critical
value $\sigma_{{\rm c}}$ Eq.\ (\ref{eq:eom}) has a fixed-point solution
$\varepsilon_{s}$, which can be obtained by setting $\dot{\varepsilon}=0$
in Eq.\ (\ref{eq:eom}) \begin{eqnarray}
\sigma_{{\rm o}}=Y\varepsilon_{s}[1-P(\varepsilon_{s})].\label{eq:stationary}\end{eqnarray}
 In this case  the strain value converges to $\varepsilon_{s}$ when 
$t\to\infty$, and no macroscopic failure occurs. But, when 
$\sigma_{{\rm o}}>\sigma_c$, no fixed-point solution exists. Here,
 $\dot{\varepsilon}$ remains always positive, that means
in this case, the strain of the system $\varepsilon(t)$
monotonically increases until the system fails globally at a finite time
$t_{f}$ \cite{hkh02}.

The solution of  the differential equation Eq.\ (\ref{eq:eom}) gives a 
complete description of the failure process. By separation
of variables, the integral becomes
 \begin{eqnarray}
t=\beta_0\int d\varepsilon\frac{1-P(\varepsilon)}{\sigma_{{\rm o}}-Y\varepsilon\left[1-P(\varepsilon)\right]}+C,\label{eq:integ}
\end{eqnarray}
where  $C$ is integration constant. 

Below the critical point $\sigma_{{\rm o}}\leq\sigma_{{\rm c}}$ the
bundle slowly relaxes to the fixed-point value $\varepsilon_{s}$. The 
characteristic time scale
of such relaxation process can be obtained by analyzing the behavior  of
 $\varepsilon(t)$  in the vicinity of $\varepsilon_{s}$. 
After introducing a new variable
 $\delta_0$ as $\delta_0(t)=\varepsilon_{s}-\varepsilon(t)$,
the differential equation  can be written as 
\begin{eqnarray}
{\displaystyle \frac{d\delta_0}{dt}=-\frac{Y}{\beta_0}\left[1-\frac{\varepsilon_{s}p(\varepsilon_{s})}{1-P(\varepsilon_{s})}\right]\delta_0.}\label{eq:delta}
\end{eqnarray}
 Clearly, the solution of Eq.\ (\ref{eq:delta}) has the form $\delta_0\sim\exp{\left[-t/\tau\right]}$,
with 
\begin{eqnarray}
{\displaystyle \tau}=\frac{\beta_0}{Y}\frac{1}{\left[1-{\displaystyle \frac{\varepsilon_{s}p(\varepsilon_{s})}{1-P(\varepsilon_{s})}}\right]},\label{eq:tau}\end{eqnarray}
where $\tau$ is the characteristic time of the relaxation process.

 The variation of the relaxation time $\tau$ 
with the external driving near the critical point $\sigma_{{\rm c}}$
is crucial for any dynamical system. 
Since $\sigma_{{\rm o}}(\varepsilon_{{\rm s}})$
has a maximum of the value $\sigma_{{\rm c}}$ at $\varepsilon_{{\rm c}}$,
in the vicinity of $\varepsilon_{{\rm c}}$  one  can use the  approximation:
 \begin{eqnarray}
\sigma_{{\rm o}}\approx\sigma_{{\rm c}}-A(\varepsilon_{{\rm c}}-\varepsilon_{s})^{2},\label{eq:series}
\end{eqnarray}
 where the multiplication factor $A$ depends on the cumulative distribution
$P$. Using the approximation Eq.\ (\ref{eq:series}), it can be  shown from 
Eq.\ (\ref{eq:tau}),that
 
\begin{eqnarray}
\tau\sim\left(\sigma_{c}-\sigma_{{\rm o}}\right)^{-1/2},\ \ \ \mbox{for}\ \ \ \sigma_{o}<\sigma_{{\rm c}},\label{eq:tau_crit}
\end{eqnarray}
 Therefore, the relaxation
time  diverges following  a universal power law with
an exponent $-1/2$. Note that a dry fiber bundle model, under constant load 
shows  similar power law divergence (see Section III.A).

\begin{figure}
\includegraphics[width=6cm,height=5cm]{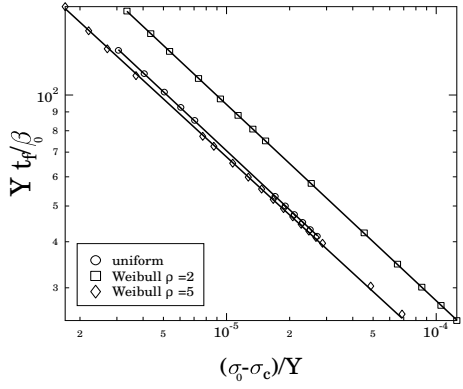}
\caption{The behavior of the time to failure $t_{f}$
for uniform and Weibull distributions with two different Weibull index
for the ELS case. All the three curves are parallel to each other
on a double logarithmic plot with an exponent close to $0.5$, in agreement
with the general result Eq.\ (\ref{critical}). From  \textcite{hkh02}. }
\label{fig:sigma}
\end{figure}

How does the system behave above the  critical point? The behavior 
can be analyzed in the same way when $\sigma_{{\rm o}}$ is close to 
 $\sigma_{{\rm c}}$. Then one can write
 $\sigma_{{\rm o}}=\sigma_{{\rm c}}+\Delta\sigma_{{\rm o}}$,
where $\Delta\sigma_{{\rm o}}<<\sigma_{{\rm c}}$.  It is obvious
that the relaxation steps are too many when $\varepsilon(t)$ becomes close to $\varepsilon_{{\rm c}}$.
Therefore, the integral in Eq.\ (\ref{eq:integ}),
is dominated by the region close to $\varepsilon_{{\rm c}}$. Using Eq.\ (\ref{eq:series}), the integral in Eq.\ (\ref{eq:integ})
becomes
 \begin{eqnarray}
t_{f}\approx\beta_0\int d\varepsilon\frac{1-P(\varepsilon)}{\Delta\sigma_{{\rm o}}-A(\varepsilon_{{\rm c}}-\varepsilon)^{2}}.\label{eq:time}
\end{eqnarray}
Evaluating the integration over a small $\varepsilon$ interval in the 
vicinity of $\varepsilon_{{\rm c}}$, one gets

\begin{eqnarray}
t_{f}\approx(\sigma_{{\rm o}}-\sigma_{{\rm c}})^{-1/2},\qquad\mbox{for}\qquad\sigma_{{\rm o}}>\sigma_{{\rm c}}.\label{critical}\end{eqnarray}
 Thus, $t_{f}$ has a two-sided power law divergence at $\sigma_{{\rm c}}$
with a universal exponent $-\frac{1}{2}$ independent of the specific
form of the disorder distribution $P(\varepsilon)$, similarly to
$\tau$ in case of dry fiber bundle (see Section III.A).

To check the validity of the universal power law behavior of $t_{f}$,
simulations were performed \cite{hkh02} with various disorder distributions, \textit{i.e.}
uniform distribution and the  Weibull distribution of the form
$P(\varepsilon)=1-\exp{\left[-\left(\varepsilon/\lambda\right)^{\rho}\right]}$, 
where $\lambda$ is the characteristic strain and $\rho$ is the shape parameter. 
The simulation results (Fig.\ \ref{fig:sigma}) are in
excellent agreement with the analytic results.

\subsubsection{Creep rupture in a bundle of slowly relaxing fibers}
\begin{figure}
\includegraphics[width=5cm,height=4cm]{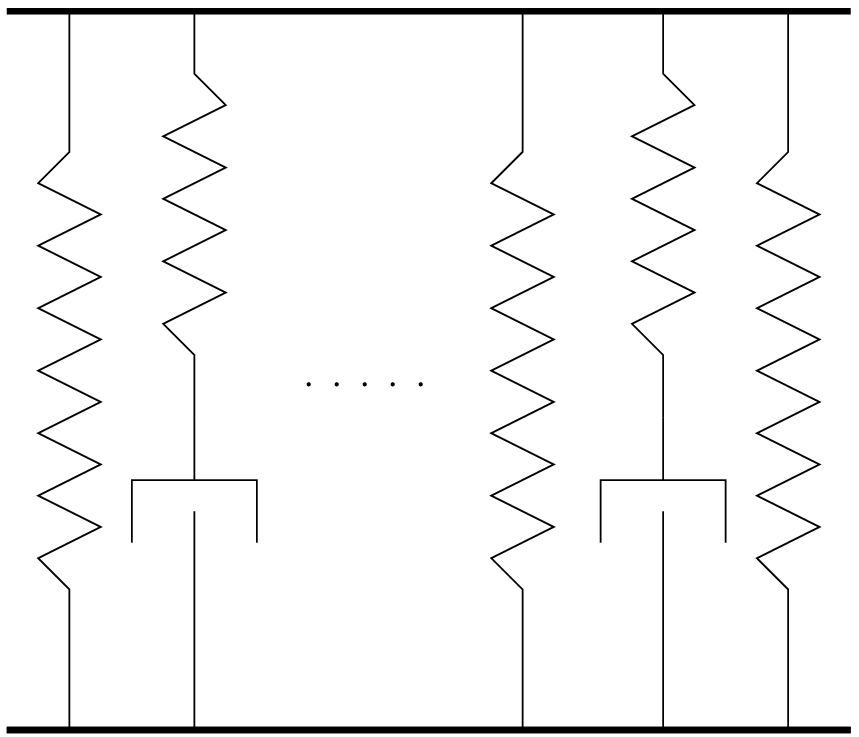}
\caption{The fibers are modeled by Maxwell elements. From \textcite{khhp03}}
\label{fig:maxwell}
\end{figure}

A slow relaxation following fiber failure can also lead to 
creep  behavior \cite{hkh01,khhp03}. In this case,
the fibers are linearly elastic until they break,
but after breaking they undergo a slow relaxation process. 
Therefore when a fiber breaks, its load does not drop to zero instantaneously.
Instead it undergoes a slow relaxation process and thus introduces a time
scale into the system. 
As the intact fibers are assumed to be linearly
elastic, the deformation rate is
 \begin{eqnarray}
\dot{\varepsilon}_{{\rm f}}=\frac{\dot{x}}{Y},\label{eq:intact}
\end{eqnarray}
where $x$ denotes the stress, $\varepsilon_{f}$ denotes the strain and $Y$ is 
the Young modulus of the fibers. 
In addition,  to capture the slow relaxation  effect, the broken
fibers with the surrounding matrix material are modeled by Maxwell
elements (Fig.\ \ref{fig:maxwell}), \textit{i.e.}
they are assumed as a serial coupling of a spring and a dashpot.
Such arrangement results in the following non-linear response \begin{eqnarray}
\dot{\varepsilon}_{{\rm b}}=\frac{\dot{x}_{{\rm b}}}{S_{{\rm b}}}+B x_{{\rm b}}^{{\rm m}}, &  & \label{eq:broken}\end{eqnarray}
where $x_{b}$ is the time
dependent stress and $\varepsilon_{{\rm b}}$ is the time
dependent  deformation of a broken fiber. The
relaxation of broken fibers is characterized by few parameters,
$S_{b},B,$ and $m$, where $S_{b}$ is the effective stiffness of
a broken fiber, the exponent $m$ characterizes the strength of
non-linearity and $B$ is a constant.

In equal load sharing mode, when external stress $\sigma_0$ is applied, 
the macroscopic elastic behavior of the composite
can be represented by the constitutive equation \cite{hkh01,khhp03}
  \begin{eqnarray}
\sigma_{{\rm o}}=x(t)\left[1-P(x(t))\right]+x_{{\rm b}}(t)P(x(t)).\label{eq:macro}\end{eqnarray}
Here $x_b (t)$ is the amount of stress carried by the  broken fibers
and  $P(x(t))$
and $1-P(x(t))$ denote the fraction of broken and intact fibers
at time $t$, respectively.

By construction (Fig.\ \ref{fig:maxwell}),
the two time derivatives have to be always equal \begin{eqnarray}
\dot{\varepsilon}_{{\rm f}}=\dot{\varepsilon}_{{\rm b}}.\label{eq:condit}\end{eqnarray}
Now the differential equation for the time evolution of the system
can be obtained using  Eq.\ (\ref{eq:macro}), Eq.\ (\ref{eq:broken})
and Eq.\ (\ref{eq:condit}) as
 \begin{eqnarray}
 &  & \dot{x}\left\{ \frac{1}{Y}-\frac{1}{S_{{\rm b}}}\left[1-\frac{1}{P(x)}+\frac{p(x)}{P(x)^{2}}(x-\sigma_{{\rm o}})\right]\right\} \nonumber\\
 &  & = B\left[\frac{\sigma_{{\rm o}}-x\left[1-P(x)\right]}{P(x)}\right]^{m}.\label{eq:eom_max} 
\end{eqnarray}
 Similar to the viscoelastic model, two different regimes of $x(t)$ 
can be distinguished depending on the value of $\sigma_{o}$: If the external
load is below the critical value $\sigma_{{\rm c}}$ a fixed-point
solution $x_{s}$  exists which can
be obtained by setting $\dot{x}=0$ in Eq.\ (\ref{eq:eom_max})
\begin{eqnarray}
\sigma_{o}=x_{s}\left[1-P(x_{s})\right].\label{eq:statio_max}\end{eqnarray}
 This means that the solution $x(t)$ of Eq.\ (\ref{eq:eom_max})
converges asymptotically to $x_{s}$ resulting in an infinite
lifetime $t_{f}$ of the system. When the external load is
above the critical value, the deformation rate 
$\dot{\varepsilon}=\dot{x}/Y$
remains always positive, resulting in a macroscopic failure in a finite
time $t_{f}$. 
Now we focus on the universal behavior of the model in the vicinity
of the critical point. Below the critical point the relaxation of
$x(t)$ to the stationary solution $x_{s}$ can be presented
by a differential equation of the form \begin{eqnarray}
\frac{d\delta_0}{dt}\sim\delta_0^{m},\end{eqnarray}
 where $\delta_o$ denotes the difference $\delta_0(t)=x_{s}-x(t)$.
Hence, the characteristic time scale $\tau$ of the relaxation process
only emerges if $m=1$. Also,  in this case  relaxation time goes as  $\tau\sim(\sigma_{c}-\sigma_{o})^{-1/2}$
 when approaching the critical point from below. 
However, for $m>1$ the
situation is  different:
relaxation process is characterized by $\delta_0(t)=at^{1/1-m}$, where
$a\rightarrow0$ with $\sigma_{0}\rightarrow\sigma_{c}$.

Again, close to the critical point, it can also be shown that the lifetime
$t_{f}$ shows  power law divergence when the external
load approaches the critical point from above \begin{eqnarray}
t_{f}\sim\left(\sigma_{0}-\sigma_{{\rm c}}\right)^{-(m-1/2)},\ \ \ \mbox{for}\ \ \ \sigma_{o}>\sigma_{{\rm c}}.\label{eq:tau_max}\end{eqnarray}
 The exponent is universal in the sense that it does not depend on
the disorder distribution. However it depends on the exponent
$m$, which characterizes the non-linearity of broken fibers.

\begin{figure}
\includegraphics[width=6cm,height=5cm]{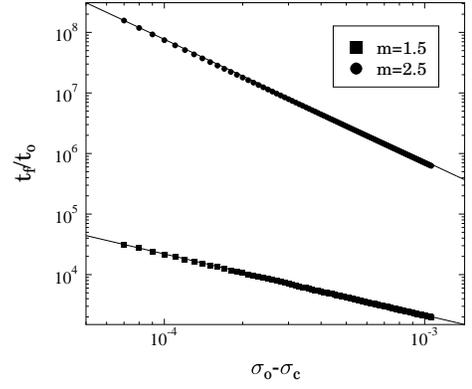}
\caption{Lifetime $t_{f}$ as a function of the distance from the critical 
point, $\sigma_{o}-\sigma_{c}$, for two different values of the parameter
$m$. The number of fibers in the bundle is $N=10^{7}$. From \textcite{khhp03}.}
\label{fig:sigma_maxwell}
\end{figure}

As a check, numerical simulations have been  performed \cite{hkh01,khhp03} 
for several
different values of the exponent $m$ (Fig.\ \ref{fig:sigma_maxwell}).
The slope of the fitted straight lines agrees  well with the analytic 
predictions (Eq.\ \ref{eq:tau_max}).

\subsubsection{Fatigue-failure experiment}

\begin{figure}
\centering
\includegraphics[width=9cm]{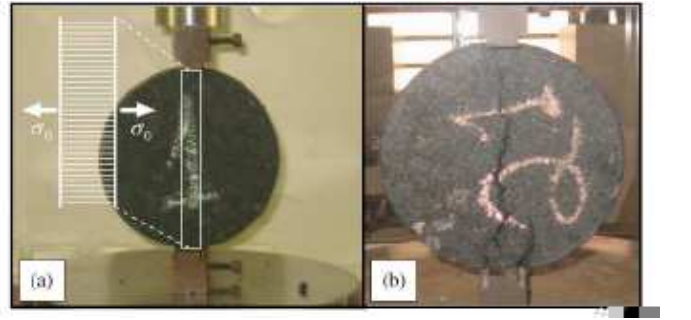}
\caption{Asphalt samples set up for experimental testing of fatigue failure.
Figure (a) shows how fatigue failure under these experimental conditions
are modeled using a fiber bundle model and (b) shows a post-failure sample. 
From \textcite{kcfaszh07}.}
\label{ah-fig13}
\end{figure}

An interesting experimental and theoretical study of fatigue failure in asphalt
was performed by \textcite{kcfaszh07}.  The experimental set-up is shown
in Fig.\ \ref{ah-fig13}.  The cylindrical sample was subjected to cyclic
diametric compression at constant load amplitude $\sigma_0$, and  the deformation
$\epsilon$ as a function of the number of cycles $N_{cycle}$ was recorded together
with the number of cycles $N_f$ at which catastrophic failure occurs.

\begin{figure}
\centering
\includegraphics[width=8cm]{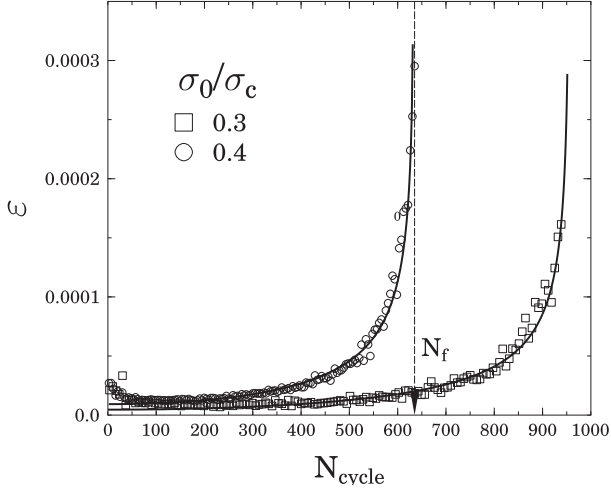}
\caption{Deformation $\varepsilon$ as a function of number of cycles
$N_{cycle}$, showing both experimental data and theoretical curves
based on the fiber bundle model. From \textcite{kcfaszh07}.}
\label{ah-fig15}
\end{figure}

\begin{figure}
\centering
\includegraphics[width=9cm]{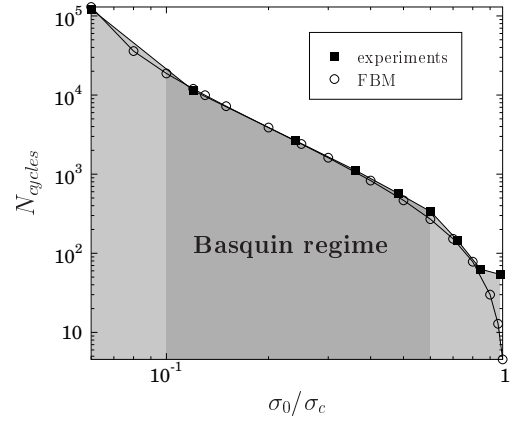}
\caption{Number of cycles at catastrophic failure, $N_f$ as a function
of load amplitude $\sigma_0/\sigma_c$.  The two curves show experimental and
fiber bundle model data respectively.
From \textcite{kcfaszh07}.}
\label{ah-fig16}
\end{figure}

Fig.\ \ref{ah-fig15} shows deformation $\epsilon$ as a function of number of
cycles $N_{cycle}$ for two different load amplitudes $\sigma_0/\sigma_c=0.3$
and 0.4.  Here $\sigma_c$ is the tensile strength of the asphalt.  Fig.\
\ref{ah-fig16} shows the number of cycles to catastrophic failure as a function
of the load amplitude $\sigma_0/\sigma_c$.  This curve shows three regimes, the
middle one being characterizable by a power law,
\begin{equation}
\label{kun_basquin}
N_f \sim \left(\frac{\sigma_0}{\sigma_c}\right)^{-\alpha'}\;.
\end{equation}
This is the {\it Basquin regime\/} \cite{s91,smb92,lm02,sll02}. 
\textcite{kcfaszh07} find $\alpha'=2.2\pm0.1$ for the asphalt system.

In order to model the behavior found in Figs.\ \ref{ah-fig15} and 
\ref{ah-fig16}, \textcite{kcfaszh07} introduce equal load-sharing fiber bundle
 model as illustrated in Fig.\ \ref{ah-fig13}a.  
 Each fiber $1 \le i \le N$ is subjected
to a time dependent load $x_i(t)$.  There are two failure mechanisms present.
Fiber $i$ fails instantaneously at time $t$ when $x_i(t)$ for the
first time reaches its failure threshold $t_i$.  However, there is also a
damage accumulation mechanism characterized by the parameter $c_i(t)$.  In
the time interval $dt$, fiber $i$ accumulates a damage
\begin{equation}
\label{kun_damage}
dc_i(t)=ax_i(t)^\gamma\ dt\;,
\end{equation}
where $a > 0$ is a scale parameter and $\gamma>0$ is an exponent to be determined.
Hence, the accumulated damage is
\begin{equation}
\label{kun_cumulative}
c_i(t)=a \int_0^t x_i(t')^\gamma\ dt'\;.
\end{equation}
When $c_i(t)$ for the first time exceeds the damage accumulation threshold,
$s_i$, fiber $i$ fails.  The thresholds $t_i$ and  $s_i$ are chosen from a
joint probability distribution $p_{t,s}(t,s)$. \textcite{kcfaszh07}
make the assumption that this distribution may be factorized,
$p_{t,s}(t,s)=p_t(t)p_s(s)$.

In addition to damage accumulation, there is yet another important mechanism
that needs to be incorporated in the model: damage {\it healing\/} 
\cite{sll02,jkccy08}.
A time scale $\tau$ is associated with this mechanism, and the ELS
average force-load equation
\begin{equation}
\label{kun_f_x}
\sigma_0(t)[1-P_t(x(t))]x(t)\;,
\end{equation}
where $P_t(t)=\int_0^t p_t(t')\ dt'$ is the cumulative instantaneous breaking
threshold probability, is generalized to
\begin{equation}
\label{kun_general_break}
\sigma_0(t)=\left[1-P_s\left(a\int_0^t e^{-(t-t')/\tau} x(t')^\gamma 
dt'\right)\right] \left[1-P_t(x(t))\right]x(t)\;,
\end{equation}
where $P_s(s)=\int_0^s p_s(s')\ ds'$. \textcite{kcfaszh07} show that
the Basquin law (\ref{kun_basquin}) may be derived analytically from Eq.\
(\ref{kun_general_break}) leading to  $\alpha'=\gamma$.  The solid curves in
Fig.\ \ref{ah-fig15} are fits of the theoretical curves based on Eq.\
(\ref{kun_general_break}) to the experimental data for the $\epsilon$ vs.\
$N_{cycle}$.  Likewise, Fig.\ \ref{ah-fig16} shows the fit of the theoretical
$N_f$ vs.\ $\sigma_0/\sigma_c$ to the experimental data.  For this fit,
$\gamma=2.0$ and $\tau=15 000$.

\subsection{Precursors of global failure}

A fundamental question in strength considerations of materials is when
does it fail.  Are there signals that can warn of imminent failure?  This
is of uttermost importance in e.g.\ the diamond mining industry where sudden
failure of the mine can be extremely costly in terms of lives.  These mines
are under continuous acoustic surveillance, but at present there are no
tell-tale acoustic signature of imminent catastrophic failure.  The same type
of question is of course also central to earthquake prediction and initiates 
the search for precursors of  global (catastrophic) failure events, see e.g., 
\textcite{sa92,sa96}, \textcite{pc05}, \textcite{pc06}. 
The precursor parameters  essentially reflect
the growing correlations within a dynamic system as it approaches 
the failure point. As we show, sometimes it is possible  to predict
the global failure points in advance.
Needless to mention that the existence of any such precursors and
detailed knowledge about their behavior for major catastrophic failures
like earthquakes\index{earthquake}, landslides, mine/bridge collapses, 
would be of supreme value for our civilization. In this sub-section we  
discuss some precursors of global failure in ELS models. We also comment on 
how can one predict the critical point (global failure point) from the 
precursor parameters.
  
\vskip.1in
\noindent \emph{(a)  Divergence of susceptibility  and relaxation time}
\vskip.1in

\begin{figure}
\includegraphics[width=6cm,height=5cm]{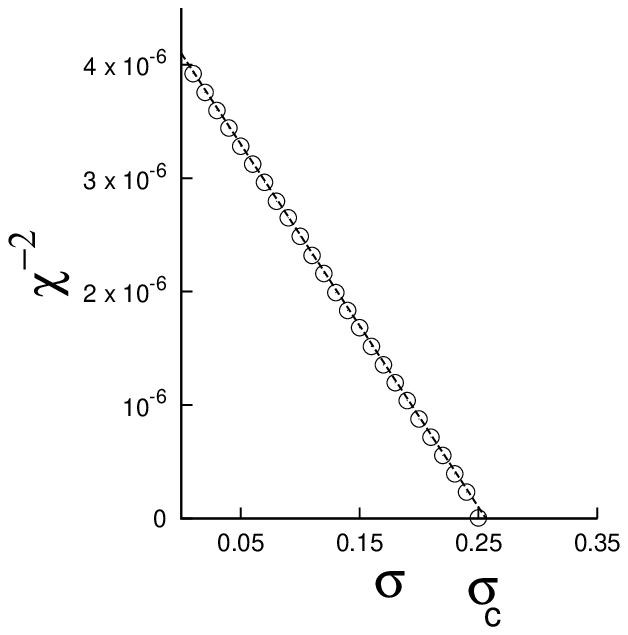} 
\includegraphics[width=6cm,height=5cm]{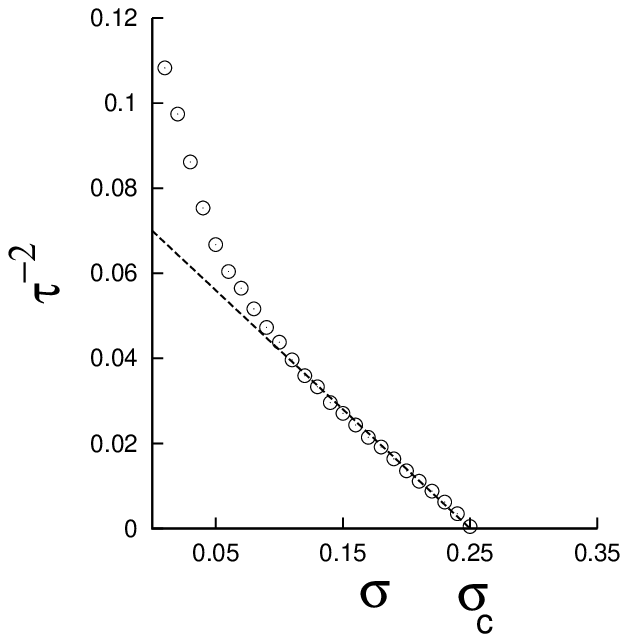} 
\caption{ Variation of $\chi^{-2}$ and $\tau^{-2}$
with applied stress for a bundle having $N=50000$ fibers.
Uniform distribution of fiber thresholds is considered and averages are taken 
over $1000$ sample.
The dotted straight lines are the best linear fits near the critical point.}
\label{fig:xi-tau}
\end{figure}
As we discussed earlier (Section III.A) in case of ELS fiber bundles the 
susceptibility ($\chi$) and the relaxation time ($\tau$)
follow power laws (exponent $= -1/2$) with external stress and both  diverge at
the critical stress. Therefore if we plot $\chi^{-2}$ and $\tau^{-2}$
with external stress, we expect a linear fit near critical point and
the straight lines should touch $X$ axis at the critical stress.
We indeed found similar behavior (Fig. \ref{fig:xi-tau}) in simulation 
experiments after taking averages over many sample.

\begin{figure}
\includegraphics[width=6cm,height=5cm]{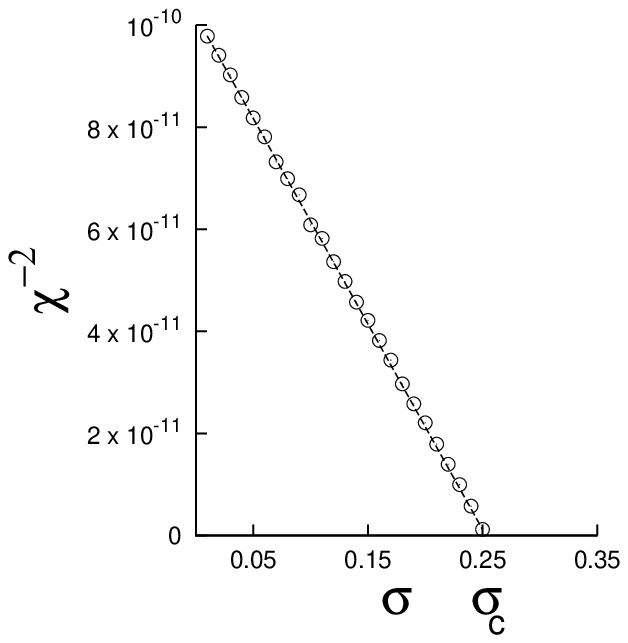} 
\includegraphics[width=6cm,height=5cm]{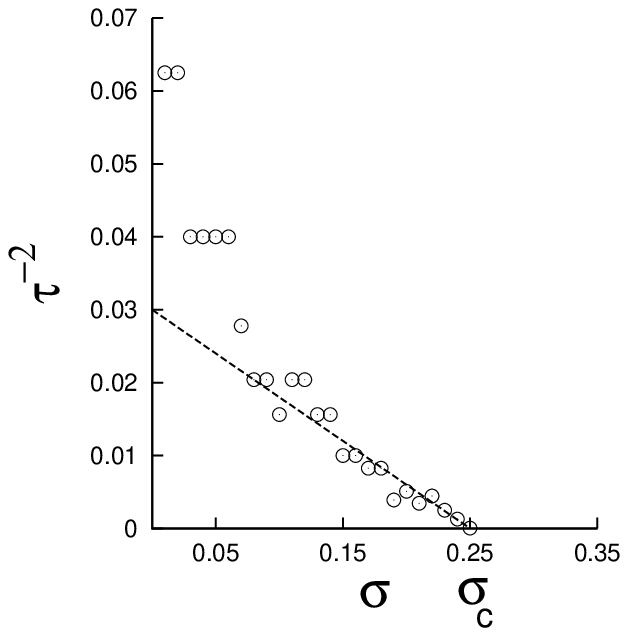} 
\caption{ Variation of $\chi^{-2}$ and $\tau^{-2}$
with applied stress for a  single bundle having $N=10000000$ fibers
with uniform distribution of fiber thresholds. Straight lines represent the 
best linear fits near the critical point. } 
\label{fig:xi-tau-single}
\end{figure}

For application, it is always important that such prediction can be
done in a single sample. For a
single bundle having very large number of fibers, similar
response of $\chi$ and $\tau$ have been observed . 
The estimation (through extrapolation)  of the failure 
point is also quite satisfactory (Fig. \ref{fig:xi-tau-single}).

\vskip.1in
\noindent \emph{(b) Pattern of  breaking-rate}
\vskip .1in

\begin{figure}
\includegraphics[width=6cm,height=5cm]{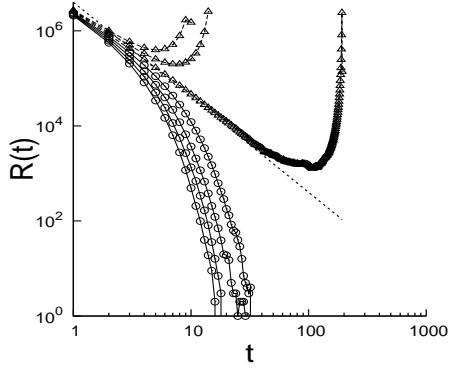} 
\caption{ 
Log-log plot of breaking-rate
with step of load redistribution for $7$ different stress values. Circles are 
for stresses below the critical stress and triangles are for stresses above the 
critical stress.
 The simulation has been performed for a single bundle with
$N=10000000$ fibers having uniform distribution of fiber thresholds. The dotted 
straight line has a slope $-2$.}
\label{fig:breaking-rate}
\end{figure}
\begin{figure}
\includegraphics[width=6cm,height=6cm]{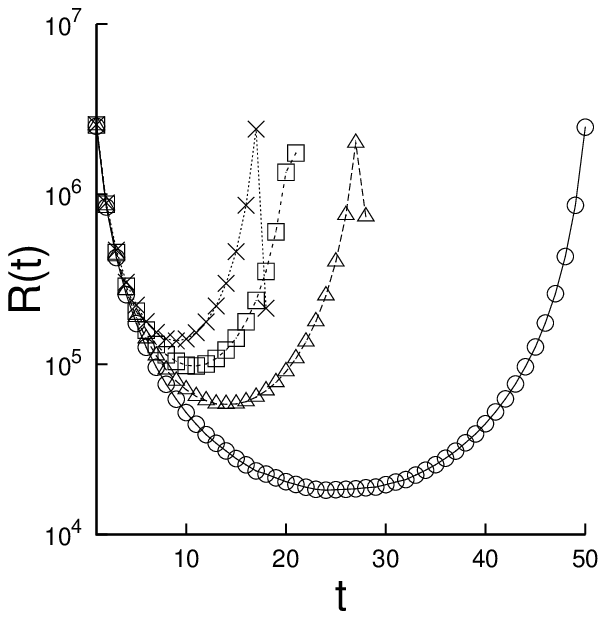}
\includegraphics[width=6cm,height=6cm]{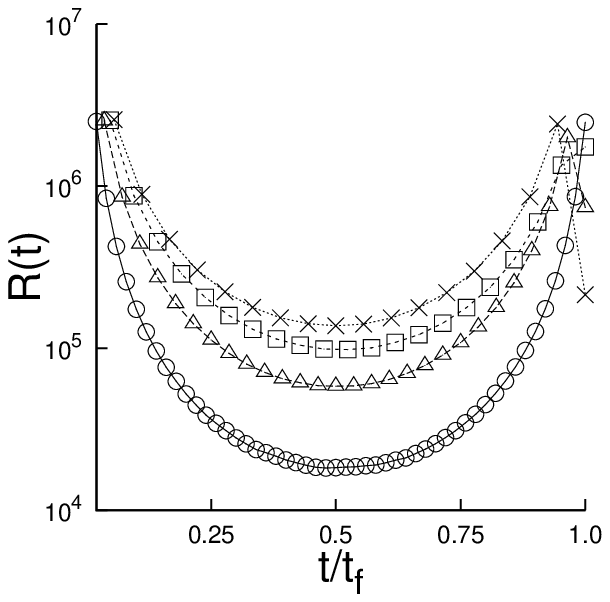}

\caption{ 
 The breaking rate $R(t)$ vs. step $t$ (upper plot)
and vs.\ the rescaled step variable $t/t_f$ (lower plot) for the uniform
threshold distribution for a bundle of $N= 10^7$ fibers.
Different symbols are used for different excess stress levels
$\sigma -\sigma _c$: 0.001  (circles), 0.003 (triangles), 0.005 (squares)
and 0.007 (crosses).
}
\label{fig:breaking-rate-min}
\end{figure}

When we apply load on a material body,  it is important
to know whether the body can support that load or not. The similar
question can be asked in fiber bundle model. We found that if we record the 
breaking-rate, 
i.e, the amount of failure in each load redistribution
-then the pattern of breaking-rate clearly shows whether the bundle
is going to fail or not. For any stress below the critical state,
the breaking-rate follows exponential decay (Fig. \ref{fig:breaking-rate})
 with the step of load redistribution and for
stress values above critical stress it is a power law followed by
a gradual rise (Fig. \ref{fig:breaking-rate}). Clearly, at critical stress 
it follows a robust power law with exponent value $-2$ that can be obtained
 analytically from Eq. (\ref{eq:fbm-omori}). As we can see from Fig.  
\ref{fig:breaking-rate-min} that when the applied stress value is above the 
critical stress, 
breaking-rate initially goes down with step number, then at some point it 
starts going up and continues till the complete breakdown. That means if 
breaking-rate changes from downward trend to upward trend - the bundle will 
fail surely -but not immediately after the change occurs -it takes few more 
steps and number of these steps decreases as we apply bigger external stress 
(above the critical value). Therefore, if we can locate this minimum 
in the breaking-rate pattern, we can save the system (bundle) from 
breaking down by withdrawing the applied load immediately. We have 
another important question here: Is there any relation between the breaking 
rate minimum and the failure time (time to collapse) of the bundle? There 
is indeed a universal 
relationship which has been explored recently \cite{ph09} through numerical and
 analytical studies:  For slightly overloaded bundle we can rewrite
Eq. (\ref{finaluni}) as 
\begin{equation}
U_t =  {\textstyle \frac{1}{2}} -\sqrt{\epsilon}\tan (A^* t-B^*),
\label{un}\end{equation}
where
\begin{equation}
 A^*= \tan^{-1}(2\sqrt{\epsilon})\hspace{1cm}\mbox{and}\hspace{1cm}B^*=\tan^{-1}(1/2\sqrt{\epsilon}.
\end{equation}
From Eq. (\ref{un}) follows the breaking rate
\begin{equation}
R(t)=- \frac{dU_t}{dt} = \sqrt{\epsilon}A^*\cos^{-2}(A^*t-B^*).
\end{equation}

$R(t)$ has a minimum when
\begin{equation}
0=\frac{dR}{dt} \propto \sin(2A^*t-2B^*),
\end{equation}
which corresponds to
\begin{equation}
t_0=\frac{B^*}{A^*}.
\end{equation}
When criticality is approached, i.e.\ when $\epsilon \rightarrow 0$, we have $A^*\rightarrow 0$, and thus $t_0\rightarrow \infty$, as expected.

We see from Eq. (\ref{un}) that $U_t=0$ for 
\begin{equation}
t_f = \left(B^*+\tan^{-1}(1/2\sqrt{\epsilon}\right)/A^*= 2B^*/A^*.
\end{equation}
This is an excellent approximation to the integer value at which the fiber 
bundle collapses completely.
Thus with very good approximation we have the simple connection
$t_f = 2 t_0$.  When the breaking rate starts increasing we are halfway 
(see Fig. \ref{fig:breaking-rate-min}) to complete collapse!\\

\vskip.1in
\noindent \emph{(c) Crossover signature in  avalanche  distribution}
\vskip.1in

\begin{figure}
\includegraphics[width=6cm,height=6cm]{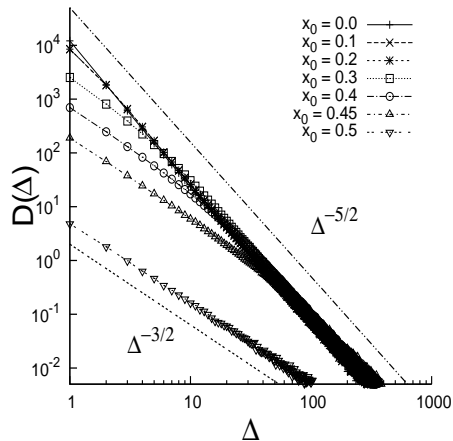}
\caption{ 
The avalanche size distributions 
for different values of $x_{0}$ in ELS model with uniform
 fiber strength distribution. Here bundle size $N=50000$ and averages are
taken over $10000$ sample. Two power laws (dotted lines)
have been drawn as reference lines to compare the numerical results. 
}
\label{fig:els-cut-dist}
\end{figure}
\begin{figure}
\centering
\includegraphics[width=8cm]{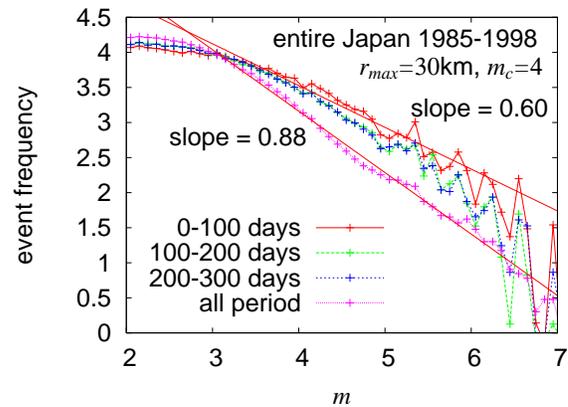}
\caption{Crossover signature in
the local magnitude distributions of earthquakes in Japan. The exponent
of the distribution during 100 days before a mainshock is about
$0.60$, much smaller than the average value $0.88$. From \textcite{k06}.}
\label{fig:kawamura}
\end{figure}

The bursts or avalanches can be 
recorded from outside -without disturbing the ongoing failure process. 
Therefore, any signature in burst statistics that can warn of imminent system 
failure would be very useful in the sense of wide scope of applicability.    
As discussed in Section III.B, when the avalanches are recorded 
close to the global failure point, the distribution shows 
(Fig. \ref{fig:els-cut-dist}) a different power 
law ($\xi=3/2$) than the one ($\xi=5/2$) characterizing the 
size distribution of all avalanches. This crossover behavior has been analyzed 
analytically in case of ELS fiber bundle model and similar crossover behavior is also seen \cite{phh06} in the burst distribution and energy distribution of 
the fuse model which is an established model for studying fracture and 
breakdown phenomena in disordered systems. The crossover length becomes bigger 
and bigger as the failure point is approached and it diverges at the failure
 point (Eq. \ref{eq-3b-16}). In some sense, the magnitude of the crossover 
length  tells us how far the 
system is from the global failure point.  Most important is that this 
crossover signal does not hinge on observing rare events and is seen also in 
a single system (see Fig. \ref{fig3b-3}). Therefore, such crossover signature 
has a strong potential to
be used as a useful detection tool. It should be mentioned that a recent 
observation \cite{k06} suggests a clear change in exponent values of the local 
magnitude distributions of earthquakes in Japan, before the onset of a 
mainshock (Fig. \ref{fig:kawamura}). This observation has definitely 
strengthened the possibility of using crossover signals in burst statistics
as a criterion for imminent failure.
       
\subsection{Fiber reinforced composites}
\label{Sec:5B}

\begin{figure}
\centering
\includegraphics[width=8cm]{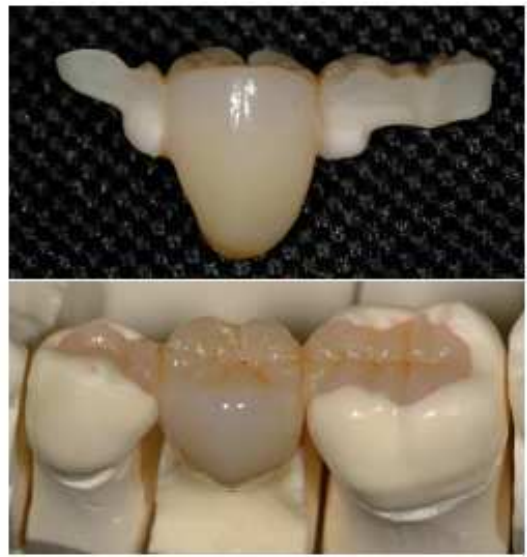}
\caption{This dental bridge is made from a fiber reinforced composite
with braided fibers made from polyethylene.  Due to the braided structure,
this composite is four times tougher than a composite made from the
same materials, but without the braiding \cite{ks07}
(Courtesy H.\ Strassler, Univ.\ of Maryland Dental School and V.\ Karbhari,
Univ.\ of Alabama, Huntsville).}
\label{ah-fig11}
\end{figure}

As we have seen, fiber bundle models provide a fertile ground for
studying a wide range of breakdown phenomena.  In some sense, they
correspond to the Ising model in the study of magnetism.  In this Section,
we will review how the fiber bundle models are generalized to
describe composites containing fibers.  Such composites are of increasing
practical importance, see e.g., Fig.\ \ref{ah-fig11}.

\begin{figure}
\centering
\includegraphics[width=6cm]{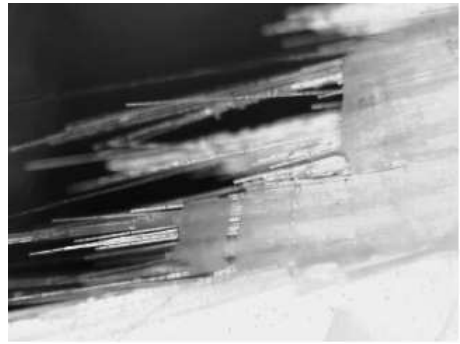}
\caption{Post mortem micrograph showing a fiber reinforced composite where
the matrix has undergone brittle failure followed by failure of the bridging
fibers through debonding. From \textcite{ks07}.}
\label{ah-fig12}
\end{figure}

The status of modeling fiber reinforced composites has recently been reviewed 
\cite{m07,mb09}. These materials consist of fibers embedded in a matrix. 
 During tensile loading the main part of the load is carried
by the fibers and the strength of the composite is governed to a large extent
by the strength of the fibers themselves.  The matrix material is chosen
so that its yield threshold is lower than that of the fibers which are
embedded in it.  Common materials used for the
fibers are aluminum, aluminum oxide, aluminum silica, asbestos,
beryllium, beryllium carbide, beryllium oxide, carbon (graphite),
glass (E-glass, S-glass, D-glass), molybdenum, polyamide (aromatic polyamide,
aramid), Kevlar 29 and Kevlar 49, polyester, quartz (fused silica),
steel, tantalum, titanium, tungsten or tungsten monocarbide.
Most matrix materials are resins as a result of their wide variation
in properties and relatively low cost. Common resin materials are epoxy,
phenolic, polyester, polyurethane and vinyl ester. When the composite is
to be used under adverse conditions such as high temperature, metallic
matrix materials such as aluminum, copper, lead, magnesium, nickel, silver or
titanium, or non-metallic matrix materials such as ceramics may be used.
When the matrix material is brittle, cracks open up in the
matrix perpendicular to the fiber direction at roughly equal spacing. In
metallic matrix materials, plasticity sets in at sufficient load.  Lastly,
in polymer matrix composites, the matrix typically responds linearly, but
still the fibers carry most of the load due to the large compliance of the
matrix.    When a fiber fails, the forces
it carried are redistributed among the surviving fibers and the matrix.
If the matrix-fiber interface is weak compared to the strength of
the fibers and the matrix themselves, fractures develop along the fibers.
When the matrix is brittle, the fibers bridging the developing crack in the
matrix will, besides binding the crack together, lead to stress alleviation
at the matrix crack front.  This leads to the energy necessary to propagate
a crack further increases with the length of the crack, \cite{sj98,sj00} i.e.,
so-called {\it R-curve behavior\/} \cite{l93}.  When the bridging fibers
fail, they typically do so through debonding at the fiber-matrix interface.
This is followed by pull-out, see Fig.\ \ref{ah-fig12}.

The Cox shear lag model forms the basis for the standard tools used for
analyzing breakdown in fiber reinforced composites \cite{c52,c92}.
It considers
the elastic response of a single fiber in a homegeous matrix
only capable of transmitting shear stresses. By treating the
properties of the matrix as effective and due to the self-consistent response
by the matrix material and the rest of the fibers, the Cox model becomes a
mean-field model \cite{rann97}.  Extensions of the Cox single-fiber model to
debonding and slip at the fiber-matrix interface have been published
\cite{ak73,bhe86,h90,h92}. In 1961 the single-fiber calculation of Cox
was extended to two-dimensional unidirectional fibers in a compliant matrix,
i.e., a matrix incapable of carrying tensile stress, by \textcite{h61}.
In 1967, this calculation was followed up by \textcite{hd67} for
three-dimensional unidirectional fibers placed in
a square or hexagonal pattern.  They found the average
{\it stress intensity factor\/} (i.e., the ratio between local stress in an
intact fiber and the applied stress) to be
\begin{equation}
\label{hedgepeth}
K_k=\prod_{i=1}^k\ \frac{2i+2}{2i+1}
\end{equation}
after $k$ fibers failing.  In his Ph.D.\ thesis, \textcite{f69} extended
these calculations to aligned arrays of broken fibers mixed with intact fibers. 
This approach was subsequently generalized
to systems where the matrix has a non-zero stiffness and hence is able to
transmit stress \cite{lm99,bl99}.  Viscoelasticity of the matrix has been
included by \textcite{lph89} and \textcite{bp98}.

\textcite{c91} demonstrated that when the fibers respond under global
load-sharing conditions, a mean-field theory may be constructed where
the breakdown of the composite is reduced to that of the failure of a single
fiber in an effective matrix \cite{c93,hdel94,hf97,rh02}.  
\textcite{we93} studied the redistribution of forces onto the neighbors of a 
single
failing fiber within a two-dimensional uni-directional composite using the
shear-lag model, finding that within this scheme, the stress-enhancement is
less pronounced than earlier calculations had shown. 
\textcite{zw99,zw00} introduced a multi-fiber failure model including debonding
and frictional effects at the fiber-matrix interface, finding that the
stress intensity factor would decrease with increasing interfiber distance.
An important calculational principle, the {\it Break Influence Superposition
Technique\/} was introduced by \textcite{sp93} based on
the method of \textcite{k85} in order to handle models with multiple
fiber failures.  The technique consists in determining  the transmission
factors, which give the load at a given position along a given fiber due to
a unit negative load at the single break point in the fiber bundle.  The
multiple failure case is then constructed through superposition of these
single-failure transmission factors.  This method has proven very efficient
from a numerical point of view, and has been generalized through a series
of later papers, see \cite{bps96,lbm00,bp97a,bp97b,ljgwa06}.

\textcite{ic97a,ic97b}, \textcite{c98}, \textcite{xc01} and \textcite{xco02} 
 analyzed the interaction between multiple breaks in
uni-directional fibers embedded in a matrix using a lattice Green function
technique \cite{zc95} to calculate the load transfer from broken to
unbroken fibers including fiber-matrix sliding with a constant
interfacial shear resistance $\tau$, given by either a debonded sliding
interface or by matrix shear yielding.  The differential load
carrying capacity of the matrix is assumed to be negligible.
In the following we describe the Zhou and Curtin approach in some detail.
The load-bearing
fibers have a strength distribution given by the cumulative probability
\begin{equation}
\label{zhou_curtin_cumulative}
P(\sigma,L)=1-e^{-\phi(\sigma,L)}\;
\end{equation}
of failure over a length of fiber $L$ experiencing a stress $\sigma$,
where
\begin{equation}
\label{zhou_curtin_phi}
\phi(\sigma,L)=\frac{L}{L_0}\ \left(\frac{\sigma}{\sigma_0}\right)^\rho\;,
\end{equation}
where $\rho$ is the Weibull index.
When a fiber breaks, the load is transferred to the unbroken fibers.  We will
return to the details henceforth.  The newly broken fiber slides relatively
to the matrix.  The shear resistance $\tau$ provides an average axial fiber
stress along the single broken fiber
\begin{equation}
\label{zhou_curtin_slide}
\sigma(z)=\min\left(\frac{2\tau z}{r},\sigma^0(z)\right)\equiv p(z)
\end{equation}
at a distance $z$ from the break, where $r$ is radius of the fibers
and $\sigma^0(z)$ is the axial fiber stress prior to the failure at point $z$.
 This defines a length scale
\begin{equation}
\label{zhou_curtin_length}
l_s=\frac{r\sigma^0(l_s)}{2\tau}\;.
\end{equation}
The total stress change within a distance $\pm l_s$ of the break is
distributed to the other fibers.  A key assumption in what now follows
is that the total stress in each plane $z$ is conserved:  The stress
difference $\sigma^0(z)-\sigma(z)$ is distributed among the other
intact fibers at the same $z$-level.

\begin{figure}
\centering
\includegraphics[width=8cm]{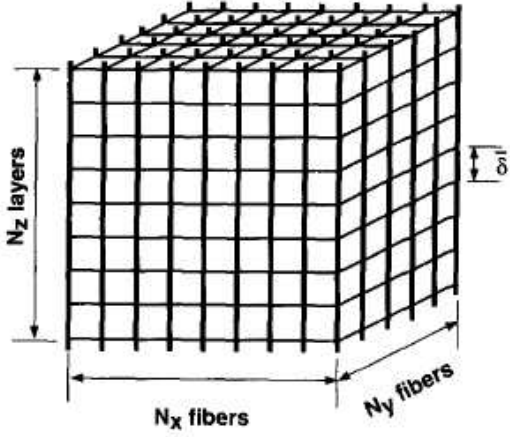}
\caption{The discretization of a three-dimensional unidirectional
fiber reinforced composite used by \textcite{zc95}.}
\label{ah-fig14}
\end{figure}

In order to set up the lattice Green function approach, the system must be
discretized.  Each fiber, oriented in the $z$ direction, of length $L_z$ is
divided into $N_z$ elements of length $\overline{\delta}=L_z/N_z$.
The fibers are arranged on the nodes of a square lattice in the
$xy$ plane so that there is a total of $N_f=N_x\times N_y$ fibers.
The lattice constants in the $x$ and $y$ directions are $a_x$
and $a_y$, respectively.  Each fiber
is labeled by $n$ where $1\le n \le N_f$.  This is shown in Fig.\
\ref{ah-fig14}.
The stress on fiber $n$ in layer $m$ along the $z$ direction is given by
$\sigma_{n,m}$.  Fiber $n$ at layer $m$ may be intact.  It then acts as
a Hookean spring with spring constant $k_t$ responding to the stress
$\sigma_{n,m}$.  If fiber $n$ has broken at layer $m$, it carries a stress
equal to zero.  The third possibility is that fiber $n$ has broken
elsewhere at $m'$, and layer $m$ is within the slip zone.  It then carries
a stress
\begin{equation}
\label{zhou_curtin_disc_slip}
\sigma_{n,m}=\min\left(\frac{2\tau\overline{\delta}}{r}|m-m'|,
\sigma^0_{n,m}\right)\equiv p_{n,m}\;,
\end{equation}
which is the discretization of Eq.\ (\ref{zhou_curtin_slide}) with zero
spring constant.

Each element $m$ of fiber $n$ has two end nodes associated with it.
At all such nodes, springs parallel to the $xy$ plane are placed linking
fiber $n$ with its nearest neighbors.  These springs have spring constant
$k_s$.  The displacement of the nodes is assumed confined to the $z$
direction only.  Zhou and Curtin denote the displacement of node connecting
element $m$ with element $m+1$ of fiber $n$, $u^+_{n,m}$, and the displacement
of node linking element $m$ with element $m-1$ of fiber $n$, $u^-_{n,m}$.
The force on element $m$ of fiber $n$ from element $m$ of fiber $n+1$ is
\begin{equation}
\label{zhou_curtin_force_on_n}
f_m(n;n+1)=k_s(u^+_{n+1,m}-u^+_{n,m})+k_s(u^-_{n+1,m}-u^-_{n,m})\;.
\end{equation}

The reader should compare the following discussion with that which was
presented in Section \ref{Sec:4C}. We now assume that it is only layer
$m=0$ that carries any damaged or slipped elements, the rest of the layers
$m\neq 0$ are perfect.  Let ${\bf u}=\{ u^\pm_{n,m}\}$.
If a force ${\bf f}=\{f^\pm_{n,m}\}$ is
applied to the nodes, the response is
\begin{equation}
\label{zhou_curtin_response}
{\bf u} = {\bf G}\ {\bf f}\;,
\end{equation}
where $\bf G$ is the lattice Green function.  Given the displacements
from solving this equation combined with Eq.\
(\ref{zhou_curtin_force_on_n}), the force carried by each broken element
is found.  The inverse of the lattice Green function is
${\bf D} = {\bf G}^{-1}$.  The elements of $\bf D$ are either zero, $k_s$
or $k_t$, reflecting the status of the springs; undamaged, slipping or broken.
When there are no breaks in layer $m=0$,
we define ${\bf D}^0 = ({\bf G}^0)^{-1}$, and
$\delta{\bf D}={\bf D}^0-{\bf D}$.  Hence, $\delta{\bf D}$ plays a r{\^o}le
somewhat similar to the matrix $\bf K$ defined in Eq.\ (\ref{M4}). By
combining these definitions, Zhou and Curtin find
\begin{equation}
\label{zhou_curtin_g_and_d}
{\bf G} = ({\bf 1} - {\bf G}^0\delta{\bf D})^{-1}\ {\bf G}^0\;.
\end{equation}
The matrices $\bf G$ and $\bf D$ have dimension $N\times N$ where
$N=N_x\times N_y\times N_z$.  By appropriately labeling the rows and
columns, the matrices ${\bf D}$ and $\bf G$ may be written
\begin{equation}
\label{zhou_curtin_g_matrix}
{\bf G} = \left(\begin{array}{cc}
                 G_{dd}& G_{dp}\\
 G_{pd}& G_{pp}\\
                \end{array}
          \right)\;,
\end{equation}
where the $(2N_x N_y)\times (2N_x N_y)$ matrix $G_{dd}$ couples elements
{\it within\/} the layer $m=0$, where all the damage is located.  The
matrix $G_{pp}$ couples elements within the rest of the layers.  These
are undamaged --- ``perfect".  The two matrices $G_{dp}$ and $G_{pd}$
provide the cross couplings.  The matrix $\delta {\bf D}$ becomes in this
representation
\begin{equation}
\label{zhou_curtin_layer_d0}
\delta {\bf D} = \left( \begin{array}{cc}
                 \delta D_{dd}&0\\
                             0&0\\
                        \end{array}
                 \right)\;.
\end{equation}
Combining this equation with Eq.\ (\ref{zhou_curtin_g_and_d}) gives
\begin{equation}
\label{zhou_curtin_dyson}
G_{dd}=(I-\delta D_{dd})^{-1} G^0_{dd}\;,
\end{equation}
where the intact Green function $G^0_{dd}$ may be found analytically
by solving Eq.\ (\ref{zhou_curtin_response}) for the intact lattice in
Fourier space $\{\vec q\}$,
\begin{eqnarray}
\label{zhou_curtin_fourier_g}
&{\bf F}{\bf G}^0{\bf F}^{-1}(\vec q) = \frac{1}{4}\nonumber\\
&\left[k_s\sin^2\left(\frac{q_xa_x}{2}\right)
     +k_s\sin^2\left(\frac{q_ya_y}{2}\right)
     +k_t\sin^2\left(\frac{q_z\overline{\delta}}{2}\right)\right]^{-1}\;.
\nonumber\\
\end{eqnarray}

By using that $f^+_{n,0}=-f^-{n,0}$, Zhou and Curtin find that
\begin{equation}
\label{zhou_curtin_u_u}
u^+_{n,0}-u^{-}_{n,0} 
= \sum_{n'}[G_{dd}(n'^+;n^+) -G_dd(n'^+;n^-)] f^+_{n',0}\;,
\end{equation}
where $n^+$ and $n^-$ refer to the upper and lower node attached to
element $n$ is layer $m=0$.
Before completing the model, the Weibull strength distribution, Eqs.\
(\ref{zhou_curtin_cumulative}) and (\ref{zhou_curtin_phi}), must be
discretized.  Each element $(n,m)$ is given a maximum sustainable load
$s_{n,m}$ from the cumulative probability
\begin{equation}
\label{zhou_curtin_discrete_cumulative}
P_f(s)=1-e^{-(s/\overline{\sigma})^\rho}\;,
\end{equation}
where $\overline{\sigma}=(L_0/\overline{\delta})^{1/\rho}\sigma_0$.

The breakdown algorithm proceeds as follows:
\begin{enumerate}
\item A force per fiber set equal to the smallest breaking
threshold, $f_0=\min_{n,m} s_{n,m}$, is applied to the system.
\item The weakest fiber or fibers  are broken by setting their spring
constants to zero.
\item Decrease the stresses in the element below and above the just
broken fibers according to Eq.\ (\ref{zhou_curtin_disc_slip}).
\item Solve Eq.\ (\ref{zhou_curtin_dyson}) for the layers in which the
breaks occurred.
\item Calculate the spring displacements in the layers where the breaks
occurred using Eq.\ (\ref{zhou_curtin_u_u}) and an effective applied
force $f^+_{n,m}=f_0-p_{n,m}$.  The force on each intact spring in such a
layer is then $\sigma_{n,m}=k_t(u^+_{n,m}-u^-_{n,m})$.
\item With the new stresses, search for other springs that carry
a force beyond their thresholds $s_{n,m}$.  If such springs are found, break
these and return to (2).  Otherwise proceed.
\item Search for the spring which is closest to it breaking threshold.  This
spring is the one with $\lambda=\min_{n,m} (s_{n,m}/f_{n,m})$.  Increase the
load by a factor $\lambda\eta$, where $\eta$ is equal to or somewhat
larger than unity.  This factor is present to take into account the
non-linearities introduced in the system due to the slip of the fibers.
\item Proceed until the system no longer can sustain a load.
\end{enumerate}

By changing the ratio $k_t/k_s$ between the moduli of the springs
in the discretized lattice, it is possible to go from fiber bundle
behavior essentially evolving according to equal load sharing (ELS) to
local load sharing (LLS).

Whereas the computational cost of finite-element calculations on fiber
reinforced composites scales with the volume of the composite, the Break
Influence Superposition Technique and the Lattice Green function technique
scale with the number of fiber breaks in the sample. This translates into
systems studied by the latter two techniques can be orders of magnitude
larger than the former \cite{ic97a}.

After this rather sketchy tour through the use of fiber bundle models as tools
for describing the increasing important fiber reinforced composite materials,
we now turn to the use of fiber bundle models in non-mechanical settings.


\subsection{Failure phenomena in networks, traffic and earthquake.}
The typical failure dynamics of the fiber bundle model captures quite
faithfully the failure behavior of several multicomponent systems like the
communication or traffic networks. Similar to the elastic networks
considered here, as the local stress or load (transmission rate or traffic
currents) at any part of the network goes beyond the sustainable limit,
that part of the system or the network fails or gets jammed, and the excess
load gets redistributed over the other intact parts. This, in turn, may
induce further failure or jamming in the system. Because of the tectonic
motions stresses develop at the  crust - tectonic plate resting (contact)
regions and the failure at any of  these supports induces additional
stresses elsewhere. Apart from the healing  phenomena in geological
faults, the fiber bundle models have built-in  features to capture the
earthquake dynamics. Naturally, the statistically established laws for
earthquake dynamics can be easily recast into the  forms derived here for
the fiber bundle models.

We will consider here in some more details these three applications
specifically.

\subsubsection{Modelling network failures}

The fiber bundle model has been applied \cite{kkj05} to study the 
cascading failures  of network structures, like Erdos-Renyi networks \cite{er59}, known as ER networks and
 Watts-Strogatz networks \cite{ws98}, known as WS networks, to model the overloading failures in 
power grids, etc. Here, the 
nodes or the individual power stations are modelled as fibers and the 
transmission links between these nodes are utilized to transfer the excess 
load (from one broken fiber or station to another).    
 
The load transfer of broken fibers or nodes through the edges or links 
of the underlying network is governed by the  LLS 
rule \cite{hp78,p79,sp81}. Under a non-zero external
load $N\sigma$, the actual stress $\sigma_{i}$ of the intact
fiber $i$ is given by the sum of $\sigma$ and the transferred
load from neighboring broken fibers. The
local load transfer, from broken fibers to intact fibers, depends on
the load concentration factor $K_{i}\equiv\sigma_{i}/{\sigma}$
with $K_{i}=1+{\sum_{j}}^{\prime}m_{j}/k_{j}$, where the primed summation
is over the cluster of broken fibers directly connected to $i$, $m_{j}$
is the number of broken fibers in the cluster $j$, and $k_{j}$ is
the number of intact fibers directly connected to $j$. 

Let the external stress  $\sigma$ be increased by an infinitesimal 
amount $\delta \sigma$ starting  from $\sigma=0$. 
Fibers for which  strength $<K_{i}\sigma$ break  iteratively
until no more fibers break. For each increment of $\sigma$,
the size $s(\sigma)$ of the avalanche is defined as the number
of broken fibers triggered by the increment. The surviving fraction
$U(\sigma)$ of fibers  can be  written as
 \begin{equation}
U(\sigma)=1-\frac{1}{N}\sum_{\sigma'<\sigma}s(\sigma').\label{eq:x}
\end{equation}
One can also measure directly the response function $\chi$, or the generalized
susceptibility, denoted as 
\begin{equation}
\chi(\sigma)=\left|\frac{dU}{d\sigma}\right|.
\end{equation}
The critical value $\sigma_{c}$ of the external
load, can be  defined from
the condition of the global breakdown $U(\sigma_{c})=0$ .

\begin{figure}
\centering{\resizebox*{0.5\textwidth}{!}{\includegraphics{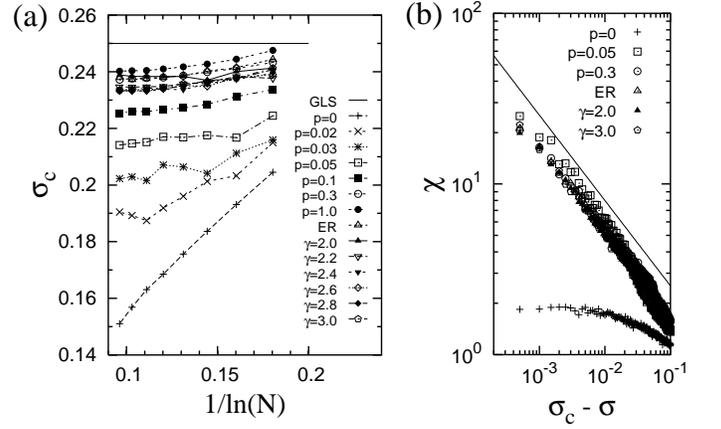}}} 

\caption{ \label{fig:sigmac} (a) The system size ($N$) dependence of critical
points ($\sigma_{c}$) for various networks with $N=2^{8},2^{9},\ldots,2^{15}$
vertices. (b) The susceptibility for the networks with $N=2^{14}$.
$p$ and $\gamma$ are the rewiring probability in the WS networks
and the exponent of degree distribution $P(k)\sim k^{-\gamma}$ 
 respectively. The data points
are obtained from the averages over $10^{4}$ ($10^{3}$ for $N=2^{15}$)
ensembles. From  \textcite{kkj05}.}

\end{figure}

The critical value $\sigma_{c}$
and the susceptibility $\chi$ have been calculated numerically 
for the model under the LLS rule on various
network structures, such as the local regular network, the WS
 network, the ER network, and the scale-free networks 
(Fig. ~\ref{fig:sigmac}). 
The results suggest that
the critical behavior of the model on complex networks is  completely
different from that on a regular lattice. More specifically, while
$\sigma_{c}$ for the FBM on a local regular network vanishes
in the thermodynamic limit and is described by $\sigma_{c}\sim1/\ln(N)$
for finite-sized systems (see the curve for $p=0$ in Fig.~\ref{fig:sigmac},
corresponding to the WS network with the rewiring probability $p=0$),
$\sigma_{c}$ for {\em all} networks except for the local
regular one does not diminish but converges to a nonzero value as
$N$ is increased. Moreover, the susceptibility diverges at the critical
point as $\chi\sim(\sigma_{c}-\sigma)^{-1/2}$,
regardless of the networks, which is again in a sharp contrast to
the local regular network [see Fig.~\ref{fig:sigmac}(b)]. The
critical exponent $1/2$ clearly indicates that the FBM under the
LLS rule on complex networks belongs to the same universality class
as that of the ELS regime~\cite{pbc02} although the load-sharing
rule is strictly local. The observed  variation for $\sigma_c$ is only 
 natural for
LLS model. In an LLS model if $n$ successive fibers fail (each with a finite 
probability $\rho_f$), then the total probability of such an event is 
$N \rho_f^{n} (1-\rho_f)^2$ as the probability is proportional to the bundle 
size $N$. If this probability is finite, then $n \sim \ln N$ (for any finite 
$\rho_f$, the failure probability of any fiber in the bundle). For a failure 
of $n$ successive fibers, the neighboring intact fibers get the transferred 
load $\sim n\sigma$ which if becomes greater than or equal to their 
strength, it surely fails giving 
$\sigma_c \sim 1/n \sim 1/\ln N$ (see Section IV.A). 

%


The evidence that the LLS model on complex networks belongs to the
universality class of the ELS model is also found \cite{kkj05} in the avalanche
size distribution $D(\Delta)$: Unanimously observed power-law behavior
$D(\Delta)\sim \Delta^{-5/2}$  (a) for all networks
except for the local regular one (the WS network with $p=0$) is in
perfect agreement with the behavior for the ELS case~\cite{hh92}.
On the other hand, the LLS model for a regular lattice has been shown
to exhibit completely different avalanche size distribution~\cite{hh94,khh97}.
Also one can observe a clear difference in terms of the failure 
probability $F(\sigma)$ defined as the probability of failure of the 
whole system at an external stress $\sigma$. While $F(\sigma)$ 
 values for LLS on complex networks fall on a common line, LLS 
on regular network shows a distinctly  different trend \cite{kkj05}, see also 
\textcite{dd07c}.   

\subsubsection{Modelling traffic jams }

 One can  apply the equal load sharing fiber bundle model
to study the traffic failure in a system of
parallel road network in a city. For some special distributions, like the 
uniform distribution,  of
traffic handling capacities (thresholds) of the roads, the critical
behavior of the jamming transition can be studied analytically. This, in fact,
is exactly comparable with that for the asymmetric simple exclusion process
in a single channel or road \cite{c06}.

Traffic jams or congestions occur essentially due to the excluded
volume effects (among the vehicles) in a single road and due to the
cooperative (traffic) load sharing by the (free) lanes or roads in
multiply connected road networks (see e.g., \textcite{d99,css00}).
Using FBM for the traffic network, it has been shown \cite{c06} that the
generic equation for the approach of the jamming transition in FBM
corresponds to that for the Asymmetric Simple Exclusion Processes
(ASEP) leading to the transport failure transition in a single channel
or lane (see e.g., \textcite{s05}).

\begin{figure}
\includegraphics[width=2.5in,height=2.5in]{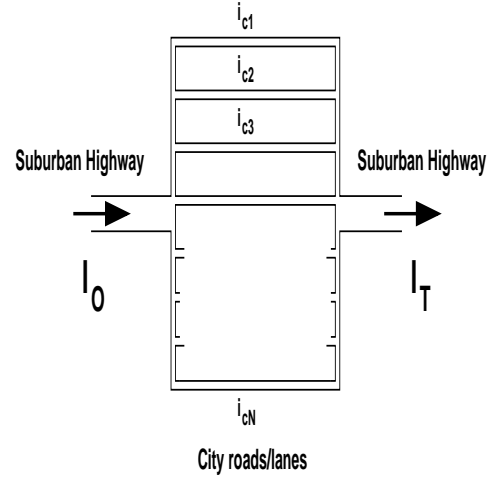} 
\caption{ The highway traffic current $I_{O}$ gets fragmented into uniform 
currents $i$ in each of the  
 narrower roads and the roads having threshold current $i_{c_{n}}\le i$
get congested or blocked. This results in extra load for the uncongested
roads. We assume that this extra load per unconjested roads gets equally 
redistributed and gets added to the existing load, causing further blocking 
of some more roads. }
\label{BKC-traffic-model}
\end{figure}

Let the suburban highway traffic, while entering the city, get fragmented
equally through the various narrower streets within the city and get
combined again outside the city (see Fig. \ref{BKC-traffic-model}). 
If $I_{O}$ denotes the input traffic current and  $I_{T}$  is the total
 output traffic current, then at the steady state, without any global traffic
jam, $I_{T}=I_{O}$. In case $I_{T}$ falls below $I_{O}$, the global
jam starts and soon $I_{T}$ drops to zero. This occurs if $I_{O}>I_{c}$,
the critical traffic current of the network, beyond which global traffic jam
occurs. Let  the parallel roads within the city have
different thresholds for traffic handling capacity: $i_{c_{1}},i_{c_{2}},\ldots,i_{c_{N}}$
for the $N$ different roads (the $n$-th road gets jammed if the
traffic current $i$ per road exceeds $i_{c_{n}}$). Initially $i=I_{O}/N$
and increases as some of the roads get jammed and the same traffic
load $I_{O}$ has to be shared equally by a lower number of unjammed
roads. Next, we assume that the distribution $p(i_{c})$ of these
thresholds is uniform  up to a maximum threshold current
(corresponding to the widest road traffic current capacity), which
is normalized to unity (sets the scale for $I_{c}$).

The jamming dynamics in this model starts from the $n$-th road
(say in the morning)
 when the traffic load  \(i\)
 per city roads exceeds the threshold $i_{c_n}$ of that road.
Due to this jam, the total
number of unconjested roads  decreases and the rest of these roads
have to bear the entire traffic load in  the system. Hence the effective
traffic load or stress
on the uncongested roads increases and this compels some more roads to 
 get jammed. These two sequential operations, namely the stress or traffic load
 redistribution and further failure in service of roads continue till an 
equilibrium is reached, where
either the surviving roads are strong (wide) enough to share equally
and  carry the entire traffic load on the system (for $I_O < I_c$)
 or all the roads fail (for $I_O \ge I_c$)  and a (global)
traffic jam occurs in the entire  road network system.

This jamming dynamics can be represented by recursion relations in
discrete time steps. Let  $U_{t}(i)$  be the fraction
of unconjested roads in the network that survive after (discrete)
time step $t$, counted from the time $t=0$ when the load (at the
level $I_{O}=iN$) is put in the system (time step indicates the number
of stress redistributions). As such, $U_{t}(i=0)=1$ for all $t$
and $U_{t}(i)=1$ for $t=0$ for any $i$; $U_{t}(i)=U^{*}(i)\ne0$
for $t\to\infty$ if $I_{O}<I_{c}$, and $U_{t}(i)=0$ for $t\to\infty$
if $I_{O}>I_{c}$.

Here $U_{t}(i)$ follows a simple recursion relation 
 \[U_{t+1}=1-i_{t};\ \ i_{t}=\frac{I_{O}}{U_{t}N}\]
 \begin{equation}
{\rm or,}\ \ U_{t+1}=1-\frac{i}{U_{t}}.\label{recU_t}\end{equation}
 The critical behavior of this model remains the same as discussed
in Section III.A  in terms of $i$ and the exponent values remain unchanged:
$\alpha=1/2=\beta=\theta$, $\eta=1$ for all these equal (traffic)
load sharing models.

\begin{figure}
\includegraphics[width=2.5in,height=1.5in]{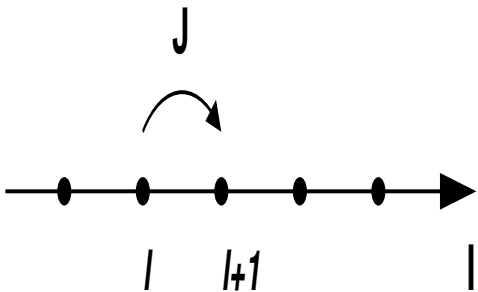} 
\caption{ The transport current $I$ in the one dimensional lane or road is
possible if, say, the $l$-th site is occupied and the $(l+1)$-th
site is vacant. The inter-site hopping probability is indicated by
$J$. }
\label{ASPE-model}
\end{figure}

In the simplest version of the asymmetric simple exclusion
process transport in a chain (see Fig. \ref{ASPE-model}), the transport 
 corresponds to movement of vehicles, which is possible only when a vehicle at
site $l$, say, moves to the vacant site $l+1$. The transport current
$I$ is then given by \cite{s05} \begin{equation}
I=J\rho'_{l}(1-\rho'_{l+1}),\label{add1}\end{equation}
 where $\rho'_{l}$ denotes the site occupation density at site $l$
and $J$ denotes the inter-site hopping probability. The above equation
can be easily recast in the form \begin{equation}
\rho'_{l+1}=1-{\frac{\sigma}{\rho'_{l}}},\label{add2}\end{equation}
 where $\sigma=I/J$. Formally it is the same as the recursion relation
 for the density of uncongested roads in the FBM model discussed
above; the site index here in ASEP plays the role of time index in
FBM. Such exact correspondence indicates
identical critical behavior in both the cases. The same universality
for different cases (different threshold distributions) in FBM suggests 
similar behavior for
other equivalent ASEP cases as well \cite{b07}. For extension of the model 
to scale free Traffic Networks, see \textcite{zgzf08}.

\subsubsection{Modelling earthquake dynamics}

The Earth's outer crust, several tens of kilometers in thickness,  
rests
on tectonic shells. Due to the high temperature-pressure phase changes
and consequent ionizations of the metallic ores, powerful
magneto-hydrodynamic convective flows occur in the earth's mantle at
several hundreds of kilometers in depth. The tectonic shell, divided into
about ten mobile plates, have got relative velocities of the order of a
few centimeters per year, see e.g., \textcite{s02}. 

The stresses developed at the interfaces between the crust and the
tectonic shells during the (long) \emph{sticking} periods get released during
the (very short) \emph{slips}, causing the releases of the stored elastic
energies (at the fault asperities) and consequent earthquakes.

Two well known phenomenologically established  laws governing
the earthquake statistics are (a) the Gutenberg-Richter law
\begin{equation} 
\mathcal{N}(E) \sim E^{-\xi'},  
\label{GR-law}
\end{equation} 
relating the number density ($\mathcal{N}$) of earthquakes with the 
released energy greater than or equal to $E$; and (b) the Omori law

\begin{equation} 
d(\tilde \mathcal{N} (t))/dt = 1/t^{\eta ^{'}}, 
\label{Omori-law}
\end{equation} 
where $\tilde \mathcal{N}$ denotes the number of aftershocks having
magnitude or released energy larger than a preassigned small  but otherwise 
arbitrary threshold value.

As mentioned already, because of the tectonic motions, stresses develop
at the  crust-tectonic plate contact regions and the entire load is
supported by such regions. Failure at any of  these supports  necessitates
load
redistributions and induces additional stresses elsewhere. 
In fact the avalanche
statistics discussed in Section III.B can easily explain the 
Gutenberg-Richter law (Eq. \ref{GR-law}) with the identification 
$\xi' = \xi - 1$. Similarly,
the decay of the number of failed fibers $N(t) = N(1 - U_t)$ at the critical
point, given by $N(t) \sim t^{-\eta}$ (see e.g., Eq. \ref{eq:fbm-omori}), 
crudely speaking,  gives in turn the
 Omori law behavior (Eq. \ref{Omori-law}) for the fiber bundle model, with the
identification $\eta ^{'} = 1 + \eta$. 

For some recent discussions on further 
studies along these lines, see e.g., \textcite{tg04}. 
The stick-slip motion in the Burridge-Knopoff model 
(see e.g., \textcite{cls94}),
 where the blocks (representing  portion of the solid crust) connected with 
springs (representing the elastic strain developed due to tectonic motion) are
pulled uniformly on a rough surface, has the same feature of stress 
redistribution as one or more blocks slip and the dynamics  
was mapped onto a ELS fiber bundle  model by \textcite{s92}. The power law 
distributions of the fluctuation driven bursts  around the critical points 
have been interpreted as the above two statistical laws for 
earthquakes.

\section{Summary and concluding remarks}

The fiber bundle model enjoys a rare double position in that it is both
useful in a practical setting for describing a class of real materials under 
real working conditions, and at the same time being abstract enough to 
function as a model for exploring fundamental breakdown mechanisms from
a general point of view.  Very few models are capable of such a double life.
This means that a review of the fiber bundle model may take on very different
character depending on the point of view.  We have in this review emphasized
the fiber bundle model as a model for exploring fundamental breakdown
mechanisms.    

Failure in loaded disordered materials is a collective phenomenon.    
It proceeds through a competition between disorder and stress distribution.  
The disorder implies a 
distribution of local strength.  If the stress distribution were uniform in
the material, it would be the weakest spot that would fail first.  Suppose now
that there has been a local failure at a given spot in the material.  The
further away we go from this failed region, the weaker the weakest region 
within this distance will be.  Hence, the disorder makes local failures 
repel each other: They will occur as far as possible from each other.  
However, as the material fails locally, the stresses are 
redistributed.  This redistribution creates hot spots where local failure is 
likely due to high stresses.  Since these hot spots occur at the boundaries of
the failed regions, the effect of the stress field is an attraction between 
the local failures \cite{rh90}.  Hence, disorder and stress has opposite 
effect on the breakdown process; repulsion vs.\ attraction, and this leads to 
competition between them.  Since the disorder in the strength of the material,
leading to repulsion, dimishes throughout the breakdown process, whereas the
stresses create increasingly important hot spots, it is the stress distribution
that ends up dominating towards the end of the process.  

The fiber bundle model catches this essential aspect of the failure process.
Depending on the load redistribution mechanism, the quantitative aspects
change.  However, qualitatively it remains the same.  In the ELS case, there
are no {\it localized\/} hot spots, all surviving fibers are loaded the same 
way.  Geometry does not enter into the redistribution of forces, and we may 
say that the ``hot spots" include all surviving fibers.  This aspect gives
the ELS fiber bundle model its mean field character, even though all other 
fluctuations are present, such as those giving rise to bursts. 
  
As shown in Section III, the lack of geometrical aspects in the redistribution 
of forces in the ELS model
enables us to construct the recursion relations (e.g., Eq. \ref{rec-U}) which
capture well the failure dynamics. We find that the eventual statistics,
governed by the fixed points for $\sigma < \sigma_c$, the average strength of
the bundle, essentially shows a normal critical behavior: order parameter
$O \sim (\sigma_c -\sigma)^{\alpha}$, breakdown susceptibility 
$\chi \sim (\sigma_c -\sigma)^{-\beta}$ and relaxation time
 $\tau = \kappa |\sigma_c -\sigma|^{-\theta}$ with 
$\alpha = 1/2 = \beta = \theta$ and 
$\kappa_{-}/\kappa_{+} = \ln N/2 \pi$ for a bundle of $N$ fibers, where 
subscripts $+$ and $-$ refers to post and pre critical cases respectively. 
The statistics of fluctuations over these average behavior,
given by the avalanche size distributions $D(\Delta) \sim \Delta^{-\xi}$
with $\xi = 3$ for discrete load increment and $= 5/2$ for quasi-static load 
increment in such ElS cases. The critical
stress $\sigma_c$ of the bundle is of course nonuniversal and its magnitude
depends on the fiber strength distribution.

For the LLS model, we essentially find (see Section IV.A), the critical 
strength of the fiber bundle $\sigma_c \sim 1/\ln N$  which vanishes in the 
macroscopic system
size limit. The avalanche size distribution is exponential for such cases.
For range-dependent redistribution of load (see Section IV.B)
one recovers the finite value of $\sigma_c$ and the ELS-like mean field
behavior for its failure statistics.

Extensions of the model to capture creep and fatigue behavior of composite
materials are discussed in Section V.A. Precursors of global failure
are discussed in Section V.B. It appears, a detailed knowledge of the critical
behavior of the model can help very precise determination of the global
failure point from the well defined precursors. Section V.C provides a
rather cursory review of models of fiber reinforced composites.  These
models go far beyond the simple fiber bundle model in complexity
and represent the state of the art of theoretical approaches to this
important class of materials. However, as complicated as these models are,
the philosophy of the fiber bundle model is still very much present.
Finally we discussed a few extensions of the model to failures in
communication networks, traffic jams and earthquakes in Section V.D.

As discussed here in details, the fiber bundle model not only gives an elegant 
and profound solution of the dynamic
critical phenomena of failures in disordered systems, with the associated
universality classes etc, but also offers the first solution
to the entire linear and non-linear stress-strain behavior for any
material up to its fracture or rupture point.
Although the model had been introduced at about the same time
($1926$) as the Ising model for static critical phenomena, it is
only now that the full (mean-field) critical dynamics in the fiber bundle
model is solved. Apart from these, as already discussed, several aspects of 
the fluctuations in this model are now well understood.  
Even from this specific point of view, the model
is not only intuitively very attractive, its behavior is
extremely rich and intriguing.  It would be surprising if it did not offer
new profound insights into failure phenomena also in the future.    

\begin{acknowledgments}
 We thank  P. Bhattacharyya and P. C. Hemmer for important 
collaborations at different parts of this work. We acknowledge the financial 
support from Norwegian Research Council through 
grant no. NFR 177958/V30. S.P.\ thanks SINTEF Petroleum Research for providing 
partial financial help and moral support toward this work.

\end{acknowledgments}

\bibliographystyle{apsrmp}


\begin{thebibliography}{39}

\expandafter\ifx\csname natexlab\endcsname\relax\def\natexlab#1{#1}\fi
\expandafter\ifx\csname bibnamefont\endcsname\relax
  \def\bibnamefont#1{#1}\fi
\expandafter\ifx\csname bibfnamefont\endcsname\relax
  \def\bibfnamefont#1{#1}\fi
\expandafter\ifx\csname citenamefont\endcsname\relax
  \def\citenamefont#1{#1}\fi
\expandafter\ifx\csname url\endcsname\relax
  \def\url#1{\texttt{#1}}\fi
\expandafter\ifx\csname urlprefix\endcsname\relax\def\urlprefix{URL }\fi
\providecommand{\bibinfo}[2]{#2}
\providecommand{\eprint}[2][]{\url{#2}}

\bibitem[{\citenamefont{Aharony}(1976)}]{a76}
\bibinfo{author}{\bibnamefont{Aharony}, \bibfnamefont{A.}},
  \bibinfo{year}{1976}, in \emph{\bibinfo{booktitle}{Phase Transition and 
Critical Phenomena}},
 edited by \bibinfo{editor}{\bibfnamefont{C.}~\bibnamefont{Domb}} and
\bibinfo{editor}{\bibfnamefont{M.}~\bibnamefont{Green}}
  (\bibinfo{publisher}{Academic Press, New Yourk}), \textbf{\bibinfo{volume}{17}}, p. \bibinfo{pages}{357}.

\bibitem[{\citenamefont{Alava} \emph{et~al.} (2006)}]{anz06}
\bibinfo{author}{\bibnamefont{Alava}, \bibfnamefont{M.~J.}}, 
\bibinfo{author}{\bibfnamefont{P.~K.~V.~V.}~\bibnamefont{Nukala}}, and
\bibinfo{author}{\bibfnamefont{S.}~\bibnamefont{Zapperi}},
\bibinfo{year}{2006},
\bibinfo{journal}{Adv. Phys.} \textbf{\bibinfo{volume}{55}},
\bibinfo{pages}{349}.

\bibitem[{\citenamefont{Andersen} \emph{et~al.} (1997)}]{asl97}
\bibinfo{author}{\bibnamefont{Andersen}, \bibfnamefont{J.~V.}}, 
\bibinfo{author}{\bibfnamefont{D.}~\bibnamefont{Sornette}} and
\bibinfo{author}{\bibfnamefont{K.}~\bibnamefont{Leung}},
\bibinfo{year}{1997},
\bibinfo{journal}{Phys. Rev. Lett.} \textbf{\bibinfo{volume}{78}},
\bibinfo{pages}{2140}.

\bibitem[{\citenamefont{Aveston and Kelly}(1973)}]{ak73}
\bibinfo{author}{\bibnamefont{Aveston}, \bibfnamefont{J.}} and
\bibinfo{author}{\bibfnamefont{A.}~\bibnamefont{Kelly}}, 
\bibinfo{year}{1973},
\bibinfo{journal}{J. Mater. Sci.} \textbf{\bibinfo{volume}{8}},
\bibinfo{pages}{352}.

\bibitem[{\citenamefont{Bak} \emph{et~al.} (1987)}]{btw87}
\bibinfo{author}{\bibnamefont{Bak}, \bibfnamefont{P.}}, 
\bibinfo{author}{\bibfnamefont{C.}~\bibnamefont{Tang}}, and
\bibinfo{author}{\bibfnamefont{K.}~\bibnamefont{Wiesenfeld}},
\bibinfo{year}{1987},
\bibinfo{journal}{Phys. Rev. Lett.} \textbf{\bibinfo{volume}{59}},
\bibinfo{pages}{381}.

\bibitem[{\citenamefont{Banerjee and Chakrabarti}(2001)}]{bc01}
\bibinfo{author}{\bibnamefont{Banerjee}, \bibfnamefont{R.}} and
\bibinfo{author}{\bibfnamefont{B.~K.}~\bibnamefont{Chakrabarti}}, 
\bibinfo{year}{2001},
\bibinfo{journal}{Bull. Mater. Sci.} \textbf{\bibinfo{volume}{24}},
\bibinfo{pages}{161}.


\bibitem[{\citenamefont{Batrouni} \emph{et~al.} (1986)}]{bhn86}
\bibinfo{author}{\bibnamefont{Batrouni}, \bibfnamefont{G.~G.}}, 
\bibinfo{author}{\bibfnamefont{A.}~\bibnamefont{Hansen}} and
\bibinfo{author}{\bibfnamefont{M.}~\bibnamefont{Nelkin}},
\bibinfo{year}{1986},
\bibinfo{journal}{Phys. Rev. Lett.} \textbf{\bibinfo{volume}{57}},
\bibinfo{pages}{1336}.

\bibitem[{\citenamefont{Batrouni and Hansen}(1988)}]{bh88}
\bibinfo{author}{\bibnamefont{Batrouni}, \bibfnamefont{G.~G.}} and
\bibinfo{author}{\bibfnamefont{A.}~\bibnamefont{Hansen}}, 
\bibinfo{year}{1988},
\bibinfo{journal}{J. Stat. Phys.} \textbf{\bibinfo{volume}{52}},
\bibinfo{pages}{747}.

\bibitem[{\citenamefont{Batrouni} \emph{et~al.} (2002)}]{bhs02}
\bibinfo{author}{\bibnamefont{Batrouni}, \bibfnamefont{G.~G.}}, 
\bibinfo{author}{\bibfnamefont{A.}~\bibnamefont{Hansen}} and
\bibinfo{author}{\bibfnamefont{J.}~\bibnamefont{Schmittbuhl}},
\bibinfo{year}{2002},
\bibinfo{journal}{Phys. Rev. E} \textbf{\bibinfo{volume}{65}},
\bibinfo{pages}{036126}.


\bibitem[{\citenamefont{Bernardes and Moreira}(1994)}]{bm94}
\bibinfo{author}{\bibnamefont{Bernardes}, \bibfnamefont{A.~T.}} and
\bibinfo{author}{\bibfnamefont{J.~G.}~\bibnamefont{Moreira}}, 
\bibinfo{year}{1994},
\bibinfo{journal}{Phys. Rev. B} \textbf{\bibinfo{volume}{49}},
\bibinfo{pages}{15035}.

\bibitem[{\citenamefont{Beyerlein and Landis}(1999)}]{bl99}
\bibinfo{author}{\bibnamefont{Beyerlein}, \bibfnamefont{I.~J.}} and
\bibinfo{author}{\bibfnamefont{C.~M.}~\bibnamefont{Landis}}, 
\bibinfo{year}{1999},
\bibinfo{journal}{Mech. Mater.} \textbf{\bibinfo{volume}{31}},
\bibinfo{pages}{331}.

\bibitem[{\citenamefont{Beyerlein and Phoenix}(1997a)}]{bp97a}
\bibinfo{author}{\bibnamefont{Beyerlein}, \bibfnamefont{I.~J.}} and
\bibinfo{author}{\bibfnamefont{S.~L.}~\bibnamefont{Phoenix}}, 
\bibinfo{year}{1997a},
\bibinfo{journal}{Eng. Fract. Mech.} \textbf{\bibinfo{volume}{57}},
\bibinfo{pages}{241}.

\bibitem[{\citenamefont{Beyerlein and Phoenix}(1997b)}]{bp97b}
\bibinfo{author}{\bibnamefont{Beyerlein}, \bibfnamefont{I.~J.}} and
\bibinfo{author}{\bibfnamefont{S.~L.}~\bibnamefont{Phoenix}}, 
\bibinfo{year}{1997b},
\bibinfo{journal}{Eng. Fract. Mech.} \textbf{\bibinfo{volume}{57}},
\bibinfo{pages}{267}.

\bibitem[{\citenamefont{Beyerlein and Phoenix}(1998)}]{bp98}
\bibinfo{author}{\bibnamefont{Beyerlein}, \bibfnamefont{I.~J.}} and
\bibinfo{author}{\bibfnamefont{S.~L.}~\bibnamefont{Phoenix}}, 
\bibinfo{year}{1998},
\bibinfo{journal}{Int. J. Solids and Struct.} \textbf{\bibinfo{volume}{35}},
\bibinfo{pages}{3177}.

\bibitem[{\citenamefont{Beyerlein} \emph{et~al.} (1996)}]{bps96}
\bibinfo{author}{\bibnamefont{Beyerlein}, \bibfnamefont{I.~J.}},
\bibinfo{author}{\bibfnamefont{S.~L.}~\bibnamefont{Phoenix}} and
\bibinfo{author}{\bibfnamefont{A.~M.} \bibnamefont{Sastry}},
\bibinfo{year}{1996},
\bibinfo{journal}{Int. J. Solids and Struct.} \textbf{\bibinfo{volume}{33}},
\bibinfo{pages}{2543}.

\bibitem[{\citenamefont{Bhattacharjee} (2007)}]{b07}
\bibinfo{author}{\bibnamefont{Bhattacharjee}, \bibfnamefont{S.~M.}},
\bibinfo{year}{2007},
\bibinfo{journal}{J. Phys. A: Math Theor.} \textbf{\bibinfo{volume}{40}},
\bibinfo{pages}{1703}.

\bibitem[{\citenamefont{Bhattacharyya} \emph{et~al.}(2003) \citenamefont{Bhattacharyya, Pradhan and Chakrabarti}}]{bpc03}
\bibinfo{author}{\bibnamefont{Bhattacharyya} \bibfnamefont{P.}},   
\bibinfo{author}{\bibfnamefont{S.} \bibnamefont{Pradhan}} and
\bibinfo{author}{\bibfnamefont{B.~K.} \bibnamefont{Chakrabarti}},
\bibinfo{year}{2003}, \bibinfo{journal}{Phys. Rev. E}
\textbf{\bibinfo{volume}{67}}, \bibinfo{pages}{046112}.

\bibitem[{\citenamefont{Bonn} \emph{et~al.}(1998) \citenamefont{Bonn, Kellay, Prochnow, Ben-Djemiaa and Meunier}}]{bkpdm98}
\bibinfo{author}{\bibnamefont{Bonn} \bibfnamefont{D.}},   
\bibinfo{author}{\bibfnamefont{H.} \bibnamefont{Kellay}},
\bibinfo{author}{\bibfnamefont{M.} \bibnamefont{Prochnow}},
\bibinfo{author}{\bibfnamefont{K.} \bibnamefont{Ben-Djemiaa}} and 
\bibinfo{author}{\bibfnamefont{J.} \bibnamefont{Meunier}},
\bibinfo{year}{1998}, \bibinfo{journal}{Science}
\textbf{\bibinfo{volume}{280}}, \bibinfo{pages}{265}.

\bibitem[{\citenamefont{Budiansky} \emph{et~al.}(1986) \citenamefont{Budiansky, Hutchinson and Evans}}]{bhe86}
\bibinfo{author}{\bibnamefont{Budiansky} \bibfnamefont{B.}},   
\bibinfo{author}{\bibfnamefont{J.~W.} \bibnamefont{Hutchinson}} and
\bibinfo{author}{\bibfnamefont{A.~G.} \bibnamefont{Evans}},
\bibinfo{year}{1986}, \bibinfo{journal}{J. Mech. Phys. Solids}
\textbf{\bibinfo{volume}{342}}, \bibinfo{pages}{167}.

\bibitem[{\citenamefont{Carlson} \emph{et~al.}(1994) \citenamefont{Carlson, Langer and Shaw}}]{cls94}
\bibinfo{author}{\bibnamefont{Carlson} \bibfnamefont{J.~M.}},   
\bibinfo{author}{\bibfnamefont{J.~S.} \bibnamefont{Langer}} and
\bibinfo{author}{\bibfnamefont{B.~E.} \bibnamefont{Shaw}},
\bibinfo{year}{1994}, \bibinfo{journal}{Rev. Mod. Phys.}
\textbf{\bibinfo{volume}{66}}, \bibinfo{pages}{657}.



\bibitem[{\citenamefont{Chakrabarti}(1994)}]{c94}
\bibinfo{author}{\bibnamefont{Chakrabarti}, \bibfnamefont{B.~K.}},
  \bibinfo{year}{1994}, in \emph{\bibinfo{booktitle}{Nonlinearity and Breakdown in Soft Condensed Matter}},
  edited by \bibinfo{editor}{\bibfnamefont{K.~K.}~\bibnamefont{Bardhan}},
  \bibinfo{editor}{\bibfnamefont{B.~K.}~\bibnamefont{Chakrabarti}} and
  \bibinfo{editor}{\bibfnamefont{A.}~\bibnamefont{Hansen}}
  (\bibinfo{publisher}{Springer-Verlag, Heidelberg}), p. \bibinfo{pages}{171}.

\bibitem[{\citenamefont{Chakrabarti}(2006)}]{c06}
\bibinfo{author}{\bibnamefont{Chakrabarti}, \bibfnamefont{B. K.}},
\bibinfo{year}{2006}, \bibinfo{journal}{Physica A}
\textbf{\bibinfo{volume}{372}}, \bibinfo{pages}{162}.

\bibitem[{\citenamefont{Chakrabarti and Benguigui}(1997)}]{cb97}
\bibinfo{author}{\bibnamefont{Chakrabarti}, \bibfnamefont{B. K.}}, and 
\bibinfo{author}{\bibfnamefont{L. G.} \bibnamefont{Benguigui}}, 
\bibinfo{year}{1997}, \emph{\bibinfo{title}{Statistical Physics of Fracture and Breakdown in Disordered Systems}} 
(\bibinfo{publisher}{Oxford University Press, Oxford}).

\bibitem[{\citenamefont{Chiao and Moore}(1971)}]{cm71}
\bibinfo{author}{\bibnamefont{Chiao}, \bibfnamefont{T.~T.}} and
\bibinfo{author}{\bibfnamefont{R.~L.}~\bibnamefont{Moore}}, 
\bibinfo{year}{1971},
\bibinfo{journal}{J. Comp. Materials} \textbf{\bibinfo{volume}{5}},
\bibinfo{pages}{2}.

\bibitem[{\citenamefont{Chou}(1997)}]{c92}
\bibinfo{author}{\bibnamefont{Chou}, \bibfnamefont{T. W.}},  
\bibinfo{year}{1992}, \emph{\bibinfo{title}{Microstructural Design of Fiber Reinforced Composites}} 
(\bibinfo{publisher}{Cambridge University Press, Cambridge}).

\bibitem[{\citenamefont{Chowdhury} \emph{et~al.}(2000) \citenamefont{Chowdhury, Santen and Schadschnider}}]{css00}
\bibinfo{author}{\bibnamefont{Chowdhury} \bibfnamefont{D.}},   
\bibinfo{author}{\bibfnamefont{L.} \bibnamefont{Santen}} and
\bibinfo{author}{\bibfnamefont{A.} \bibnamefont{Schadschnider}},
\bibinfo{year}{2000}, \bibinfo{journal}{Phys. Rep.}
\textbf{\bibinfo{volume}{329}}, \bibinfo{pages}{199}.

\bibitem[{\citenamefont{Ciliberto} \emph{et~al.}(2001) \citenamefont{Ciliberto,Guarino and Scorretti }}]{cgs01}
\bibinfo{author}{\bibnamefont{Ciliberto}, \bibfnamefont{S.}},
\bibinfo{author}{\bibfnamefont{A.} \bibnamefont{Guarino}} and  
\bibinfo{author}{\bibfnamefont{R.} \bibnamefont{Scorretti}},
\bibinfo{year}{2001}, \bibinfo{journal}{Physica D}
\textbf{\bibinfo{volume}{158}}, \bibinfo{pages}{83}.

\bibitem[{\citenamefont{Coleman}(1956)}]{c56}
\bibinfo{author}{\bibnamefont{Coleman}, \bibfnamefont{B. D.}},
\bibinfo{year}{1956}, \bibinfo{journal}{J. App. Phys.}
\textbf{\bibinfo{volume}{27}}, \bibinfo{pages}{862}.

\bibitem[{\citenamefont{Coleman}(1957a)}]{c57a}
\bibinfo{author}{\bibnamefont{Coleman}, \bibfnamefont{B. D.}},
\bibinfo{year}{1957a}, \bibinfo{journal}{J. App. Phys.}
\textbf{\bibinfo{volume}{28}}, \bibinfo{pages}{1058}.

\bibitem[{\citenamefont{Coleman}(1957b)}]{c57b}
\bibinfo{author}{\bibnamefont{Coleman}, \bibfnamefont{B. D.}},
\bibinfo{year}{1957b}, \bibinfo{journal}{J. App. Phys.}
\textbf{\bibinfo{volume}{28}}, \bibinfo{pages}{1065}.

\bibitem[{\citenamefont{Cox}(1952)}]{c52}
\bibinfo{author}{\bibnamefont{Cox}, \bibfnamefont{H. L.}},
\bibinfo{year}{1952}, \bibinfo{journal}{Br. J. App. Phys.}
\textbf{\bibinfo{volume}{3}}, \bibinfo{pages}{72}.

\bibitem[{\citenamefont{Curtin}(1991)}]{c91}
\bibinfo{author}{\bibnamefont{Curtin}, \bibfnamefont{W. A.}},
\bibinfo{year}{1991}, \bibinfo{journal}{J. Am. Cearam. Soc.}
\textbf{\bibinfo{volume}{74}}, \bibinfo{pages}{2837}.

\bibitem[{\citenamefont{Curtin}(1993)}]{c93}
\bibinfo{author}{\bibnamefont{Curtin}, \bibfnamefont{W. A.}},
\bibinfo{year}{1993}, \bibinfo{journal}{J. Mech. Phys. Solids}
\textbf{\bibinfo{volume}{41}}, \bibinfo{pages}{217}.

\bibitem[{\citenamefont{Curtin}(1998)}]{c98}
\bibinfo{author}{\bibnamefont{Curtin}, \bibfnamefont{W. A.}},
\bibinfo{year}{1998}, \bibinfo{journal}{Phys. Rev. Lett.}
\textbf{\bibinfo{volume}{80}}, \bibinfo{pages}{1445}.

\bibitem[{\citenamefont{Curtin and Scher}(1997)}]{cs97}
\bibinfo{author}{\bibnamefont{Curtin}, \bibfnamefont{W. A.}} and 
\bibinfo{author}{\bibfnamefont{H.} \bibnamefont{Scher}},
\bibinfo{year}{1997}, \bibinfo{journal}{Phys. Rev. B}
\textbf{\bibinfo{volume}{55}}, \bibinfo{pages}{12038}.

\bibitem[{\citenamefont{Daniels}(1945)}]{d45}
\bibinfo{author}{\bibnamefont{Daniels}, \bibfnamefont{H. E.}},
\bibinfo{year}{1945}, \bibinfo{journal}{Proc. Roy. Soc. London}
\textbf{\bibinfo{volume}{A 183}}, \bibinfo{pages}{405}.

\bibitem[{\citenamefont{Daniels and Skyrme}(1985)}]{ds85}
\bibinfo{author}{\bibnamefont{Daniels}, \bibfnamefont{H. E.}} and 
\bibinfo{author}{\bibfnamefont{T.~H.~R.} \bibnamefont{Skyrme}},
\bibinfo{year}{1985}, \bibinfo{journal}{Adv. Appl. Prob.}
\textbf{\bibinfo{volume}{17}}, \bibinfo{pages}{85}.

\bibitem[{\citenamefont{Daniels}(1989)}]{d89}
\bibinfo{author}{\bibnamefont{Daniels}, \bibfnamefont{H. E.}},
\bibinfo{year}{1989}, \bibinfo{journal}{Adv. Appl. Prob.}
\textbf{\bibinfo{volume}{21}}, \bibinfo{pages}{315}.

\bibitem[{\citenamefont{de Silveira}(1999)}]{d98}
\bibinfo{author}{\bibnamefont{de Silveira}, \bibfnamefont{R.}},
\bibinfo{year}{1998}, \bibinfo{journal}{Phys. Rev. Lett. (Comment)}
\textbf{\bibinfo{volume}{80}}, \bibinfo{pages}{3157}.
\bibitem[{\citenamefont{de Silveira}(1999)}]{d99}
\bibinfo{author}{\bibnamefont{de Silveira}, \bibfnamefont{R.}},
\bibinfo{year}{1999}, \bibinfo{journal}{Am. J. Phys.}
\textbf{\bibinfo{volume}{67}}, \bibinfo{pages}{1177}.

\bibitem[{\citenamefont{Dill-Langer} \emph{et~al.}(2003) \citenamefont{Dill-Langer, Hidalgo, Kun, Moreno, Aicher and Herrmann}}]{dhkmah03}
\bibinfo{author}{\bibnamefont{Dill-Langer}, \bibfnamefont{G.}},
\bibinfo{author}{\bibfnamefont{R.~C.} \bibnamefont{Hidalgo}},  
\bibinfo{author}{\bibfnamefont{F.} \bibnamefont{Kun}}, 
\bibinfo{author}{\bibfnamefont{Y.} \bibnamefont{Moreno}}, 
\bibinfo{author}{\bibfnamefont{S.} \bibnamefont{Aicher}} and 
\bibinfo{author}{\bibfnamefont{H.~J.} \bibnamefont{Herrmann}}, 
\bibinfo{year}{2003},
\bibinfo{journal}{Physica A} \textbf{\bibinfo{volume}{325}},
\bibinfo{pages}{547}.

\bibitem[{\citenamefont{Diodati} \emph{et~al.}(1997) \citenamefont{Diodati, Marchesoni and Piazza }}]{dmp91}
\bibinfo{author}{\bibnamefont{Diodati}, \bibfnamefont{P.}},
\bibinfo{author}{\bibfnamefont{F.} \bibnamefont{Marchesoni}} and
\bibinfo{author}{\bibfnamefont{S.} \bibnamefont{Piazza}}, 
\bibinfo{year}{1991},
\bibinfo{journal}{Phys. Rev. Lett.} \textbf{\bibinfo{volume}{67}},
\bibinfo{pages}{2239}.

\bibitem[{\citenamefont{Divakaran and Dutta}(2007a)}]{dd07a}
\bibinfo{author}{\bibnamefont{Divakaran}, \bibfnamefont{U.}} and
\bibinfo{author}{\bibfnamefont{A.}~\bibnamefont{Dutta}}, 
\bibinfo{year}{2007a},
\bibinfo{journal}{Phys. Rev. E} \textbf{\bibinfo{volume}{75}},
\bibinfo{pages}{011109}.

\bibitem[{\citenamefont{Divakaran and Dutta}(2007b)}]{dd07b}
\bibinfo{author}{\bibnamefont{Divakaran}, \bibfnamefont{U.}} and
\bibinfo{author}{\bibfnamefont{A.}~\bibnamefont{Dutta}}, 
\bibinfo{year}{2007b},
\bibinfo{journal}{Phys. Rev. E} \textbf{\bibinfo{volume}{75}},
\bibinfo{pages}{011117}.

\bibitem[{\citenamefont{Divakaran and Dutta}(2007c)}]{dd07c}
\bibinfo{author}{\bibnamefont{Divakaran}, \bibfnamefont{U.}} and
\bibinfo{author}{\bibfnamefont{A.}~\bibnamefont{Dutta}}, 
\bibinfo{year}{2007c},
\bibinfo{journal}{Int. J. Mod. Phys.} \textbf{\bibinfo{volume}{18}},
\bibinfo{pages}{919}.

\bibitem[{\citenamefont{Divakaran and Dutta}(2008)}]{dd08}
\bibinfo{author}{\bibnamefont{Divakaran}, \bibfnamefont{U.}} and
\bibinfo{author}{\bibfnamefont{A.}~\bibnamefont{Dutta}}, 
\bibinfo{year}{2008},
\bibinfo{journal}{Phys. Rev. E} \textbf{\bibinfo{volume}{78}},
\bibinfo{pages}{021118}.

\bibitem[{\citenamefont{Duxbury and Leath}(1994)}]{dl94}
\bibinfo{author}{\bibnamefont{Duxbury}, \bibfnamefont{P.~M.}} and
\bibinfo{author}{\bibfnamefont{P. M.}~\bibnamefont{Leath}}, 
\bibinfo{year}{1994},
\bibinfo{journal}{Phys. Rev. B} \textbf{\bibinfo{volume}{49}},
\bibinfo{pages}{12676}.

\bibitem[{\citenamefont{Erd\"{o}s and  R\'{e}nyi}(1959)}]{er59}
\bibinfo{author}{\bibnamefont{Erd\"{o}s}, \bibfnamefont{P.}} and
\bibinfo{author}{\bibfnamefont{A.}~\bibnamefont{R\'{e}nyi}}, 
\bibinfo{year}{1959},
\bibinfo{journal}{Publ. Math. (Debrecen)} \textbf{\bibinfo{volume}{6}},
\bibinfo{pages}{290}.

\bibitem[{\citenamefont{Fazzini}(1991)}]{f91}
\bibinfo{author}{\bibnamefont{Fazzini}, \bibfnamefont{P.}},
\bibinfo{year}{1991}, 
 \emph{Basic Acoustic Emission}
in  \emph{\bibinfo{title}{Nondestructive Testing Monographs and Track Vol 6}} 
(\bibinfo{publisher}{Gordon and Breach Science Publishers, New York}).
i
\bibitem[{\citenamefont{Feller}(1966)}]{f66}
\bibinfo{author}{\bibnamefont{Feller}, \bibfnamefont{W.}}, 
\bibinfo{year}{1966}, \emph{\bibinfo{title}{An Introduction to Probability Theory and Its
applications}}, Second Edition, Vol. 1
(\bibinfo{publisher}{J. Wiley, New York}).

\bibitem[{\citenamefont{Fichter}(1969)}]{f69}
\bibinfo{author}{\bibnamefont{Fichter}, \bibfnamefont{W.~B.}}, 
\bibinfo{year}{1969},
\bibinfo{journal}{Ph. D. Thesis, North Caroline State University}

\bibitem[{\citenamefont{Fisher}(1974)}]{f74}
\bibinfo{author}{\bibnamefont{Fisher}, \bibfnamefont{M.~E.}}, 
\bibinfo{year}{1974},
\bibinfo{journal}{Rev. Mod. Phys.} \textbf{\bibinfo{volume}{46}},
\bibinfo{pages}{597}.

\bibitem[{\citenamefont{Garcimart\'{i}n} \emph{et~al.}(1997) \citenamefont{Garcimartin, Guarino, Bellon and Ciliberto}}]{ggbc97}
\bibinfo{author}{\bibnamefont{Garcimart\'{i}n}, \bibfnamefont{A.}},
\bibinfo{author}{\bibfnamefont{A.} \bibnamefont{Guarino}}, 
\bibinfo{author}{\bibfnamefont{L.} \bibnamefont{Bellon}} and
\bibinfo{author}{\bibfnamefont{S.} \bibnamefont{Ciliberto}}, 
\bibinfo{year}{1997},
\bibinfo{journal}{Phys. Rev. Lett.} \textbf{\bibinfo{volume}{79}},
\bibinfo{pages}{3202}.

\bibitem[{\citenamefont{Golubovi\'{c} and Feng}(1991)}]{gf91}
\bibinfo{author}{\bibnamefont{Golubovi\'{c}}, \bibfnamefont{L.}} and
\bibinfo{author}{\bibfnamefont{S.}~\bibnamefont{Feng}}, 
\bibinfo{year}{1991},
\bibinfo{journal}{Phys. Rev. A} \textbf{\bibinfo{volume}{43}},
\bibinfo{pages}{5223}.

\bibitem[{\citenamefont{Gomez} \emph{et~al.}(1993) \citenamefont{Gomez, Iniguez, and Pacheco}}]{gip93}
\bibinfo{author}{\bibnamefont{Gomez}, \bibfnamefont{J.~B.}},
\bibinfo{author}{\bibfnamefont{D.} \bibnamefont{Iniguesz}} and 
\bibinfo{author}{\bibfnamefont{A.~F.} \bibnamefont{Pacheco}}, 
\bibinfo{year}{1993},
\bibinfo{journal}{Phys. Rev. Lett.} \textbf{\bibinfo{volume}{71}},
\bibinfo{pages}{380}.


\bibitem[{\citenamefont{Guarino} \emph{et~al.}(1999) \citenamefont{Guarino, Scorretti and Ciliberto}}]{gsc99}
\bibinfo{author}{\bibnamefont{Guarino}, \bibfnamefont{A.}},
\bibinfo{author}{\bibfnamefont{R.} \bibnamefont{Scorretti}} and
\bibinfo{author}{\bibfnamefont{S.} \bibnamefont{Ciliberto}}, 
  \bibinfo{year}{1999}, \eprint{cond-mat/9908329}.

\bibitem[{\citenamefont{Guarino} \emph{et~al.}(1998) \citenamefont{Guarino, Garcimart\'{i}n and Ciliberto}}]{ggc98}
\bibinfo{author}{\bibnamefont{Guarino}, \bibfnamefont{A.}},
\bibinfo{author}{\bibfnamefont{A.} \bibnamefont{Garcimart\'{i}n}} and 
\bibinfo{author}{\bibfnamefont{S.} \bibnamefont{Ciliberto}}, 
\bibinfo{year}{1998},
\bibinfo{journal}{Europhys. J. B} \textbf{\bibinfo{volume}{6}},
\bibinfo{pages}{13}.

\bibitem[{\citenamefont{Guarino} \emph{et~al.}(1999) \citenamefont{Guarino, Garcimart\'{i}n and Ciliberto}}]{ggc99}
\bibinfo{author}{\bibnamefont{Guarino}, \bibfnamefont{A.}},
\bibinfo{author}{\bibfnamefont{A.} \bibnamefont{Garcimart\'{i}n}} and 
\bibinfo{author}{\bibfnamefont{S.} \bibnamefont{Ciliberto}}, 
\bibinfo{year}{1999},
\bibinfo{journal}{Europhys. Lett.} \textbf{\bibinfo{volume}{47}},
\bibinfo{pages}{456}.

\bibitem[{\citenamefont{Guarino} \emph{et~al.}(2002) \citenamefont{Guarino, Ciliberto, Garcimart\'{i}n, Zei and Scorretti }}]{gcgzs02}
\bibinfo{author}{\bibnamefont{Guarino}, \bibfnamefont{A.}},
\bibinfo{author}{\bibfnamefont{S.} \bibnamefont{Ciliberto}}, 
\bibinfo{author}{\bibfnamefont{A.} \bibnamefont{Garcimart\'{i}n}}, 
\bibinfo{author}{\bibfnamefont{M.} \bibnamefont{Zei}} and
\bibinfo{author}{\bibfnamefont{R.} \bibnamefont{Scorretti}},
  \bibinfo{year}{2002}, \eprint{cond-mat/0201257}.

\bibitem[{\citenamefont{Guarino} \emph{et~al.}(2006) \citenamefont{Guarino, Vanel, Scorretti and Ciliberto }}]{gvsc06}
\bibinfo{author}{\bibnamefont{Guarino}, \bibfnamefont{A.}},
\bibinfo{author}{\bibfnamefont{L.} \bibnamefont{Zei}},
\bibinfo{author}{\bibfnamefont{R.} \bibnamefont{Scorretti}} and
\bibinfo{author}{\bibfnamefont{S.} \bibnamefont{Ciliberto}}, 
\bibinfo{year}{2006},
\bibinfo{journal}{J. Stat. Mech.} \textbf{\bibinfo{volume}{P06020}}.

\bibitem[{\citenamefont{Hansen and Hemmer}(1994)}]{hh94b}
\bibinfo{author}{\bibnamefont{Hansen}, \bibfnamefont{A.}} and
\bibinfo{author}{\bibfnamefont{P.~C.}~\bibnamefont{Hemmer}}, 
\bibinfo{year}{1994},
\bibinfo{journal}{Trends in Stat. Phys.} \textbf{\bibinfo{volume}{1}},
\bibinfo{pages}{213}.

\bibitem[{\citenamefont{Hansen and Hemmer}(1994)}]{hh94}
\bibinfo{author}{\bibnamefont{Hansen}, \bibfnamefont{A.}} and
\bibinfo{author}{\bibfnamefont{P.~C.}~\bibnamefont{Hemmer}}, 
\bibinfo{year}{1994},
\bibinfo{journal}{Phys. Lett. A} \textbf{\bibinfo{volume}{184}},
\bibinfo{pages}{394}.

\bibitem[{\citenamefont{Harlow}(1985)}]{h85}
\bibinfo{author}{\bibnamefont{Harlow}, \bibfnamefont{D. G.}},
\bibinfo{year}{1985}, \bibinfo{journal}{Proc. Roy. Soc. London}
\textbf{\bibinfo{volume}{A397}}, \bibinfo{pages}{211}.

\bibitem[{\citenamefont{Harlow and Phoenix}(1978)}]{hp78}
\bibinfo{author}{\bibnamefont{Harlow}, \bibfnamefont{D.~G.}} and
\bibinfo{author}{\bibfnamefont{S.~L.}~\bibnamefont{Phoenix}}, 
\bibinfo{year}{1978},
\bibinfo{journal}{J. Compos. Matter.} \textbf{\bibinfo{volume}{12}},
\bibinfo{pages}{314}.

\bibitem[{\citenamefont{Harlow and Phoenix}(1981)}]{hp81}
\bibinfo{author}{\bibnamefont{Harlow}, \bibfnamefont{D.~G.}} and
\bibinfo{author}{\bibfnamefont{S.~L.}~\bibnamefont{Phoenix}}, 
\bibinfo{year}{1981},
\bibinfo{journal}{Int. J. Fracture} \textbf{\bibinfo{volume}{17}},
\bibinfo{pages}{601}.

\bibitem[{\citenamefont{Harlow and Phoenix}(1991)}]{hp91}
\bibinfo{author}{\bibnamefont{Harlow}, \bibfnamefont{D.~G.}} and
\bibinfo{author}{\bibfnamefont{S.~L.}~\bibnamefont{Phoenix}}, 
\bibinfo{year}{1991},
\bibinfo{journal}{J. Mech. Phys. Solids} \textbf{\bibinfo{volume}{39}},
\bibinfo{pages}{173}.



\bibitem[{\citenamefont{Harlow} \emph{et~al.}(1983) \citenamefont{Harlow, Smith and Taylor}}]{hst83}
\bibinfo{author}{\bibnamefont{Harlow}, \bibfnamefont{D.~G.}},
\bibinfo{author}{\bibfnamefont{R.~L.} \bibnamefont{Smith}} and
\bibinfo{author}{\bibfnamefont{H.~M.} \bibnamefont{Taylor}}, 
\bibinfo{year}{1983}, \bibinfo{journal}{J. Appl. Prob.}
\textbf{\bibinfo{volume}{20}}, \bibinfo{pages}{358}.

\bibitem[{\citenamefont{Hedgepeth}(1961)}]{h61}
\bibinfo{author}{\bibnamefont{Hedgepeth}, \bibfnamefont{J.~M.}},
\bibinfo{year}{1961}, \bibinfo{journal}{Tech. Rep. TND-882 NASA}

\bibitem[{\citenamefont{Hedgepeth and Dyke}(1967)}]{hd67}
\bibinfo{author}{\bibnamefont{Hedgepeth}, \bibfnamefont{J.~M.}} and
\bibinfo{author}{\bibfnamefont{P.~Van} \bibnamefont{Dyke}}, 
\bibinfo{year}{1967}, \bibinfo{journal}{J. Composite Mat.}
\textbf{\bibinfo{volume}{1}}, \bibinfo{pages}{294}.

\bibitem[{\citenamefont{Hemmer and Hansen}(1992)}]{hh92}
\bibinfo{author}{\bibnamefont{Hemmer}, \bibfnamefont{P.~C.}} and
\bibinfo{author}{\bibfnamefont{A.}~\bibnamefont{Hansen}}, 
\bibinfo{year}{1992},
\bibinfo{journal}{ASME J. Appl. Mech.} \textbf{\bibinfo{volume}{59}},
\bibinfo{pages}{909}.


\bibitem[{\citenamefont{Hemmer} \emph{et~al.}(2006) \citenamefont{Hemmer, Hansen and Pradhan}}]{hhp06}
\bibinfo{author}{\bibnamefont{Hemmer}, \bibfnamefont{P.~C.}},
\bibinfo{author}{\bibfnamefont{A.} \bibnamefont{Hansen}} and
\bibinfo{author}{\bibfnamefont{S.} \bibnamefont{Pradhan}}, 
  \bibinfo{year}{2006}, in \emph{\bibinfo{booktitle}{Modelling Critical and Catastrophic Phenomena in Geoscience: A Statistical Physics Approach}},
 edited by \bibinfo{editor}{\bibfnamefont{P.}~\bibnamefont{Bhattacharyya}} and
\bibinfo{editor}{\bibfnamefont{B.~K.}~\bibnamefont{Chakrabarti}}
  (\bibinfo{publisher}{Springer-Verlag, Berlin}), \textbf{\bibinfo{volume}{705}}, p. \bibinfo{pages}{27}.

\bibitem[{\citenamefont{Hemmer and Pradhan}(2007)}]{hp07}
\bibinfo{author}{\bibnamefont{Hemmer}, \bibfnamefont{P. C.}} and
\bibinfo{author}{\bibfnamefont{S.}~\bibnamefont{Pradhan}}, 
\bibinfo{year}{2007},
\bibinfo{journal}{Phys. Rev. E} \textbf{\bibinfo{volume}{75}},
\bibinfo{pages}{046101}.

\bibitem[{\citenamefont{Herrmann and Roux}(1990)}]{hr90}
\bibinfo{editor}{\bibnamefont{Herrmann}, \bibfnamefont{H. J. }} and 
\bibinfo{editor}{\bibfnamefont{S.} \bibnamefont{Roux }} (Eds.), 
\bibinfo{year}{1990}, \emph{\bibinfo{title}{Statistical Models for the  Fracture of  Disordered Media}} 
(\bibinfo{publisher}{North-Holland, Amsterdam}).

\bibitem[{\citenamefont{Hidalgo} \emph{et~al.}(2001) \citenamefont{Hidalgo, Kun and Herrmann}}]{hkh01}
\bibinfo{author}{\bibnamefont{Hidalgo}, \bibfnamefont{R.~C.}},   
\bibinfo{author}{\bibfnamefont{F.} \bibnamefont{Kun}} and
\bibinfo{author}{\bibfnamefont{H.~J.} \bibnamefont{Herrmann}},
\bibinfo{year}{2001}, \bibinfo{journal}{Phys. Rev. E}
\textbf{\bibinfo{volume}{64}}, \bibinfo{pages}{066122}.

\bibitem[{\citenamefont{Hidalgo} \emph{et~al.}(2002) \citenamefont{Hidalgo, Kun and Herrmann}}]{hkh02}
\bibinfo{author}{\bibnamefont{Hidalgo}, \bibfnamefont{R.~C.}},   
\bibinfo{author}{\bibfnamefont{F.} \bibnamefont{Kun}} and
\bibinfo{author}{\bibfnamefont{H.~J.} \bibnamefont{Herrmann}},
\bibinfo{year}{2002}, \bibinfo{journal}{Phys. Rev. E}
\textbf{\bibinfo{volume}{65}}, \bibinfo{pages}{032502}.

\bibitem[{\citenamefont{Hidalgo} \emph{et~al.}(2002) \citenamefont{Hidalgo, Moreno, Kun and Herrmann}}]{hmkh02}
\bibinfo{author}{\bibnamefont{Hidalgo}, \bibfnamefont{R.~C.}},   
\bibinfo{author}{\bibfnamefont{Y.} \bibnamefont{Moreno}},
\bibinfo{author}{\bibfnamefont{F.} \bibnamefont{Kun}} and
\bibinfo{author}{\bibfnamefont{H.~J.} \bibnamefont{Herrmann}},
\bibinfo{year}{2002}, \bibinfo{journal}{Phys. Rev. E}
\textbf{\bibinfo{volume}{65}}, \bibinfo{pages}{046148}.

\bibitem[{\citenamefont{Hidalgo} \emph{et~al.}(2008) \citenamefont{Hidalgo, 
Kovacs, Pagonbarraga and Kun}}]{hkpk08}
\bibinfo{author}{\bibnamefont{Hidalgo}, \bibfnamefont{R.~C.}},   
\bibinfo{author}{\bibfnamefont{K.} \bibnamefont{Kovacs}},
\bibinfo{author}{\bibfnamefont{I.} \bibnamefont{Pagonbarraga}} and
\bibinfo{author}{\bibfnamefont{F.} \bibnamefont{Kun}},
\bibinfo{year}{2008}, \bibinfo{journal}{Europhys. Lett.}
\textbf{\bibinfo{volume}{81}}, \bibinfo{pages}{54005}.

\bibitem[{\citenamefont{Hidalgo} \emph{et~al.}(2008) \citenamefont{Hidalgo, 
Zapperi and Herrmann}}]{hzh08}
\bibinfo{author}{\bibnamefont{Hidalgo}, \bibfnamefont{R.~C.}},   
\bibinfo{author}{\bibfnamefont{S.} \bibnamefont{Zapperi}}, and
\bibinfo{author}{\bibfnamefont{H.~J.} \bibnamefont{Herrmann}},
\bibinfo{year}{2008}, \bibinfo{journal}{J.\ Stat.\ Mech.\ Theor.\ Exp.}
\bibinfo{pages}{P01004}.

\bibitem[{\citenamefont{Hild} \emph{et~al.}(1994) \citenamefont{Hild, Domergue, Evans and Leckie}}]{hdel94}
\bibinfo{author}{\bibnamefont{Hild}, \bibfnamefont{F.}},   
\bibinfo{author}{\bibfnamefont{J.~M.} \bibnamefont{Domergue}},
\bibinfo{author}{\bibfnamefont{A.~G.} \bibnamefont{Evans}} and
\bibinfo{author}{\bibfnamefont{F.~A.} \bibnamefont{Leckie}},
\bibinfo{year}{1994}, \bibinfo{journal}{Int. J. Solids Struct.}
\textbf{\bibinfo{volume}{31}}, \bibinfo{pages}{1035}.
\bibitem[{\citenamefont{Hild and Feillard}(1997)}]{hf97}
\bibinfo{author}{\bibnamefont{Hild}, \bibfnamefont{F.}} and   
\bibinfo{author}{\bibfnamefont{P.} \bibnamefont{Feillard}},
\bibinfo{year}{1997}, \bibinfo{journal}{Rel. Eng. Sys. Saf.}
\textbf{\bibinfo{volume}{56}}, \bibinfo{pages}{225}.

\bibitem[{\citenamefont{Hsueh}(1990)}]{h90}
\bibinfo{author}{\bibnamefont{Hsueh}, \bibfnamefont{C.~H.}},
\bibinfo{year}{1990}, \bibinfo{journal}{Mater. Sci. Engng. A}
\textbf{\bibinfo{volume}{123}}, \bibinfo{pages}{1}.

\bibitem[{\citenamefont{Hsueh}(1992)}]{h92}
\bibinfo{author}{\bibnamefont{Hsueh}, \bibfnamefont{C.~H.}},
\bibinfo{year}{1992}, \bibinfo{journal}{Mater. Sci. Engng. A}
\textbf{\bibinfo{volume}{154}}, \bibinfo{pages}{125}.

\bibitem[{\citenamefont{Ibnabdeljalil and Curtin}(1997a)}]{ic97a}
\bibinfo{author}{\bibnamefont{Ibnabdeljalil}, \bibfnamefont{M.}} and
\bibinfo{author}{\bibfnamefont{W.~A.}~\bibnamefont{Curtin}}, 
\bibinfo{year}{1997a},
\bibinfo{journal}{Acta Mater.} \textbf{\bibinfo{volume}{9}},
\bibinfo{pages}{3641}.

\bibitem[{\citenamefont{Ibnabdeljalil and Curtin}(1997b)}]{ic97b}
\bibinfo{author}{\bibnamefont{Ibnabdeljalil}, \bibfnamefont{M.}} and
\bibinfo{author}{\bibfnamefont{W.~A.}~\bibnamefont{Curtin}}, 
\bibinfo{year}{1997b},
\bibinfo{journal}{Int.\ J.\ Solid Struct.} \textbf{\bibinfo{volume}{21}},
\bibinfo{pages}{2649}.

\bibitem[{\citenamefont{Jo} \emph{et~al.}(2008) \citenamefont{Jo, Kang, Choi,
Choi and Yoon}}]{jkccy08}
\bibinfo{author}{\bibnamefont{Jo}, \bibfnamefont{J.}},
\bibinfo{author}{\bibnamefont{H.}, \bibfnamefont{Kang}}, 
\bibinfo{author}{\bibnamefont{M.\ Y.}, \bibfnamefont{Choi}}, 
\bibinfo{author}{\bibnamefont{J.}, \bibfnamefont{Choi}}, 
and \bibinfo{author}{\bibfnamefont{B.-G.}~\bibnamefont{Yoon}}, 
\bibinfo{year}{2008},
\bibinfo{journal}{J.\ Phys.\ A} \textbf{\bibinfo{volume}{41}},
\bibinfo{pages}{145101}.

\bibitem[{\citenamefont{Johnson}(1985)}]{j85}
\bibinfo{author}{\bibnamefont{Johnson}, \bibfnamefont{K.~L.}}, 
\bibinfo{year}{1985}, \emph{\bibinfo{title}{Contact Mechanics}},
(\bibinfo{publisher}{Cambridge University Press, Cambridge}).

\bibitem[{\citenamefont{Kachanov}(1985)}]{k85}
\bibinfo{author}{\bibnamefont{Kachanov}, \bibfnamefont{M.}},
\bibinfo{year}{1985},
\bibinfo{journal}{Int. J. Solids Fract.} \textbf{\bibinfo{volume}{28}},
\bibinfo{pages}{R11}.


\bibitem[{\citenamefont{Karbhari and Strassler}(2007)}]{ks07}
\bibinfo{author}{\bibnamefont{Karbhari}, \bibfnamefont{V.}} and
\bibinfo{author}{\bibfnamefont{H.}~\bibnamefont{Strassler}}, 
\bibinfo{year}{2007},
\bibinfo{journal}{Dental Materials} \textbf{\bibinfo{volume}{23}},
\bibinfo{pages}{960}.

\bibitem[{\citenamefont{Kawamura}(2006)}]{k06}
\bibinfo{author}{\bibnamefont{Kawamura}, \bibfnamefont{H.}},
  \bibinfo{year}{2006}, \eprint{cond-mat/0603335}.


\bibitem[{\citenamefont{Kim}(2004)}]{k04}
\bibinfo{author}{\bibnamefont{Kim}, \bibfnamefont{B.~J.}},
\bibinfo{year}{2004}, \bibinfo{journal}{Europhys. Lett.}
\textbf{\bibinfo{volume}{66}}, \bibinfo{pages}{819}.

\bibitem[{\citenamefont{Kim} \emph{et~al.}(2005) \citenamefont{Kim, Kim and Jeong}}]{kkj05}
\bibinfo{author}{\bibnamefont{Kim}, \bibfnamefont{D.~-H.}},
\bibinfo{author}{\bibfnamefont{B.~J.} \bibnamefont{Kim}} and
\bibinfo{author}{\bibfnamefont{H.} \bibnamefont{Jeong}}, 
\bibinfo{year}{2005}, \bibinfo{journal}{Phys. Rev. Lett.}
\textbf{\bibinfo{volume}{94}}, \bibinfo{pages}{025501}.

\bibitem[{\citenamefont{Kloster} \emph{et~al.}(1997) \citenamefont{Kloster, Hansen and Hemmer}}]{khh97}
\bibinfo{author}{\bibnamefont{Kloster}, \bibfnamefont{M.}},
\bibinfo{author}{\bibfnamefont{A.} \bibnamefont{Hansen}} and
\bibinfo{author}{\bibfnamefont{P. C.} \bibnamefont{Hemmer}}, 
\bibinfo{year}{1997}, \bibinfo{journal}{Phys. Rev. E}
\textbf{\bibinfo{volume}{56}}, \bibinfo{pages}{2615}.

\bibitem[{\citenamefont{Kun} \emph{et~al.}(2007) \citenamefont{Kun, Costa, 
Filho,Andrade, Soares, Zapperi and Herrmann}}]{kcfaszh07}
\bibinfo{author}{\bibnamefont{Kun}, \bibfnamefont{F.}},
\bibinfo{author}{\bibfnamefont{M.~H.} \bibnamefont{Costa}},
\bibinfo{author}{\bibfnamefont{R.~N.} \bibnamefont{Costa Filho}},
\bibinfo{author}{\bibfnamefont{J.~S.} \bibnamefont{Andrade Jr}},
\bibinfo{author}{\bibfnamefont{J.~B.} \bibnamefont{Soares}},
\bibinfo{author}{\bibfnamefont{S.} \bibnamefont{Zapper}} and
\bibinfo{author}{\bibfnamefont{H.~J.} \bibnamefont{Herrmann}},
\bibinfo{year}{2007}, \bibinfo{journal}{J. Stat. Mech.}
\textbf{\bibinfo{volume}{02}}, \bibinfo{pages}{P02003}.

\bibitem[{\citenamefont{Kun} \emph{et~al.}(2003) \citenamefont{Kun, Hidalgo, Herrmann and Pal}}]{khhp03}
\bibinfo{author}{\bibnamefont{Kun}, \bibfnamefont{F.}},
\bibinfo{author}{\bibfnamefont{R.~C.} \bibnamefont{Hidalgo}},
\bibinfo{author}{\bibfnamefont{H.~J.} \bibnamefont{Herrmann}} and
\bibinfo{author}{\bibfnamefont{K.~F.} \bibnamefont{Pal}},
\bibinfo{year}{2003}, \bibinfo{journal}{Phys. Rev. E}
\textbf{\bibinfo{volume}{67}}, \bibinfo{pages}{061802}.

\bibitem[{\citenamefont{Kun} \emph{et~al.}(2006) \citenamefont{Kun, Hidalgo, Raischel and  Herrmann}}]{khrh06}
\bibinfo{author}{\bibnamefont{Kun}, \bibfnamefont{F.}},
\bibinfo{author}{\bibfnamefont{R.~C.} \bibnamefont{Hidalgo}},
\bibinfo{author}{\bibfnamefont{F.} \bibnamefont{Raischel}} and
\bibinfo{author}{\bibfnamefont{H.~J.} \bibnamefont{Herrmann}}, 
  \bibinfo{year}{2006}, in \emph{\bibinfo{booktitle}{Modelling Critical and Catastrophic Phenomena in Geoscience: A Statistical Physics Approach}},
 edited by \bibinfo{editor}{\bibfnamefont{P.}~\bibnamefont{Bhattacharyya}} and
\bibinfo{editor}{\bibfnamefont{B.~K.}~\bibnamefont{Chakrabarti}}
  (\bibinfo{publisher}{Springer-Verlag, Berlin}), \textbf{\bibinfo{volume}{705}}, p. \bibinfo{pages}{57}.

\bibitem[{\citenamefont{Kun and Nagy}(2008)}]{kn08}
\bibinfo{author}{\bibnamefont{Kun}, \bibfnamefont{F.}} and
\bibinfo{author}{\bibfnamefont{S.}~\bibnamefont{Nagy}}, 
\bibinfo{year}{2008},
\bibinfo{journal}{Phys. Rev. E} \textbf{\bibinfo{volume}{77}},
\bibinfo{pages}{016608}.

\bibitem[{\citenamefont{Kun} \emph{et~al.}(2000) \citenamefont{Kun, Zapperi and Herrmann}}]{kzh00}
\bibinfo{author}{\bibnamefont{Kun}, \bibfnamefont{F.}},
\bibinfo{author}{\bibfnamefont{S.} \bibnamefont{Zapperi}} and
\bibinfo{author}{\bibfnamefont{H.~J.} \bibnamefont{Herrmann}},
\bibinfo{year}{2000}, \bibinfo{journal}{Europhys. J. B}
\textbf{\bibinfo{volume}{17}}, \bibinfo{pages}{269}.

\bibitem[{\citenamefont{Kuo and Phoenix}(1987)}]{kp87}
\bibinfo{author}{\bibnamefont{Kuo}, \bibfnamefont{C.~C.}} and
\bibinfo{author}{\bibfnamefont{S.~L.}~\bibnamefont{Phoenix}}, 
\bibinfo{year}{1987},
\bibinfo{journal}{J. Appl. Prob.} \textbf{\bibinfo{volume}{24}},
\bibinfo{pages}{137}.

\bibitem[{\citenamefont{Lagoudas} \emph{et~al.}(1989) \citenamefont{Lagoudas, Phoenix and Hui}}]{lph89}
\bibinfo{author}{\bibnamefont{Lagoudas}, \bibfnamefont{D.~C.}},
\bibinfo{author}{\bibfnamefont{S.~L.} \bibnamefont{Phoenix}} and
\bibinfo{author}{\bibfnamefont{C.~Y.} \bibnamefont{Hui}},
\bibinfo{year}{1989}, \bibinfo{journal}{Int. J. Solids and Struct.}
\textbf{\bibinfo{volume}{25}}, \bibinfo{pages}{45}.

\bibitem[{\citenamefont{Landau and Litshitz}(1958)}]{ll58}
\bibinfo{author}{\bibnamefont{Landau}, \bibfnamefont{L.}} and
\bibinfo{author}{\bibfnamefont{E.~M.} \bibnamefont{Lifshitz}},
\bibinfo{year}{1958}, \emph{\bibinfo{title}{Theory of Elasticity }}
(\bibinfo{publisher}{Clarendon Press, Oxford}).

\bibitem[{\citenamefont{Landis and McMeeking}(1999)}]{lm99}
\bibinfo{author}{\bibnamefont{Landis}, \bibfnamefont{C.~M.}} and
\bibinfo{author}{\bibfnamefont{R.~M.} \bibnamefont{McMeeking}},
\bibinfo{year}{1999}, \bibinfo{journal}{Composites Scie. Tech.}
\textbf{\bibinfo{volume}{59}}, \bibinfo{pages}{447}.

\bibitem[{\citenamefont{Landis} \emph{et~al.} (2000) \citenamefont{Landis, Beyerlein and McMeeking}}]{lbm00}
\bibinfo{author}{\bibnamefont{Landis}, \bibfnamefont{C.~M.}}, 
\bibinfo{author}{\bibfnamefont{I.~J.} \bibnamefont{Beyerlein}} and
\bibinfo{author}{\bibfnamefont{R.~M.} \bibnamefont{McMeeking}},
\bibinfo{year}{2000}, \bibinfo{journal}{J. Mech. Phys. Solids}
\textbf{\bibinfo{volume}{48}}, \bibinfo{pages}{621}.

\bibitem[{\citenamefont{Lawn}(1993)}]{l93}
\bibinfo{author}{\bibnamefont{Lawn}, \bibfnamefont{B. R.}}, 
\bibinfo{year}{1993}, \emph{\bibinfo{title}{Fracture of Brittle Solids}} 
(\bibinfo{publisher}{Cambridge University Press, Cambridge}).

\bibitem[{\citenamefont{Layton and Sastry}(2004)}]{ls04}
\bibinfo{author}{\bibnamefont{Layton}, \bibfnamefont{B.~E.}} and
\bibinfo{author}{\bibfnamefont{A.~M.} \bibnamefont{Sastry}},
\bibinfo{year}{2004}, \bibinfo{journal}{Biomech. Engng.}
\textbf{\bibinfo{volume}{126}}, \bibinfo{pages}{803}.

\bibitem[{\citenamefont{Lee}(1994)}]{l94}
\bibinfo{author}{\bibnamefont{Lee}, \bibfnamefont{W.}},
\bibinfo{year}{1994}, \bibinfo{journal}{Phys. Rev. B}
\textbf{\bibinfo{volume}{50}}, \bibinfo{pages}{3797}.

\bibitem[{\citenamefont{Li} \emph{et~al.} (2006)}]{ljgwa06}
\bibinfo{author}{\bibnamefont{Li}, \bibfnamefont{H.}},
\bibinfo{author}{\bibfnamefont{X.~J.} \bibnamefont{Jia}}, 
\bibinfo{author}{\bibfnamefont{M.} \bibnamefont{Geni}}, 
\bibinfo{author}{\bibfnamefont{J.} \bibnamefont{Wei}} and 
\bibinfo{author}{\bibfnamefont{L.~J.} \bibnamefont{An}}, 
\bibinfo{year}{2006}, \bibinfo{journal}{Mater. Sci. and Eng.}
\textbf{\bibinfo{volume}{425}}, \bibinfo{pages}{178}.


\bibitem[{\citenamefont{Li and Metcalf}(2002)}]{lm02}
\bibinfo{author}{\bibnamefont{Li}, \bibfnamefont{Y.}} and
\bibinfo{author}{\bibfnamefont{J.} \bibnamefont{Metcalf}},
\bibinfo{year}{2002}, \bibinfo{journal}{J. Mater. Civ. Eng.}
\textbf{\bibinfo{volume}{14}}, \bibinfo{pages}{303}.

\bibitem[{\citenamefont{Lund and Byrne}(2001)}]{lb01}
\bibinfo{author}{\bibnamefont{Lund}, \bibfnamefont{J.~R.}} and
\bibinfo{author}{\bibfnamefont{J.~P.} \bibnamefont{Byrne}},
\bibinfo{year}{2001}, \bibinfo{journal}{Civ. Eng. Env. Systems}
\textbf{\bibinfo{volume}{18}}, \bibinfo{pages}{243}.


\bibitem[{\citenamefont{Maes} \emph{et~al.}(1998) \citenamefont{Maes, Moffaeret,  Frederix and Strauven}}]{mmfs98}
\bibinfo{author}{\bibnamefont{Maes}, \bibfnamefont{C.}},
\bibinfo{author}{\bibfnamefont{A.~V.} \bibnamefont{Moffaeret}} 
\bibinfo{author}{\bibfnamefont{H.} \bibnamefont{Frederix}} and 
\bibinfo{author}{\bibfnamefont{H.} \bibnamefont{Strauven}}, 
\bibinfo{year}{1998}, \bibinfo{journal}{Phys. Rev. B}
\textbf{\bibinfo{volume}{57}}, \bibinfo{pages}{9}.

\bibitem[{\citenamefont{Mishnaevsky}(2007)}]{m07}
\bibinfo{author}{\bibnamefont{Mishnaevsky Jr.}, \bibfnamefont{L.}}, 
\bibinfo{year}{2007}, \emph{\bibinfo{title}{Computational Mesomechanics of
Composites}} 
(\bibinfo{publisher}{Wiley, New York}).

\bibitem[{\citenamefont{Mishnaevsky and Br{\o}ndsted}(2009)}]{mb09}
\bibinfo{author}{\bibnamefont{Mishnaevsky Jr.}, \bibfnamefont{L.}} and
\bibinfo{author}{\bibfnamefont{Br{\o}ndsted}~\bibnamefont{P.}}, 
\bibinfo{year}{2009},
\bibinfo{journal}{Comp. Mater. Sci.} 
\textbf{\bibinfo{volume}{44}}, \bibinfo{pages}{1351}.

\bibitem[{\citenamefont{McCartney and Smith}(1983)}]{ms83}
\bibinfo{author}{\bibnamefont{McCartney}, \bibfnamefont{L.~N.}} and
\bibinfo{author}{\bibfnamefont{R.~L.}~\bibnamefont{Smith}}, 
\bibinfo{year}{1983},
\bibinfo{journal}{J. Appl. Mech.} \textbf{\bibinfo{volume}{105}},
\bibinfo{pages}{601}.

\bibitem[{\citenamefont{Monette}(1994)}]{m94}
\bibinfo{author}{\bibnamefont{Monette}, \bibfnamefont{L.}},
\bibinfo{year}{1994}, \bibinfo{journal}{Int. J. Mod. Phys.}
\textbf{\bibinfo{volume}{B 8}}, \bibinfo{pages}{1417}.

\bibitem[{\citenamefont{Moore} \emph{et~al.}(1974) \citenamefont{Moore, Hamstad and Chiao}}]{mhc74}
\bibinfo{author}{\bibnamefont{Moore}, \bibfnamefont{R.~L.}},
\bibinfo{author}{\bibfnamefont{M.~A.} \bibnamefont{Hamstad}} and
\bibinfo{author}{\bibfnamefont{T.~T.} \bibnamefont{Chiao}}, 
\bibinfo{year}{1974}, \bibinfo{journal}{Composite Materials and Structure}
\textbf{\bibinfo{volume}{3}}, \bibinfo{pages}{19}.

\bibitem[{\citenamefont{Moral} \emph{et~al.}(2001) \citenamefont{Moral, G\'{o}mez and Moreno}}]{mgm01}
\bibinfo{author}{\bibnamefont{Moral}, \bibfnamefont{L.}},
\bibinfo{author}{\bibfnamefont{J.~B.} \bibnamefont{G\'{o}mez}} and
\bibinfo{author}{\bibfnamefont{Y.} \bibnamefont{Moreno}}, 
\bibinfo{year}{2001}, \bibinfo{journal}{J. Phys. A-Math. Gen.}
\textbf{\bibinfo{volume}{34}}, \bibinfo{pages}{9983}.

\bibitem[{\citenamefont{Moral} \emph{et~al.}(2001) \citenamefont{Moral, Moreno, G\'{o}mez and Pacheco}}]{mmgp01}
\bibinfo{author}{\bibnamefont{Moral}, \bibfnamefont{L.}},
\bibinfo{author}{\bibfnamefont{Y.} \bibnamefont{Moreno}}, 
\bibinfo{author}{\bibfnamefont{J.~B.} \bibnamefont{G\'{o}mez}} and
\bibinfo{author}{\bibfnamefont{A.~F.} \bibnamefont{Pacheco}}, 
\bibinfo{year}{2001}, \bibinfo{journal}{Phys. Rev. E}
\textbf{\bibinfo{volume}{63}}, \bibinfo{pages}{066106}.

\bibitem[{\citenamefont{Moreno} \emph{et~al.}(2001) \citenamefont{Moreno, Correig, G\'{o}mez and Pacheco}}]{mcgp01}
\bibinfo{author}{\bibnamefont{Moreno}, \bibfnamefont{Y.}},
\bibinfo{author}{\bibfnamefont{A.~M.} \bibnamefont{Correig}}, 
\bibinfo{author}{\bibfnamefont{J.~B.} \bibnamefont{G\'{o}mez}} and
\bibinfo{author}{\bibfnamefont{A.~F.} \bibnamefont{Pacheco}}, 
\bibinfo{year}{2001}, \bibinfo{journal}{J. Geophys. Res.}
\textbf{\bibinfo{volume}{B 106}}, \bibinfo{pages}{6609}.


\bibitem[{\citenamefont{Moreno} \emph{et~al.}(2000) \citenamefont{Moreno, G\'{o}mez and Pacheco}}]{mgp00}
\bibinfo{author}{\bibnamefont{Moreno}, \bibfnamefont{Y.}},
\bibinfo{author}{\bibfnamefont{J.~B.} \bibnamefont{G\'{o}mez}} and
\bibinfo{author}{\bibfnamefont{A.~F.} \bibnamefont{Pacheco}}, 
\bibinfo{year}{2000}, \bibinfo{journal}{Phys. Rev. Lett.}
\textbf{\bibinfo{volume}{85}}, \bibinfo{pages}{2865}.

\bibitem[{\citenamefont{Moreno} \emph{et~al.}(2001) \citenamefont{Moreno, G\'{o}mez and Pacheco}}]{mgp01}
\bibinfo{author}{\bibnamefont{Moreno}, \bibfnamefont{Y.}},
\bibinfo{author}{\bibfnamefont{J.~B.} \bibnamefont{G\'{o}mez}} and
\bibinfo{author}{\bibfnamefont{A.~F.} \bibnamefont{Pacheco}},
\bibinfo{year}{2001}, \bibinfo{journal}{Physica A}
\textbf{\bibinfo{volume}{296}}, \bibinfo{pages}{9}.

\bibitem[{\citenamefont{Moreno} \emph{et~al.}(2002) \citenamefont{Moreno, G\'{o}mez and Pacheco}}]{mgp02}
\bibinfo{author}{\bibnamefont{Moreno}, \bibfnamefont{Y.}},
\bibinfo{author}{\bibfnamefont{J.~B.} \bibnamefont{G\'{o}mez}} and
\bibinfo{author}{\bibfnamefont{A.~F.} \bibnamefont{Pacheco}},
\bibinfo{year}{2002}, \bibinfo{journal}{Europhys. Lett.}
\textbf{\bibinfo{volume}{58}}, \bibinfo{pages}{630}.



\bibitem[{\citenamefont{M{\aa}l{\o}y and Schmittbuhl}(1997)}]{ms97}
\bibinfo{author}{\bibnamefont{M{\aa}l{\o}y}, \bibfnamefont{K.~J.}} and
\bibinfo{author}{\bibfnamefont{J.}~\bibnamefont{Schmittbuhl}}, 
\bibinfo{year}{1997},
\bibinfo{journal}{Phys. Rev. Lett.} \textbf{\bibinfo{volume}{78}},
\bibinfo{pages}{3888}.

\bibitem[{\citenamefont{M{\aa}l{\o}y and Schmittbuhl}(2001)}]{ms01}
\bibinfo{author}{\bibnamefont{M{\aa}l{\o}y}, \bibfnamefont{K.~J.}} and
\bibinfo{author}{\bibfnamefont{J.}~\bibnamefont{Schmittbuhl}}, 
\bibinfo{year}{2001},
\bibinfo{journal}{Phys. Rev. Lett.} \textbf{\bibinfo{volume}{87}},
\bibinfo{pages}{105502}.

\bibitem[{\citenamefont{Nechad} \emph{et~al.}(2005) \citenamefont{Nechad, Helmstetter,Guerjouma and Sornette}}]{nhgs05}
\bibinfo{author}{\bibnamefont{Nechad}, \bibfnamefont{H.}},
\bibinfo{author}{\bibfnamefont{A.} \bibnamefont{Helmstetter}},
\bibinfo{author}{\bibfnamefont{R.} \bibnamefont{El Guerjouma}} and
\bibinfo{author}{\bibfnamefont{D.} \bibnamefont{Sornette}},
\bibinfo{year}{2002}, \bibinfo{journal}{J. Mech. Phys. Solids}
\textbf{\bibinfo{volume}{53}}, \bibinfo{pages}{1099}.

\bibitem[{\citenamefont{Newman and Gabrielov}(1991)}]{ng91}
\bibinfo{author}{\bibnamefont{Newman}, \bibfnamefont{W.~I.}} and
\bibinfo{author}{\bibfnamefont{A.~M.}~\bibnamefont{Gabrielov}}, 
\bibinfo{year}{1991},
\bibinfo{journal}{Int. J. Fract.} \textbf{\bibinfo{volume}{50}},
\bibinfo{pages}{1}.

\bibitem[{\citenamefont{Newman} \emph{et~al.}(1994) \citenamefont{Newman, Gabrielov, Durand, Phoenix and Turcotte}}]{ngdpt94}
\bibinfo{author}{\bibnamefont{Newman}, \bibfnamefont{W.~I.}},
\bibinfo{author}{\bibfnamefont{A.~M.}~\bibnamefont{Gabrielov}}, 
\bibinfo{author}{\bibfnamefont{T.~A.}~\bibnamefont{Durand}}, 
\bibinfo{author}{\bibfnamefont{S.~L.}~\bibnamefont{Phoenix}} and 
\bibinfo{author}{\bibfnamefont{D.~T.}~\bibnamefont{Turcotte}}, 
\bibinfo{year}{1994},
\bibinfo{journal}{Physica D} \textbf{\bibinfo{volume}{77}},
\bibinfo{pages}{200}.


\bibitem[{\citenamefont{Newman and Phoenix}(2001)}]{np01}
\bibinfo{author}{\bibnamefont{Newman}, \bibfnamefont{W.~I.}} and
\bibinfo{author}{\bibfnamefont{S.~L.}~\bibnamefont{Phoenix}}, 
\bibinfo{year}{2001},
\bibinfo{journal}{Phys. Rev. E} \textbf{\bibinfo{volume}{63}},
\bibinfo{pages}{021507}.


\bibitem[{\citenamefont{Pauchard and Meunier}(1993)}]{pm93}
\bibinfo{author}{\bibnamefont{Pauchard}, \bibfnamefont{L.}} and
\bibinfo{author}{\bibfnamefont{J.}~\bibnamefont{Meunier}}, 
\bibinfo{year}{1993},
\bibinfo{journal}{Phys. Rev. Lett.} \textbf{\bibinfo{volume}{70}},
\bibinfo{pages}{3565}.


\bibitem[{\citenamefont{Peirce}(1926)}]{p26}
\bibinfo{author}{\bibnamefont{Peirce}, \bibfnamefont{F. T.}},
\bibinfo{year}{1926}, \bibinfo{journal}{J. Text. Ind.}
\textbf{\bibinfo{volume}{17}}, \bibinfo{pages}{355}.

\bibitem[{\citenamefont{Petri} \emph{et~al.}(1994) \citenamefont{Petri, Paparo, Vespignani, Alippi and Constantini}}]{ppvac94}
\bibinfo{author}{\bibnamefont{Petri}, \bibfnamefont{A.}},
\bibinfo{author}{\bibfnamefont{G.}~\bibnamefont{Paparo}}, 
\bibinfo{author}{\bibfnamefont{A.}~\bibnamefont{Vespignani}}, 
\bibinfo{author}{\bibfnamefont{A.}~\bibnamefont{Alippi}} and 
\bibinfo{author}{\bibfnamefont{M.}~\bibnamefont{Constantini}}, 
\bibinfo{year}{1994},
\bibinfo{journal}{Phys. Rev. Lett.} \textbf{\bibinfo{volume}{73}},
\bibinfo{pages}{3423}.


\bibitem[{\citenamefont{Phoenix}(1978)}]{p78}
\bibinfo{author}{\bibnamefont{Phoenix}, \bibfnamefont{S. L.}},
\bibinfo{year}{1978}, \bibinfo{journal}{Int. J. Fracture}
\textbf{\bibinfo{volume}{14}}, \bibinfo{pages}{327}.

\bibitem[{\citenamefont{Phoenix}(1979)}]{p79}
\bibinfo{author}{\bibnamefont{Phoenix}, \bibfnamefont{S. L.}},
\bibinfo{year}{1978}, \bibinfo{journal}{Adv. Appl. Prob.}
\textbf{\bibinfo{volume}{11}}, \bibinfo{pages}{153}.

\bibitem[{\citenamefont{Phoenix and Smith}(1983)}]{ps83}
\bibinfo{author}{\bibnamefont{Phoenix}, \bibfnamefont{S.~L.}} and
\bibinfo{author}{\bibfnamefont{R.~L.}~\bibnamefont{Smith}}, 
\bibinfo{year}{1983},
\bibinfo{journal}{Int. J. Sol. Struct.} \textbf{\bibinfo{volume}{19}},
\bibinfo{pages}{479}.

\bibitem[{\citenamefont{Phoenix and Taylor}(1973)}]{pt73}
\bibinfo{author}{\bibnamefont{Phoenix}, \bibfnamefont{S.~L.}} and
\bibinfo{author}{\bibfnamefont{H.~M.}~\bibnamefont{Taylor}}, 
\bibinfo{year}{1973},
\bibinfo{journal}{Adv. Appl. Prob.} \textbf{\bibinfo{volume}{5}},
\bibinfo{pages}{200}.

\bibitem[{\citenamefont{Phoenix and Tierney}(1983)}]{pt83}
\bibinfo{author}{\bibnamefont{Phoenix}, \bibfnamefont{S.~L.}} and
\bibinfo{author}{\bibfnamefont{L.~J.}~\bibnamefont{Tierney}}, 
\bibinfo{year}{1983},
\bibinfo{journal}{Eng. Fract. Mech.} \textbf{\bibinfo{volume}{18}},
\bibinfo{pages}{193}.

\bibitem[{\citenamefont{Politi} \emph{et~al.}(2001) \citenamefont{Politi, Ciliberto and Scorretti}}]{pcs02}
\bibinfo{author}{\bibnamefont{Politi}, \bibfnamefont{A.}},
\bibinfo{author}{\bibfnamefont{S.} \bibnamefont{Ciliberto}} and 
\bibinfo{author}{\bibfnamefont{R.} \bibnamefont{Scorretti}},
\bibinfo{year}{2002},
\bibinfo{journal}{Phys. Rev. E} \textbf{\bibinfo{volume}{66}},
\bibinfo{pages}{026107}.

\bibitem[{\citenamefont{Pomeau}(1992)}]{p92}
\bibinfo{author}{\bibnamefont{Pomeau}, \bibfnamefont{Y.}},
\bibinfo{year}{1992},
\bibinfo{journal}{C. R. Acad. Sci. Paris} \textbf{\bibinfo{volume}{314 II}},
\bibinfo{pages}{553}.





\bibitem[{\citenamefont{Pradhan} \emph{et~al.}(2002) \citenamefont{Pradhan, Bhattacharyya and Chakrabarti}}]{pbc02}
\bibinfo{author}{\bibnamefont{Pradhan}, \bibfnamefont{S.}},
\bibinfo{author}{\bibfnamefont{P.} \bibnamefont{Bhattacharyya}} and  
\bibinfo{author}{\bibfnamefont{B.~K.} \bibnamefont{Chakrabarti}},
\bibinfo{year}{2002}, \bibinfo{journal}{Phys. Rev. E}
\textbf{\bibinfo{volume}{66}}, \bibinfo{pages}{016116}.

\bibitem[{\citenamefont{Pradhan and Chakrabarti}(2001)}]{pc01}
\bibinfo{author}{\bibnamefont{Pradhan}, \bibfnamefont{S.}} and
\bibinfo{author}{\bibfnamefont{B.~K.}~\bibnamefont{Chakrabarti}}, 
\bibinfo{year}{2001},
\bibinfo{journal}{Phys. Rev. E} \textbf{\bibinfo{volume}{71}},
\bibinfo{pages}{016113}.

\bibitem[{\citenamefont{Pradhan and Chakrabarti}(2003a)}]{pc03a}
\bibinfo{author}{\bibnamefont{Pradhan}, \bibfnamefont{S.}} and
\bibinfo{author}{\bibfnamefont{B.~K.}~\bibnamefont{Chakrabarti}}, 
\bibinfo{year}{2003a},
\bibinfo{journal}{Int. J. Mod. Phys. B} \textbf{\bibinfo{volume}{17}},
\bibinfo{pages}{5565}.

\bibitem[{\citenamefont{Pradhan and Chakrabarti}(2003b)}]{pc03b}
\bibinfo{author}{\bibnamefont{Pradhan}, \bibfnamefont{S.}} and
\bibinfo{author}{\bibfnamefont{B.~K.}~\bibnamefont{Chakrabarti}}, 
\bibinfo{year}{2003b},
\bibinfo{journal}{Phys. Rev. E} \textbf{\bibinfo{volume}{67}},
\bibinfo{pages}{046124}.

\bibitem[{\citenamefont{Pradhan and Chakrabarti}(2005)}]{pc05}
\bibinfo{author}{\bibnamefont{Pradhan}, \bibfnamefont{S.}} and
\bibinfo{author}{\bibfnamefont{B.~K.}~\bibnamefont{Chakrabarti}}, 
  \bibinfo{year}{2005}, in \emph{\bibinfo{booktitle}{Nonequilibrium Phenomena 
in Plasmas}},
 edited by \bibinfo{editor}{\bibfnamefont{A.~S.}~\bibnamefont{Sharma}} and
\bibinfo{editor}{\bibfnamefont{P.~K.}~\bibnamefont{Kaw}}
  (\bibinfo{publisher}{Springer, Dordrecht}), 
 p. \bibinfo{pages}{293}.

\bibitem[{\citenamefont{Pradhan and Chakrabarti}(2006)}]{pc06}
\bibinfo{author}{\bibnamefont{Pradhan}, \bibfnamefont{S.}} and
\bibinfo{author}{\bibfnamefont{B.~K.}~\bibnamefont{Chakrabarti}}, 
  \bibinfo{year}{2006}, in \emph{\bibinfo{booktitle}{Modelling Critical and Catastrophic Phenomena in Geoscience: A Statistical Physics Approach}},
 edited by \bibinfo{editor}{\bibfnamefont{P.}~\bibnamefont{Bhattacharyya}} and
\bibinfo{editor}{\bibfnamefont{B.~K.}~\bibnamefont{Chakrabarti}}
  (\bibinfo{publisher}{Springer-Verlag, Berlin}), \textbf{\bibinfo{volume}{705}}, p. \bibinfo{pages}{459}.


\bibitem[{\citenamefont{Pradhan} \emph{et~al.}(2005) \citenamefont{Pradhan, Chakrabarti and Hansen}}]{pch05}
\bibinfo{author}{\bibnamefont{Pradhan}, \bibfnamefont{S.}},
\bibinfo{author}{\bibfnamefont{B.~K.} \bibnamefont{Chakrabarti}} and
\bibinfo{author}{\bibfnamefont{A.} \bibnamefont{Hansen}}, 
\bibinfo{year}{2005}, \bibinfo{journal}{Phys. Rev. E}
\textbf{\bibinfo{volume}{71}}, \bibinfo{pages}{036149}.

\bibitem[{\citenamefont{Pradhan and Hansen}(2005)}]{ph05}
\bibinfo{author}{\bibnamefont{Pradhan}, \bibfnamefont{S.}} and
\bibinfo{author}{\bibfnamefont{A.}~\bibnamefont{Hansen}}, 
\bibinfo{year}{2005},
\bibinfo{journal}{Phys. Rev. E} \textbf{\bibinfo{volume}{72}},
\bibinfo{pages}{026111}.

\bibitem[{\citenamefont{Pradhan} \emph{et~al.}(2005) \citenamefont{Pradhan, Hansen and Hemmer}}]{phh05}
\bibinfo{author}{\bibnamefont{Pradhan}, \bibfnamefont{S.}},
\bibinfo{author}{\bibfnamefont{A.} \bibnamefont{Hansen}} and
\bibinfo{author}{\bibfnamefont{P. C.} \bibnamefont{Hemmer}}, 
\bibinfo{year}{2005}, \bibinfo{journal}{Phys. Rev. Lett.}
\textbf{\bibinfo{volume}{74}}, \bibinfo{pages}{125501}.

\bibitem[{\citenamefont{Pradhan} \emph{et~al.}(2006) \citenamefont{Pradhan, Hansen and Hemmer}}]{phh06}
\bibinfo{author}{\bibnamefont{Pradhan}, \bibfnamefont{S.}},
\bibinfo{author}{\bibfnamefont{A.} \bibnamefont{Hansen}} and
\bibinfo{author}{\bibfnamefont{P. C.} \bibnamefont{Hemmer}}, 
\bibinfo{year}{2006}, \bibinfo{journal}{Phys. Rev. E}
\textbf{\bibinfo{volume}{74}}, \bibinfo{pages}{016122}.

\bibitem[{\citenamefont{Pradhan and Hemmer}(2007)}]{ph07}
\bibinfo{author}{\bibnamefont{Pradhan}, \bibfnamefont{S.}} and
\bibinfo{author}{\bibfnamefont{P.~C.}~\bibnamefont{Hemmer}}, 
\bibinfo{year}{2007},
\bibinfo{journal}{Phys. Rev. E} \textbf{\bibinfo{volume}{75}},
\bibinfo{pages}{056112}.

\bibitem[{\citenamefont{Pradhan and Hemmer}(2008)}]{ph08}
\bibinfo{author}{\bibnamefont{Pradhan}, \bibfnamefont{S.}} and
\bibinfo{author}{\bibfnamefont{P.~C.}~\bibnamefont{Hemmer}}, 
\bibinfo{year}{2008},
\bibinfo{journal}{Phys. Rev. E} \textbf{\bibinfo{volume}{77}},
\bibinfo{pages}{031138}.

\bibitem[{\citenamefont{Pradhan and Hemmer}(2009)}]{ph09}
\bibinfo{author}{\bibnamefont{Pradhan}, \bibfnamefont{S.}} and
\bibinfo{author}{\bibfnamefont{P.~C.}~\bibnamefont{Hemmer}}, 
\bibinfo{year}{2009},
\bibinfo{journal}{Phys. Rev. E.} \textbf{\bibinfo{volume}{79}},
\bibinfo{pages}{041148}.

\bibitem[{\citenamefont{Press} \emph{et~al.} (1992)\citenamefont{Press, Teukolsky, Vetterling and Flannery}}]{ptvf92}
\bibinfo{author}{\bibnamefont{Press}, \bibfnamefont{W.~H.}}, 
\bibinfo{author}{\bibfnamefont{S.~A.}~\bibnamefont{Teukolsky}}, 
\bibinfo{author}{\bibfnamefont{W.~T.}~\bibnamefont{Vetterling}} and 
\bibinfo{author}{\bibfnamefont{B.~P.}~\bibnamefont{Flannery}}, 
\bibinfo{year}{1992}, \emph{\bibinfo{title}{Numerical Recipes in Fortran 77: The Art of Scientific Computing}},
(\bibinfo{publisher}{Cambridge University Press, Cambridge}).

\bibitem[{\citenamefont{Pride and Toussaint}(2002)}]{pt02}
\bibinfo{author}{\bibnamefont{Pride}, \bibfnamefont{S.~R.}} and
\bibinfo{author}{\bibfnamefont{R.}~\bibnamefont{Toussaint}}, 
\bibinfo{year}{2002}, \bibinfo{journal}{Physica A}
\textbf{\bibinfo{volume}{312}}, \bibinfo{pages}{159}.

\bibitem[{\citenamefont{Raischel} \emph{et~al.}(2005) \citenamefont{Raischel, Kun and  Herrmann}}]{rkh05}
\bibinfo{author}{\bibnamefont{Raischel}, \bibfnamefont{F.}},
\bibinfo{author}{\bibfnamefont{F.} \bibnamefont{Kun}} and
\bibinfo{author}{\bibfnamefont{H.~J.} \bibnamefont{Herrmann}}, 
\bibinfo{year}{2005}, \bibinfo{journal}{Phys. Rev. E}
\textbf{\bibinfo{volume}{72}}, \bibinfo{pages}{046126}.

\bibitem[{\citenamefont{Raischel} \emph{et~al.}(2006) \citenamefont{Raischel, Kun and  Herrmann}}]{rkh06}
\bibinfo{author}{\bibnamefont{Raischel}, \bibfnamefont{F.}},
\bibinfo{author}{\bibfnamefont{F.} \bibnamefont{Kun}} and
\bibinfo{author}{\bibfnamefont{H.~J.} \bibnamefont{Herrmann}}, 
\bibinfo{year}{2006}, \bibinfo{journal}{Phys. Rev. E}
\textbf{\bibinfo{volume}{74}}, \bibinfo{pages}{035104}.

\bibitem[{\citenamefont{Roux}(2000)}]{r00}
\bibinfo{author}{\bibnamefont{Roux}, \bibfnamefont{S.}},
\bibinfo{year}{2000}, \bibinfo{journal}{Phys. Rev. E}
\textbf{\bibinfo{volume}{62}}, \bibinfo{pages}{6164}.

\bibitem[{\citenamefont{Roux and Hansen}(1990)}]{rh90}
\bibinfo{author}{\bibnamefont{Roux}, \bibfnamefont{S.}},
\bibinfo{author}{\bibnamefont{Hansen}, \bibfnamefont{A.}},
\bibinfo{year}{2000}, \bibinfo{journal}{europhys. Lett.}
\textbf{\bibinfo{volume}{11}}, \bibinfo{pages}{37}.

\bibitem[{\citenamefont{Roux and Hild}(2002)}]{rh02}
\bibinfo{author}{\bibnamefont{Roux}, \bibfnamefont{S.}} and
\bibinfo{author}{\bibfnamefont{F.}~\bibnamefont{Hild}}, 
\bibinfo{year}{2002}, \bibinfo{journal}{Int. J. Fract.}
\textbf{\bibinfo{volume}{116}}, \bibinfo{pages}{219}.

\bibitem[{\citenamefont{R{\"a}is{\"a}nen} \emph{et~al.} (1997)\citenamefont{R{\"a}is{\"a}nen, Alava, Niskanen and Nieminen}}]{rann97}
\bibinfo{author}{\bibnamefont{R{\"a}is{\"a}nen}, \bibfnamefont{V.~I.}}, 
\bibinfo{author}{\bibfnamefont{M.~J.}~\bibnamefont{Alava}}, 
\bibinfo{author}{\bibfnamefont{K.~J.}~\bibnamefont{Niskanen}} and 
\bibinfo{author}{\bibfnamefont{R.~M.}~\bibnamefont{Nieminen}}, 
\bibinfo{year}{1997}, \bibinfo{journal}{J. Mater. Res.}
\textbf{\bibinfo{volume}{12}}, \bibinfo{pages}{2725}.


\bibitem[{\citenamefont{Sahimi}(2003)}]{s03}
\bibinfo{author}{\bibnamefont{Sahimi}, \bibfnamefont{M.}},
\bibinfo{year}{2003}, \emph{\bibinfo{title}{Heterogeneous Materials II: Nonlinear and Breakdown Properties}} 
(\bibinfo{publisher}{Springer-Verlag, Berlin}).

\bibitem[{\citenamefont{Sahimi and Arbabi}(1992)}]{sa92}
\bibinfo{author}{\bibnamefont{Sahimi}, \bibfnamefont{M.}} and
\bibinfo{author}{\bibfnamefont{S.}~\bibnamefont{Arbabi}}, 
\bibinfo{year}{1992}, \bibinfo{journal}{Phys. Rev. Lett.}
\textbf{\bibinfo{volume}{68}}, \bibinfo{pages}{608}.

\bibitem[{\citenamefont{Sahimi and Arbabi}(1996)}]{sa96}
\bibinfo{author}{\bibnamefont{Sahimi}, \bibfnamefont{M.}} and
\bibinfo{author}{\bibfnamefont{S.}~\bibnamefont{Arbabi}}, 
\bibinfo{year}{1996}, \bibinfo{journal}{Phys. Rev. Lett.}
\textbf{\bibinfo{volume}{77}}, \bibinfo{pages}{3689}.

\bibitem[{\citenamefont{Sastry and Phoenix}(1993)}]{sp93}
\bibinfo{author}{\bibnamefont{Sastry}, \bibfnamefont{A.~M.}} and
\bibinfo{author}{\bibfnamefont{S.~L.}~\bibnamefont{Phoenix}}, 
\bibinfo{year}{1993},
\bibinfo{journal}{Mat. Sci. Lett.} \textbf{\bibinfo{volume}{12}},
\bibinfo{pages}{1596}.

\bibitem[{\citenamefont{Scholz}(2002)}]{s02}
\bibinfo{author}{\bibnamefont{Scholz}, \bibfnamefont{C.~H.}},
\bibinfo{year}{2002}, \emph{\bibinfo{title}{
The mechanics of Earthquakes and Faulting}} 
(\bibinfo{publisher}{ Cambridge Univ. Press, Cambridge}).

\bibitem[{\citenamefont{Scorretti} \emph{et~al.}(2001) \citenamefont{Scorretti, Ciliberto and Guarino}}]{scg01}
\bibinfo{author}{\bibnamefont{Scorreti}, \bibfnamefont{R.}},
\bibinfo{author}{\bibfnamefont{S.} \bibnamefont{Ciliberto}} and  
\bibinfo{author}{\bibfnamefont{A.} \bibnamefont{Guarino}},
\bibinfo{year}{2001}, \bibinfo{journal}{Europhys. Lett.}
\textbf{\bibinfo{volume}{55}}, \bibinfo{pages}{626}.

\bibitem[{\citenamefont{Scott}(1991)}]{s91a}
\bibinfo{author}{\bibnamefont{Scott}, \bibfnamefont{I.~G.}},
\bibinfo{year}{1991}, 
 \emph{Basic Acoustic Emission}
in  \emph{\bibinfo{title}{Nondestructive Testing Monographs and Track Vol 6}} 
(\bibinfo{publisher}{Gordon and Breach Science Publishers, New York}).



\bibitem[{\citenamefont{Si} \emph{et~al.}(2002) \citenamefont{Si, Little and Lytton}}]{sll02}
\bibinfo{author}{\bibnamefont{Si}, \bibfnamefont{Z.}},
\bibinfo{author}{\bibfnamefont{D.~N.} \bibnamefont{Little}} and  
\bibinfo{author}{\bibfnamefont{R.~L.} \bibnamefont{Lytton}},
\bibinfo{year}{2002}, \bibinfo{journal}{J. Mater. Civ. Eng.}
\textbf{\bibinfo{volume}{14}}, \bibinfo{pages}{461}.


\bibitem[{\citenamefont{Smith}(1980)}]{s80}
\bibinfo{author}{\bibnamefont{Smith}, \bibfnamefont{R. L.}},
\bibinfo{year}{1980}, \bibinfo{journal}{Proc. R. Soc. London}
\textbf{\bibinfo{volume}{A 372}}, \bibinfo{pages}{539}.

\bibitem[{\citenamefont{Smith}(1982)}]{s82}
\bibinfo{author}{\bibnamefont{Smith}, \bibfnamefont{R. L.}},
\bibinfo{year}{1982}, \bibinfo{journal}{Ann. Prob.}
\textbf{\bibinfo{volume}{10}}, \bibinfo{pages}{137}.

\bibitem[{\citenamefont{Smith and Phoenix}(1981)}]{sp81}
\bibinfo{author}{\bibnamefont{Smith}, \bibfnamefont{R.~L.}} and
\bibinfo{author}{\bibfnamefont{S.~L.}~\bibnamefont{Phoenis}}, 
\bibinfo{year}{1981},
\bibinfo{journal}{J. Appl. Mech.} \textbf{\bibinfo{volume}{48}},
\bibinfo{pages}{75}.

\bibitem[{\citenamefont{Sornette}(1989)}]{s89}
\bibinfo{author}{\bibnamefont{Sornette}, \bibfnamefont{D.}},
\bibinfo{year}{1989}, \bibinfo{journal}{J. Phys. A}
\textbf{\bibinfo{volume}{22}}, \bibinfo{pages}{L 243}.

\bibitem[{\citenamefont{Sornette}(1992)}]{s92}
\bibinfo{author}{\bibnamefont{Sornette}, \bibfnamefont{D.}},
\bibinfo{year}{1992}, \bibinfo{journal}{J. Phys. I France}
\textbf{\bibinfo{volume}{2}}, \bibinfo{pages}{2089}.

\bibitem[{\citenamefont{Sornette}(2000)}]{s00}
\bibinfo{author}{\bibnamefont{Sornette}, \bibfnamefont{D.}},
\bibinfo{year}{2000}, \emph{\bibinfo{title}{Critical Phenomena in Natural Sciences}} 
(\bibinfo{publisher}{Springer-Verlag, Berlin}).


\bibitem[{\citenamefont{Sornette} \emph{et~al.}(2002) \citenamefont{Sornette, Magnin and Brechet}}]{smb92}
\bibinfo{author}{\bibnamefont{Sornette}, \bibfnamefont{D.}},
\bibinfo{author}{\bibfnamefont{T.} \bibnamefont{Magnin}} and  
\bibinfo{author}{\bibfnamefont{Y.} \bibnamefont{Brechet}},
\bibinfo{year}{1992}, \bibinfo{journal}{Europhys. Lett.}
\textbf{\bibinfo{volume}{20}}, \bibinfo{pages}{433}.


\bibitem[{\citenamefont{Stanley}(1987)}]{s87}
\bibinfo{author}{\bibnamefont{Stanley}, \bibfnamefont{H. E.}}, 
\bibinfo{year}{1987}, \emph{\bibinfo{title}{Introduction to Phase Transition and Critical Phenomena}} 
(\bibinfo{publisher}{Oxford University Press, Oxford}).

\bibitem[{\citenamefont{Stauffer  and Aharony}(1994)}]{sa94}
\bibinfo{author}{\bibnamefont{Stauffer}, \bibfnamefont{D.}}  and 
\bibinfo{author}{\bibfnamefont{A.} \bibnamefont{Aharony}}, 
\bibinfo{year}{1994}, \emph{\bibinfo{title}{Introduction to Percolation
Theory}} 
(\bibinfo{publisher}{Taylor and Francis, London}).

\bibitem[{\citenamefont{Stinchcombe}(2005)}]{s05}
\bibinfo{author}{\bibnamefont{Stinchcombe}, \bibfnamefont{R. B}},
\bibinfo{year}{2005}, \bibinfo{journal}{Physica A}
\textbf{\bibinfo{volume}{346}}, \bibinfo{pages}{1}.


\bibitem[{\citenamefont{Suresh}(1991)}]{s91}
\bibinfo{author}{\bibnamefont{Suresh}, \bibfnamefont{S.}}, 
\bibinfo{year}{1991}, \emph{\bibinfo{title}{Fatigue of Materials}} 
(\bibinfo{publisher}{Cambridge University Press, Cambridge}).

\bibitem[{\citenamefont{S{\o}rensen and  Jacobsen}(1998)}]{sj98}
\bibinfo{author}{\bibnamefont{S{\o}rensen}, \bibfnamefont{B.~F.}} and
\bibinfo{author}{\bibfnamefont{T.~K.}~\bibnamefont{Jacobsen}}, 
\bibinfo{year}{1998},
\bibinfo{journal}{Composites A} \textbf{\bibinfo{volume}{29}},
\bibinfo{pages}{1443}.

\bibitem[{\citenamefont{S{\o}rensen and  Jacobsen}(2000)}]{sj00}
\bibinfo{author}{\bibnamefont{S{\o}rensen}, \bibfnamefont{B.~F.}} and
\bibinfo{author}{\bibfnamefont{T.~K.}~\bibnamefont{Jacobsen}}, 
\bibinfo{year}{2000},
\bibinfo{journal}{Plastics, Rubber and Composites} \textbf{\bibinfo{volume}{29}},
\bibinfo{pages}{119}.
\bibitem[{\citenamefont{Toffoli and  Lehman}(2001)}]{tl01}
\bibinfo{author}{\bibnamefont{Toffoli}, \bibfnamefont{S.~M.}} and
\bibinfo{author}{\bibfnamefont{R.~L.}~\bibnamefont{Lehman}}, 
\bibinfo{year}{2001},
\bibinfo{journal}{J. Am. Ceram. Soc.} \textbf{\bibinfo{volume}{84}},
\bibinfo{pages}{123}.

\bibitem[{\citenamefont{Toland} \emph{et~al.}(1978-79) \citenamefont{Toland, Sanchez, Freeman, Chiao and Barlow}}]{tsfcb78}
\bibinfo{author}{\bibnamefont{Toland} \bibfnamefont{R.~H.}},   
\bibinfo{author}{\bibfnamefont{R.~J.} \bibnamefont{Sanchez}},
\bibinfo{author}{\bibfnamefont{D.}  \bibnamefont{Freeman}}, 
\bibinfo{author}{\bibfnamefont{T.~T.} \bibnamefont{Chiao}} and
\bibinfo{author}{\bibfnamefont{R.~E.} \bibnamefont{Barlow}},
\bibinfo{year}{1978-79}, \bibinfo{journal}{Lawrence Livermore Laboratory Reports, UCID-17755 Parts 1-3}.

\bibitem[{\citenamefont{Turcotte and Glassco}(2004)}]{tg04}
\bibinfo{author}{\bibnamefont{Turcotte}, \bibfnamefont{D.~L.}} and
\bibinfo{author}{\bibfnamefont{M.~T.}~\bibnamefont{Glassco}}, 
\bibinfo{year}{2004},
\bibinfo{journal}{Tectonophysics} \textbf{\bibinfo{volume}{383}},
\bibinfo{pages}{71}.

\bibitem[{\citenamefont{Wagner and Eitan}(1993)}]{we93}
\bibinfo{author}{\bibnamefont{Wagner}, \bibfnamefont{H.~D.}} and
\bibinfo{author}{\bibfnamefont{A.}~\bibnamefont{Eitan}}, 
\bibinfo{year}{1993},
\bibinfo{journal}{Comp. Sci. and Tech.} \textbf{\bibinfo{volume}{46}},
\bibinfo{pages}{353}.

\bibitem[{\citenamefont{Watts and Strogatz}(1994)}]{ws98}
\bibinfo{author}{\bibnamefont{Watts}, \bibfnamefont{D.~J.}} and
\bibinfo{author}{\bibfnamefont{S.~H.}~\bibnamefont{Strogatz}}, 
\bibinfo{year}{1998},
\bibinfo{journal}{Nature (London)} \textbf{\bibinfo{volume}{393}},
\bibinfo{pages}{440}.

\bibitem[{\citenamefont{Xia and Curtin}(2001)}]{xc01}
\bibinfo{author}{\bibnamefont{Xia}, \bibfnamefont{Z.~H.}} and
\bibinfo{author}{\bibfnamefont{W.~A.}~\bibnamefont{Curtin}}, 
\bibinfo{year}{2001},
\bibinfo{journal}{Compos. Sci. and Tech. } \textbf{\bibinfo{volume}{61}},
\bibinfo{pages}{2247}.

\bibitem[{\citenamefont{Xia} \emph{et~al.} (2002)}]{xco02}
\bibinfo{author}{\bibnamefont{Xia}, \bibfnamefont{Z.~H.}},
\bibinfo{author}{\bibfnamefont{W.~A.}~\bibnamefont{Curtin}} and 
\bibinfo{author}{\bibfnamefont{T.}~\bibnamefont{Okabe}}, 
\bibinfo{year}{2002},
\bibinfo{journal}{Compos. Sci. and Tech. } \textbf{\bibinfo{volume}{62}},
\bibinfo{pages}{1279}.

\bibitem[{\citenamefont{Yoshioka} \emph{et~al.} (2008)}]{yki08}
\bibinfo{author}{\bibnamefont{Yoshioka}, \bibfnamefont{N.}},
\bibinfo{author}{\bibfnamefont{F.}~\bibnamefont{Kun}} and 
\bibinfo{author}{\bibfnamefont{N.}~\bibnamefont{Ito}}, 
\bibinfo{year}{2008},
\bibinfo{journal}{Phys. Rev. Lett.} \textbf{\bibinfo{volume}{101}},
\bibinfo{pages}{145502}.

\bibitem[{\citenamefont{Zapperi} \emph{et~al.}(1997) \citenamefont{Zapperi, Ray, Stanley and Vespignani}}]{zrsv97}
\bibinfo{author}{\bibnamefont{Zapperi} \bibfnamefont{S.}},   
\bibinfo{author}{\bibfnamefont{P.} \bibnamefont{Ray}},
\bibinfo{author}{\bibfnamefont{H.~E.}, \bibnamefont{Stanley}} and
\bibinfo{author}{\bibfnamefont{A.} \bibnamefont{Vespignani}},
\bibinfo{year}{1997}, \bibinfo{journal}{Phys. Rev. Lett.}
\textbf{\bibinfo{volume}{78}}, \bibinfo{pages}{1408}.

\bibitem[{\citenamefont{Zapperi} \emph{et~al.}(1999a) \citenamefont{Zapperi, Ray, Stanley and Vespignani}}]{zrsv99a}
\bibinfo{author}{\bibnamefont{Zapperi} \bibfnamefont{S.}},   
\bibinfo{author}{\bibfnamefont{P.} \bibnamefont{Ray}},
\bibinfo{author}{\bibfnamefont{H.~E.}, \bibnamefont{Stanley}} and
\bibinfo{author}{\bibfnamefont{A.} \bibnamefont{Vespignani}},
\bibinfo{year}{1999a}, \bibinfo{journal}{Phys. Rev. E.}
\textbf{\bibinfo{volume}{59}}, \bibinfo{pages}{5049}.

\bibitem[{\citenamefont{Zapperi} \emph{et~al.}(1999b) \citenamefont{Zapperi, Ray, Stanley and Vespignani}}]{zrsv99b}
\bibinfo{author}{\bibnamefont{Zapperi} \bibfnamefont{S.}},   
\bibinfo{author}{\bibfnamefont{P.} \bibnamefont{Ray}},
\bibinfo{author}{\bibfnamefont{H.~E.}, \bibnamefont{Stanley}} and
\bibinfo{author}{\bibfnamefont{A.} \bibnamefont{Vespignani}},
\bibinfo{year}{1999b}, \bibinfo{journal}{Phyica A}
\textbf{\bibinfo{volume}{270}}, \bibinfo{pages}{57}.

\bibitem[{\citenamefont{Zhang and Ding}(1994)}]{zd94}
\bibinfo{author}{\bibnamefont{Zhang}, \bibfnamefont{S.~D.}} and
\bibinfo{author}{\bibfnamefont{E.~J.}~\bibnamefont{Ding}}, 
\bibinfo{year}{1994},
\bibinfo{journal}{Phys. Lett. A} \textbf{\bibinfo{volume}{193}},
\bibinfo{pages}{425}.

\bibitem[{\citenamefont{Zhang and Ding}(1995)}]{zd95}
\bibinfo{author}{\bibnamefont{Zhang}, \bibfnamefont{S.~D.}} and
\bibinfo{author}{\bibfnamefont{E.~J.}~\bibnamefont{Ding}}, 
\bibinfo{year}{1995},
\bibinfo{journal}{J. Phys. A} \textbf{\bibinfo{volume}{28}},
\bibinfo{pages}{4323}.

\bibitem[{\citenamefont{Zhang and Ding}(1995)}]{zd96}
\bibinfo{author}{\bibnamefont{Zhang}, \bibfnamefont{S.~D.}} and
\bibinfo{author}{\bibfnamefont{E.~J.}~\bibnamefont{Ding}}, 
\bibinfo{year}{1996},
\bibinfo{journal}{Phys. Rev. B} \textbf{\bibinfo{volume}{53}},
\bibinfo{pages}{646}.

\bibitem[{\citenamefont{Zheng} \emph{et~al.}(2008) \citenamefont{Zheng, Gao, Zhao and Fu}}]{zgzf08}
\bibinfo{author}{\bibnamefont{Zheng} \bibfnamefont{J.-F.}},   
\bibinfo{author}{\bibfnamefont{Z. -Y.} \bibnamefont{Gao}},
\bibinfo{author}{\bibfnamefont{X. -M.}, \bibnamefont{Zhao}} and
\bibinfo{author}{\bibfnamefont{B. -B.} \bibnamefont{Fu}},
\bibinfo{year}{2008}, \bibinfo{journal}{Int. J. Mod. Phys. C}
\textbf{\bibinfo{volume}{19}}, \bibinfo{pages}{1727}.

\bibitem[{\citenamefont{Zhou and Curtin}(2001)}]{zc95}
\bibinfo{author}{\bibnamefont{Zhou}, \bibfnamefont{S.~J.}} and
\bibinfo{author}{\bibfnamefont{W.~A.}~\bibnamefont{Curtin}}, 
\bibinfo{year}{1995},
\bibinfo{journal}{Acta Metallurgica Mater.} \textbf{\bibinfo{volume}{43}},
\bibinfo{pages}{3093}.

\bibitem[{\citenamefont{Zhou and Wagner}(1999)}]{zw99}
\bibinfo{author}{\bibnamefont{Zhou}, \bibfnamefont{X.~F.}} and
\bibinfo{author}{\bibfnamefont{H.~D.}~\bibnamefont{Wagner}}, 
\bibinfo{year}{1999},
\bibinfo{journal}{Comp. Sci. and Tech.} \textbf{\bibinfo{volume}{59}},
\bibinfo{pages}{1063}.

\bibitem[{\citenamefont{Zhou and Wagner}(2000)}]{zw00}
\bibinfo{author}{\bibnamefont{Zhou}, \bibfnamefont{X.~F.}} and
\bibinfo{author}{\bibfnamefont{H.~D.}~\bibnamefont{Wagner}}, 
\bibinfo{year}{2000},
\bibinfo{journal}{Comp. Sci. and Tech.} \textbf{\bibinfo{volume}{60}},
\bibinfo{pages}{367}.

 
\end{thebibliography}

\end{document}